%% file: Thesis_master_v1.tex
\documentclass[12pt,a4paper,oneside,openright]{book}

\usepackage{amsmath}
\usepackage{amsfonts}
\usepackage{amssymb}
\usepackage{amscd}

\usepackage{setspace}
\usepackage{color}
\usepackage[usenames,dvipsnames]{xcolor}
\definecolor{dblue}{rgb}{0, 0, 139}
\usepackage[pdftex,hyperindex,bookmarks=true,colorlinks=true,citecolor=red,linkcolor=dblue]{hyperref}

\usepackage[page]{appendix}

\usepackage{graphicx}
\DeclareGraphicsExtensions{.pdf}
\usepackage[margin=10mm]{caption}

\usepackage{lettrine}
\usepackage{type1cm}

\usepackage{fancyhdr}
\usepackage{sectsty}
\allsectionsfont{\sffamily}

\usepackage[defernumbers=true, sorting=nyt, style=numeric-comp, sortcites=true, isbn=false, url=false, doi=false, eprint=false, firstinits=true, maxbibnames=10, style=numeric, backend=bibtex]{biblatex}

\DeclareFieldFormat[article]{journaltitle}{\mkbibbold{\mkbibemph{#1\isdot}},}
\DeclareFieldFormat[article]{volume}{\textbf{#1}\addcolon\space}
\renewbibmacro{in:}{%
  \ifentrytype{article}{}{%
  \printtext{\bibstring{in}\intitlepunct}}}

\nocite{*}
\appto{\bibsetup}{\sloppy}
\bibliography{refs,mypub}

\author{Debraj Roy}
\title{Effects of curvature and gravity from flat spacetime}

\begin{document}

\frontmatter
\include{titlepage}
\setlength\bibitemsep{1.2em}
\printbibliography[title={\vspace{-3em}~\\Publications included in thesis}, keyword=drpub]
\addcontentsline{toc}{chapter}{Publications included in thesis\vspace*{-1em}}

\include{Acknowledgement}

%
\newpage
\thispagestyle{empty}
\begin{center}
\vspace*{3em}
{\sffamily EFFECTS OF CURVATURE AND GRAVITY FROM FLAT SPACETIME}
\end{center}
\tableofcontents
\listoffigures
%
%

\mainmatter
\begin{spacing}{1.4}
\include{C1_introduction}
\include{C2_unruh}

\include{C3_pgt}

\include{C4_mb}

\include{C5_bht}

\include{C6_lagrangian}

\include{C7_trivial}

\include{C8_conclusion}
\include{Capp}
\end{spacing}

\backmatter
\addcontentsline{toc}{chapter}{\refname}
\printbibliography[notkeyword=drpub,resetnumbers=true]

\end{document}

%% file: titlepage.tex
\begin{titlepage}
\begin{center}

\hrule\vspace{1em}
{\huge \textsf{EFFECTS OF CURVATURE AND GRAVITY\\[1ex] FROM FLAT SPACETIME}}\vspace{2em}
\hrule

\vfill

{\sffamily \Large \sl Thesis submitted for the degree of}\\[1em]
\textbf{\sffamily\Large Doctor of Philosophy (Science)}\\[1ex]
{\sffamily\Large \sl in}\\[1ex]
\textbf{\sffamily\Large Physics (Theoretical)}\\[2em]
{\sffamily\Large \sl By}\\[1em]
\textbf{\sffamily\Large Debraj Roy}

\vfill

{\sffamily\sl \Large Department of Physics}\\[1ex]
{\sffamily\sl \Large University of Calcutta}\\[1em]
{\sffamily\Large \sl 2013}

\end{center}
\end{titlepage}

%% file: Acknowledgement.tex
\chapter*{Acknowledgements}
\addcontentsline{toc}{chapter}{Acknowledgements}
\label{P:ack}

\hspace*{\parindent}To my supervisor Prof. Rabin Banerjee, for supervision of this thesis work.

To S. N. Bose National Centre for Basic Sciences (SNBNCBS), Kolkata (India), my affiliating institute.

To Prof. Sayan Kar (Indian Institute of Technology, Kharagpur), Prof. Pradip Mukherjee (Barasat Govt. College, West Bengal), Prof. Subir Ghosh (Indian Statistical Institute, Kolkata), Prof. Manu Mathur (SNBNCBS), Prof. Amitabha Lahiri (SNBNCBS), Prof. Partha Guha (SNBNCBS), Prof. Samir Kumar Paul (SNBNCBS) and Prof. Priya Mahadevan (SNBNCBS) for numerous assistance.

To Prof. Marc Henneaux (Universit\'{e} Libre de Bruxelles), Prof. Andreas Buchleitner (Albert-Ludwigs-Universit\"{a}t Freiburg), Prof. Josep M. Pons (Universitat de Barcelona), Prof. T. Padmanbhan (IUCAA, Pune) and Prof. M. Siva Kumar (Univ. of Hyderabad) for hosting short visits; To Prof. Ghanashyam Date (IMSc, Chennai) for both organising a one month school and later hosting short visits.

To various conference organisers for allowing contribution and participation: To Charles University, Prague and the organisers of the meeting ``Relativity and Gravitation, 100 Years after Einstein in Prague;'' To the organisers of the ``Thirteenth Marcel Grossmann Meeting,'' held at Stockholm University; To the organisers of ``COSGRAV-12'' at the Indian Statistical Institute, Kolkata; To the organisers of ``IAGRG-27'' at HNBGU Srinagar Garhwal. 

To my group, both at SNBNCBS current and former graduates: Saurav Samanta, Shailesh Kulkarni, Bibhas R. Majhi, Sujoy K. Modak, Dibakar Roychowdhury, Biswajit Paul, Arindam Lala, Arpan K. Mitra, Shirsendu Dey.

To my friends at SNBNCBS (including former graduates): Abhijit Chakraborty, Abhinav Kumar, Amartya Sarkar, Ambika P. Jena, Arghya Datta, Biswajit Das, Debmalya Mukhopadhyay, Himadri Ghosh, Indrakshi Raychowdhury, Kapil Gupta, Mitali Banerjee, Prashant Singh, Rajiv K. Chouhan, Rudranil Basu, Sourav Bhattacharya, Sandeep Agarwal, Sandeep Singh, Sudip Garain, Sumit Ghosh, Sunandan Gangopadhyay, and Tanmoy Ghosh.

\thispagestyle{plain}

To my friends who have known me for a long time: Suman Majumdar and Suchandra Ghatak.

To my brother, mother and father for everything.

And to the cosmos $\ldots$ for being there.


%% file: C1_introduction.tex

\chapter{Introduction}
\label{C:intro} 

\lettrine[lraise=0.0, loversize=0.3, nindent=0pt]{O}{ur} knowledge of nature being derived from observations of natural phenomena makes the role of an observer itself an important aspect -- to be analysed and understood as part of our concepts about the natural world. Each observer sees and measures the world from their own frame, perceiving different versions of the `same' world. And our understanding of the mutual relation between these observations, and of the nature of the act of an observation itself has undergone several revolutions during the development of Physics. In-fact, the start of the scientific renaissance was triggered by discard of the geocentric view as The Canonical view and subsequent adoption of a heliocentric coordinates for planetary calculations. Much later, special relativity overthrew the seeming necessity of an absolute rest frame of {\ae}ther. This was through a revolution in our understanding of the relation between measurements of time (and space) among different reference frames, moving with constant velocities with respect to each other. However, till recently, the concept of such inertial frames was still sacrosanct for the validity of physical laws. This was finally overcome in general relativity with the equivalence principle relating accelerating frames to gravity -- bringing them within the ambit of valid physical reference frames through the principle of `general covariance.'

Apart from being a refinement in our concepts of reference frames, the principle of equivalence can also inspire the setting up of frameworks to understand gravity in connection with flat-spacetime (in the Minkowskian sense), the idea forming a preamble to our thesis. We will investigate two lines of thought starting from this. First, we study the Unruh effect \cite{Unruh:1976db} where an accelerated observer in flat spacetime sees the Minkowski vacuum (of some matter field) as a thermal background, from his/her own perspective -- the Rindler frame. This is in direct relation to the formulation of the principle of equivalence dealing with uniform accelerations. Secondly,  we study certain gauge symmetry aspects of Poincar\'{e} gauge theoretic models of gravity. In the formulation of such theories, one starts from the global Poincar\'{e} symmetry of observed matter fields in flat spacetime and then goes on to localising this symmetry. This gauging of the Poincar\'{e} group gives rise to additional gauge potentials and field-strengths that describe gravity. If the principle of equivalence is meant to understand the invariance of general relativity under local Poincar\'{e} transformations, the Poincar\'{e} gauge procedure has been seen as the recovery of a principle of equivalence from the gauge principle \cite{Blagojevic:2002du, Blagojevic:2003cg}. We now outline the specific problems that we identify and address during the course of this thesis.


\section{Rindler space, Unruh effect \& quantum tunnelling}
\label{Cintro:Eq}

\begin{figure}[t]
\centering
\includegraphics[angle=0, width=0.9\textwidth]{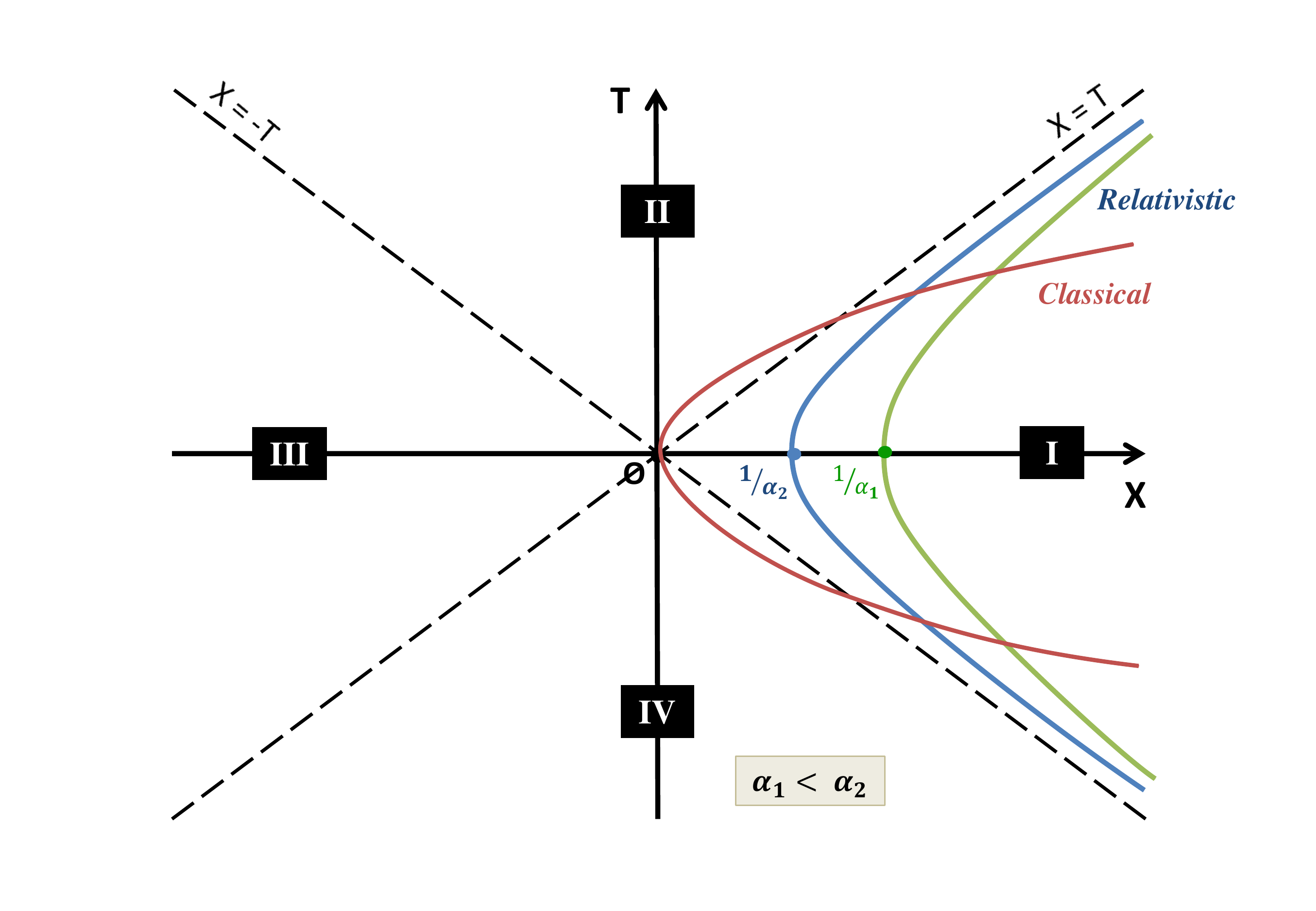}
\caption{\it Non-relativistic and relativistic accelerated observers in the Minkowski plane $(T,X)$.}
\label{D:accobs}
\end{figure}

The trajectory of a uniformly accelerating particle (acceleration $\alpha$) in special relativity is a hyperbola in spacetime (see Chapter 6, \cite{Misner:1974qy}) $$X^2 - T^2 = \frac{1}{\alpha^2}.$$ The speed is bounded by the maximum attainable speed: equal to the speed of light $c$, here taken to be unity by measuring lengths in \textit{light-seconds}. To go into the reference frame adapted to such an observer, a family of observers having all physically possible accelerations between $\pm \infty$ is taken. Their hyperbolic world lines cover only a single wedge of the Minkowski spacetime (wedge $\mathtt{I}$ in Figure \ref{D:accobs}).\footnote{The wedge $\mathtt{III}$ is a time reversed world \cite{Mould:1995gj} and hence is un-physical. The conjugate hyperbolae form a totally different world in wedges $\mathtt{II}$ (can't emit a signal back to the physical world) \& wedge $\mathtt{IV}$ (can't receive any signal from the physical world). The latter two are thus analogues of black hole and white hole regions, respectively.} Now the Rindler metric is usually written as some parametrisation of such world lines. In fact, different authors use various different parametrisations resulting in algebraically different metrics \cite{Rindler:1966zz, Unruh:1976db, Carroll:2004st}. It would be simpler to have a single metric which we can adapt to any parametrisation -- a generalised Rindler metric. This may be implemented by allowing someone to choose their observers' accelerations as some function of space $F(x)^2 = \frac{1}{\alpha(x)^2}$, and keeping this arbitrary function in the metric. We present such a programme in \S \ref{Cunruh:Rindler}. Note that the asymptotes given by $X=\pm T, \ F(x)=0$ represent horizons seen by the accelerated observers: no communication is allowed from wedge $\mathtt{II}$ to wedge $\mathtt{I}$, a fact easily seen by drawing light-cones along the physical trajectories.

Once having written the Rindler metric, it is usual to quantize matter fields in both coordinates and show their inequivalence \cite{Fulling:1972md}. This is because the Bogolyubov coefficients between the two sets of modes mix the annihilation and creation operators, so that neither of the two different sets of vacua are compatible with both the number operators. Unruh showed that the Rindler observer's number operator gives a thermal interpretation to the Minkowski vacuum, characterised by a temperature proportional to the acceleration \cite{Unruh:1976db}. However, it is not easy to understand the origin of this thermalisation of the quantum field \cite{Unruh:1995gn, Kiefer:2012boa}. Also, Unruh's result was significant since he demonstrated its close correspondence with the phenomenon of Hawking radiation from black holes \cite{Hawking:1974rv, Hawking:1974sw}.

An alternative derivation of black hole radiation has been proposed recently based on the phenomenon of quantum tunnelling \cite{Srinivasan:1998ty, Parikh:1999mf}. This method considers the quantum tunnelling of particles across the horizon in a classically forbidden process. The original method is appealing since it makes the role of the horizon relevant, and presents an alternate conceptual picture. However there were problems in understanding why it was needed to take an imaginary factor of time in the calculations. Moreover, the method could only be used to calculate the Hawking temperature, but not the black-body spectrum. Subsequently, within the ambit of the tunnelling approach, a method was proposed to calculate the spectrum in \cite{Banerjee:2009wb}. This was an interesting development, and we apply this procedure in the case of the Unruh effect.

It should be clear that since the wedge $\mathtt{I}$ is the representative of our physical world and the wedge $\mathtt{II}$ represents the black hole world (see Fig. \ref{D:accobs}), the tunnelling must occur between them across the accelerated horizon $X=T,\ T>0$. Following \cite{Banerjee:2009wb}, we adopt a statistical approach by considering a collection of particles. The in-falling ones will be classically trapped inside. But the quantum mechanical boundary matching of modes at the horizon will result in a set of modes to tunnel out, with an exponentially suppressed factor. This is because the nature of time and space coordinates get reversed across the horizon with analytically continuing them across requiring imaginary factors, both in time and space.\footnote{This also happens in relating coordinates across the Schwarzschild black-hole horizon in a Kruskally extended set of coordinates.} Now considering a {\it reduced } density matrix of only outgoing modes leads to a thermalisation. We will show in this thesis (Chapter \ref{C:unruh}) how this method gives us both the Unruh black-body spectrum and temperature, for both bosons and fermions.

As a further interest, considering our aim of seeing gravitational effects from flat spacetime, we note an interesting method to link the Unruh effect in Minkowski space with the Hawking radiation from black holes. This is following Deser and Levin's \cite{Deser:1998bb, Deser:1998xb} idea (coined `GEMS' -- Global Embedding Minkowski Spacetime) of using the embedding of the Schwarzschild geometry in a higher dimensional Minkowski-flat metric. It was known earlier \cite{Fronsdal:1959No} that such embeddings can be found; for the Schwarzschild black-hole, we need a six-dimensional flat space. This maps static observers in the Schwarzschild spacetime to hyperbolic trajectories in the higher dimensional embedding space. Carrying out a tunnelling analysis in this extended flat spacetime gives us back the Unruh temperature, and in this case, it is identical with the Hawking temperature when considering observers who were at asymptotic Schwarzschild infinity. Observers at other positions see temperatures which can be explained through the phenomenon of gravitational red-shifting \cite{Tolman:1934}.


\section{Poincar\'{e} gauge symmetries}
\label{Cintro:PGT}

\subsection{Poincar\'{e} gauge theory}
\label{Cintro:PGT:PGT}

Let us consider a Minkowski flat spacetime $M_4$ with global Poincar\'{e} symmetry. Special relativity holds everywhere. It is possible to set up the global coordinates $x^\mu$ and local laboratory frames $x^i$ at each point such that both are perfectly aligned. Cartesian coordinates at both global and local levels suffice to implement this scheme. A vector in the local coordinates will then be represented in the global frame through $$A^i = \delta^i_\mu \, A^\mu.$$ The Poincar\'{e} symmetry is implemented at the infinitesimal level through infinitesimal parameters of rotation $\theta^{\mu\nu}$ and translation $\varepsilon^\mu$ such that any matter field transforms as
\begin{align*}
\delta_0\phi &= \left(\frac{1}{2}\theta^{\mu\nu}M_{\mu\nu} + \varepsilon^\mu P_\mu\right) \phi\\
\delta x^\mu &= \theta^\mu_{\ \nu}x^\nu + \varepsilon^\mu,
\end{align*}
where $M_{\mu\nu}$ and $P_\mu$ are the rotation/spin and translation generators respectively, obeying the usual Poincar\'{e} algebra. This means that at each point we are free to transform our coordinates (or fields in an active view) by a rotation through $\theta^{\mu\nu}$ and a translation $\varepsilon^\mu$, without bringing change in any physical observables. The global nature implies that the parameters $\theta$ and $\varepsilon$ are constant, i.e. we would be required to implement the same transformations everywhere in spacetime.

Inspired by special relativity and with a finite speed of propagation of information, we can go forward and relax the procedure by allowing arbitrary rotations and translations at each point \cite{Hehl:1976kj}. The parameters $\theta$ and $\varepsilon$ then become functions of spacetime. The local frames are still Minkowski flat, but a vector in the local frame is related to its representation in the global space by non-trivial vielbien fields $$A^i = b^i_{\ \mu} A^\mu.$$ The local frames may also have a relative rotation and the derivative operator changes as $$\partial_\mu \rightarrow \nabla_\mu = \partial_\mu + \frac{1}{2}\omega^{ij}_{\ \ \mu}\Sigma_{ij},$$ where $\Sigma_{ij}$ is the spin part of $M_{\mu\nu}$ obeying the same algebra, and $\omega^{ij}_{\ \ \mu}$ are spin connection fields. So we see that localising (or `gauging') of the Poincar\'{e} symmetries require introduction of additional fields, sometimes known as gauge potentials. The geometry of the global space is now no longer $M_4$, but a more general $U_4$ Riemann-Cartan spacetime, which has both curvature and torsion. The vielbien fields can be used to define a metric on the global space through $$g_{\mu\nu}=b^i_{\ \mu}b^j_{\ \nu}\eta_{ij},$$ where $\eta_{ij}$ is the Minkowski metric of the local frames. The curvature and torsion get introduced in this formalism as field strengths, due to the changed covariant derivative operator $\nabla_\mu$. This procedure was originally developed by Utiyama \cite{Utiyama:1956sy}, Sciama \cite{Sciama:1964wt} and Kibble \cite{Kibble:1961ba} and is known as the Poincar\'{e} gauge theory. We have detailed the setting up of the formalism in Chapter \ref{C:pgt}, including a section (\S \ref{Cpgt:geometrical}) containing some points on the relation of the Poincar\'{e} gauge transformations with the known diffeomorphisms of relativity.

Numerous models of gravity have been studied in the literature (see \S \ref{Cpgt:3Dactions} for a selective outline of the literature), written down using the field strengths mentioned above. In general, the presence of torsion has not yet been detected experimentally. The difficulty, as summarised by Hehl et al. \cite{Hehl:1976kj, Hehl:2013qga} lies in the fact that torsion couples only to the intrinsic spin, not orbital angular momentum. And typically, spin gets averaged out in any significantly large distribution of particles, as against mass and energy-momentum which is a source of the observable curvature in our universe. However, from the point of view of gauge theory, Poincar\'{e} gravity models present some remarkable features which by itself are important enough to study. We now turn to one such issue.

\subsection{Canonicity of Poincar\'{e} gauge symmetries}
\label{Cintro:PGT:canon}

It is not straightforward to write down a hamiltonian for a given action which has gauge symmetries. The Legendre transform is then inexact and some velocities are left undetermined. This is a manifestation of the fact that the equations of motion cannot be used to fully determine the time evolution of a given physical state. There occur arbitrary functions of time in the solutions, which is indicative of presence of gauge symmetries. Such symmetries or transformations leave the action invariant across all paths in phase space, and additional gauge-fixing conditions have to be imposed to yield unique time evolution of physical states.

Following Dirac's procedure for handling such systems \cite{Dirac:Lectures}, we first need to identify the constraints corresponding to the inexact Legendre transform. A complete set of all such constraints, consistent with the dynamics, is to be then identified. The dynamics itself is generated by a `total hamiltonian' that contains a linear combination of all primary constraints, those which arose directly from the definition of momenta adopted while implementing the Legendre transform. These are added into the hamiltonian weighted by arbitrary Lagrange multipliers. Now, some constraints lead to a fixing of some of these multipliers and are classified as second class. The rest are known as first-class and generate gauge symmetries. They form a closed algebra (under Poisson or Dirac brackets, as the case may be) among themselves. However not all of these generate indpendent gauge symmetries. There are restrictions such that the number of independent gauge symmetries (and hence gauge parameters) must be equal to the number of independent primary first-class constraints \cite{Henneaux:1990au, Banerjee:1999hu, Banerjee:1999yc}.

Now, with regard to the canonical structure of Poincar\'{e} gauge symmetry models, a wide number of studies by various authors \cite{Blagojevic:2004hj, Blagojevic:2008bn, Blagojevic:2010ir, Grumiller:2008pr, Park:2008yy, Carlip:2005zn} are available. And while discussing the gauge symmetries recovered via a first-class gauge generator, it was noted that the Poincare\'{e} symmetries were recoverable only {\em on-shell}. So, there appeared to be two different sets of symmetries corresponding to the same action, off-shell. From the point of view of gauge symmetries, in the back drop of the theorem mentioned before as well as looking towards the maximum number of parameters admissible in the Poincar\'{e} group, this represented a puzzle. And the major part of our endeavour in this portion of our thesis, was to resolve this.

At first, we made sure that both sets of symmetries (the geometrical/gauge ones and the canonically generated) were off-shell by checking explicitly. This was done for the Mielke-Baekler type 3-dimensional model of gravity with torsion \cite{Mielke:1991nn}, presented in Chapter \ref{C:mb}. Then, since most works employed an on-shell procedure \cite{Castellani:1981us} of constructing the gauge generator, we went forward to repeat the canonical procedure in a clean off-shell manner and re-construct the gauge generator following an explicitly off-shell algorithm \cite{Henneaux:1990au, Banerjee:1999hu, Banerjee:1999yc}. We carried out our analysis both for the Mielke-Baekler model (Chapter \ref{C:mb}) and the `New Massive Gravity' model proposed recently \cite{Bergshoeff:2009hq, Bergshoeff:2009aq} (Chapter \ref{C:bht}). To stress that our algorithm for constructing the gauge generator is indeed off-shell, we also present in this thesis (\S \ref{Cmb:ConstraintsRev}) a new derivation, simpler than the original one. Our results however showed that -- atleast algebraically -- indeed the two sets of symmetries were distinct.

To understand the results better, we carried out an extensive lagrangian analysis. Using the formalism of \cite{Shirzad:1998af} we constructed a set of `lagrangian generators' which could reproduce the Poincar\'{e} gauge symmetries directly. This is presented in Chapter \ref{C:lag}. In addition, we explicitly constructed the Noether identities corresponding to both sets of symmetries. By comparing these, we found a procedure to construct a map between the two sets of gauge parameters. The mapping is an important step since to compare the geometric Poincar\'{e} gauge symmetries with the canonical ones, we first have to ensure that both symmetries are parametrised by the same set of infinitesimal parameters. The mapping itself was often used in the literature based on inspection and intuition. However it was instructive to find an algorithm to generate it, even more so as it is a non-trivial field-dependent map.

This lagrangian analysis through gauge/Noether identities reveals that instead of two different sets of Noether identities, we have exactly one set and their number is consistent with the expected number of independent gauge symmetries/parameters. Thus the apparent algebraic difference in the two sets of symmetries is not significant at a canonical level. Subsequently, a study of the canonical symmetries in the hamiltonian formalism also supports this fact and we show how the two sets of symmetries are {\em canonically equivalent} off-shell: their difference not being responsible for introduction of any arbitrariness in the time evolution of physical states. The role of `trivial gauge' transformations \cite{Henneaux:1992ig} is explicitly highlighted as a result of our work, alongside the need for a synergistic canonical treatment adopting both hamiltonian and lagrangian procedures that complement each other. This part of our work is presented in Chapter \ref{C:tr}.

We will present a detailed summary of our results and some further discussions in the concluding Chapter \ref{C:conc}.


%% file: C2_unruh.tex

\chapter{The Unruh effect through quantum tunneling}
\label{C:unruh}

\lettrine[lraise=0.0, loversize=0.3, findent=-1pt]{A}{} route to studying effects of gravity from flat spacetime is enshrined in the `{\it principle of equivalence}.' It states that an uniform gravitational field may be transformed away by adopting the point of view of a uniformly accelerating observer: the effect of gravity is mimicked by passing to a non-inertial frame. This principle is also discussed as an equivalence between inertial and gravitational mass.

So, one may ask, what are the effects of gravity that one may study in such settings? There are many, with some results coming from consideration of quantum fields on curved general relativistic spacetimes. One such result is that black holes do radiate and have a particular temperature. This phenomenon is now known as Hawking radiation, after Hawking \cite{Hawking:1974rv, Hawking:1974sw} who showed that a collapsing geometry with formation of a horizon results in particle creation. This is made evident through non-trivial Bogolyubov coefficients between the ingoing and outgoing modes of a quantum field in such a geometry. There are several facets of this derivation that one could highlight, with the crucial role played by the black hole horizon being one of them. A horizon is, in general, a boundary in spacetime which restricts the flow of {\it all} information across it, to a single direction -- much like a valve.

Unruh tried to understand this phenomenon and correctly encapsulated the idea by studying an accelerated observer in flat spacetime, the corresponding horizon being the accelerated-horizon restricting the field of view of such an observer. After earlier such works by Fulling \cite{Fulling:1972md} and Davies \cite{Davies:1974th}, he showed \cite{Unruh:1976db} that similar non-trivial Bogolyubov coefficients result in the Minkowski vacuum to be seen as a warm thermal state by an accelerated observer. This explicitly demonstrated the in-equivalence of the positive and negative frequency mode decomposition of quantum fields between different observers.

However, such demonstrations were still captured in the language of field theory, which has its own notion of definition of particles. A different and ordinary wave-packet quantum mechanics has also been used by Padmanabhan \cite{Srinivasan:1998ty} and Parikh-Wilczek \cite{Parikh:1999mf} in recent times to study the Hawking radiation. This method is based on a `quantum tunneling' perspective and clearly demonstrates that presence of quantum fields across a horizon forces quantum penetration of particles across the classically forbidden direction. This is dictated by basic requirements of continuity of quantum modes across boundaries.

Originally, quantum tunneling was employed to calculate the temperature corresponding to Hawking radiation from black holes. However, the method has been extended to include the calculation of the black-body energy spectrum \cite{Banerjee:2009wb}. Here, we will adopt this method to discuss both the temperature and spectrum in the case of an observer accelerating in flat spacetime, i.e. the case of the Unruh effect.

Having at hand the Unruh effect, one can now go back and relate it to the Hawking effect. The two effects are closely related, but are not quite the same. The Hawking radiation at the horizon (a process occurring locally with infinite energy) is red-shifted to the predicted value to an observer at asymptotic infinity. In the case of the Unruh effect however, the temperature is measured by an observer throughout his uniformly accelerated motion. Now a method to envision the Hawking radiation more closely as an Unruh effect lies in embedding the blackhole geometry into a flat spacetime. The required flat space thus is of higher dimensions than the black hole spacetime and the Blackhole Hawking effect may now be mapped into the flatspace Unruh effect. This idea was introduced by Deser and Levin \cite{Deser:1998xb} and is known as GEMS (global embedding Minkowski spacetime) method. We will finally relate the Hawking and Unruh effects through this approach. It presents another interesting route to understand curved space effects of gravity through flat spacetime.


\section{Non-inertial observers: Rindler coordinates}
\label{Cunruh:Rindler}

The world seen by a particle in special relativity differs from the Newtonian particle through the introduction of a maximal allowed speed of propagation of signals: the speed of light `$c$.' The Newtonian parabolic distance formula $X=\frac{1}{2} \alpha T^2$ is changed to a hyperbolic law
\begin{align}
\label{hyper}
X^2 - T^2 = \frac{1}{\alpha^2},
\end{align}
where $\alpha$ is the constant acceleration of the particle along the Minkowski space direction $X$, w.r.t the Minkowski time $T$. If we allow all possible accelerations from zero to infinity, the hyperbolae cover the full region $X>0$ between the lines $X=\pm T$. This region is known as a `wedge' and is one among four such wedges which cover the full Minkowski spacetime. Thus a particle with any conceivable acceleration gets trapped in a sub-space of the Minkowski spacetime -- the `physical wedge.' We can now construct an appropriate coordinate system with the worldlines of the accelerated observer through a parametrization of the hyperbolic worldlines
\begin{align}
\label{transformationsI}
\begin{aligned}
T &= F(x_\mathtt{\scriptscriptstyle I}) \sinh(at_\mathtt{\scriptscriptstyle I})\\
X &= F(x_\mathtt{\scriptscriptstyle I}) \cosh(at_\mathtt{\scriptscriptstyle I})\\
Y &= y_\mathtt{\scriptscriptstyle I}~, \quad Z = z_\mathtt{\scriptscriptstyle I},
\end{aligned}
\end{align}
where $F(x)^2=1/\alpha^2$ and $a$ is a constant. The subscript $\mathtt{I}$ indicates that this transformation is for the physical wedge only. Adopting this transformation $(X, T) \longrightarrow (x, t)$, the Minkowski metric $ds^2 = -dT^2 + dX^2 + dY^2 + dZ^2$ changes to
\begin{align}
\label{genRind}
ds^2 = -a^2F(x)^2dt^2 + F'(x)^2dx^2 + dy^2 + dz^2,
\end{align}
known as the `Rindler metric.' The new Rindler time and space coordinates are given by $t$ and $x$ respectively.

\begin{figure}[t]
\centering
\includegraphics[angle=0, width=0.9\textwidth]{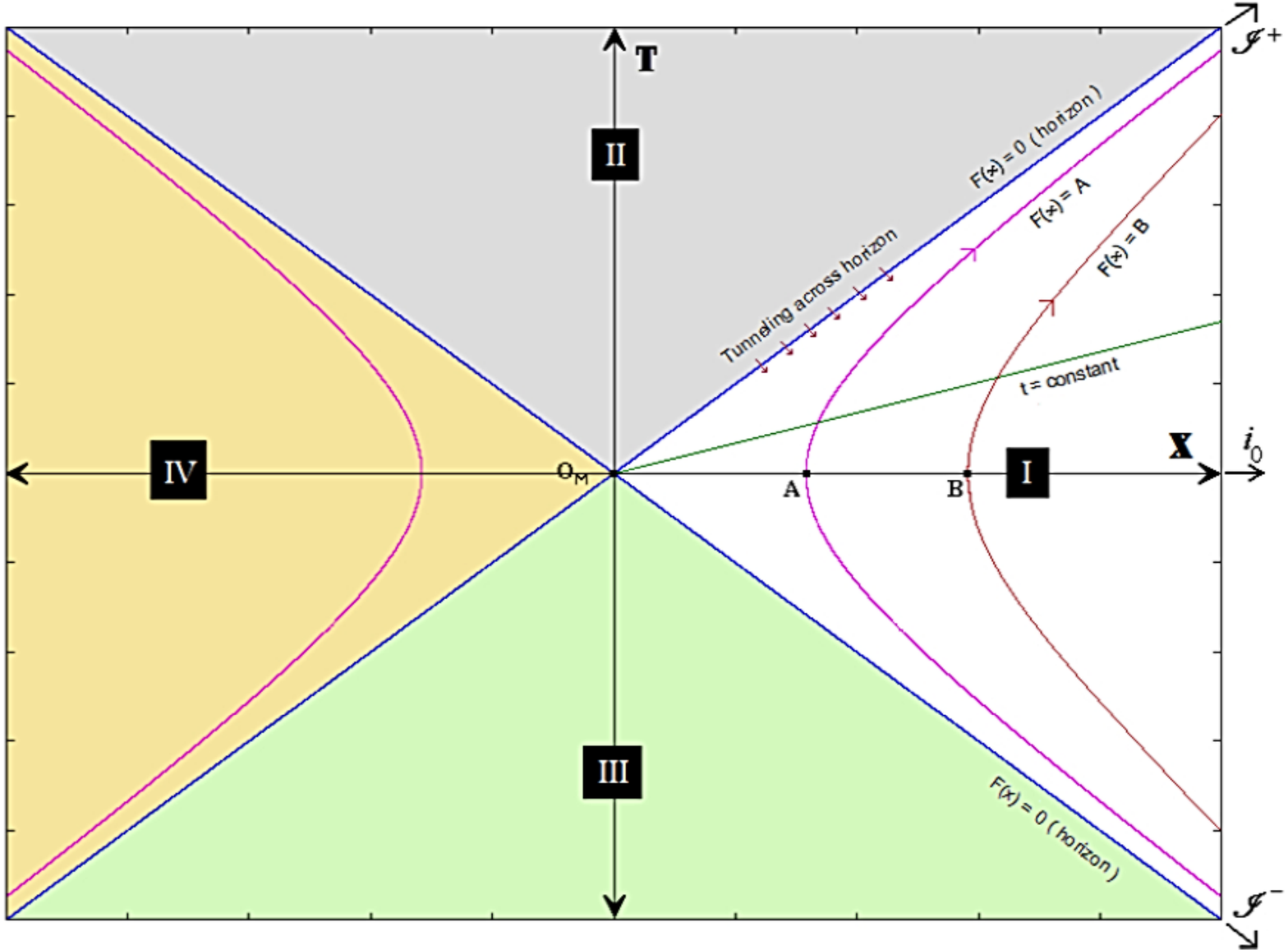}
\caption{{\it Four Rindler wedges covering the Minkowski plane. {\em (}{\bf T},{\bf X}{\em )} and {\em(}$t,x${\em)} are Minkowski and Rindler coordinates. The infinities lie outside the diagram, towards the directions shown. Tunneling occurs from region $\textrm{II}$ to region $\textrm{I}$, across the horizon $X=T$.}}
\label{D:wedge}
\end{figure}


The three other regions or wedges (Fig \ref{D:wedge}) have to be covered by transformations similar to this but having different signs or reversal of roles between time and space. For example the region $T>0$ and bounded by the lines $X=\pm T$ is covered by coordinate transformations
\begin{align}
\label{transformationsII}
\begin{aligned}
T &= F(x_\mathtt{\scriptscriptstyle II}) \cosh(at_\mathtt{\scriptscriptstyle II})\\
X &= F(x_\mathtt{\scriptscriptstyle II}) \sinh(at_\mathtt{\scriptscriptstyle II})\\
Y &= y_\mathtt{\scriptscriptstyle II}~, \quad Z = z_\mathtt{\scriptscriptstyle II}.
\end{aligned}
\end{align}
Comparing between the two transformations \eqref{transformationsI} and \eqref{transformationsII} we see that the role of Rindler time $t$ and space $x$ get interchanged between the two regions $\mathtt{I}$ and $\mathtt{II}$. This is because the transformations \eqref{transformationsII} parametrise the hyperbolae $$T^2 - X^2 = \frac{1}{\alpha^2},$$ the conjugate hyperbolae to \eqref{hyper}. The boundary marked by the worldline of an observer with infinite acceleration, i.e. the lines $X=\pm T$ (0r $F(x)=0$ in Rindler coordinates) act as horizons -- the `accelerated horizons.' The region $\mathtt{II}$ is the known as the black hole like region as no information can leave from it and reach any accelerating observer, though all such observers can send information inside. This can be seen by imagining a light cone attached to any observer travelling along a hyperbole with constant acceleration, i.e. constant $F(x)$. The past of all such light cones will always never overlap with region $\mathtt{II}$, while their future will definitely overlap.

The other two regions are the time reversed world (region $\mathtt{IV}$) and the white hole like world (region $\mathtt{III}$). Corresponding to every forward moving path in the physical wedge, there is a time reversed path in wedge $\mathtt{IV}$ which makes it a time reversed copy of the physical world. And in case of the region $\mathtt{III}$, information can come out and reach any accelerated observer though no information can ever be sent inside as it is always in the past of any accelerated observer.

The metric presented in \eqref{genRind} is a generalised Rindler metric and various other forms seen in literature can all be attained through specific choices of the parametrizing function $F(x)$. This is demonstrated in appendix \ref{App:Unruh}. However for our convenience, we now change to a tortoise-like coordinate $x_\star$ appropriate for the Rindler metric \eqref{genRind} through the transformation
\begin{align}
\label{tortoise}
\begin{aligned}
dx_\star = \frac{F'(x)}{a F(x)} ~dx\\
x_\star = \frac{1}{a} \ln F(x),
\end{aligned}
\end{align}
after which the metric (\ref{genRind}) becomes
\begin{align}
\label{RinTor}
ds^2 = \left(a~e^{ax\star}\right)^2\left(-dt^2 + dx_\star^2\right) + dy^2 + dz^2.
\end{align}

Now a maximal extension of the Rindler coordinate system $(t,x,\ldots)$ turns out to be the Minkowski coordinates $(T,X,\ldots)$ which are defined in the entire Minkowski plane, irrespective of accelerated horizons which the Rindler observer encounters \cite{Rindler:1966zz}. The Rindler coordinates however undergo a finite shift through the horizon, as can easily be seen through a comparison of \eqref{transformationsI} and \eqref{transformationsII}. This induces a relation between the $t-x_\star$ coordinates defined in wedges $\mathtt{I}$ and $\mathtt{II}$ as
\begin{align}
\label{1_2relation}
\begin{aligned}
t_{\mathtt{\scriptscriptstyle II}}&=t_{\mathtt{\scriptscriptstyle I}}-\frac{i\pi}{2a}\\
x_{\star}{}_{\mathtt{\scriptscriptstyle II}}
&=x_{\star}{}_{\mathtt{\scriptscriptstyle I}}+\frac{i\pi}{2a}.
\end{aligned}
\end{align}
The directions perpendicular to the acceleration ($y, z$) remain unchanged. Such transforms were reported earlier in \cite{Akhmedov:2008ru, Banerjee:2008sn}. It is to be noted that the transformation $t_{\mathtt{\scriptscriptstyle II}} = t_{\mathtt{\scriptscriptstyle I}}+\frac{i\pi}{2a};\ \ x_{\star}{}_{\mathtt{\scriptscriptstyle II}} = x_{\star}{}_{\mathtt{\scriptscriptstyle I}}-\frac{i\pi}{2a}$ also suffice in relating the coordinates $\mathtt{II}$ and $\mathtt{I}$. This second pair leads to some problems in taking the classical limit of an outgoing tunneling probability, and so is not considered at the accelerated horizon. However they are required at cosmological horizons for incomming radiations \cite{Modak:2008tg}.


\section{Quantum tunneling}
\label{Cunruh:qtun}

The method of quantum tunneling captures the intuitive picture of radiation tunneling across the horizon in a classically forbidden process. Originally demonstrated for scalar particles in a Schwarzschild spacetime, the formalism since then has both been refined and extended to fermions and various other black hole spacetimes \cite{Parikh:2004ih, Shankaranarayanan:2000qv, Vagenas:2001qw, Akhmedov:2006un, Akhmedov:2008ru, Akhmedova:2008dz, DiCriscienzo:2008dm, Mitra:2006qa, Banerjee:2008gc, Banerjee:2008fz, Majhi:2008gi, Majhi:2009uk, Medved:2005yf, Chowdhury:2006sk, Pilling:2007cn, Li:2008ws, Jiang:2008gq, Kerner:2007rr, Banerjee:2009wb, Banerjee:2009pf, Kaul:2000kf, Modak:2008tg, Banerjee:2009tz, Banerjee:2009sz, Banerjee:2009xx, Angheben:2005rm, Kerner:2006vu, Stotyn:2008qu, Sarkar:2007sx, Nozari:2009nr}.

The starting point of the calculation is to find the quantum modes, either Klein-Gordon or Dirac as the case demands, in the background geometry being considered. In general, this is a difficult task and so the WKB approximation is adopted with the ansatz for the wave function as $\phi = \exp[-\frac{i}{\hbar} S(x,t)]$, where $S(x,t)$ is the single particle action. Classically the the ingoing probability (always given by $|\phi|^2$) is unity while the outgoing probability is zero. Now, the ingoing single particle action is real and the outgoing action is complex with the tunneling rate being proportional to the exponential of the imaginary part of the outgoing action. This is compared with a Boltzmann factor through the `principle of detailed balance' \cite{Srinivasan:1998ty} leading to the Unruh temperature. Therefore the important part of this method lies in the calculation of the imaginary outgoing one-particle action.

In the original exposition outlined above, use is made of a single-particle picture and the only contact with statistics is brought through the principle of detailed balance, a thermodynamical relation. This could only yield the temperature. To be able to calculate the spectrum, we need to adopt a more statistical viewpoint. Following \cite{Banerjee:2009wb}, we here use a density matrix of many such particles tunneling out, to find the spectrum as well as the corresponding temperature. The role of reversal of time and space across the horizon in determining the quantum radiation is highlighted through use of the coordinate matching relations \eqref{1_2relation} to maintain continuity of modes across the horizon. Since we are discussing the Unruh effect for Rindler spacetime, which is essentially flat, we make a full determination of the quantum modes without going into any WKB approximation. Below, we start with the determination of scalar and fermionic modes in the Rindler background.


\section{Scalar particles in Rindler spacetime}
\label{Cunruh:KGmodes}

Scalar particle modes are obtained from the Klein Gordon (massless) equation $\square \Phi = 0$, written in the Rindler metric \eqref{RinTor}
\begin{align}
\label{kg}
\frac{e^{-2 a x_\star}}{a^2}\left[-\partial_t^2\Phi + \partial_{x_\star}^2\Phi\right] + \partial_y^2\Phi + \partial_z^2\Phi = 0.
\end{align}
The metric being independent of the coordinates $t, y$ \& $z$, we adopt an ansatz for $\Phi$ as
\begin{align}
\label{sansatz}
\Phi(t,x_\star,y,z)=\phi(x_\star)~e^{-\frac{i}{\hbar}\left(\Omega t+k_y y+k_z z\right)},
\end{align}
where $\Omega$ is a constant. It is related to the local energy $\omega$ at some Rindler spacetime point $x_\star$ through the Tolman red-shift relation \cite{Tolman:1934, Carroll:2004st} $E_1~V_1=E_2~V_2=\Omega$ connecting the observed energies $E_1$ and $E_2$ at two different points in a gravitating system at equilibrium. The result ensures that though the observed energies $E$ and the \emph{Tolman red-shift} factor $V$ vary locally (as functions of $x_\star$), their product $\Omega$ is a constant. In a gravitational system in equilibrium, this condition of the constancy of $\Omega$ characterizes the equilibrium, just as the bare temperature is in a flat space thermodynamic system. In Rindler spacetime this locally observed energy $E$ is the energy of the tunneling-particle ($\omega$) and the red-shift factor $V$ equals $\sqrt{|g_{00}|}$. So
\begin{align}
\label{redshift}
\Omega = \omega\ a e^{a x_\star} = \frac{a\,\omega}{\alpha}
\end{align}
where appropriate definitions $F(x_\star)=1/\alpha=e^{a x_\star}$ (see \ref{genRind} and \ref{RinTor}) have been used. Substituting the ansatz \eqref{sansatz} back in the Klein Gordon equation \eqref{kg} leads to the following differential equation
\begin{align}
\label{reducedkg}
\phi''(x_\star)+\left(\frac{\Omega^2}{\hbar^2}-a^2 e^{2 a
x\star}\frac{k_\perp^2}{\hbar^2}\right)\phi(x_\star)=0,
\end{align}
with $k_\perp = \sqrt{k_y^2+k_z^2}$.

We pause to make some observations from equation governing the scalar modes \eqref{reducedkg}. Near the horizon $x_\star \rightarrow -\infty$ and the term containing $k_\perp$ drops out resulting in a simple harmonic type equation with plane wave solutions. Again at large spatial distances, $x_\star \rightarrow \infty$ and thus the term containing $\Omega^2$ becomes negligible. This leaves an equation with exponentially increasing and decreasing solutions $I_0\left(\frac{k_\perp}{\hbar}e^{a x_\star}\right)$ and $K_0\left(\frac{k_\perp}{\hbar}e^{a x_\star}\right)$, where $I_0$ and $K_0$ are the zero-th order modified Bessel functions of first and second types respectively. Thus throwing away the $I_0$ solutions, we have an exponentially vanishing solution at infinity, in $K_0$. The resulting modes are similar to those found by Boulware \cite{Boulware:1974dm}.

The solution of the full equation (\ref{reducedkg}) that is well defined through the horizon is,
\begin{align}
\label{phix}
\phi(x_\star) =& A_-\ e^{\frac{\pi\Omega}{2a\hbar}} \Gamma\!\left(1-\frac{i \Omega} {a\hbar}\right) I_{-\frac{i \Omega}{a\hbar}}\!\left(\frac{k_\perp}{\hbar} e^{a x_\star}\right) + \nonumber\\
&\quad A_+\ e^{-\frac{\pi\Omega}{2a\hbar}}\Gamma\!\left(1+\frac{i \Omega}{a\hbar}\right) I_{\frac{i \Omega}{a\hbar}}\! \left(\frac{k_\perp}{\hbar}e^{a x_\star}\right)
\end{align}
where $A_\mp$ are arbitrary integration constants. For small arguments, the appropriate expansion of the modified Bessel function is $I_\nu(z)\simeq\frac{\left(z/2\right)^\nu} {\Gamma(1+\nu)}$. This holds if $k_\perp\ll \Omega$ and also especially near the horizon where $x_\star\rightarrow-\infty$. Therefore (\ref{phix}) simplifies to
\begin{align*}
\phi(x_\star)\simeq A_\mp\ e^{\pm\frac{\pi\Omega}{2a\hbar}}
\left(\frac{k_\perp}{2\hbar}\right)^{\mp\frac{i\Omega}{a\hbar}}\
e^{\mp\frac{i}{\hbar}\Omega x_\star}.
\end{align*}
The total wave function $\Phi(t,x_\star,y,z)$ near the horizon is then
\begin{align}
\label{totswave}
\Phi(t,x_\star,y,z)=&B_{\mathsf{\scriptscriptstyle in}}e^{-\frac{i}{\hbar}
\left[\Omega(t+x_\star) + k_y y+k_z z \right]} + B_{\mathsf{\scriptscriptstyle out}} e^{-\frac{i}{\hbar}\left[\Omega(t-x_\star) + k_y y+k_z z \right]},
\end{align}
where all the constants have been absorbed within $B_{\mathsf{in/out}}$. The subscript `${\mathsf{in}}$' here stands for the ingoing mode which travels toward the accelerated horizon at $x_\star=-\infty$, while the subscript `${\mathsf{out}}$' stands for the outgoing mode traveling away from horizon, i.e. towards $x_\star=\infty$.


\section{Dirac particles in Rindler spacetime}
\label{Cunruh:Dirmodes}

In a curved spacetime spinors are dealt with as objects in the local frames at each spacetime point \cite{Birrell:1982ix}. Usually in a Riemannian space with metric $g$, the local frames are chosen to be Minkowskian with metric $\eta$. The local tetrad frame fields constitute a map between tensors on tangent space to the local Lorentz frame and vice-versa through relations as
\begin{align*}
A^\mu = V_a^\mu A^a.
\end{align*}
Here $A^\mu$ is a tensor in the tangent space, $A^a$ is the corresponding vector in the local frame and $V^a_\mu$ is the tetrad field. Our convention is Latin ($a,b,\ldots$) and Greek ($\mu,\nu,\ldots$) letters for local frame and curved space indices respectively. The tetrad relates the metrics between curved and local frames through the relation
\begin{align}
\label{metrictetrad}
g_{\mu\nu}=V^a_\mu\ V^b_\nu\ \eta_{ab}.
\end{align}
In our case of the Rindler metric \eqref{RinTor}, the choice of the tetrad field $V^a_\mu$ may be
\begin{align}
\label{tetrad}
V_\mu^a={\rm diag}(a e^{a x_\star}, a e^{a x_\star}, 1, 1)
\end{align}
and the metric signature, both global and local, is kept same as ($-,+,+,+$). It is to be noted that the relation \eqref{metrictetrad} cannot uniquely specify all of the $4 \times 4=16$ (in $4$-D) components of the tetrad. The specific choice \eqref{tetrad} is one of several possible choices possible. We adopt it here in order to work with a simple diagonal object.

The massless Dirac equation may be written as \cite{Giammatteo:2004wp, Crispino:2007eb}
\begin{align}
\label{DiracEq}
\left[\gamma^a \ V_a^\mu\ (\partial _\mu+\Gamma_\mu)\right]\Psi=0
\end{align}
where $\gamma^a $ are the usual Dirac matrices obeying $\left[\gamma^a,\gamma^b\right]=2\eta^{ab}$ and $\Gamma_\mu$ are connection coefficients given by
\begin{align*}
\Gamma_\mu &= \frac{1}{2}\ \Sigma^{ab}\ V_a^\nu\ V_{b\nu;\mu}\\
\Sigma^{ab}&=\frac{1}{4}\left[\gamma^a,\gamma^b\right].
\end{align*}
Covariant derivatives over the curved space indices is defined in the usual way $V_{b\nu;\mu} = \partial_{\mu}V_{b\nu} - \Gamma^{\alpha}_{\mu\nu} V_{b\alpha}$, where $\Gamma^{\alpha}_{\mu\nu}$ is the Christoffel symbol. On using the properties of $\gamma$ matrices and the diagonal choice of the tetrad \eqref{tetrad}, the spin-connection becomes $\Gamma_\mu = -\frac{1}{2}\ \Sigma^{ab}\ \Gamma^\lambda_{\mu\nu}\ V^\nu_a\ V_{b \lambda}$. Substituting all this, the Dirac equation \eqref{DiracEq} becomes
\begin{align}
\label{DiracEq2}
\big[\left(\partial_t-a\Sigma^{01}\right)-\gamma^0\gamma^1 \partial_{x_\star} - ae^{ax_\star}\big(\gamma^0\gamma^2 \partial_y + \gamma^0\gamma^3\partial_z \big)\big]~\Psi=0.
\end{align}
The equation is independent of all coordinates except $x_\star$. So we the an ansatz for the total solution $\Psi$ as a spinor depending on $x_\star$ modulated by a phase factor
\begin{align}
\label{DansatzPsi}
\Psi(t,x_\star,y,z)&=\psi(x_\star)e^{-\frac{i}{\hbar}
 \left(\Omega t+k_y y+k_z z\right)}\nonumber\\
\psi(x_\star)&=\left[\begin{matrix}A(x_\star)\\0\\B(x_\star)\\0\end{matrix}\right].
\end{align}
Upon using this ansatz, equation (\ref{DiracEq2}) can be cast into a Schr\"{o}dinger like equation
\begin{align}
\label{Schrod}
\hat{H}\ \psi(x_\star)=\Omega\ \psi(x_\star)
\end{align}
with a Hamiltonian given by
\begin{align}
\label{hamilt}
\hat{H}=i\hbar\big(a\Sigma^{01} + \gamma^0\gamma^1\partial_{x_\star}\big) + ae^{a x_\star} \big(k_y\gamma^0\gamma^2+k_z\gamma^0\gamma^3\big).
\end{align}
Squaring the above equation we get $\hat{H}^2\psi=\Omega^2\psi$. Subsequent use of\eqref{DansatzPsi} along with the usual definition of gamma matrices $\gamma^0=\left(\begin{smallmatrix}-i&0\\0&i\end{smallmatrix}\right)$, $\gamma^j=\left(\begin{smallmatrix}0&-i\sigma^j\\i\sigma^j&0\end{smallmatrix}\right)$ ($j=1,2,3$ and $\sigma^j$ the Pauli matrices), the following equation governing the spinor component functions is obtained
\begin{align}
\label{ABEqns}
\blacklozenge''+a\,\blacklozenge'+\Biggl[\frac{\Omega^2}{\hbar^2}-a^2\,e^{2ax_\star}\frac{k_\perp^2}{\hbar^2}+\frac{a^2}{4} \Biggl]\blacklozenge=0.
\end{align}
Here $\blacklozenge$ (blacklozenge) is a place-holder for the functions $A(x_\star)$ and $B(x_\star)$. Similar to the case of scalar modes in \eqref{reducedkg}, a study of asymptotic behaviour of this equation show oscillatory behaviour near the horizon and vanishing modes near infinity. From above considerations, the solution for the full equation \eqref{ABEqns} may be written as
\begin{align}
\label{ABSolns}
\blacklozenge = e^{-\frac{ax_\star}{2}} &\biggl[M_-^{(\blacklozenge)}\ e^{\frac{\pi\Omega}{2a\hbar}}\ \Gamma \left(1-\frac{i \Omega}{a\hbar}\right) I_{-\frac{i \Omega}{a\hbar}}\left(\frac{k_\perp}{\hbar}e^{a x_\star}\right)\ \nonumber\\
{}& \quad + M_+^{(\blacklozenge)}\ e^{-\frac{\pi\Omega}{2a\hbar}}\ \Gamma \left(1+\frac{i \Omega}{a\hbar}\right) I_{\frac{i\Omega}{a\hbar}}\left(\frac{k_\perp}{\hbar}e^{a x_\star}\right)\biggl].
\end{align}
Note that both \eqref{ABSolns} and \eqref{phix} are full solutions of the respective differential equations \eqref{ABEqns} and \eqref{reducedkg}, without using any approximation. However, since the phenomenon of tunneling occurs near the horizon, we can study the near horizon forms of these complete solutions by using the expansion of the modified Bessel function for small arguments, $I_\nu(z)\simeq\frac{\left(z/2\right)^\nu} {\Gamma(1+\nu)}$. This holds if $k_\perp\ll \Omega$ and also especially near the horizon where $x_\star\rightarrow-\infty$. So near to the horizon the total spinor $\Psi$ can finally be written as
\begin{align}
\label{DiracFinSol}
\Psi(t,x_\star,y,z)=\xi_{\mathsf{\scriptscriptstyle in}}e^{-\frac{ax_\star}{2}} e^{-\frac{i}{\hbar}\left[\Omega(t+x_\star) + k_y y+k_z z \right]} + \xi_{\mathsf{\scriptscriptstyle out}}e^{-\frac{ax_\star}{2}} e^{-\frac{i}{\hbar} \left[\Omega(t-x_\star) + k_y y+k_z z \right]},
\end{align}
with $\xi_{\mathsf{\scriptscriptstyle in/out}}$ being constant spinors and subscripts `$\mathsf{in}$'/`$\mathsf{out}$' denoting ingoing/outgoing modes.


\section{Tunneling: temperature and thermal spectrum}
\label{Cunruh:unruhtunn}

We now have single-particle modes for bosons \eqref{totswave} and fermions \eqref{DiracFinSol} in the Rindler background. These solutions are valid in both Rindler wedges $\mathtt{I}$ and $\mathtt{II}$, but in their respective native coordinates. Now let a virtual pair of particles be formed just inside the horizon, in wedge $\mathtt{II}$, by some pair production process. Classically, both the ingoing and outgoing modes inside the horizon are trapped as nothing can travel out across the horizon. However the modes being quantum in nature, an outgoing particle can quantum-mechanically tunnel out from wedge $\mathtt{II}$ to wedge $\mathtt{I}$. This process occurs with a Maxwellian probability $e^{-\frac{2 \pi \omega}{\hbar a}}$, that appropriately goes to zero in the classical ($\hbar \rightarrow 0$) limit. Now to find the energy distribution of a collection of such particles, we will construct a suitable density matrix for both bosons and fermions, and find out the average number of particles tunneling out at some particular energy $\omega$. This will give us the spectrum of the radiation.

Starting first with bosonic particles, the relation between inside and outside wave-functions is dictated by the connection between coordinates \eqref{1_2relation} in equation \eqref{totswave} for the modes.
\begin{align}
\label{scalarOutIn}
\begin{aligned}
B_{\mathsf{\scriptscriptstyle in}}e^{-\frac{i}{\hbar}\left[\Omega(t_{\mathtt{\scriptscriptstyle II}} + x_\star{}_{\mathtt{\scriptscriptstyle II}}) + k_y y_{\mathtt{\scriptscriptstyle II}}+ k_z z_{\mathtt{\scriptscriptstyle II}}\right]} &= B_{\mathsf{\scriptscriptstyle in}} e^{-\frac{i}{\hbar} \left[\Omega(t_{\mathtt{\scriptscriptstyle I}} + x_\star{}_{\mathtt{\scriptscriptstyle I}}) + k_y y_{\mathtt{\scriptscriptstyle I}} + k_z z_{\mathtt{\scriptscriptstyle I}}\right]} \\%
B_{\mathsf{\scriptscriptstyle out}}e^{-\frac{i}{\hbar} \left[\Omega(t_{\mathtt{\scriptscriptstyle II}}-x_\star{}_{\mathtt{\scriptscriptstyle II}}) + k_y y_{\mathtt{\scriptscriptstyle II}}+ k_z z_{\mathtt{\scriptscriptstyle II}} \right]} &= \left(e^{-\frac{\pi\Omega}{a\hbar}}\right) B_{\mathsf{\scriptscriptstyle out}}e^{-\frac{i}{\hbar}\left[
\Omega(t_{\mathtt{\scriptscriptstyle I}}-x_\star{}_{\mathtt{\scriptscriptstyle I}}) + k_y y_{\mathtt{\scriptscriptstyle I}}+ k_z z_{\mathtt{\scriptscriptstyle I}} \right]}
\end{aligned}
\end{align}

Let there be `$n$' pair of free particles (ingoing and outgoing) just inside the horizon in wedge $\mathtt{II}$,each being described by the modes \eqref{totswave}. The total state of this system of particles can be written as a direct product of the single particle states
\begin{align}
\label{scalTotn1}
|\chi_{\scriptscriptstyle B}\rangle &= N_B \displaystyle\sum_{n=0}^\infty |n_\mathtt{\scriptscriptstyle II}^\mathsf{\scriptscriptstyle in}\rangle \otimes |n_\mathtt{\scriptscriptstyle II}^\mathsf{\scriptscriptstyle out}\rangle\nonumber\\
&= N_B \displaystyle\sum_{n=0}^\infty \left(e^{-\frac{n\pi\Omega}{a\hbar}} \right)|n_\mathtt{\scriptscriptstyle I}^\mathsf{\scriptscriptstyle in}\rangle \otimes |n_\mathtt{\scriptscriptstyle I}^\mathsf{\scriptscriptstyle out}\rangle
\end{align}
where the normalization $\langle\chi_{\scriptscriptstyle B}|\chi_{\scriptscriptstyle B}\rangle=1$ determines the constant $N_B$. Here in the case of bosons, the sum over $n$ runs from $0$ to $\infty$. But in case of fermions, which will be considered later, $n$ is limited to $0$ and $1$ by Pauli's exclusion principle. The normalization of $|\chi_{\scriptscriptstyle B}\rangle$ leads to
\begin{align*}
N_B^2\displaystyle\sum_{n,m=0}^\infty &e^{-\frac{(n+m)\pi\Omega}{a\hbar}} \Big(\langle m_\mathtt{\scriptscriptstyle I}^\mathsf{\scriptscriptstyle out}| \otimes \langle m_\mathtt{\scriptscriptstyle I}^\mathsf{\scriptscriptstyle in}|\Big) \Big(| n_\mathtt{\scriptscriptstyle I}^\mathsf{\scriptscriptstyle in}\rangle \otimes |n_\mathtt{\scriptscriptstyle I}^\mathsf{\scriptscriptstyle out}\rangle\Big)=1\nonumber\\
\Rightarrow N_B^2&=\left[\displaystyle\sum_{n=0}^\infty e^{-\frac{2\pi n \Omega}{a\hbar}}\right]^{-1},
\end{align*}
and finally we have
\begin{align}
\label{scalN}
N_B=\left( 1 - e^{-\frac{2\pi\Omega}{a\hbar}} \right)^{\frac{1}{2}}.
\end{align}
The density matrix operator for this system of bosons is defined as
\begin{align}
\label{scalDens}
\hat{\rho}_{\scriptscriptstyle B}=\Big(1-e^{-\frac{2\pi\Omega}{a\hbar}}\Big) \displaystyle\sum_{n,m=0}^\infty e^{-\frac{(n+m)\pi\Omega}{a\hbar}} |n_\mathtt{\scriptscriptstyle I}^\mathsf{\scriptscriptstyle in}\rangle \otimes |n_\mathtt{\scriptscriptstyle I}^\mathsf{\scriptscriptstyle out}\rangle \langle m_\mathtt{\scriptscriptstyle I}^\mathsf{\scriptscriptstyle out}| \otimes\langle m_\mathtt{\scriptscriptstyle I}^\mathsf{\scriptscriptstyle in}|.
\end{align}
Since ingoing waves are trapped within the horizon and only the outgoing particles contribute to spectrum, we trace out the ingoing particles in the density matrix to get the density matrix for outgoing modes,
\begin{align}
\label{scalDensOut}
\hat{\rho}_{\scriptscriptstyle B}^\mathsf{\scriptscriptstyle out}=\Big(1-e^{-\frac{2\pi\Omega}{a\hbar}}\Big)\displaystyle\sum_{n=0}^\infty e^{-\frac{2\pi n\Omega}{a\hbar}} |n_\mathtt{\scriptscriptstyle I}^\mathsf{\scriptscriptstyle out}\rangle \langle n_\mathtt{\scriptscriptstyle I}^\mathsf{\scriptscriptstyle out}|.
\end{align}
The spectrum, given by the average number of outgoing particles is then calculated as
\begin{align}
\label{BoseSpect}
\langle\hat{n}_{\scriptscriptstyle B}\rangle &= \text{Tr}^\mathsf{\scriptscriptstyle out}\left[\hat{n}_{\scriptscriptstyle B}\hat{\rho}_{\scriptscriptstyle B}^\mathsf{\scriptscriptstyle out}\right]=\Big(1-e^{-\frac{2\pi\Omega}{a\hbar}}\Big)\displaystyle\sum_{n=0}^\infty n e^{-\frac{2\pi n\Omega}{a\hbar}}\nonumber\\
&=\frac{1}{e^{\frac{2\pi\Omega}{a \hbar}}-1}\nonumber\\
&=\frac{1}{e^{\frac{2\pi\omega}{\hbar \alpha}}-1}
\end{align}
where in the last step, the red-shift definition of $\Omega$ given in \eqref{redshift} was used. The above spectrum is clearly the Bose-Einstein distribution for a black body at a temperature $T_{\scriptscriptstyle U}$, the Unruh temperature \cite{Unruh:1976db}, given as
\begin{align}
\label{UnruhTemp}
T_{\scriptscriptstyle U}=\frac{\hbar \alpha}{2 \pi}
\end{align}
with $\alpha$ being the local acceleration.

For fermions, the same algorithm is to be adopted starting with the fermionic modes \eqref{DiracFinSol}. The connection between spinorial wave-functions in wedges $\mathtt{II}$ and $\mathtt{I}$ is obtained by using \eqref{1_2relation} in \eqref{DiracFinSol}.
\begin{align}
\label{fermOutIn}
\xi_{\mathsf{\scriptscriptstyle in}}e^{-\frac{i}{\hbar}\left[\Omega(t_{\mathtt{\scriptscriptstyle II}} + x_\star{}_{\mathtt{\scriptscriptstyle II}}) + k_y y_{\mathtt{\scriptscriptstyle II}}+ k_z z_{\mathtt{\scriptscriptstyle II}}\right]} = \xi_{\mathsf{\scriptscriptstyle in}} e^{-\frac{i}{\hbar} \left[\Omega(t_{\mathtt{\scriptscriptstyle I}}+x_\star{}_{\mathtt{\scriptscriptstyle I}}) + k_y y_{\mathtt{\scriptscriptstyle I}}+k_z z_{\mathtt{\scriptscriptstyle I}}\right]} \nonumber\\%
\xi_{\mathsf{\scriptscriptstyle out}}e^{-\frac{i}{\hbar} \left[\Omega(t_{\mathtt{\scriptscriptstyle II}}-x_\star{}_{\mathtt{\scriptscriptstyle II}}) + k_y y_{\mathtt{\scriptscriptstyle II}}+ k_z z_{\mathtt{\scriptscriptstyle II}} \right]} = \left(e^{-\frac{\pi\Omega}{a\hbar}}\right) \xi_{\mathsf{\scriptscriptstyle out}} e^{-\frac{i}{\hbar}\left[ \Omega(t_{\mathtt{\scriptscriptstyle I}}-x_\star{}_{\mathtt{\scriptscriptstyle I}}) + k_y y_{\mathtt{\scriptscriptstyle I}}+ k_z z_{\mathtt{\scriptscriptstyle I}} \right]}
\end{align}
The normalization of the total state for fermions
\begin{align}
\label{fermTotn1}
|\chi_{\scriptscriptstyle F}\rangle &= N_F \displaystyle\sum_{n=0}^1 |n_\mathtt{\scriptscriptstyle II}^\mathsf{\scriptscriptstyle in}\rangle \otimes |n_\mathtt{\scriptscriptstyle II}^\mathsf{\scriptscriptstyle out}\rangle\nonumber\\
&= N_F \displaystyle\sum_{n=0}^1 \left(e^{-\frac{n\pi\Omega}{a\hbar}} \right)|n_\mathtt{\scriptscriptstyle I}^\mathsf{\scriptscriptstyle in}\rangle \otimes |n_\mathtt{\scriptscriptstyle I}^\mathsf{\scriptscriptstyle out}\rangle
\end{align}
is again done through $\langle\chi_{\scriptscriptstyle F}|\chi_{\scriptscriptstyle F}\rangle=1$. Fermions being governed by Pauli's exclusion principle, the sum over number of particles in a given fermionic state always run from $0$ to $1$. The normalization constant $N_F$ turns out to be
\begin{align}
\label{fermN}
N_F=\frac{1}{\sqrt{1+e^{-\frac{2\pi \Omega}{a\hbar}}}}.
\end{align}
The density operator for fermions $\hat{\rho}_{\scriptscriptstyle F}$ is defined as $|\chi_{\scriptscriptstyle F}\rangle  \langle\chi_{\scriptscriptstyle F}|$. Tracing over the ingoing modes and using the resultant outgoing fermionic density operator $\hat{\rho}_{\scriptscriptstyle F}^\mathsf{\scriptscriptstyle out}$, the average number of outgoing particles is calculated to obtain the spectrum.
\begin{align}
\label{Fermspect}
\langle\hat{n}_{\scriptscriptstyle F}\rangle &= \text{Tr}^\mathsf{\scriptscriptstyle out}\left[\hat{n}_{\scriptscriptstyle F}\hat{\rho}_{\scriptscriptstyle F}^\mathsf{\scriptscriptstyle out}\right]=\frac{1}{1+e^{-\frac{2\pi \Omega}{a\hbar}}}\displaystyle\sum_{n=0}^1 n e^{-\frac{2\pi n\Omega}{a\hbar}}\nonumber\\
&=\frac{1}{e^{\frac{2\pi\Omega}{ah}}+1}\nonumber\\
&=\frac{1}{e^{\frac{2\pi\omega}{\hbar \alpha}}+1}
\end{align}
where equation \eqref{redshift} was used in the last step. The spectrum obtained is the Fermi-Dirac distribution, at the Unruh temperature $T_{\scriptscriptstyle U}$ defined in \eqref{UnruhTemp}. This completes the demonstration of the Unruh effect for fermions, through the method of quantum tunneling.


\section{Black holes embedded in flat space}
\label{Cunruh:gems}

An interesting way of studying curved spaces is by describing them as an embedding in a (higher dimensional) flat space. Such embeddings can always be constructed for black holes in 4-D \cite{Goenner:1980aa, Friedman:1965Ja}. Several examples of this mapping have been illustrated in \cite{Rosen:1965Ja}. Once we can construct an appropriate embedding, the black hole observers (ones at fixed $(r,\theta,\phi)$ in Schwarzschild coordinates) become accelerated Rindler observers, thereby leading to a mapping of the Hawking effect into the Unruh effect. The Hawking temperature can then be found corresponding to the Schwarzschild observer at infinity embedded in the flat space. This approach was introduced by \cite{Deser:1997ri, Deser:1998bb, Deser:1998xb}, and the procedure is known as `{\em global embedding Minkowski spacetime}' (GEMS).

For a flat space embedding of the Schwarzschild black hole
\begin{align}
\label{schwarz}
ds^2 = -f(r) dt^2 + \frac{dr^2}{f(r)} + r^2d\theta^2 + r^2\sin^2\theta d\phi^2
\end{align}
where $f(r)=1-\frac{2M}{r}$, we need a 6 dimensional Minkowski spacetime \cite{Fronsdal:1959No}
\begin{align}
\label{GEMSmetric}
ds^2 = -\left(dz^0\right)^2 + \left(dz^1\right)^2 + \left(dz^2\right)^2 + \left(dz^3\right)^2 + \left(dz^4\right)^2 + \left(dz^5\right)^2
\end{align}
The particular hypersurface in this space which embeds the Schwarzschild solution is given through
\begin{align}
\label{GEMStrans}
\begin{aligned}
z^0 =& 4M\sqrt{1-2M/r}~\sinh(t/4M),\\
z^1 =& 4M\sqrt{1-2M/r}~\cosh(t/4M),\\
z^2 =& \int dr\sqrt{(2Mr^2+4M^2r+8M^3)/r^3},\\
z^3 =& r\sin \theta \sin \phi ,\\
z^4 =& r \sin \theta \cos \phi ,\\
z^5 =& r \cos \theta .
\end{aligned}
\end{align}
To check this, one can plug this transformation back in the GEMS flat metric \eqref{GEMSmetric} to get back the 4-dimensional Schwarzschild metric.

Observers in the Schwarzschild spacetime located at some fixed spatial point with constant coordinates $(r, \theta, \phi)$ is represented in the GEMS space by observers having fixed $(z^2, z^3, z^4, z^5)$ coordinates, with the other two coordinates related as 
\begin{align}
\label{GEMShyper}
\left(z^1\right)^2-\left(z^0\right)^2 = 16M^2\left(1-\frac{2M}{r}\right)\quad = \quad \frac{1}{\alpha_6^2}.
\end{align}
Comparing with the hyperbolic trajectories \eqref{hyper} of an accelerated observer in Min-kowski space, we see that \eqref{GEMShyper} represent similar trajectories in the $z^0-z^1$ plane with local acceleration $\alpha_6$. So different observers located at different values of the Schwarzschild radial coordinate $r$ represent different hyperbolae in the $z^0-z^1$ plane. This fact and the form of the first two transformations in \eqref{GEMStrans} indicate that the coordinates $(t,r)$ behave as Rindler coordinates modelling an accelerated observer in the $z^0-z^1$ plane. This is verified by explicitly transforming the metric of the $z^0-z^1$ sector into the corresponding metric in the coordinates $(t, r)$
\begin{align}
\label{GEMSrtrind}
ds^2=-\left(1-\frac{2M}{r}\right)~dt^2+\frac{16 M^4}{r^4\left(1-\frac{2M}{r}\right)}~dr^2.
\end{align}
This metric is in Rindler form as can be seen using our generalised Rindler metric \eqref{genRind} with the choice
\begin{align}
\label{choiceGEMSrind}
\begin{aligned}
F(r) := \frac{1}{\alpha_6} &= 4M~\sqrt{1-\frac{2M}{r}}\\
a &= \frac{1}{4M}.\\
\end{aligned}
\end{align}
Then, the analysis carried out in Section \ref{Cunruh:unruhtunn} can be carried out straight away to lead to a Unruh temperature \eqref{UnruhTemp} observed by the accelerated observer in the $z^0 - z^1$ plane
\begin{align}
\label{gemsUtemp}
T_U = \frac{\hbar\alpha_6}{2\pi},
\end{align} 
with $\alpha_6$ defined in \eqref{choiceGEMSrind}.

The Unruh temperature \eqref{gemsUtemp} varies from one hyperbolae to the other in $z^0-z^1$ plane, depending upon the value of $r$, and each hyperbola represents accelerated observers in $(z^0,z^1)$ coordinates having acceleration $\alpha_6 = \frac{1}{4M}\left(1-\frac{2M}{r}\right)^{-1/2}$. The Hawking observer is the observer who is situated at the Schwarzschild infinity $r=\infty$. So here in this case the Hawking temperature is
\begin{align}
\label{gemsUnruhHawk}
\begin{aligned}
T_H &= \lim_{r \rightarrow \infty} T_U\\
 &= \frac{\hbar}{2\pi}\ \lim_{r \rightarrow \infty}\ \frac{1}{4M}\left(1-\frac{2M}{r}\right)^{-1/2}\\
 &= \frac{\hbar}{8\pi M}.
\end{aligned}
\end{align}
This is precisely the Hawking temperature for a Schwarzschild black hole.


\section{Discussions}

The study in this chapter illustrates the crucial role played by a horizon in the Unruh effect. The classical field theoretic result encapsulated by the Unruh effect was meant to stress how different observers have different concepts of particles corresponding to quantum states. And the derivation discussed here stresses the important role played by the horizon in this process. Use was made only of simple quantum mechanics where the notion of a particle is encoded in the wave packet, and is well accepted. It was seen that the horizon, across which the time and space coordinates interchange their nature, forces a quantum tunnelling of particle-modes in a classically forbidden direction. This is based on quantum mechanical requirements of continuous matching of wave modes across boundaries. Finally, we also saw another approach of relating the Unruh effect for flat-spacetime with the Hawking effect seen in black holes, through GEMS.


%% file: C3_pgt.tex
\chapter{Gauging Poincar\'{e} symmetries}
\label{C:pgt}

\lettrine[lraise=0.0, loversize=0.3, findent=3pt, nindent=0pt]{S}{ymmetries} are ubiquitous in nature and play a guiding role in our understanding of physical interactions \cite{Gross:1996}. Noether's theorem shows how continuous infinitesimal symmetries of the action give rise to conservation laws. Invariance under spatial translations lead to conservation of momentum, invariance under time translations lead to energy conservations and the $U(1)$ phase symmetry is related to electromagnetic charge conservation. In the following parts of this thesis we adopt a framework, the Poincar\'{e} gauge theory of Utiyama-Kibble-Sciama \cite{Utiyama:1956sy, Kibble:1961ba,  1962rdgr.book..415S} and others \cite{Hehl:1976kj, Blagojevic:2002du}, where gravity is constructed from basic symmetry considerations, starting from a {\em flat spacetime}. Eventually this leads to curvature and torsion of spacetime upon localising or `gauging' the symmetries.

While working with field theories it is seen that certain field configurations, linked by (infinitesimal) symmetry transformations, keep physical observables unchanged. Such symmetry transformations usually form a (Lie) group. This is much like rotation of a vector around an axis leaving its length unchanged. Thus solutions of Maxwell's equations remain unchanged under a transformation of the electromagnetic vector potential $A_\mu$ by addition of a total derivative $A_\mu \rightarrow A_\mu + \partial_\mu \Lambda$. There are equivalent classes of solutions consistent with the same initial conditions and given set of sources and currents. To get unique time evolution, one has to resort to additional conditions like $\partial^\mu A_\mu = 0$ (Lorentz gauge) to weed out the extra unphysical degrees of freedom.\footnote{A correct choice of such a condition is necessary so that physical results derived do not depend on this arbitrary gauge choice. Also, not all correct gauge choices are easy to handle at the level of actual computations. But these considerations require detailed attention and since we do not make any gauge choice, lies outside the scope of our present discussion.} It was later understood that this degeneracy can be more than a mathematical artefact arising from an effort in making the theory explicitly invariant under the symmetries. With the advent of Yang-Mills theory \cite{Yang:1954ek} in 1954 it was seen \cite{Utiyama:1956sy} that symmetries of fields under general Lie groups could severely constrain the form of any new field and its interactions. In fact, gauge theory today helps us in gaining a unified quantum description of three of the four fundamental interactions known in nature: electromagnetism, the weak force and the strong force, through what is known as the Standard Model. It has the internal/gauge symmetry group $SU(3) \times SU(2) \times U(1)$.


\section[Gauge symmetries]{Gauge symmetries\footnote{We mainly follow the work of Utiyama \cite{Utiyama:1956sy} here.}}
\label{Cpgt:gaugeth}

We consider a system of fields $\phi^A(x)\ (A=1,2,\ldots,N)$ with a lagrangian density $\mathcal{L}(\phi, \partial_\mu\phi)$ such that the action $$ S = \int d^4x ~\mathcal{L}(\phi, \partial_\mu\phi)$$ remains invariant under infinitesimal linear transformations described by
\begin{align}
\label{globalgauge}
\delta \phi = \varepsilon^a T_a \,\phi.
\end{align}
The field index $A$ has been suppressed for simplicity but, we then remember, that the object $\phi$ is a column matrix with $N$ entities. The index `$a$' denotes $n$ infinitesimal {\em constant} parameters $\varepsilon^a$. Also, we assume that $T_a$ are $n$ matrices satisfying commutation relations
\begin{align}
\label{Tcommutator}
[T_a,T_b] = f_{a\ b}^{\ c} \,T_c,
\end{align}
thus describing the generators of a Lie group with structure coefficients $f_{a\ b}^{\ c}$, antisymmetric in the lower indices $a$ and $b$. These commutators then follow the usual Jacobi identity. Now, the derivatives of fields will also follow the same transformations, viz.
\begin{align}
\delta\partial_\mu\phi = \varepsilon^a T_a \,\partial_\mu\phi
\end{align}
which comes about due to the commutativity of arbitrary variations with coordinate differentials $\delta\circ\partial=\partial\circ\delta$, and the constancy of $\varepsilon^a$. The condition of invariance of the action, upto total derivatives, is found to lead to
\begin{align}
\label{deltaLglobal}
&\quad\,\delta \mathcal{L} = 0\nonumber\\
\Rightarrow &\left(\frac{\partial\mathcal{L}}{\partial\phi}\right)T_a\phi + \left(\frac{\partial\mathcal{L}}{\partial\partial_\mu\phi}\right)T_a\,\partial_\mu\phi = 0.
\end{align}
These are identities, $n$ in number, each corresponding to an arbitrary parameter $\varepsilon^a$. Also, the condition of invariance of the action leads to
\begin{align}
\label{currentNeomForGlobal}
\left[ \frac{\partial\mathcal{L}}{\partial\phi} - \partial_\mu\left(\frac{\partial\mathcal{L}}{\partial\partial_\mu \phi}\right) \right] \delta\phi + \partial_\mu\left[\frac{\partial\mathcal{L}}{\partial\partial_\mu\phi}\delta\phi\right] = 0.
\end{align}
The first term within square brackets is the `Euler derivative' or the equation of motion for field $\phi$. We define the second term to be a current
\begin{align}
\label{globalCurrent}
J^\mu_a = \frac{\partial\mathcal{L}}{\partial\partial_\mu\phi} T_a \,\phi,
\end{align}
which is conserved on-shell, when equations of motion hold, in the sense that $\partial_\mu J^\mu_a = 0$. Of course in all these we have used the form of the field variations \eqref{globalgauge} and the fact that the $\varepsilon$'s are constants.

Now, in relativistic field theories, fields at different spatial points are independent. So it becomes meaningful to relax the global nature of the symmetries considered in \eqref{globalgauge} and make them spacetime dependent by promoting the hitherto constant arbitrary parameters $\varepsilon$ to arbitrary functions of spacetime, i.e. $\varepsilon(x)$. This process of localising symmetries is referred to as `gauging' the symmetry. The parameters $\varepsilon$ are now `gauge parameters'.

But then, the field derivatives are no longer covariant under the transformations \eqref{globalgauge} and there occurs an  additional term containing derivative of the gauge parameter,
\begin{align}
\label{NoncovDeltaPartial}
\delta\,\partial_\mu\phi = \partial_\mu\,\delta\phi = \varepsilon^a T_a \,\phi + \partial_\mu\varepsilon^a \,T_a \,\phi.
\end{align}
This term destroys the invariance of the lagrangian, which now becomes
\begin{align}
\label{gaugeNonCovLagrangianNCurrent}
\delta \mathcal{L} = J^\mu_a \,\partial_\mu \varepsilon^a.
\end{align}
To `cure' this problem, the gauge idea proposes introduction of $4n$ new fields $A^a_\mu$ which will effectively cancel the extra terms and render the lagrangian invariant. But this new lagrangian $\mathcal{L}'(\phi,\partial_\mu\phi,A)$ is now a functional of the original system of fields and the new fields, which are known as gauge potentials or gauge connections. The new fields occur only in conjunction with the derivative terms such that we can define a new `gauge covariant derivative'
\begin{align}
\label{gaugeCovDeriv}
\nabla_\mu \phi = \partial_\mu \phi - T_a\,\phi\,A^a_\mu,
\end{align}
covariant in the sense that
\begin{align}
\label{CovTransfGaugeDeriv}
\delta \,\nabla_\mu\phi = \varepsilon^a T_a \,\nabla_\mu\phi.
\end{align}
This in turn fixes the transformation of the gauge potential as
\begin{align}
\label{GaugePotTransf}
\delta \,A^a_\mu = \varepsilon^b f^{\phantom{b}a}_{b\phantom{a}c}\,A^c_\mu + \partial_\mu\varepsilon^a.
\end{align}
That the new fields $A^a_\mu$ appear in the lagrangian only through covariant derivatives \cite{Utiyama:1956sy} mean that  the lagrangian $\mathcal{L}'(\phi,\partial_\mu\phi,A) \equiv \mathcal{L}'(\phi,\nabla_\mu\phi)$. It allows us to define covariant currents analogous to \eqref{globalCurrent} as
\begin{align}
\label{MatterGaugeCurrent}
J'^\mu_a = \left(\frac{\partial\mathcal{L}}{\partial A^a_\mu}\right) = \left(\frac{\partial\mathcal{L}}{\partial\nabla_\mu \phi}\right)T_a \phi,
\end{align}
which are covariantly conserved on-shell, i.e.
\begin{align}
\label{MatterGaugeCurrConsv}
\nabla_\mu J'^\mu_a = \partial_\mu J'^\mu_a + A^b_\mu\,f^{\phantom{b}c}_{b\phantom{c}a}\,J'^\mu_c = 0.
\end{align}

Covariant derivatives are not commutative in nature, rather their commutator is used to define a `field strength'
\begin{align}
\label{GaugeFieldStrength}
\left[\nabla_\mu, \nabla_\nu\right]\phi = F^a_{\mu\nu} T_a \phi,
\end{align}
where
\begin{align}
\label{GenFieldStrength}
F^a_{\mu\nu} = \partial_\mu A^a_\nu - \partial_\nu A^a_\mu - f^{\phantom{b}a}_{b\phantom{a}c} A^b_\mu A^c_\nu.
\end{align}
The field strength is covariant under the gauge transformations \eqref{globalgauge}.

The gauge principle introduced thus far even allows us to constrain the form of any kinetic term in $A^a_\mu$ \cite{Utiyama:1956sy}. The free lagrangian term $\mathcal{L}_0$  can contain $A^a_\mu$ only in the specific combination occurring through the covariant field strength $F^a_{\mu\nu}$ defined in \eqref{GenFieldStrength}, and must be gauge invariant. One of the simplest such choices is
\begin{align}
\label{YMgaugeLagrangian}
\mathcal{L}_0 = -\frac{1}{4} F^a_{\mu\nu} F_a^{\mu\nu}.
\end{align}
This is nothing but the celebrated Yang-Mills lagrangian. Thus the gauge principle, through fundamental symmetry arguments, allows us knowledge \cite{Utiyama:1956sy} of the five following basic points:
\begin{itemize}
\item Specify the nature of new fields to be introduced -- the gauge potentials $A^a_\mu$.
\item Transformation of these new fields under the symmetry \eqref{GaugePotTransf}.
\item Form of interaction \eqref{gaugeCovDeriv} between the new field $A^a_\mu$ and the original fields $\phi$.
\item The form of the new lagrangian $\mathcal{L}'$ containing the original and new fields $A^a_\mu$ -- $\mathcal{L}'(\phi,\nabla_\mu\phi)$.
\item Lastly, the form of the kinetic terms in the new field $A^a_\mu$ in the free lagrangian \eqref{YMgaugeLagrangian}, and hence the equation of motion of $A^a_\mu$.
\end{itemize}

\paragraph*{Lorentz symmetry as gauge symmetry:}

The general apparatus discussed above applies well to `internal' symmetries like $U(1)$ which doesn't affect the spacetime and acts only on the fields being considered. Utiyama went from here to try \cite{Utiyama:1956sy} to describe gravity starting from the Lorentz symmetry and working through the gauge principle. The Lorentz symmetry is fundamental to relativity and the causal structure, as we know, and thus it was a good start. However, complicacies arise due to the fact that the Lorentz group acts both on the fields and the coordinates. In fact general relativity stands on full diffeomorphism symmetry, allowing arbitrary infinitesimal change of coordinates.

To allow for this change in coordinates and resultant change in the measure of the action integral, Utiyama started with two coordinate systems at the onset. A curvilinear global `$u$' coordinate system (indexed by Greek alphabets $\mu, \nu, \ldots$) with arbitrary metric $g_{\mu\nu}$ and a local Lorentz frame `$x$' (indexed in Latin $i, j, \ldots$) with Minkowskian flat metric $\eta_{\mu\nu}$ where he defined the system of fields $\phi^A(x)$. This local Lorentz coordinate frame is also necessary to accommodate spinor fields that are well defined only w.r.t. Lorentz frames. To interchange objects like vectors and tensors between the two coordinates, he introduced a tetrad system $b^k_{\ \mu}$, and its inverse $b_k^{\ \mu}$. Having all this apparatus, he introduced the Lagrangian $\mathfrak{L}=b\,\mathcal{L}\left(\phi(u),b_k^{\ \mu}(u)\partial_\mu\phi(u)\right)$, where $b=\text{det}(b^k_{\ \mu})=\sqrt{-g}$ which is actually part of the curvilinear measure in the action
\begin{align}
\label{UtiyamaAction}
S = \int d^4u\, \mathfrak{L} = \int d^4u\, b\,\mathcal{L}\left(\phi(u),b_k^{\ \mu}(u)\partial_\mu\phi(u)\right).
\end{align}
The action \eqref{UtiyamaAction} is invariant under Lorentz transformations
\begin{align}
\label{UtiyamaLorentzGlobal}
\begin{aligned}
\delta\phi &= \frac{1}{2}\theta^{kl}T_{kl} \,\phi, \qquad \theta^{kl}=-\theta^{lk}\\
\delta b^k_{\ \mu} &= \theta^k_{\ l} \,b^l_{\ \mu}\\
u^\mu &= \text{unchanged}
\end{aligned}
\end{align}
with $6$ constant antisymmetric parameters $\theta^{kl}$ and Lorentz generators $T_{kl}$ which follow the algebra
\begin{align}
\label{UtiyamaLorAlgebra}
[T_{kl}, T_{mn}] = \frac{1}{2} f^{\phantom{kl}ab}_{kl\phantom{ab}mn} T_{ab}.
\end{align}
From here Utiyama went on to localising the Lorentz symmetry by making the parameters $\theta^{ik}$ functions of the curvilinear coordinates $u$ as $\theta^{ik}(u)$. He introduced $24$ compensating fields $\omega^{ij}_{\phantom{ij}\mu}$, covariant derivatives, field strength which was identified with the Riemann tensor and finally wrote down the free lagrangian of the $\omega^{ij}_{\phantom{ij}\mu}$ fields as a gravitational action.

This derivation was however not satisfactory from many considerations \cite{Kibble:1961ba, Hehl:1976kj, Hehl:2012pi}. He considered the tetrad fields $h^k_{\ \mu}$ to be pre-supplied functions and somewhat arbitrarily identified the connection $\omega^{ij}_{\phantom{ij}\mu}$ with the usual symmetric Christoffel connection of general relativity, ending up with a theory in the Riemann spacetime. However, it may be argued, that the whole point of a gauge theoretic formulation of gravity should have been to produce the gravitational fields (both the metric and the connection) from a gauge principle. But the most striking shortcomming \cite{Hehl:2012pi} was that the current associated with the Lorentz gauge was the antisymmetric angular momentum current alone. However, we know that gravity in general relativity is sourced from the symmetric energy momentum tensor.

These shortcommings were later overcome through the work of Sciama \cite{1962rdgr.book..415S} and Kibble \cite{Kibble:1961ba}. This was through gauging the full Poincar\'{e} or the inhomogeneous Lorentz group with both boosts and translations, instead of just the Lorentz group. We now turn towards this `Poincar\'{e} gauge theory.'


\section{Flat spacetime, Poincar\'{e} symmetries, gauging}
\label{Cpgt:Psymms}

In implementing Poincar\'{e} symmetry \cite{Kibble:1961ba, Blagojevic:2002du}, we have to consider infinitesimal transformations of both the fields $\phi^A(x)$ and the coordinates $x^\mu$. So, at this point, we segregate two different kinds of variations. Form variations `$\delta_0$' which consider change in functional form at the same position, $\delta_0 \phi(x)=\phi'(x)-\phi(x)$, and the total variations `$\delta$' which take into consideration change of functional form {\em and} spacetime point, $\delta\phi(x) = \phi'(x')-\phi(x)$. According to this refinement, the variations considered in \S \ref{Cpgt:gaugeth} dealing with internal symmetries, were form variations or $\delta_0$. They did not affect spacetime.

We start on a 3+1 dimensional flat Minkowskian space with metric $\eta_{\mu\nu}$. The Poincar\'{e} group generators are composed of angular momentum $L_{\mu\nu} = x_\mu\partial_\nu - x_\nu\partial_\mu$, spin angular momentum $\Sigma_{\mu\nu}$ whose representation depends on the species of field being acted upon, and the translations $P_\mu = -\partial_\mu$. The first two are often written together as $M_{\mu\nu} = L_{\mu\nu} + \Sigma_{\mu\nu}$. The generators form the Poincar\'{e} algebra:
\begin{align}
\label{PoincareAlgebra}
\begin{aligned}
\left[M_{\mu\nu},M_{\lambda\rho}\right] &= \eta_{\nu\lambda}M_{\mu\rho} - \eta_{\mu\lambda}M_{\nu\rho} + \eta_{\mu\rho}M_{\nu\lambda} - \eta_{\nu\rho}M_{\mu\lambda}\\
\left[M_{\mu\nu},P_\lambda\right] &= \eta_{\nu\lambda}P_\mu - \eta_{\mu\lambda}P_\nu\\
\left[P_\mu,P_\nu\right] &= 0.
\end{aligned}
\end{align}

Under the Poincar\'{e} group, both coordinates and fields undergo infinitesimal transformations as follows:
\begin{align}
\label{PoincareGlobalTransfms}
\begin{aligned}
\delta_0\phi &= \left( \frac{1}{2}\theta^{\mu\nu}M_{\mu\nu} + \varepsilon^\mu P_\mu \right)\phi \\
\delta \phi &= \delta_0\phi + \delta x^\mu \partial_\mu\phi = \frac{1}{2}\theta^{\mu\nu}\Sigma_{\mu\nu} \phi \\
\delta x^\mu &= \left( \frac{1}{2}\theta^{\lambda\nu}L_{\lambda\nu} + \varepsilon^\nu P_\nu \right)x^\mu = \theta^\mu_{\ \nu}x^\nu + \varepsilon^\mu, \qquad \theta^{\mu\nu}=-\theta^{\nu\mu}.
\end{aligned}
\end{align}
Here $\theta^{\mu\nu}$ and $\varepsilon^\mu$ are the infinitesimal parameters corresponding to Lorentz transformations and translations, with the antisymmetry of $\theta^{\mu\nu}$ linked to the fact that the coordinate transformations in  \eqref{PoincareGlobalTransfms} are isometries of the flat Minkowski metric $\eta_{\mu\nu}$.

The derivatives commute with form variations $\delta_0$, but not with total variations. Thus $\delta_0 \,\partial_\mu\phi = \partial_\mu \,\delta_0\phi$, and remembering that $\delta\phi = \delta_0\phi + \delta x^\mu \,\partial_\mu\phi$, we have
\begin{align}
\label{TderivPartialCommutator}
\delta \,\partial_\mu\phi &= \delta_0 \,\partial_\mu\phi + \delta x^\lambda\,\partial_\lambda\partial_\mu\phi \nonumber\\
&= \partial_\mu \,\delta_0 \phi + \partial_\mu(\delta x^\lambda \,\partial_\lambda\phi) - \partial_\mu\delta x^\lambda \,\partial_\lambda\phi \nonumber\\
&= \partial_\mu \,\delta\phi - \partial_\mu\delta x^\lambda\,\partial_\lambda\phi.
\end{align}
In the case of the Poincar\'{e} transformations \eqref{PoincareGlobalTransfms} being considered here, this boils down to
\begin{align}
\label{PoincareGlobalDeriv}
\delta \,\partial_\mu\phi = \frac{1}{2}\theta^{\lambda\rho}\Sigma_{\lambda\rho}\,\partial_\mu\phi - \theta^\lambda_{\ \mu}\,\partial_\lambda\phi.
\end{align}

Turning towards the invariance of the action $S=\int \,d^4x \,\mathcal{L}$, we see that since the coordinates $x^\mu$ also changes under transformations, we have to take into consideration the change in the measure under transformations $x^\mu \rightarrow x'^\mu$. This is taken care of by the Jacobian $\frac{\partial(x')}{\partial(x)} \simeq 1+\partial_\mu\delta x^\mu$. Thus the condition of invariance of the action implies that the lagrangian must change as a density
\begin{align}
\label{LagrangianDensityVariation}
\triangle \mathcal{L} = \delta_0 \mathcal{L} + \delta x^\mu \,\partial_\mu\mathcal{L} + \partial_\mu\delta x^\mu \,\mathcal{L},
\end{align}
where the first two terms come from total variation and the last term from the Jacobian. However, in the particular case of rigid Poincar\'{e} symmetries \eqref{PoincareGlobalTransfms}, $\partial_\mu\delta x^\mu =0$ and the change in Jacobian doesn't really have an effect; the action remains invariant simply if the lagrangian itself is (ignoring any surface term). This condition can be finally written as (compare: \ref{currentNeomForGlobal})
\begin{align}
\label{PoincareGlobalCanonicalConsv}
\triangle\mathcal{L} = \left[\frac{\partial\mathcal{L}}{\partial\phi}-\partial_\mu\left(\frac{\partial\mathcal{L}}{\partial\partial_\mu\phi}\right)\right]\,\delta \phi + \partial_\mu J^\mu = 0,
\end{align}
where the first term within square brackets is the Euler derivative of the field $\phi$ which vanishes on-shell. This gives the canonical conservation law $\partial_\mu J^\mu = 0$, and the current may be expressed as
\begin{align}
\label{PoincareCanCurrent}
J^\mu = \frac{1}{2} \theta^{\nu\lambda}M^\mu_{\ \nu\lambda} - \varepsilon^\nu \mathcal{T}^\mu_{\ \nu}.
\end{align}
Here $M^\mu_{\ \nu\lambda}$ is the canonical angular-momentum tensor, $\mathcal{T}^\mu_{\ \nu}$ is the energy-momentum tensor and $\Sigma^\mu_{\ \nu\lambda}$ specifically is the spin tensor. They are defined as follows:
\begin{align}
\label{PoincareCanTensorsDef}
\begin{aligned}
\mathcal{T}^\mu_{\ \nu} &= \frac{\partial\mathcal{L}}{\partial\partial_\mu\phi}\partial_\nu\phi - \delta^\mu_\nu\,\mathcal{L} \\
M^\mu_{\ \nu\lambda} &= (x_\nu\mathcal{T}^\mu_{\ \lambda} - x_\lambda\mathcal{T}^\mu_{\ \nu}) - \Sigma^\mu_{\ \nu\lambda} \\
\Sigma^\mu_{\ \nu\lambda} &= \frac{\partial\mathcal{L}}{\partial\partial_\mu\phi}\,\Sigma_{\nu\lambda}\phi
\end{aligned}
\end{align}
Comparing the expression of $\mathcal{T}^\mu_{\ \nu}$ with the typical current \eqref{globalCurrent}, we see that apart from terms proportional to derivatives of the lagrangian, we also have a term linear in the lagrangian. This has come from considering total variations as $\delta x^\mu \partial_\mu\mathcal{L}$. But the important difference that has arisen out of considering the Poincar\'{e} group with the translations $\varepsilon^\mu$ is the appearance of the energy-momentum tensor $\mathcal{T}^\mu_{\ \nu}$ itself, which is the usual source of curvature in general relativity. This is in addition to the spin and angular momentum tensors coupled to the Lorentz parameter $\theta^{\mu\nu}$ and was absent in Utiyama's Lorentz gauge structure \cite{Utiyama:1956sy} as a `naturally occurring' canonical current. Also, we may remind the reader that we are still now on a flat Minkowskian space in a single coordinate system with only infinitesimal changes induced by the symmetry group. And with the symmetries still rigid, no new fields have been erected yet.

We can now go on with localisation of the Poincar\'{e} symmetries \eqref{PoincareGlobalTransfms} by making the parameters $\theta^{\mu\nu}$ and $\varepsilon^\mu$ functions of spacetime. However we wish to be able to separate coordinate and field transformations independently. To do this \cite{Kibble:1961ba}, we set $\xi^\mu$ in $\delta x'^\mu = \xi^\mu = \theta^\mu_{\ \nu} + \varepsilon^\mu$ as the independent parameter instead of $\varepsilon^\mu$. This leaves us with the freedom to consider generalised transformations with $\xi^\mu = 0$ but still keeping non-zero $\theta^{\mu\nu}$. And having done this, we index the field transformations through $\theta^{ij}$ in Latin and the coordinate transformations  through $\xi^\mu$ in Greek. The localised Poincar\'{e} transformations now reads:
\begin{align}
\label{PoincareLocalTransfms}
\begin{aligned}
\delta_0\phi &= \frac{1}{2}\theta^{ij}(x)\,\Sigma_{ij}\phi - \xi^\mu(x) \,\partial_\mu\phi \\
\delta \phi &= \frac{1}{2}\theta^{ij}(x) \,\Sigma_{ij} \phi \\
\delta x^\mu &= \xi^\mu(x).
\end{aligned}
\end{align}
This segregation and separate indices will take on the meaning of local coordinate frames ($x^i$) that support matter fields like spinors, and the global possibly curved coordinates ($x^\mu$). This will be further discussed in the next section. Here, we continue with the gauge procedure.

The localisation of the gauge parameters expectedly render the derivatives non-covariant
\begin{align}
\label{PoincareLocalNonCovDeriv}
\delta \,\partial_\mu\phi = \frac{1}{2}\theta^{ij}\Sigma_{ij} \,\partial_\mu\phi + \frac{1}{2}\partial_\mu\theta^{ij} \Sigma_{ij}\phi - \partial_\mu\xi^\nu\,\partial_\nu\phi.
\end{align}
The lagrangian no longer fits the condition \eqref{LagrangianDensityVariation} for invariance of the actions and similar to \eqref{gaugeNonCovLagrangianNCurrent} gives residual terms related to the canonical currents
\begin{align}
\label{PoincareGaugeNonCovLagNCurrent}
\triangle \mathcal{L} &= -\frac{1}{2}\partial_\mu\theta^{ij}\,\Sigma^\mu_{\ ij} - \partial_\mu\xi^\rho \,\mathcal{T}^\mu_{\ \rho}.
\end{align}
We would however want to make \eqref{PoincareLocalNonCovDeriv} covariant, so that the invariance of the lagrangian can be maintained. So, we adopt a two stage procedure to remove the two non-vanishing terms in \eqref{PoincareGaugeNonCovLagNCurrent} following the spirit og gauge theory. Let us first note that the terms with $\partial_\mu \theta^{ij}$, $24$ in number, are much like the generic terms in \eqref{NoncovDeltaPartial}; in \eqref{PoincareLocalNonCovDeriv} they are proportional to $\phi$. This is taken care of by introducing $24$ new fields $\omega^{ij}_{\phantom{ij}\mu}= -\omega^{ji}_{\phantom{ij}\mu}$ and a covariant derivative in $x^\mu$
\begin{align}
\label{PoincareCovD1}
\nabla_\mu \phi = \partial_\mu \phi + \frac{1}{2}\omega^{ij}_{\phantom{ij}\mu}\Sigma_{ij}\phi.
\end{align}
We require the new derivative to transform as
\begin{align}
\label{PoincareCovD1transfms}
\delta \,\nabla_\mu \phi = \frac{1}{2}\theta^{ij}\Sigma_{ij} \,\nabla_\mu\phi - \partial_\mu\xi^\nu\,\nabla_\nu\phi,
\end{align}
which fixes the transformation of the connections to (in $\delta_0$ form),
\begin{align}
\label{PoincareAconnTransfms}
\delta_0 \omega^{ij}_{\phantom{ij}\mu} = \theta^i_{\ k} \omega^{kj}_{\phantom{kj}\mu} + \theta^j_{\ k} \omega^{ik}_{\phantom{ik}\mu}  - \partial_\mu\theta^{ij} - \partial_\mu\xi^\lambda \,\omega^{ij}_{\phantom{ij}\lambda} - \xi^\lambda\,\partial_\lambda \omega^{ij}_{\phantom{ij}\mu}.
\end{align}
Looking at \eqref{PoincareCovD1transfms} we see that we are still left with the $\partial_\mu\xi^\nu$ term. This term contains the derivative $\nabla_\nu\phi$ itself, instead of just the field $\phi$. We recognise this feature to have come from the last term in \eqref{PoincareLocalNonCovDeriv}, $\partial_\mu\xi^\nu\,\partial_\nu\phi$, which still is of the form $\partial\xi \,P\, \phi$, where the Poincar\'{e} group translation generator is a derivative operator $P_\nu = \partial_\nu$. This is unlike the case of usual gauge theory with internal symmetry groups, and is another peculiarity of taking translations along with Lorentz transformations. To get rid of this term from \eqref{PoincareCovD1transfms}, we have to add a term in $\nabla_\mu \phi$ itself, which boils down to multiplying by a new set of fields $b_k^{\ \mu}$, $16$ in number,
\begin{align}
\label{PoincareCovD2}
\nabla_k\phi = \delta^\mu_k \nabla_\mu \phi - \bar{\omega}^{\ \mu}_{k}\nabla_\mu\phi = b_k^{\ \mu}\nabla_\mu\phi.
\end{align}
Here, the $\bar{\omega}^{\ \mu}_{k}$ fields were new fields which have been absorbed and redefined as $b_k^{\ \mu}$. The derivatives $\nabla_k$ are now fully covariant under Poincar\'{e} transformations
\begin{align}
\label{PoincareCovD2transfms}
\delta_0 \,\nabla_k \phi = \left( \frac{1}{2}\theta^{ij}M_{ij} + \varepsilon^\mu P_\mu \right)\nabla_k \phi + \theta_k^{\ i} \nabla_i\phi,
\end{align}
and this fixes the transformations of the new fields $b_k^{\ \mu}$ as
\begingroup						 
\setlength\belowdisplayskip{0pt} 
\begin{align}
\label{Poincarebframetransfms}
\delta_0 b_k^{\ \mu} = \theta_k^{\ j}b_j^{\ \mu} + \partial_\lambda\xi^\mu b_k^{\ \lambda} - \xi^\lambda\,\partial_\lambda b_k^{\ \mu}
\end{align}
\endgroup

Having achieved the covariance of derivatives, we now require an invariant lagrangian {\em density} so that the action is invariant \eqref{LagrangianDensityVariation}. For this, we multiply $\mathcal{L}$ by some suitable function of the new fields, say $\Lambda$, such that $\Lambda$ itself is an invariant density $(\delta \Lambda + \partial_\mu\xi^\mu\,\Lambda=0)$.  The most suitable choice is \cite{Kibble:1961ba, Blagojevic:2002du} $\Lambda = b = \text{det}(b^i_{\ \mu})$ where $b^i_{\ \mu}$ is the inverse of the field $b^i_{\ \mu}$ defined by $b^i_{\ \mu}b_i^{\ \nu}=\delta^\mu_\nu$ and $b^i_{\ \mu}b_j^{\ \mu}=\delta^i_j$. So the general form of the Poincar\'{e} gauge theory invariant lagrangian is
\begingroup						 
\setlength\abovedisplayskip{6pt} 
\setlength\belowdisplayskip{1pt} 
\begin{align}
\label{PoincareGeneralInvLag}
\mathfrak{L} = b\,\mathcal{L}(\phi,\nabla_k\phi).
\end{align}
\endgroup

As we had seen in \S \ref{Cpgt:gaugeth}, gauge invariant kinetic terms of gauge fields in the lagrangian are written in terms of field strengths, defined through the commutator of the covariant derivatives. In the case of Poincar\'{e} gauge theory, we have two types of derivatives $\nabla_\mu$ and $\nabla_k$. They give rise to two distinct fields strengths.
\begin{align}
\label{PoincareCovDGcommu}
\left[\nabla_\mu, \nabla_\nu\right] \phi &= \frac{1}{2}R^{ij}_{\ \ \mu\nu}\Sigma_{ij}\phi\\
\left[\nabla_k, \nabla_l\right] \phi &= \frac{1}{2}R^{ij}_{\ \ kl}\Sigma_{ij}\phi - T^s_{\ kl}\nabla_s\phi.
\end{align}
The relevant quantities are defined below:
\begin{align}
\label{PoincareFieldStrengths}
\begin{aligned}
R^{ij}_{\ \ \mu\nu} &= \partial_\mu \omega^{ij}_{\ \ \nu} - \partial_\nu \omega^{ij}_{\ \ \mu} + \omega^i_{\ l\mu}\omega^{lj}_{\ \ \nu} - \omega^i_{\ l\nu}\omega^{lj}_{\ \ \mu} \\
T^s_{\ kl} &= b_k^{\ \mu} b_l^{\ \nu} T^s_{\ \mu\nu} = b_k^{\ \mu} b_l^{\ \nu} \left(\nabla_\mu b^s_{\ \nu} - \nabla_\nu b^s_{\ \mu}\right).
\end{aligned}
\end{align}
Jacobi identities for the commutators would imply the following Bianchi identities \cite{Blagojevic:2002du}
\begin{align}
\label{PoincareFieldStrengthBianchi}
\begin{aligned}
\epsilon^{\rho\mu\lambda\nu}\nabla_\mu T^s_{\ \lambda\nu} &= \epsilon^{\rho\mu\lambda\nu} R^s_{\ k\lambda\rho}b^k_{\ \mu} \\
\epsilon^{\rho\mu\lambda\nu}\nabla_\lambda R^{ij}_{\ \ \mu\nu} &= 0.
\end{aligned}
\end{align}
Using these field strengths, various actions have been investigated. We will review some of them later in this chapter. But before that, we would like to discuss the geometric picture that may be attached to this theory.


\section{Geometrical interpretation}
\label{Cpgt:geometrical}

A general curved manifold obeying metricity, i.e. vanishing covariant derivative of the metric $(D_\mu(\Gamma)g_{\nu\lambda}=0)$, and equipped with a connection $\tilde{\Gamma}$ that is linear $(\delta A^\mu = -\tilde{\Gamma}^\mu_{\lambda\rho}A^\lambda \text{d}x^\rho)$ is the Riemann-Cartan spacetime $U_4$ (in $4$-D). It has both curvature, characterised by the Riemann tensor 
\begin{align}
\label{RiemannGenRiemCartan}
R^\mu_{\ \nu\lambda\rho} = \partial_\lambda \tilde{\Gamma}^\mu_{\nu\rho} + \tilde{\Gamma}^\mu_{\sigma\lambda}\tilde{\Gamma}^\sigma_{\nu\rho} - \partial_\rho \tilde{\Gamma}^\mu_{\nu\lambda} - \tilde{\Gamma}^\mu_{\sigma\rho}\tilde{\Gamma}^\sigma_{\nu\lambda},
\end{align}
and torsion, characterised by the antisymmetric part of the linear connection 
\begin{align}
\label{TorsionGenRiemCartan}
T^\mu_{\ \lambda\rho} = \tilde{\Gamma}^\mu_{\rho\lambda} - \tilde{\Gamma}^\mu_{\lambda\rho}.
\end{align}
General relativity is formulated on a Riemannian spacetime $V_4$ where torsion is set to zero, and the connection is the symmetric Christoffel connection $\Gamma$, determined uniquely from the metric through
\begin{align}
\label{ChristoffelRiem}
\Gamma^\mu_{\nu\rho} = \frac{1}{2}g^{\mu\lambda}\left(\partial_\rho g_{\lambda\nu} + \partial_\nu g_{\lambda\rho} - \partial_\lambda g_{\nu\rho}\right).
\end{align}
Another parallel formulation exists \cite{Blagojevic:2002du}, using the Weitzenb\"{o}ck connection in what is known as teleparallel gravity, where the curvature is set to zero ($T_4$) and gravity is sourced by torsion. Finally, when both the torsion and curvature (in whichever order) is set to zero, we get back to our flat Minskowsi spacetime $M_4$.

In $U_4$, we can always set up a local Lorentz (Minskowskian) $L$-frame through vectors $e_i$ at any spacetime point $x^\mu$ \cite{Hehl:1976kj}. Then a quantity (say a vector $A^\mu$) in the global coordinate $C$-frame $e_\mu \equiv \partial_\mu$ is represented in the $L$-frame through $$A^i = e^i_{\ \mu}A^\mu,$$ where $e^i_{\ \mu}$ are frame fields\footnote{These are known as triads in 3D, tetrads in 4D and vielbeins in general $n$ dimensions.} relating the two coordinate systems.

We now turn to the fields $b^i_{\ \mu}$. They relate the invariant $k$-covariant derivative to the $\mu$-covariant derivative through $$\nabla_k = b_k^{\ \mu}\nabla_\mu.$$ What about general gauge fields, say $A_\mu$ and $A_i$, if we consider them as fields $\phi$ transforming as \eqref{PoincareLocalTransfms}? Let us check. We may recall here our different indices convention adopted in \S \ref{Cpgt:Psymms}. Quantities with Latin indices transform through both Poincar\'{e} generators of Lorentz rotation $\theta^{ij}$ and translations  $\xi^\rho$; Greek indices transform like a density with a $\partial_\mu \xi$ term; and quantities with mixed indices transform in both ways.
\begin{align}
\label{btimesAiTransfms}
\delta_0 (b^i_{\ \mu}A_i) &= (\delta_0 b^i_{\ \mu}) A_i + b^i_{\ \mu}(\delta_0 A_i) \nonumber\\
&= \left(\theta^i_{\ k}b^k_{\ \mu} - \xi^\rho\,\partial_\rho b^i_{\ \mu} - \partial_\mu\xi^\rho \,b^i_{\ \rho} \right) A_i + b^i_{\ \mu}\left(\theta_i^{\ k} A_k - \xi^\rho\,\partial_\rho A_i \right) \nonumber\\
&= - \xi^\rho\,\partial_\rho A_\mu - \partial_\mu\xi^\rho \,A_\rho,
\end{align}
which is nothing but the transformation for the field $A_\mu$. Thus the fields $b^i_{\ \mu}$ and its inverse $b_i^{\ \mu}$ may be regarded as a tetrad system $e^i_{\ \mu}$. They can be used to define a metric $g_{\mu\nu}$ through the relations
\begin{align}
\label{PoincareTransMetric}
\begin{aligned}
g_{\mu\nu} &= b^i_{\ \mu}b_{i\nu} = \eta_{ij}b^i_{\ \mu}b^j_{\ \nu} \\
\eta_{ij} &= b_i^{\ \mu}b_{j\mu} = g_{\mu\nu}b_i^{\ \mu}b_j^{\ \nu}.
\end{aligned}
\end{align}
Now, the notion of parallel transport in the $L$-frame works through the spin connection $\omega^{ij}_{\ \ \mu}$ and that in the $C$-frame uses the manifold connection $\Gamma^\mu_{\nu\rho}$. However the two must match in the sense \cite{Blagojevic:2002du} $$A^i + \delta A^i = e^i_{\ \mu}\,(A^\mu + \delta A^\mu).$$  This connects $\omega$ and $\Gamma$ through the {\em `vielbein postulate'}:
\begin{align}
\label{cpgtVielbeinPostulate}
D_\mu(\omega+\Gamma)e^i_{\ \nu} = \partial_\mu e^i_{\ \nu} + \omega^i_{s\mu}e^s_{\ \nu} - \Gamma^\lambda_{\nu\mu}e^i_{\ \lambda} = 0.
\end{align}
Solving this for $\Gamma$ in terms of the connection $\omega$ and using them in the geometric definitions of the Riemann \eqref{RiemannGenRiemCartan} and torsion \eqref{TorsionGenRiemCartan}, we get \cite{Blagojevic:2002du}
\begin{align}
\label{GeometricTensorsFieldStrengths}
\begin{aligned}
T^\mu_{\ \nu\lambda}(\Gamma) &= e_i^{\ \mu}T^i_{\ \nu\lambda}(\omega) \\
R^\mu_{\ \nu\lambda\rho}(\Gamma) &= e_i^{\ \mu}e_{j\nu}R^{ij}_{\ \ \lambda\rho}
\end{aligned}
\end{align}
Thus we see that the field strengths $T^i_{\ \nu\lambda}$ and $R^{ij}_{\ \ \lambda\rho}$ defined earlier in \S \ref{Cpgt:Psymms} are counterparts of the torsion and the Riemann respectively. Also, using \eqref{cpgtVielbeinPostulate} and \eqref{PoincareTransMetric} we can recover the {\em `metricity condition'}:
\begin{align}
\label{metricity}
D_\mu(\Gamma)g_{\nu\lambda} = D_\mu(\omega+\Gamma)g_{\nu\lambda} = D_\mu(\omega+\Gamma)\eta_{ij}b^i_{\ \nu}b^j_{\ \lambda} = 0.
\end{align}

Uptil now, we showed the correspondence of the Poincar\'{e} gauge structure and the geometrical manifold picture. We saw that the Poincar\'{e} gauge theory lives on a Riemann-Cartan manifold which has both curvature and torsion. Our work in this thesis will focus largely on the symmetry aspects of this gauge theoretic formulation. So in addition to the gauge picture of the symmetries of the triad and connection fields, which is an attempt to create general coordinate transformations out of a gauge inspired idea, we now try to find \cite{Banerjee:2009vf} the appropriate symmetries, starting with the diffeomorphism symmetry (diff) inherent in gravitational structures. This throws a new light on the symmetries of $b^i_{\ \mu}$ and $\omega^{ij}_{\ \ \mu}$ fields; we see that their transformations can be deduced starting from the usual diff
\begin{align}
\label{metricvar}
\delta g_{\mu\nu} =  -\partial_\mu\xi^\rho g_{\rho\nu}- \partial_\nu\xi^\rho g_{\mu\rho} - \xi^\rho\partial_\rho g_{\mu\nu}.
\end{align}
Using \eqref{PoincareTransMetric}, the lhs of \eqref{metricvar} can be expressed in terms of the variation $\delta b^i_{\ \mu}$ as 
\begin{align}
\label{metricvarlhs}
\delta g_{\mu\nu} = \delta\eta_{ij}b^i_{\ \mu}b^j_{\ \nu}+\eta_{ij}\,\delta b^i_{\ \mu}\,b^j_{\ \nu} + \eta_{ij}b^i_{\ \mu}\,\delta b^j_{\ \nu},
\end{align}
where $\delta\eta_{ij}$ is the variation of $\eta_{ij}$ under local Lorentz transformations (LLT), given by
\begin{align}
\label{metricidentity}
\delta\eta_{ij} = \theta_i^{\ k}\eta_{kj} + \theta_j^{\ k}\eta_{ik} = 0.
\end{align}
Equating \eqref{metricvar} with \eqref{metricvarlhs} we get:
\begin{align}
\label{metricvarstep1}
\eta_{ij} \,\delta b^i_{\ \mu}\,b^j_{\ \nu} + \eta_{ij}b^i_{\ \mu}\,\delta b^j_{\ \nu} &+ \theta_i^{\ k}\eta_{kj}b^i_{\ \mu} b^j_{\ \nu} + \theta_j^{\ k}\eta_{ik}b^i_{\ \mu} b^j_{\ \nu} \\ \nonumber
&= -\partial_\nu\xi^\rho \eta_{ij}b^i_{\ \mu} b^j_{\ \rho}-\partial_\mu\xi^\rho \eta_{ij}b^i_{\ \rho} b^j_{\ \nu} - \xi^\rho\partial_\rho \left(b^i_{\ \mu}b^j_{\ \nu}\right) \eta_{ij}.
\end{align}
Simplification yields
\begin{align}
\label{metricvarstep2}
b_{i\nu}[\delta b^i_{\ \mu} + \partial_\mu\xi^\rho b^i_{\ \rho} &+ \xi^\rho\partial_\rho b^i_{\ \mu} + \theta_k^{\ i}b^k_{\ \mu}] \nonumber\\
&+ b_{j\mu}[\delta b^j_{\ \nu} + \partial_\nu\xi^\rho b^j_{\ \rho} + \xi^\rho\partial_\rho b^j_{\ \nu} + \theta_k^{\ j}b^k_{\ \nu}]= 0,
\end{align}
and the last equation leads to
\begin{align}
\label{metricvarstep3}
\delta b^i_{\ \mu} = \theta^i_{\ k} b^k_{\ \mu} - \partial_\mu\xi^\rho b^i_{\ \rho} - \xi^\rho\partial_\rho b^i_{\ \mu},
\end{align}
which is identical with \eqref{Poincarebframetransfms}. Note that the variation $\delta\eta_{ij}$ of the constant tensor $\eta_{ij}$ under local Lorentz transformations (LLT), given by \eqref{metricidentity}, reproduces the expected vanishing result. Nevertheless it has to be included and split in \eqref{metricvarstep2} in order to get agreement with the corresponding Poincar\'{e} gauge transformation. Otherwise, the $\theta$-contribution in \eqref{metricvarstep3} will be lacking.

In order to reproduce the connection transformations \eqref{PoincareAconnTransfms}, consider the transformation of the affine connection $\Gamma^\mu_{\nu\lambda}$. Using the vielbein postulate \eqref{cpgtVielbeinPostulate} we write
\begin{align}
\label{affcon}
\Gamma^\mu_{\nu\lambda} = b_i^{\ \mu}\partial_\lambda b^i_{\ \nu} + \omega^i_{\ j\lambda}b_i^{\ \mu} b^j_{\ \nu}
\end{align}
It transforms under diff as 
\begin{align}
\label{conalj1}
\delta\Gamma^\mu_{\nu\lambda} = -\partial_\nu\xi^\rho~\Gamma^\mu_{\rho\lambda} - \partial_\lambda\xi^\rho~\Gamma^\mu_{\nu\rho} + \partial_\rho\xi^\mu~\Gamma^\rho_{\nu\lambda} -\partial_\nu\partial_\lambda\xi^\mu - \xi^\rho~\partial_\rho\Gamma^\mu_{\nu\lambda}
\end{align}
On the other hand, from \eqref{affcon}, we obtain
\begin{align}
\label{varGamma19}
\delta \Gamma^\mu_{\nu\lambda} = \delta b_i^{\ \mu}\,\partial_\lambda b^i_{\ \nu} + b_i^{\ \mu}\,\partial_\lambda \delta b^i_{\ \nu} + \delta\omega^i_{\ j\lambda}\,b_i^{\ \mu} b^j_{\ \nu} + \omega^i_{\ j\lambda}\,\delta b_i^{\ \mu}\,b^j_{\ \nu} + \omega^i_{\ j\lambda}b_i^{\ \mu}\,\delta b^j_{\ \nu}
\end{align}
Equating \eqref{conalj1} with \eqref{varGamma19} and using \eqref{metricvarstep3}, one finds after a long algebra,
\begin{align}
\label{conalj2}
\delta \omega^{ij}_{\ \ \mu} = \theta^i_{\ k} \omega^{kj}_{\ \ \mu} + \theta^j_{\ k} \omega^{ik}_{\ \ \mu} - \partial_\mu\theta^{ij} - \partial_\mu\xi^\rho \omega^{ij}_{\ \ \rho} - \xi^{\rho}\partial_{\rho}\omega^{ij}_{\ \ \mu}
\end{align}
which is equivalent to \eqref{PoincareAconnTransfms}. We thus find that if we identify the triad $b^i_{\ \mu}$ and the spin connection $\omega^{ij}_{\ \ \lambda}$ with the `gauge potentials' $b^i_{\ \mu}$ and  $\omega^{ij}_{\ \ \lambda}$, then spacetime symmetry transformations (namely, the LLT and diff) generate the same transformations as the Poincar\'{e} gauge transformations. 

The above correspondence may further be pursued at the level of the field strengths. From the point of view of Poincar\'{e} gauge theory, the transformations of the field strengths $R^{ij}_{\ \ \mu\nu}$ and $T^i_{\ \mu\nu}$ can easily be obtained by direct substitution of \eqref{Poincarebframetransfms} and \eqref{PoincareAconnTransfms} in the definition of field strengths \eqref{PoincareFieldStrengths},
\begin{align}
\label{lorentztrans}
\begin{aligned}
\delta R^{ij}_{\ \ \mu\nu} &= \theta^i_{\ k}\,R^{kj}_{\ \ \mu\nu} +\theta ^j_{\ k}\,R^{ik}_{\ \ \mu\nu} -\partial_{\mu}\xi^{\rho}\,R^{ij}_{\ \ \rho\nu} - \partial_{\nu}\xi^{\rho}\,R^{ij}_{\ \ \mu\rho} -\xi^{\rho}\,\partial_{\rho}R^{ij}_{\ \ \mu\nu} \\
\delta T^{i}_{\ \mu\nu} &= \theta ^i_{\ k}\,T^{k}_{\ \ \mu\nu} -\partial_{\mu}\xi^{\rho}\,T^{i}_{\ \rho\nu} - \partial_{\nu}\xi^{\rho}\,T^{i}_{\ \mu\rho} - \xi^{\rho}\,\partial_{\rho}T^{i}_{\ \mu\nu}
\end{aligned}
\end{align}
As noted earlier in discussions on the transformations (see above \eqref{btimesAiTransfms}), the Latin indices transform as under LLT with parameters $\theta^{ij}$ and the Greek indices transform as under diff with parameters $\xi^\mu$ in the above transformations. We thus get back the expected transformations under LLT and diff. This agreement lends further support to viewing Poincar\'{e} gauge theory as a theory of gravity.

At the end of this discussion on geometric correspondence, we would like to list some dual fields that can be defined in 3 dimensions to help us in simplifying the calculations
\begin{align}
\label{Poincare3Ddualities}
\begin{aligned}
\theta_i &= -\frac{1}{2}\epsilon_{ijk}\,\theta^{jk}\\
\omega^i_{\ \mu} &= -\frac{1}{2}\epsilon^i_{\ jk}\,\omega^{jk}_{\ \ \mu}\\
R_{i\mu\nu} &= -\frac{1}{2}\epsilon_{ijk}\,R^{jk}_{\ \ \mu\nu}.
\end{aligned}
\end{align}
Using these in the expressions \eqref{PoincareFieldStrengths} for the field strengths, we get the following dual strengths in 3D
\begin{align}
\label{Poincare3DdualFldStrengths}
\begin{aligned}
T_{i\nu\rho} &= \partial_\nu b_{i\rho} + \epsilon_{ijk} \omega^j_{\ \nu} b^k_{\ \rho} - \partial_\rho b_{i\nu} - \epsilon_{ijk} \omega^j_{\ \rho} b^k_{\ \nu }\\
R_{i\nu\rho} &= \partial_\nu \omega_{i\rho}-\partial_\rho \omega_{i\nu} + \epsilon_{ijk} \omega^j_{\ \nu} \omega^k_{\ \rho},
\end{aligned}
\end{align}
and the basic field transformations \eqref{Poincarebframetransfms} and \eqref{PoincareAconnTransfms} turn out to be
\begin{align}
\label{Poincare3Dfieldtrans}
\begin{aligned}
\delta b^i_{\ \mu} &= -\epsilon^i_{\ jk}b^j_{\ \mu}\theta^k - \partial_\mu\xi^\rho b^i_{\ \rho} - \xi^{\rho}\partial_{\rho} b^i_{\ \mu}\\
\delta \omega^{i}_{\ \mu} &= -\left(\partial_\mu \theta^i + \epsilon^i_{\ jk} \omega^j_{\ \mu} \theta^k  \right) - \partial_\mu\xi^\rho \omega^{i}_{\ \rho} - \xi^{\rho}\partial_{\rho}\omega^{i}_{\ \mu}.
\end{aligned}
\end{align}


\section{Poincar\'{e} gauge gravity}
\label{Cpgt:3Dactions}

We have given a construction of the Poincar\'{e} gauge theory (PGT) following the original attempts by Utiyama \cite{Utiyama:1956sy} (with the Lorentz group) and Kibble \cite{Kibble:1961ba}. However, with the success of gauge theory in describing all the other fundamental interactions except gravity, there has been a prolific activity in the direction of obtaining gravity also as a gauge theory. We list some of them here to give an overview.

The basic setup was revisited or investigated further by Hehl et al. \cite{Hehl:1976kj}, Trautman \cite{trautman1973structure}, and several others \cite{Cho:1976fr, Ne'eman:1978uk, Tseytlin:1981nu, Hennig:1981id, Szczyrba:1985pd}. The methodology was both through taking matter field lagrangians and constructing gauge invariance, as we have done, and through using standard geometric descriptions of gauging Lie groups in a fibre bundle or jet structure \cite{Sardanashvily:2002mi}. Gronwald discusses \cite{Gronwald:1998ag} the unique challenges posed by the translations within the Poincar\'{e} group, in trying to construct a pure Yang-Mills type gauge description. Our point of view of PGT starting with {\em flat spacetime} is also discussed by Wiesendanger \cite{Wiesendanger:1995eh} and Hehl et al. \cite{Hehl:1976kj}.

Several reviews and text books have been written on the subject. Hehl et al. in their review \cite{Hehl:1976kj} follow the original developments in PGT discussing the Riemann-Cartan spacetime and its consequences - a weak spin-spin interaction. In another review \cite{Hehl:1994ue} they give the ``kinematic" layout and discusses, in first-order, the inclusion of fermionic matter, Noether identities and energy momentum tensors. Ivanenko and Sardanashvily \cite{Ivanenko:1984vf} and Sardanshvily \cite{Sardanashvily:2011aa} discusses the geometric description. Obukhov \cite{Obukhov:2006gea} discusses the dynamical aspects of Poincar\'{e} gauge gravity, especially its spinless limit and correspondences with general relativity at large distances, where GR already has experimental verifications. Trautman presents the theory succinctly in a short encyclopaedia entry \cite{Trautman:2006fp}. Little more pedagogic `lectures' have been given by Hehl in \cite{Hehl:1979xk} and by Blagojevic in \cite{Blagojevic:2003cg}. Among text books, the one written by Blagojevic \cite{Blagojevic:2002du} covers a wide range of topics lucidly. We also have the very recent collection of papers with commentaries \cite{Blagojevic:2013xpa}.

As regards to the dynamics, different gauge lagrangians were studied by numerous authors. This was done in the Riemann-Cartan spacetime with both curvature and torsion, as well as in Weitzenbock spacetime with zero curvature and non-zero torsion \cite{Hehl:1976kj, Hayashi:1979qx, Blagojevic:2000pi, Blagojevic:2002du}. Now, the most direct generalisation of Einstein's general relativity is the Einstein-Cartan action, which contains just a term linear in the Riemann or Lorentz field strength \cite{Sciama:1964wt, Kibble:1961ba, Nester:1977zz, Blagojevic:2002du}
\begin{align}
\label{demoEC}
S = \int d^4x\: b\,R = \int d^4x\: b\, b_i^{\ \mu}b_j^{\ \nu} R^{ij}_{\ \ \mu\nu} = \int d^4x\: \epsilon^{\lambda\rho\mu\nu} \epsilon_{hkij} b^h_{\ \lambda} b^k_{\ \rho} R^{ij}_{\ \ \mu\nu}.
\end{align}
Various other theories were also studied, including different combinations of terms linear and quadratic in the two field strengths Riemann `$R$' and torsion `$T$.' But if the Einstein-Cartan is the most `simplest' one, the most general Poincar\'{e} gauge lagrangian, which still gives equations of motion which are not higher than second order in field derivatives, was proposed by Hayashi and Shirafuji \cite{Hayashi:1979wj}. They used a decomposition of the torsion and Riemann field strengths into tensor, vector and scalar modes. The final form of the lagrangian \cite{Blagojevic:2002du} is
\begin{align}
\label{demoHayashiShirafuji}
\begin{aligned}
S &= \int d^4x \,b\left(-\alpha R + \mathcal{L}_T + \mathcal{L}_R + \lambda \right)\\
\mathcal{L}_T &= a\left(A T_{ijk}T^{ijk} + B T_{ijk}T^{jik} + C T^m_{\ mi} T_n^{\ ni} \right)\\
\mathcal{L}_R &= b_1 R_{ijkl}R^{ijkl} + b_2 R_{ijkl}R^{klij} + b_3 R_{ij}R^{ij} + b_4 R_{ij}R^{ji} + b_5 R^2 + b_6\!\left(\epsilon_{ijkl}R^{ijkl}\right)^2\!\!,\\
\end{aligned}
\end{align}
where $\alpha$, $A$, $B$, $C$ and $b_i$ are free parameters, $\lambda$ is cosmological constant, and $a=\frac{1}{2G}$ with $G$ being the gravitational constant.

All these were in the usual four dimensions. But PGT was also studied in various lower dimensions. Complete integrability of PGT in 2 dimensions was studied by Mielke et al. \cite{Mielke:1993nc}. Solodukhin studied solutions of 2D PGT with fermions \cite{Solodukhin:1994ux} while conserved quantities and asymptotics were studied by Blagojevic et al. \cite{Blagojevic:1996aw}. In 3 dimensions, we know from Witten's work \cite{Witten:1988hc} that gravity can be written down as Chern-Simons gauge theory. This had lead to an intense development of gravity in 3D \cite{Carlip:1998uc}. In another direction, inspired by the topologically massive gravity of Deser, Jackiw and Templeton \cite{Deser:1982vy}, Mielke and Baekler \cite{Mielke:1991nn} proposed a cosmological gravity model with torsion in the PGT framework. The action contained a Chern-Simons term, a term linear in torsion (the `translational Chern-Simons' term), and the usual Einstein-Cartan term. We can also add a cosmological term \cite{Blagojevic:2004hj} so that the action finally looks like
\begin{align}
\label{demoMB}
S = \int d^3x\,\epsilon^{\mu\nu\rho}\Big[ab^i_{\ \mu}R_{i\nu\rho} &- \frac{\Lambda}{3} \epsilon_{ijk}b^i_{\ \mu}b^j_{\ \nu}b^k_{\ \rho} \nonumber\\
&+ \alpha_3\left(\omega^i_{\ \mu}\partial_\nu\omega_{i\rho} + \frac{1}{3} \epsilon_{ijk}\,\omega^i_{\ \mu}\omega^j_{\ \nu}\omega^k_{\ \rho} \right) + \frac{\alpha_4}{2}b^i_{\ \mu}T_{i\nu\rho} \Big].
\end{align}
However this action does not have any propagating degree of freedom. To really study a topologically massive gravity with propagating degrees of freedom, one has to couple the term linear in torsion through a new lagrange multiplier type field which itself has no dynamics. This type of action was studied within the PGT framework in  \cite{Park:2008yy, Grumiller:2008pr, Blagojevic:2008bn}
\begin{align}
\label{demoTMG}
S = \int d^3x\,\epsilon^{\mu\nu\rho}\Big[ab^i_{\ \mu}R_{i\nu\rho} &- \frac{\Lambda}{3} \epsilon_{ijk}b^i_{\ \mu}b^j_{\ \nu}b^k_{\ \rho} \nonumber\\
&+ \frac{a}{\mu}\left(\omega^i_{\ \mu}\partial_\nu\omega_{i\rho} + \frac{1}{3} \epsilon_{ijk}\,\omega^i_{\ \mu}\omega^j_{\ \nu}\omega^k_{\ \rho} \right) + \frac{1}{2}\lambda^i_{\ \mu}T_{i\nu\rho} \Big].
\end{align}
Other types of massive gravity models were also studied, such as the PGT version \cite{Blagojevic:2010ir} of the `new massive gravity' proposed by Bergshoeff, Hohm and Townsend \cite{Bergshoeff:2009hq}, which contains a Pauli-Fierz type mass term. The action looks like
\begin{align}
\label{demoBHT}
S = \int d^3x a\epsilon^{\mu\nu\rho}\left( \sigma b^i_{\ \mu}R_{i\nu\rho}-\frac{\Lambda}{3}\epsilon_{ijk}b^i_{\ \mu}b^j_{\ \nu}b^k_{\ \rho} \right) + \frac{a}{m^2} \mathcal{L}_K + \frac{1}{2} \epsilon^{\mu\nu\rho}\lambda^i_{\ \mu}T_{i\nu\rho},
\end{align}
where $\mathcal{L}_K$ is given by
\begin{align}
\label{demoL_K}
\begin{aligned}
\mathcal{L}_K &=\frac{1}{2} \epsilon^{\mu\nu\rho}f^i_{\ \mu}R_{i\nu\rho}-b\mathcal{V}_K\\
\mathcal{V}_K &=\frac{1}{4}\left(f_{i\mu}f^{i\mu}-f^2\right).
\end{aligned}
\end{align}

Several interesting solutions within PGT were also put forward. Kerr-Neumann type black hole solutions were investigated in \cite{Bakler:1988nq}. Study of BTZ type black hole solutions \cite{Garcia:2003nm} and asymptotics of such geometries was also extensively carried out \cite{Blagojevic:2002du}. The Cauchy problem was studied, both in first order \cite{Dimakis:1989ba} and second order formalism, \cite{Dimakis:1989az} by Dimakis. 

Now, the whole point of PGT was to construct a theory of gravity starting from localised Poincar\'{e} symmetries. Thus, it is both very interesting and imperative to study the symmetry aspects of PGT models from a rigorous canonical viewpoint. Several studies were made \cite{Blagojevic:2002du}, but it was found that the Poincar\'{e} symmetries were not recovered from the first-class generator, usually constructed through the methods of Castellani \cite{Castellani:1981us}. The symmetries could be recovered only on-shell. But we note that gauge symmetries can only be understood meaningfully from an off-shell perspective! Also, existence of two independent sets of complete symmetries for the same action would be against the counting principles of gauge symmetries. We know, that the number of independent gauge parameters must be equal to the number of independent primary first-class constraints \cite{Henneaux:1990au}. Also, each gauge symmetry is tied to a Noether identity, and the number of such identities is restricted. So, in the remaining part of this thesis, we will turn towards a resolution of this puzzling scenario. Adopting both the canonical hamiltonian and lagrangian methodologies, we provide a clean off-shell and rigorous analysis of the gauge generator and resulting gauge symmetries.


%% file: C4_mb.tex
\chapter[Topological gravity with torsion]{Topological gravity with torsion: Mielke-Baekler model}
\label{C:mb}

\lettrine[lraise=0.0, loversize=0.3, findent=3pt, nindent=0pt, lhang=0.2]{W}{e} have, till now, described the building of a Poincar\'{e} gauge theoretic description of gravity, starting from a flat spacetime. Our starting point was the Poincar\'{e} symmetries, in their local form. But, as was mentioned, a hamiltonian analysis of the symmetries through a first class generator could not recover these symmetries \cite{Blagojevic:2004hj, Blagojevic:2008bn} in general. They could be recovered only on-shell. However, a fact to be noted here is, that the gauge generator itself in \cite{Blagojevic:2008bn, Blagojevic:2004hj} is constructed by an algorithm \cite{Castellani:1981us} based on an approach that considers symmetries that maps solutions to solutions of the equations of motion. It leads to the existence of two sets of complete canonical symmetries for the same action. By complete we mean that, in each set, there are as many independent symmetries as there are independent primary first class constraints in the theory. This leads to a paradox. As a first step towards a resolution, we present \cite{Banerjee:2009vf} an off-shell canonical hamiltonain analysis in this chapter. Gauge transformations are viewed as mappings of field configurations to field configurations at the level of the action.

The model that we adopt is the Mielke-Baekler \cite{Mielke:1991nn} type topological 3D gravity with torsion and a cosmological term \cite{Blagojevic:2004hj}. This model is formulated in the first order Poincar\'{e} gauge theory (PGT) formalism and is invariant {\em off-shell} under the Poincar\'{e} gauge transformations by construction, as we also explicitly verify here. We provide a thorough canonical analysis of the model following Dirac's approach of constrained hamiltonian analysis \cite{Dirac:Lectures}. The model contains both first class and second class constraints. Our work \cite{Banerjee:2009vf} supplements the canonical analysis of \cite{Blagojevic:2004hj} in that we work out the reduced phase-space structure. This is done by eliminating the second class constraints through replacement of the Poisson brackets by Dirac brackets. An explicitly off-shell method \cite{Banerjee:1999hu, Banerjee:1999yc} is used to construct a first-class gauge generator. This algorithm has been applied to different reparametrization invariant models \cite{Banerjee:2004un, Gangopadhyay:2007gn, Banerjee:2006pi, Samanta:2007fk} including second order metric gravity \cite{Mukherjee:2007yi}. The transformation of the basic fields are obtained from their Dirac brackets with the generator. Though the gauge transformations of the basic fields obtained from this method should naturally be invariances of the action, we provide an explicit check of the off-shell invariances thus obtained.

However, despite following a completely off-shell procedure, the geometric Poincar\'{e} gauge symmetries and the ones recovered through the canonical first class generator remain algebraically distinct. We will show that this is indeed a peculiarity of the PGT. For example, we will consider the Einstein action in the second order metric gravity formalism. The spacetime invariance of the theory is the diffeomorphism (diff) transformations. In (3+1) dimensions these diff invariances have been shown to map off-shell to the hamiltonian gauge transformations \cite{Mukherjee:2007yi}. We show that the same analysis is applicable to the corresponding (2+1) dimensional theory. But when we treat the same theory in the PGT framework, the discrepancy between the Poincar\'{e} transformations and the gauge transformations comes to the fore.

Now, before we go into the canonical analysis of the model, we would like to briefly review the canonical procedure that we have adopted here \cite{Dirac:Lectures}. More importantly, we would like to explain our method of constructing the generator through an off-shell analysis. We present the same method as was given in \cite{Henneaux:1990au, Banerjee:1999hu, Banerjee:1999yc}, but offer a new derivation of it. Our derivation presented here makes its off-shell nature more transparent.



\section{An off-shell canonical generator}
\label{Cmb:ConstraintsRev}

A canonical analysis for constrained systems, leading from a lagrangian to the hamiltonian, was given by Dirac \cite{Dirac:Lectures} and Bergmann \cite{Anderson:1951ta, Bergmann:1956zz, Bergmann:1961wa}. The idea, since then,  has been presented and extended innumerable times. Some of the standard references relevant to our purposes are \cite{Henneaux:1992ig, Wipf:1993xg, Blagojevic:2002du, Rothe:2010book}. For simplicity, in this section we refrain from an action depending on field\footnote{A generalisation to fields is formally straightforward, and relevant field theory definitions will be presented as we need them, later.}, rather taking a lagrangian depending on `$n$' coordinates $q_i$, $i=1,2,\ldots,n$, described by the action
\begin{align}
\label{revChParticleaction}
S = \int dt\: L(q,\,\dot{q}).
\end{align}
Considering fixed-endpoint variation, the equations of motion are obtained by setting the Euler derivatives
\begin{align}
\label{revchEOM}
L_i := -\frac{d}{dt}\left(\frac{\partial L}{\partial \dot{q}_i}\right) + \frac{\partial L}{\partial q_i},
\end{align}
to zero $L_i=0$. The equations of motion can be further written as
\begin{align}
\label{revChEOMbr}
L_i = -\frac{\partial^2 L}{\partial \dot{q}_i\dot{q}_j} \ddot{q}_j -\frac{\partial^2 L}{\partial \dot{q}_i q_j} \dot{q}_j + \frac{\partial L}{\partial q_i} = 0,
\end{align}
where the co-efficient of the acceleration, $W_{ij} = \frac{\partial^2 L}{\partial \dot{q}_i \,\partial\dot{q}_j}$, is known as the `Hessian.' These should ideally give us equations specifying the accelerations $\ddot{q}_i$. However, the equations of motion may be solved for the accelerations only if this Hessian is invertible. If the inverse doesn't exist, or in other words if $\text{det}(W_{ij})=0$, then the accelerations are not fixed uniquely  and the time evolution of the coordinates $q_i(t)$ has some arbitrariness in it \cite{Wipf:1993xg}.

In going from the lagrangian to the hamiltonian, we choose to move to a $2n$-dimensional phase space $(q,p)$ description starting from the $n$-dimensional configuration space of coordinates. This requires defining the canonical momenta $p_i$ as
\begin{align}
\label{revChmomenta}
p_i := \frac{\partial L}{\partial \dot{q}_i}(q,\,\dot{q}).
\end{align}
The velocities $\dot{q}_i$ may be eliminated in favour of these momenta, through the Legendre transform, only if the Hessian
\begin{align*}
W_{ij} = \frac{\partial p_i}{\partial \dot{q}_j} = \frac{\partial p_j}{\partial \dot{q}_i} = \frac{\partial^2 L}{\partial \dot{q}_i\,\partial\dot{q}_j}
\end{align*}
is invertible, or non-singular. In constrained systems, it is not so. This creates an arbitrariness in the time-evolution of physical variables and the usual definition of the canonical Hamiltonian is not possible. The Dirac-Bergmann canonical treatment overcomes this by the introduction of `constraints.'

The momenta defined through \eqref{revChmomenta} can either be a function of $(q, \dot{q})$, such that in principle it can be solved for the velocities $\dot{q}$, or may turn out to be a function of the coordinates $q$ only. In the later case, the defining relation of the momenta is then a {\em `primary constraint'}, $\phi(p,q)\approx 0$. The $\approx$ sign indicates that the constraint is to be equated to zero, only after computation of all brackets. The canonical hamiltonian $H_C$ is now modified to a `total hamiltonian' $H_T$ through addition of a linear combination of all the primary constraints along with arbitrary multipliers $\lambda_a$
\begin{align}
\label{revChHamilt}
\begin{aligned}
H_T &= H_C + \lambda_a \phi_a \\
H_C &= p_i \dot{q}_i - L.
\end{aligned}
\end{align}

The primary constraints $\phi_a$ have to be consistent with time evolution, and thus we have to impose
\begin{align}
\label{revChprimaryconsistency}
\lbrace \phi_a , H_T \rbrace = 0.
\end{align}
This may be identically satisfied (upto already existing constraints, if needed), may lead to functions involving some of the multipliers $\lambda_a$ (thus fixing them), or may be non-trivial functions of phase space variables $\chi_b(p,q)\approx 0$. In the last case we get new constraints $\chi_b$, known as secondary constraints, which have to be again checked for consistency. This process may lead to further generation of tertiary, quaternary, or higher constraints and then finally close so that no new constraints are obtained. At this point, all constraints are consistent with the dynamics\footnote{We are considering only such lagrangians which do not lead to inconsistent equations of motion.} and we are (possibly) left with some undetermined multipliers $\lambda_a$, which still inject some indeterminacy in the time evolution. The determined multipliers may be notationally denoted through a overbar, as in $\bar{\lambda}_a$, and along with their corresponding primary constraint, may be absorbed into the canonical hamiltonian, leading to a re-definition $\bar{H}_C$.

The constraints $\lbrace \phi_a, \chi_b \rbrace$ can now be divided into two categories. The ones which close among themselves under Poisson brackets, with structure constants $C^{\ l}_{h\ k}$
\begin{align}
\label{revChFCCalgebra}
\lbrace\xi_h, \xi_k\rbrace = C^{\ l}_{h\ k}\,\xi_l,
\end{align}
are known as first-class constraints $\xi_f$ (FCC). Rest of the constraints are second-class $\zeta_s$ (SCC). It is the SCC sector that determines (some of) the arbitrary multipliers $\lambda_a$. The SCC may be strongly put to zero if we modify the Poisson brackets
\begin{align}
\label{revChPoissonBr}
\lbrace f,g \rbrace := \frac{\partial f}{\partial q_i} \frac{\partial f}{\partial p_i} - \frac{\partial g}{\partial q_i} \frac{\partial f}{\partial p_i},
\end{align}
and introduce a new set of brackets -- the Dirac brackets:
\begin{align}
\label{revChDiracBr}
\lbrace f,g \rbrace_* := \lbrace f,g \rbrace - \sum_{r,s} \lbrace f,\zeta_r \rbrace \triangle^{-1}_{r,s} \lbrace \zeta_s, g\rbrace.
\end{align}
Here $\triangle^{-1}_{r,s}$ is the inverse of the matrix $\triangle_{r,s} = \lbrace \zeta_r,\zeta_s \rbrace$, which represents the algebra of the second class sector.

The FCC, through appropriate brackets with the canonical coordinates and their functions, generate transformations which are invariances of the action \cite{Dirac:Lectures}. Thus $\lbrace\xi_a, q_i\rbrace = \delta_a q_i$ such that $\delta_a S = 0$, where $S$ is the action. These invariances are known as `{\em gauge symmetries}' of the action and are a reflection of the arbitrariness of the dynamics in the theory. Also, the closure property within the FCC sector ensures that the symmetries form a group -- the gauge group. The number of independent gauge symmetries of an action must be equal to the number of independent primary first-class constraints \cite{Henneaux:1990au}. But since all first-class constraints generate gauge symmetries, an ansatz for the most general gauge generator may be written down as
\begin{align}
\label{revChGansatz}
G = \varepsilon_f \xi_f \qquad \text{(summation over repeated indices implied)}
\end{align}
The multipliers $\varepsilon_f$ are arbitrary parameters known as gauge parameters. A method of eliminating the dependent gauge parameters from the above ansatz and constructing an {\em off-shell} canonical gauge generator was provided by Henneaux-Teitelboim-Zanelli (HTZ) in an extended hamiltonian formalism\footnote{The extended hamiltonian formalism uses a hamiltonian containing linear combination of all constraints instead of only the primary constraints as used in the total hamiltonian method we are using.} \cite{Henneaux:1990au} and by Banerjee-Rothe-Rothe \cite{Banerjee:1999yc, Banerjee:1999hu} in a total hamiltonian formalism, one that we use here. The crux of this method lies in introducing additional relations obeyed by the gauge parameters by noting that arbitrary variations must commute with total time derivatives
\begin{align}
\label{revChDeltaDerivCommute}
\delta \bullet \left(\frac{d}{dt} f\right) \equiv \frac{d}{dt} \bullet \left(\delta \,f\right).
\end{align}
These can then be used to precisely weed out the adequate number of dependent parameters from \eqref{revChGansatz}. Below, we give a new and concise derivation of the relevant equations which highlights the {\em off-shell} nature of this algorithm.

Let us first consider a system with only irreducible first-class constraints. We note that $H_T$ preserves all constraints by construction, and in the definition \eqref{revChHamilt}, all the $\phi_a$ are of course FCC. So $H_C$ must be $FCC$ in nature, having an algebra with other FCC as
\begin{align}
\label{revChStructureV}
\lbrace H_C, \xi_f\rbrace = V_f^{\ h}\,\xi_h.
\end{align}
The generator \eqref{revChGansatz} induces variations in phase space functions as $\delta f = \lbrace f,G \rbrace = \varepsilon_f \lbrace f, \xi_f \rbrace$, and in particular,
\begin{align}
\label{revChGaugeVarQP}
\begin{aligned}
\delta q_i &= \varepsilon_f \lbrace q_i, \xi_f \rbrace = \varepsilon_f \frac{\partial \xi_f}{\partial p_i}\\
\delta p_i &= \varepsilon_f \lbrace p_i, \xi_f \rbrace = -\varepsilon_f \frac{\partial \xi_f}{\partial q_i}.
\end{aligned}
\end{align} 
Now, the `canonical action' may be written as
\begin{align}
\label{revChCanAction}
S = \int p_i \dot{q}_i - H_C - \lambda_{f_1} \xi_{f_1},
\end{align}
where the subscript $1$ on indices denotes primary nature of the constraints. Secondary and all higher constraints will be given subscript $2$. Variation of this $S$ gives
\begin{align*}
\delta S = \int \delta p_i \,\dot{q}_i + p_i \,\delta\dot{q}_i - \delta \left( H_C - \lambda_{f_1} \xi_{f_1} \right).
\end{align*}
Using the commutativity of total derivatives and arbitrary variations \eqref{revChDeltaDerivCommute} in the second term within the action integral produces
\begin{align*}
\delta S = \int \delta p_i \,\dot{q}_i -\delta q_i \,\dot{p}_i + \frac{d}{dt}\left(p_i \,\delta q_i\right) - \delta \left( H_C - \lambda_{f_1} \xi_{f_1} \right).
\end{align*}
Now we substitute variations of the coordinates \eqref{revChGaugeVarQP} induced by the generator \eqref{revChGansatz} in the first two terms
\begin{align*}
\delta S &= \int -\varepsilon_f \frac{\partial \xi_f}{\partial q_i} \frac{d q_i}{dt} - \varepsilon_f \frac{\partial \xi_f}{\partial p_i} \frac{d p_i}{dt} + \frac{d}{dt}\left(p_i \,\delta q_i\right) - \delta \left( H_C - \lambda_{f_1} \xi_{f_1} \right) \\
&= \int \frac{d}{dt} \left( p_i \,\delta q_i - \varepsilon_f \xi_f \right) + \dot{\varepsilon}_f \xi_f - \delta \left( H_C - \lambda_{f_1} \xi_{f_1} \right).
\end{align*}
Dropping total derivatives and expanding the variation on the last term,
\begin{align*}
\delta S = \int \dot{\varepsilon}_f \xi_f - \delta H_C - \lambda_{h_1} \,\delta \xi_{h_1} - \delta\lambda_{h_1} \,\xi_{h_1},
\end{align*}
where we are ready to substitute the variations, in terms of the structure constants, of the hamiltonian \eqref{revChStructureV} and the primary FCC \eqref{revChFCCalgebra}.
\begin{align*}
\delta S = \int \dot{\varepsilon}_f \xi_f - \varepsilon_h V_h^{\ f} \xi_f - \varepsilon_k\lambda_{h_1}\, C_{{h_1}\ k}^{\ f} \,\xi_f - \delta\lambda_{f_1} \,\xi_{f_1}.
\end{align*}
Finally, we now want to segregate the primary and the secondary/higher sector in the above expression
\begin{align*}
\delta S = &\int \dot{\varepsilon}_{f_1} \xi_{f_1} - \varepsilon_h V_h^{\ {f_1}} \xi_{f_1} - \varepsilon_k\lambda_{h_1}\, C_{{h_1}\ k}^{\ {f_1}} \,\xi_{f_1} - \delta\lambda_{f_1} \,\xi_{f_1}\\
+ &\int \dot{\varepsilon}_{f_2} \xi_{f_2} - \varepsilon_h V_h^{\ {f_2}} \xi_{f_2} - \varepsilon_k\lambda_{h_1}\, C_{{h_1}\ k}^{\ {f_2}} \,\xi_{f_2},
\end{align*}
so that the variation of the action becomes
\begin{align*}
\delta S = \int \left[\dot{\varepsilon}_{f_1} - \varepsilon_h V_h^{\ {f_1}} - \varepsilon_k\lambda_{h_1}\, C_{{h_1}\ k}^{\ {f_1}} - \delta\lambda_{f_1} \right]\,\xi_{f_1} + \int \left[\dot{\varepsilon}_{f_2} - \varepsilon_h V_h^{\ {f_2}} - \varepsilon_k\lambda_{h_1}\, C_{{h_1}\ k}^{\ {f_2}} \right] \xi_{f_2}.
\end{align*}
Now, we see that {\em if} the terms within square brackets vanish, the action is invariant under the transformations induced by the generator G \eqref{revChGansatz}, subject to the conditions
\begin{align}
\label{revChMasterEqn1}
\delta\lambda_{f_1} &= \dot{\varepsilon}_{f_1} - \varepsilon_h V_h^{\ {f_1}} - \varepsilon_k\lambda_{h_1}\, C_{{h_1}\ k}^{\ {f_1}}\\
\label{revChMasterEqn2}
\dot{\varepsilon}_{f_2} &= \varepsilon_h V_h^{\ {f_2}} + \varepsilon_k\lambda_{h_1}\, C_{{h_1}\ k}^{\ {f_2}}.
\end{align}
Of course we cannot comment on whether a solution to these equations will {\em always} be possible. However, if we can show that these equations can be solved to get the dependent gauge parameters in a particular model, then on using these conditions, the generator constructed is an off-shell canonical generator of gauge symmetries for that model.

Among the two equations obtained, the first \eqref{revChMasterEqn1} gives variations of the arbitrary multipliers, and is actually not an independent equation. It can be derived from the second \eqref{revChMasterEqn2} and the definitions required for the total hamiltonian $H_T$ \cite{Banerjee:1999hu}. So, we actually have only one independent equation \eqref{revChMasterEqn2} that is imposed on the gauge parameters $\varepsilon_f$. Observing this equation carefully, we see that it is a set of $f_2$ equations, i.e. there are as many equations as there are secondary/higher FCCs. So, ultimately the number of remaining independent gauge parameters is equal to the number of independent primary FCCs. This is a result \cite{Henneaux:1990au} that we have stressed previously, and now prove explicitly.

Finally, we would like to comment on mixed systems having both FCCs and SCCs. The SCC sector fixes some of the arbitrary multipliers $\lambda_a$ in the total hamiltonian \eqref{revChHamilt}. As noted before, if we denote these determined multipliers by an overbar $\bar{\lambda}_a$, then we can absorb them along with their corresponding primary SCCs with the canonical hamiltonian to define a new quantity $\bar{H}_C = H_C + \bar{\lambda}_s \chi_s$. This leaves the total hamiltonian as $H_T = \bar{H}_C + \lambda_f \xi_f$. Noting, as in the previous case, that $H_T$ preserves all constraints and $\xi_f$'s are by definition FCC, $\bar{H}_C$ must be within FCC and will follow an algebra analogous to \eqref{revChStructureV}. Rest of the derivation follows along same lines and we reach the same set of equations. However, we also note that an alternate way to handle mixed systems is by adopting the Dirac brackets defined in \eqref{revChDiracBr}. We can then set the SCC sector strongly to zero, and all algebra is computed using Dirac brackets instead of Poisson brackets.


\section{3D gravity with torsion}
\label{Cmb:3dGravIntro}

Gravity theories in (2+1) dimensions offer an arena where one can address the problem of quantization on a simpler setting \cite{Carlip:2004ba}. Interest in 3D gravity increased a lot after Witten's discovery of the equivalence of 3D gravity with a Chern-Simons gauge theory \cite{Witten:1988hc}. Inclusion of the Chern-Simons term in the Einstein-Hilbert action leads to a theory known as `topologically massive gravity' which has a massive propagating degree of freedom \cite{Deser:1981wh, Deser:1983tn, Deser:1983nh}. These theories were studied in Riemannian space time. Later, a 3D gravity theory was formulated in the Riemann-Cartan spacetime, that is with non-zero torsion \cite{Mielke:1991nn, Baekler:1992ab}. The canonical structure of the `topological 3D gravity with torsion' plus a cosmological term was investigated in \cite{Blagojevic:2004hj} following Dirac's constrained hamiltonian analysis. Recently a surge of activity in various 3D gravity models has been witnessed \cite{Blagojevic:2008bn, Blagojevic:2004hj, Blagojevic:2002du, Maloney:2009ck, Carlip:2005zn, Park:2008yy, Grumiller:2008pr, Carlip:2008qh, Banerjee:2012kq}. Now, before we start with our canonical analysis, let us list our conventions.

\paragraph{Conventions:}We adopt the conventions of Blagojevic (see for example \cite{Blagojevic:2004hj}). Latin indices refer to the local Lorentz frame and the Greek indices refer to the coordinate frame. The first letters of both alphabets $(a,b,c,\ldots)$ and $(\alpha,\beta,\gamma,\ldots)$ run over spatial indices 1,2 while the middle alphabet letters $(i,j,k,\ldots)$ and $(\mu,\nu,\lambda,\ldots)$ run over all, i.e. 0,1,2. The totally antisymmetric tensor densities $\epsilon^{ijk}$ and  $\epsilon^{\mu\nu\rho}$ are both normalized so that $\epsilon^{012}=1$. The signature of spacetime is $(+,-,-)$.

The first-order Mielke-Baekler type \cite{Mielke:1991nn} action for the topological 3D gravity model with torsion and a cosmological term  \cite{Blagojevic:2004hj} is
\begin{align}
\label{CmbMBaction}
S = \int d^3x\ \epsilon^{\mu\nu\rho}&\left[a\,b^i_{\ \mu} R_{i\nu\rho} - \frac{\Lambda}{3} \epsilon_{ijk}\,b^i_{\ \mu} b^j_{\ \nu} b^k_{\ \rho} \right.\nonumber\\
&\phantom{+}\left. + \alpha_3 \left(\omega^i_{\ \mu} \,\partial_\nu\omega_{i\rho} + \frac{1}{3} \epsilon_{ijk}\,\omega^i_{\ \mu} \omega^j_{\ \nu} \omega^k_{\ \rho} \right) + \frac{\alpha_4}{2} b^i_{\ \mu} T_{i\nu\rho} \right].
\end{align}
Here $a$, $\Lambda$, $\alpha_3$ and $\alpha_4$ are arbitrary parameters. The first term, proportional to $a$, is the Einstein-Hilbert term written in three dimensions, using the identity
\begin{align}
\label{theidentity}
bR = -\epsilon^{\mu\nu\rho}\,b^i_{\ \mu} R_{i\nu\rho}.
\end{align}
Here $b = \text{det}\,b^i_{\ \mu}$, and $R = b_i^{\ \mu} b_j^{\ \nu} R^{ij}_{\ \ \mu\nu}$. The second term contains the cosmological constant, the third is the Chern-Simons action while the fourth includes torsion.
 
The variation of the action \eqref{CmbMBaction} w.r.t the triad $b^i_{\ \mu}$ and the spin connection $\omega^i_{\ \mu}$ are given by
\begin{align}
\label{ChmbEulerDerivs}
\begin{aligned}
\frac{\delta S}{\delta b^i_{\ \mu}} &= \epsilon^{\mu\nu\rho}\left[a R_{i\nu\rho} + \alpha_4 T_{i\nu\rho} - \Lambda\epsilon_{ijk}b^j_{\ \nu}b^k_{\ \rho}\right] \\
\frac{\delta S}{\delta \omega^i_{\ \mu}} &= \epsilon^{\mu\nu\rho}\left[\alpha_3 R_{i\nu\rho} + a T_{i\nu\rho} + \alpha_4 \epsilon_{ijk} b^j_{\ \nu}b^k_{\ \rho}\right]
\end{aligned}
\end{align}
These Euler derivatives, when equated to zero, yield the equations of motion in the usual way. The equations of motion following from the action can be simplified in the sector\footnote{Later, we will find that this condition is essential in computing the Dirac brackets.\label{fn:neq 0 condition}} $\alpha_3\alpha_4 - a^2 \neq 0$ as
\begin{align}
\label{eqnmot}
\begin{aligned}
T^i_{\ \mu\rho} - p\,\epsilon^{ijk}\,b_{j \mu}b_{k\rho} &= 0\\
R^i_{\ \mu\rho} - q\,\epsilon^{ijk}\,b_{j \mu}b_{k\rho} &= 0.
\end{aligned}
\end{align}
where $p=\dfrac{\alpha_3\Lambda + \alpha_4 a}{\alpha_3\alpha_4 - a^2}$ and $q=-\dfrac{\alpha_4^{\ 2} +a\Lambda}{\alpha_3\alpha_4 - a^2}$. This gives us two independent equations to find the two characteristics of the Riemann-Cartan manifold, curvature and torsion.

From the analysis of the geometric correspondence it is natural to expect the action \eqref{CmbMBaction} to be invariant under the Poincar\'{e} gauge transformations \eqref{Poincare3Dfieldtrans}
\begin{align}
\tag{\ref{Poincare3Dfieldtrans}}
\begin{aligned}
\delta b^i_{\ \mu} &= -\epsilon^i_{\ jk}b^j_{\ \mu}\theta^k - \partial_\mu\xi^\rho b^i_{\ \rho} - \xi^{\rho}\partial_{\rho} b^i_{\ \mu}\\
\delta \omega^{i}_{\ \mu} &= -\left(\partial_\mu \theta^i + \epsilon^i_{\ jk} \omega^j_{\ \mu} \theta^k  \right) - \partial_\mu\xi^\rho \omega^{i}_{\ \rho} - \xi^{\rho}\partial_{\rho}\omega^{i}_{\ \mu}.
\end{aligned}
\end{align}
But we may also check it explicitly. A straightforward calculation leads to the following variation $\delta S$ of the action \eqref{CmbMBaction}
\begin{align}
\label{actiontrans}
\delta S = \delta S^{(1)} + \delta S^{(2)},
\end{align}
where
\begin{align}
\label{invariance1}
\delta S^{(1)} = \int d^3x\ \partial_\lambda \left[\xi^\lambda\,\epsilon^{\mu\nu\rho} \left\lbrace - ab^i_{\ \mu} R_{i\nu\rho} -\alpha_3\left(\omega^i_{\ \mu}\partial_\nu\omega_{i\rho} + \frac{1}{3}\epsilon_{ijk} \omega^i_{\ \mu} \omega^j_{\ \nu} \omega^k_{\ \rho}\right) \right.\right. \nonumber\\
\left.\left. + \ \Lambda\epsilon_{ijk}b^i_{\ \mu}b^j_{\ \nu}b^k_{\ \rho}  - \dfrac{\alpha_4}{2} b^i_{\ \mu}T_{i\nu\rho}\right\rbrace + \theta^i\epsilon^{\lambda\nu\rho} \,\partial_\nu \omega_{i\rho} \right],
\end{align}
and
\begin{align}
\label{invariance2}
\delta S^{(2)} = \int d^3x &~\epsilon^{\mu\nu\rho}\left[a\,\left\lbrace\partial_\mu\xi^\lambda\, b^i_{\ \lambda}R_{i\nu\rho} +\partial_\nu\xi^\lambda\, b^i_{\ \mu}R_{i\lambda\rho} +\partial_\rho\xi^\lambda\, b^i_{\ \mu}R_{i\nu\lambda} -\partial_\lambda\xi^\lambda\, b^i_{\ \mu}R_{i\nu\rho}\right\rbrace\right. \nonumber\\
&- \frac{\Lambda}{3} \epsilon_{ijk} \left\lbrace -3\,\partial_\mu\xi^\lambda\, b^i_{\ \lambda}b^j_{\ \nu}b^k_{\ \rho} + \partial_\lambda\xi^\lambda\, b^i_{\ \mu}b^j_{\ \nu}b^k_{\ \rho} \right\rbrace \nonumber\\
&+ \alpha_3\left\lbrace \partial_\mu\xi^\lambda (\omega^i_{\ \lambda}\partial_\nu\omega_{i\rho} +\frac{1}{3}\epsilon_{ijk}\omega^i_{\ \lambda}\omega^j_{\ \nu}\omega^k_{\ \rho}) + \partial_\nu\xi^\lambda\omega^i_{\ \mu} \partial_\lambda\omega_{i\rho}\right. \nonumber\\
& \left. \qquad+\,\partial_\rho\xi^\lambda\omega^i_{\ \mu}\partial_\nu\omega_{i\lambda} - \partial_\lambda\xi^\lambda ( \omega^i_{\ \mu}\partial_\nu\omega_{i\rho} + \frac{1}{3}\epsilon_{ijk}\omega^i_{\ \mu}\omega^j_{\ \nu}\omega^k_{\ \rho})\right\rbrace \nonumber\\
&+ \left.\frac{\alpha_4}{2}\left\lbrace\partial_\mu\xi^\lambda b^i_{\ \lambda}T_{i\nu\rho} + \partial_\nu\xi^\lambda b^i_{\ \mu}T_{i\lambda\rho} + \partial_\rho\xi^\lambda b^i_{\ \mu}T_{i\nu\lambda}\! - \partial_\lambda\xi^\lambda b^i_{\ \mu}T_{i\nu\rho}\right\rbrace\right].
\end{align}
The piece $\delta S^{(1)}$ is a total boundary term but $\delta S^{(2)}$ is not so. The latter actually vanishes. To see this in a compact manner we use the following identity, from the transformation of the {\em constant} tensor density $\epsilon^{\mu\nu\rho}$
\begin{equation}
\delta\epsilon^{\mu\nu\rho} = \partial_\lambda\xi^\mu\epsilon^{\lambda\nu\rho} +\partial_\lambda\xi^\nu\epsilon^{\mu\lambda\rho}+\partial_\lambda\xi^\rho\epsilon^{\mu\nu\lambda}-\partial_\lambda\xi^\lambda\epsilon^{\mu\nu\rho} = 0.
\label{epstrans}
\end{equation}
Now the rhs of \eqref{invariance2} simplifies as,
\begin{align}
\delta S^{(2)} = \int d^3x\, \delta\epsilon^{\mu\nu\rho}\ &\left[ab^i_{\ \mu} R_{i\nu\rho}-\frac{\Lambda}{3} \epsilon_{ijk}b^i_{\ \mu} b^j_{\ \nu} b^k_{\ \rho} \right.\nonumber\\
&+\left. \alpha_3 \left\lbrace\omega^i_{\ \mu} \partial_\nu\omega_{i\rho} + \frac{1}{3} \epsilon_{ijk}\omega^i_{\ \mu} \omega^j_{\ \nu} \omega^k_{\ \rho} \right\rbrace + \frac{\alpha_4}{2} b^i_{\ \mu} T_{i\nu\rho} \right]
\end{align}
and hence, $\delta S^{(2)}$ vanishes on account of \eqref{epstrans}. The invariance of the theory \eqref{CmbMBaction}) under Poincare gauge transformations \eqref{Poincare3Dfieldtrans} is thus explicitly verified.


\section{Canonical analysis of the model}
\label{Cmb:CanonicalMB}

In considering the 2+1 dimensional model \eqref{CmbMBaction} with the Chern-Simons term along with the torsion, cosmological and usual Einstein-Hilbert terms, we actually get a mixed system with both first-class and second-class constraints. This calls for a more general analysis than what is done for pure gauge systems with only first-class constraints. Such mixed systems can be dealt through two different approaches.
\begin{itemize}
\item {\bf\emph{Using Poisson brackets:}} In this method, the entire algebra is computed using Poisson brackets. Second class constraints are taken care by introducing Lagrange multipliers which enforce these constraints. The multipliers can be fixed from the time conservation of the constraints.
\item {\bf\emph{Using Dirac brackets:}} The second-class constraints can be strongly eliminated by using Dirac brackets,  and we can deal with an effectively pure system having only first-class constraints. All Poisson brackets will have to be replaced by corresponding Dirac brackets.
\end{itemize}
Here in this paper, we adopt the method of using Dirac brackets due to two reasons. First, the analysis of this model  through Dirac brackets is new and provides an interesting alternative to other studies \cite{Blagojevic:2004hj} on this model. Secondly, the systematic method of computing a generator from a structured algorithm \cite{Banerjee:1999hu, Banerjee:1999yc}, which is adopted here, is technically simple for pure systems. Thus it is desirable, though not essential, to first convert our mixed system into a pure gauge system.

The action \eqref{CmbMBaction} is written in terms of the triads $b^i_{\ \mu}(x)$ and spin connections $\omega^i_{\ \mu}(x)$, which are the basic fields in this theory. The corresponding momenta $\pi_i^{\ \mu}(x)$ and $\Pi_i^{\ \mu}(x)$, defined as $\frac{\partial \mathcal{L}}{\partial(\partial_0 b^i_{\ \mu})}$ and $\frac{\partial \mathcal{L}}{\partial(\partial_0 \omega^i_{\ \mu})}$ respectively, are found to be,
\begin{align}
\label{primary const}
\begin{aligned}
\phi^{\ 0}_i &= \pi^{\ 0}_i \approx 0\\
\phi^{\ \alpha}_i &=  \pi_i^{\ \alpha}-\alpha_4 \epsilon^{0\alpha\beta}b_{i\beta}\approx 0\\
\Phi^{\ 0}_i &= \Pi^{\ 0}_i \approx 0\\
\Phi^{\ \alpha}_i &= \Pi_i^{\ \alpha} - \epsilon^{0\alpha\beta}\left(2ab_{i\beta} + \alpha_3 \omega_{i\beta} \right)\approx 0.
\end{aligned}
\end{align}
We now see that \emph{all} the momenta lead to constraints. These are the primary constraints of the theory, defined by $\phi^{\ 0}_i$, $\phi^{\ \alpha}_i$, $\Phi^{\ 0}_i$ and $\Phi^{\ \alpha}_i$. The symbol $\approx$ stands for weak equality in the sense of Dirac \cite{Dirac:Lectures} implying that the constraints can be set equal to zero only after computing all relevant brackets.

The canonical hamiltonian density $\mathcal{H}_C$ can now be written down, through a Legendre transformation $\mathcal{H}_C=\pi^{\ \mu}_i \,\partial_0 b^i_{\ \mu} + \Pi^{\ \mu}_i \,\partial_0 \omega^i_{\ \mu} - \mathcal{L}$,
\begin{align}
\label{canon Hamilt}
\begin{aligned}
\mathcal{H}_C &= b^i_{\ 0}\mathcal{H}_i + \omega^i_{\ 0} \mathcal{K}_i + \partial_\alpha D^\alpha\\
\mathcal{H}_i &= -\epsilon^{0\alpha\beta}\left(a\,R_{i\alpha\beta}+\alpha_4 \,T_{i\alpha\beta}-\Lambda \,\epsilon_{ijk} \,b^j_{\ \alpha} b^k_{\ \beta} \right)\\
\mathcal{K}_i &= -\epsilon^{0\alpha\beta}\left(a\,T_{i\alpha\beta}+\alpha_3 \,R_{i\alpha\beta}+\alpha_4 \,\epsilon_{ijk} \,b^j_{\ \alpha} b^k_{\ \beta} \right)\\
D^\alpha &= \epsilon^{0\alpha\beta} \left[\omega^i_{\ 0}\left(2a\,b_{i \beta} + \alpha_3\,\omega_{i\beta}\right)+ \alpha_4 \,b^i_{\ 0} b_{i\beta}\right];\\
\end{aligned}
\end{align}
and the total hamiltonian density $\mathcal{H}_T$ with all primary constraints
\begin{align}
\label{total Hamilt1}
\mathcal{H}_T &= \mathcal{H}_C + \lambda^i_{\ \mu} \,\phi^{\ \mu}_i + \varrho^i_{\ \mu} \,\Phi^{\ \mu}_i\nonumber\\
&= b^i_{\ 0} \,\mathcal{H}_i + \omega^i_{\ 0} \,\mathcal{K}_i + \lambda^i_{\ 0} \,\pi_i^{\ 0} + \lambda^i_{\ \alpha} \,\phi_i^{\ \alpha} + \varrho^i_{\ 0} \,\Pi_i^{\ 0} + \varrho^i_{\ \alpha} \,\Pi_i^{\ \alpha} + \partial_\alpha D^\alpha,
\end{align}
where $\lambda^i_{\ \mu}$ and $\varrho^i_{\ \mu}$ are undetermined multipliers.

The consistency conditions for the primary constraints $\phi_i^{\ \alpha}$ and $\Phi_i^{\ \alpha}$ fixes the multipliers $\lambda^i_{\ \alpha}$ and $\varrho^i_{\ \alpha}$ \cite{Blagojevic:2004hj}, which we can substitute back into \eqref{total Hamilt1} to get a modified total hamiltonian
\begin{align}
\label{total Hamilt2}
\mathcal{H}_T = b^i_{\ 0} \,\bar{\mathcal{H}}_i + \omega^i_{\ 0} \,\bar{\mathcal{K}}_i + \lambda^i_{\ 0} \,\pi_i^{\ 0} + \varrho^i_{\ 0} \,\Pi_i^{\ 0} + \partial_\alpha \bar{D^\alpha},
\end{align}
where the barred quantities are obtained by combining the terms with determined multipliers suitably with the corresponding unbarred ones. Their explicit forms are:
\begin{subequations}
\begin{align}
\label{ChmbSCC}
\begin{aligned}
\bar{\mathcal{H}_i} &:=\mathcal{H}_i - \nabla_\beta \phi_i^{\ \beta} + \epsilon_{ijk}\,b^j_{\ \beta} \left( p \phi^{k\beta}+q\Phi^{k\beta} \right) \approx 0\\
\bar{\mathcal{K}_i} &:= \mathcal{K}_i - \nabla_\beta \Phi_i^{\ \beta} - \epsilon_{ijk}\,b^j_{\ \beta} \phi^{k\beta} \approx 0\\
\end{aligned}
\end{align}\vspace*{-2.5em}
\begin{align}
\label{barDdefn}
\bar{D^\alpha} &:= D^\alpha + b^i_{\ 0} \phi_i^{\ \alpha} + \omega^i_{\ 0} \phi_i^{\ \alpha}.
\end{align}
\end{subequations}
On using the the time conservation conditions of the primary constraints $\pi_i^{\ o}$ and $\Pi_i^{\ 0}$, we find  the two secondary constraints $\bar{\mathcal{H}}_i$ and $\bar{\mathcal{K}}_i$ \eqref{ChmbSCC}. 

The consistency of the secondary constraints leads to no new constraints, ending the iterative procedure here. So, we have the complete constraint structure of the theory. On examining the Poisson algebra of the constraints ({\it see Appendix \ref{App:Poisson}}), we see that this is a mixed system, with both first-class (whose algebra close with all constraints) and second-class (whose algebra does not close among themselves) constraints.

In the table below, we give a complete classification of the constraints along with an explanation of the notation. First-class constraints will be denoted as $\Sigma$ whereas second-class constraints will be denoted as $\Omega$.
\begin{table}[ht]
\caption{Classification of Constraints\vspace*{-2em}}
\label{table:constraints}
\centering
\begin{tabular}{l  c  c}
\\[0.5ex]
\hline
\hline\\[-2ex]
 & First class $\Sigma$ & Second class $\Omega$\\[0.5ex]
\hline\\[-2ex]
Primary &\ \ $\Sigma_{(1)i}=\phi_i^{\ 0},~ \Sigma_{(2)i}=\Phi_i^{\ 0}$ &\ \ ${\Omega_{(1)}}_i^{\ \alpha}=\phi_i^{\ \alpha}, ~ {\Omega_{(2)}}_i^{\ \alpha}=\Phi_i^{\ \alpha}$\\[0.5ex]
\hline\\[-2ex]
Secondary &\ $\Sigma_{(3)i}=\bar{\mathcal{H}_i},~\Sigma_{(4)i}=\bar{\mathcal{K}_i}$ &\ \ \\[0.5ex]
\hline
\hline
\end{tabular}
\end{table}

As explained at the beginning of this section, we will now implement the method of Dirac brackets\footnote{Dirac brackets are denoted by a star $\lbrace\, , \, \rbrace^*$ to distinguish them from Poisson brackets $\lbrace\, , \,\rbrace$.}  and thus eliminate all second-class constraints from the theory. The Dirac bracket is defined in terms of Poisson brackets as,
\begin{align}
\label{Dirac defn}
\lbrace f(x),g(x') \rbrace^* := &\lbrace f(x),g(x') \rbrace \nonumber\\
& - \displaystyle{\sum_{(YZ)}} \int dy\, dz ~ \lbrace f(x),\Omega_{(Y)}(y)\rbrace ~\Delta^{-1}_{(YZ)}(y,z) ~\lbrace \Omega_{(Z)}(z), g(x') \rbrace.
\end{align}
The quantity $\Delta^{-1}_{(YZ)}(y,z)$ is the inverse of the matrix $\Delta_{(YZ)}(y,z)$, formed from the second class constraints $\Omega_{(Z)}$, with $Y,Z=1,2$. The elements of the matrix $\Delta_{(YZ)}(y,z)=\lbrace \Omega_{(Y)}, \Omega_{(Z)} \rbrace$ are given by
\begin{align}
\label{const matrix}
\left[\Delta_{(YZ)}(x,x')\right]^{\ \ \alpha\beta}_{ij} = - 2 ~\epsilon^{0\alpha\beta}~\eta_{ij} \left(\begin{array}{cc}\alpha_4 & a\\a & \alpha_3 \end{array}\right)  \delta(x-x'),
\end{align}
and the matrix $\Delta^{-1}_{(YZ)}(y,z)$ can thus be written down as
\begin{align}
\label{inv const matrix}
\left[\Delta^{-1}_{(YZ)}(x,x')\right]_{\ \ \beta\alpha}^{ij} = -\frac{1}{2\left(\alpha_3\alpha_4-a^2\right)} \epsilon_{0\beta\alpha} ~\eta^{ij} \left(\begin{array}{rr}\alpha_3 & -a\\-a & \alpha_4 \end{array}\right)  \delta(x-x').
\end{align}
Here the condition $\alpha_3\alpha_4-a^2 \neq 0$ ensures the invertibility of the matrix $\Delta_{YZ}$. This is the same condition as encountered before, (see footnote in page \pageref{fn:neq 0 condition}) and we observe that it also comes up naturally in this canonical analysis. The Dirac brackets between pairs of basic fields and momenta can now be computed, and they \emph{all} turn out to be non-zero. These brackets are listed below,
\begin{align}
\label{basicDirac}
\begin{aligned}
\lbrace b^i_{\ \mu}(x),b^j_{\ \nu}(x')\rbrace^* &= \frac{\alpha_3}{2\left(\alpha_3 \alpha_4 - a^2\right)} ~\epsilon_{0\alpha\beta} ~\delta^\alpha_\mu \delta^\beta_\nu ~\eta^{ij} ~\delta(x-x')\\
\lbrace b^i_{\ \mu}(x),\omega^j_{\ \nu}(x')\rbrace^* &= -\frac{a}{2\left(\alpha_3 \alpha_4 - a^2\right)} ~\epsilon_{0\alpha\beta}  \delta^\alpha_\mu \delta^\beta_\nu ~\eta^{ij} ~\delta(x-x')\\
\lbrace b^i_{\ \mu}(x),\pi_j^{\ \nu}(x')\rbrace^* &= \left[\delta^\nu_\mu - ~\delta^\alpha_\mu \delta^\nu_\alpha ~\frac{\left(\alpha_3 \alpha_4 - 2a^2\right)}{2\left(\alpha_3 \alpha_4 - a^2\right)} \right] ~\delta^i_j ~\delta(x-x')\\
\lbrace b^i_{\ \mu}(x),\Pi^{\ \nu}_j(x')\rbrace^* &= \frac{\alpha_3 a}{2\left(\alpha_3 \alpha_4 - a^2\right)} ~\delta^\alpha_\mu \delta^\nu_\alpha ~\delta^i_j ~\delta(x-x')\\
\lbrace \omega^i_{\ \mu}(x),\omega^j_{\ \nu}(x')\rbrace^* &= \frac{\alpha_4}{2\left(\alpha_3 \alpha_4 - a^2\right)} ~\epsilon_{0\alpha\beta} ~\delta^\alpha_\mu \delta^\beta_\nu ~\eta^{ij} ~\delta(x-x')\\
\lbrace \omega^i_{\ \mu}(x),\pi^{\ \nu}_j(x')\rbrace^* &= -\frac{\alpha_4 a}{2\left(\alpha_3 \alpha_4 - a^2\right)} ~\delta^\alpha_\mu \delta^\nu_\alpha ~\delta^i_j ~\delta(x-x')\\
\lbrace \omega^i_{\ \mu}(x),\Pi^{\ \nu}_j(x')\rbrace^* &= \left[\delta^\nu_\mu - ~\delta^\alpha_\mu \delta^\nu_\alpha ~\frac{\left(\alpha_3 \alpha_4 - 2a^2\right)}{2\left(\alpha_3 \alpha_4 - a^2\right)} \right] ~\delta^i_j ~\delta(x-x')\\
\lbrace \pi_i^{\ \mu}(x),\pi^{\ \nu}_j(x')\rbrace^* &= \left[\frac{\alpha_4^2\alpha_3 + 4a^3 - 2\alpha_4a^2 - 2\alpha_4\alpha_3a}{2\left(\alpha_3 \alpha_4 - a^2\right)} \right] \epsilon^{0\alpha\beta} \delta^\mu_\alpha \delta^\nu_\beta ~\eta_{ij} \delta(x-x')\\
\lbrace \pi_i^{\ \mu}(x),\Pi^{\ \nu}_j(x')\rbrace^* &= \frac{\alpha_3 \alpha_4a}{2\left(\alpha_3 \alpha_4 - a^2\right)} ~\epsilon^{0\alpha\beta} \delta^\mu_\alpha \delta^\nu_\beta ~\eta_{ij}~\delta(x-x')\\
\lbrace \Pi_i^{\ \mu}(x),\Pi^{\ \nu}_j(x')\rbrace^* &= \frac{\alpha_4\alpha_3^2}{2\left(\alpha_3 \alpha_4 - a^2\right)} ~\epsilon^{0\alpha\beta} \delta^\mu_\alpha \delta^\nu_\beta ~\eta_{ij}~\delta(x-x')\\
\end{aligned}
\end{align}
The Dirac brackets of the second class constraints among themselves and with all other quantities turn out to be zero, as expected. Hence these constraints can be strongly set equal to zero.

We now list the various Dirac brackets relevant to us. The involutive algebra of the first class constraints:
\begin{align}
\label{constrnsDirac}
\begin{aligned}
\lbrace\bar{\mathcal{H}_i}(x),\bar{\mathcal{H}_j}(x')\rbrace^* &=  \epsilon_{ijk}\left(p\,\bar{\mathcal{H}^k}+q\,\bar{\mathcal{K}^k}\right)\delta(x-x')\\
\lbrace\bar{\mathcal{K}_i}(x),\bar{\mathcal{K}_j}(x')\rbrace^* &= - \epsilon_{ijk}\,\bar{\mathcal{K}^k}\,\delta(x-x')\\
\lbrace\bar{\mathcal{H}_i}(x),\bar{\mathcal{K}_j}(x')\rbrace^* &= - \epsilon_{ijk}\,\bar{\mathcal{H}^k}\,\delta(x-x').
\end{aligned}
\end{align}
Next, the algebra of the first class constraints with the canonical hamiltonian $H_C=\int d^2x'  ~\mathcal{H}_C(x')$:
\begin{align}
\label{constrnsDiracHamilt}
\begin{aligned}
\lbrace H_C,\bar{\mathcal{H}_i}(x)\rbrace^* &= \left[\epsilon_{ijk}\left\lbrace\omega^j_{\ 0}(x) - p b^j_{\ 0}(x) \right\rbrace \bar{\mathcal{H}^k}(x) - q \,\epsilon_{ijk} b^j_{\ 0}(x) ~\bar{\mathcal{K}^k}(x)\right]\\
\lbrace H_C,\bar{\mathcal{K}_i}(x)\rbrace^* &= \left[\epsilon_{ijk} \,b^j_{\ 0}(x) ~\bar{\mathcal{H}^k}(x) + \epsilon_{ijk} \,\omega^j_{\ 0}(x) ~\bar{\mathcal{K}^k}(x)\right]\\
\lbrace H_C,\pi_i^{\ 0}(x)\rbrace^* &= \bar{\mathcal{H}_i}(x)\\
\lbrace H_C,\Pi_i^{\ 0}(x)\rbrace^* &= \bar{\mathcal{K}_i}(x).
\end{aligned}
\end{align}
Note that now we have a system with only first class constraints whose Dirac algebra has been given above. The second class constraints, as already stated, are strongly set equal to zero. In the next section, we will use these results to systematically find the gauge generator following \cite{Banerjee:1999hu, Banerjee:1999yc} and show that it generates \emph{off-shell} symmetries of the action \eqref{CmbMBaction}.


\section{Gauge generator and gauge symmetries}
\label{Cmb:GaugeGen}

In this section, we systematically calculate the gauge symmetry generator $G$ of the action \eqref{CmbMBaction}. We follow the method enunciated in \cite{Banerjee:1999hu, Banerjee:1999yc} to construct $G$. It is to be noted at the very onset that this method does not require any use of the equations of motion. Consequently the generated symmetries are off-shell. This may be compared to the approach \cite{Castellani:1981us} adopted in \cite{Blagojevic:2004hj}, for discussions in this model, where the generator maps solutions to solutions of the equations of motion. Since equations of motion are involved, it becomes debatable whether the generator would be able to reproduce the genuine (off-shell) symmetries of the model. In this sense our approach is conceptually cleaner.  We next outline this approach briefly.

Having eliminated all the second class constraints through introduction of Dirac brackets, we are left with a theory with only first class constraints. The set of constraints $\Sigma_{(I)}$ is now classified as
\begin{align}
\label{RB const1}
\left[\Sigma_{(I)}\right] = \left[\Sigma_{(A)};\Sigma_{(Z)}\right]
\end{align}
where $A=1,2$ belong to the set of primary (first class) constraints, $Z=3,4$ to the set of secondary (first class) constraints and $I=1,2,3,4$ refer to all (first class) constraints. The total hamiltonian is
\begin{align}
\label{RB tot hamilt}
H_{T} = H_{C} + \int d^2x ~\lambda^{(A)}\Sigma_{(A)}
\end{align}
where $H_C$ is the canonical hamiltonian and $\lambda^{(A)}$ are Lagrange multipliers enforcing the primary constraints. The most general expression for the generator of gauge transformations is obtained according to the Dirac conjecture as
\begin{align}
\label{RB gen G}
G = \int d^2x ~\varepsilon^{(I)}\Sigma_{(I)}
\end{align}
where $\varepsilon^{(I)}$ are the gauge parameters. However, not all of these are independent.  This is most simply and elegantly seen by  demanding the commutation of an arbitrary gauge variation with the total time derivative, i.e. $\frac{d}{dt}\left(\delta q \right) = \delta \left(\frac{d}{dt} q \right)$. Recalling that,
\begin{align}
\begin{aligned}
\delta q &= \lbrace q, G \rbrace^*\\
\frac{dq}{dt} &= \lbrace q, H_T \rbrace^*,
\end{aligned}
\end{align}
a little algebra, using \eqref{RB tot hamilt} and \eqref{RB gen G}, yields the following conditions (see \cite{Banerjee:1999hu, Banerjee:1999yc} and Section:\ref{Cmb:ConstraintsRev})
\begin{align}
\label{RB master 1}
\delta\lambda^{(A)}(x) = \frac{d\epsilon^{(A)}(x)}{dt}-\int d^2x' \,\varepsilon^{(I)}(x') \,&\left[ \left(V^A_{\;\;\: I}\right)(x,x')\right.\nonumber\\
&\left.\ +\int d^2x''\,\lambda^{(B)}(x'') ~\left(C^A_{\;\;\, IB}\right)(x,x',x'')\right]\\
\label{RB master 2}
0 = \frac{d\epsilon^{(Z)}(x)}{dt} - \int d^2x'~\varepsilon^{(I)}(x')\,&\left[\left(V^Z_{\;\;\: I}\right)(x,x') \right.\nonumber\\
&\left.\ +\int d^2x''\,\lambda^{(B)}(x'') \,\left(C^Z_{\;\;\, IB}\right)(x,x',x'')\right].
\end{align}
Here the coefficients $\left(V^I_{\;\;\: J}\right)(x,x')$ and $\left(C^I_{\;\;\, JK}\right)(x,x',x'')$ are the structure functions of the involutive (first-class) algebra, defined through
\begin{align}
\label{StructureConsts}
\begin{aligned}
\lbrace\Sigma_{(I) i}(x),\Sigma_{(J) j}(x')\rbrace^* &= \int d^2x''\,{\left(C^K_{\;\;\, IJ}\right)}_{ijk}(x'',x,x') ~\Sigma_{(K)}^{\quad k}(x'')\\
\lbrace H_C,\Sigma_{(I)i}(x)\rbrace^* &= \int d^2x'\,{\left(V^J_{\;\;\: I}\right)}_{ik}(x',x)~\Sigma_{(J)}^{\quad k}(x').
\end{aligned}
\end{align}
The second condition \eqref{RB master 2} makes it is possible to choose $A$ independent gauge parameters from the set $\varepsilon^{(I)}$ and express the generator $G$ of \eqref{RB gen G} entirely in terms of them. This shows that the number of independent gauge parameters is equal to the number of independent, primary first-class constraints. 

Before proceeding further let us note the following point. The derivation of \eqref{RB master 2} is based only on the relation between the velocities and the canonical momenta, namely, the first of Hamilton's equations \cite{Banerjee:1999hu, Banerjee:1999yc}. Note that the full dynamics, implemented through the second of Hamilton's equations $\left(\frac{dp}{dt}=\lbrace p, H \rbrace\right)$ is not required to impose restrictions on the gauge parameters. Since this is the only input in our method of abstraction of the independent gauge parameters, we find that our analysis will be valid off-shell. The off-shell invariance will also be demonstrated explicitly.

The structure constants defined in \eqref{StructureConsts} can now be obtained using the results of the various Dirac brackets \eqref{basicDirac}, \ref{constrnsDirac} \& \ref{constrnsDiracHamilt}). These are:
\begin{align}
\label{Cs}
\begin{aligned}
{\left(C^A_{\;\; IJ}\right)}_{ijk}(x'',x,x') &= 0.\\
{\left(C^Z_{\;\; AB}\right)}_{ijk}(x'',x,x') &= 0.\\
{\left(C^3_{\;\; 33}\right)}_{ijk}(x'',x,x') &= p \,\epsilon_{ijk} \,\delta(x-x'') \delta(x''-x')\\
{\left(C^3_{\;\; 34}\right)}_{ijk}(x'',x,x') &= - \epsilon_{ijk} \,\delta(x-x'') \delta(x''-x')\\
{\left(C^3_{\;\; 44}\right)}_{ijk}(x'',x,x') &= 0\\
{\left(C^4_{\;\; 33}\right)}_{ijk}(x'',x,x') &= q \,\epsilon_{ijk} \,\delta(x-x'') \delta(x''-x')\\
{\left(C^4_{\;\; 34}\right)}_{ijk}(x'',x,x') &= 0\\
{\left(C^4_{\;\; 44}\right)}_{ijk}(x'',x,x') &= - \epsilon_{ijk} \,\delta(x-x'') \delta(x''-x')
\end{aligned}
\end{align}
and,
\begin{align}
\label{Vs}
\begin{aligned}
{\left(V^A_{\;\;\, I}\right)}_{ik}(x',x) &= 0\\
{\left(V^3_{\;\;\, 1}\right)}_{ik}(x',x) &= \eta_{ik} \,\delta(x-x')\\
{\left(V^3_{\;\;\, 2}\right)}_{ik}(x',x) &= 0\\
{\left(V^3_{\;\;\, 3}\right)}_{ik}(x',x) &= \epsilon_{ijk} \,\left[\omega^j_{\ 0}(x') - p\,b^j_{\ 0}(x')\right] \delta(x-x')\\
{\left(V^3_{\;\;\, 4}\right)}_{ik}(x',x) &= \epsilon_{ijk} \,b^j_{\ 0}(x') \,\delta(x-x')\\
{\left(V^4_{\;\;\, 1}\right)}_{ik}(x',x) &= 0\\
{\left(V^4_{\;\;\, 2}\right)}_{ik}(x',x) &= \eta_{ik} \,\delta(x-x').\\
{\left(V^4_{\;\;\, 3}\right)}_{ik}(x',x) &= -q \,\epsilon_{ijk} \,b^j_{\ 0}(x') \,\delta(x-x')\\
{\left(V^4_{\;\;\, 4}\right)}_{ik}(x',x) &= \epsilon_{ijk} \,\omega^j_{\ 0}(x') \,\delta(x-x')
\end{aligned}
\end{align}
Now the generator \eqref{RB gen G} is expanded as,
\begin{align}
\label{generatorGen}
G=\int d^2x \left[\varepsilon^{(1)i}(x)\,\pi^{\ 0}_i(x)+\varepsilon^{(2)i}(x)\,\Pi^{\ 0}_i(x) + \varepsilon^{(3)i}(x)\,\bar{\mathcal{H}_i}(x) + \varepsilon^{(4)i}(x)\,\bar{\mathcal{K}_i}(x)\right]
\end{align}
where the parameters $\varepsilon^{(I)i}$ are not all independent, but satisfy the equation \eqref{RB master 2}, so that,
\begin{align}
\label{master eq}
\displaystyle\frac{d\epsilon^{\scriptscriptstyle{(Z)i}}(x)}{dt}-\int d^2x' \,\varepsilon^{\scriptscriptstyle{(I)k}}(x')\,{\left({V^{\scriptscriptstyle{Z}}}_{\scriptscriptstyle{\,I}}\right)}_k^{\;\;\,i}(x',x)=0
\end{align}
Using the structure constants ${\left({V^{\scriptscriptstyle{Z}}}_{\scriptscriptstyle{\,I}}\right)}_k^{\;\;\,i}(x',x)$ already determined in \eqref{Vs}, we get the following two relations among the parameters $\varepsilon^{\scriptscriptstyle{(Z)}}$
\begin{align}
\label{rel epsilons}
\begin{aligned}
\dot{\varepsilon}^{(3)i}(x) &= \varepsilon^{(1)i}(x) + \varepsilon^{(3)k}(x) \,\epsilon_k^{\;\:ij}\left[ p\,b_{j0}(x)-\omega_{j0}(x) \right] - \varepsilon^{(4)k}(x) \,\epsilon_k^{\;\:ij}\,b_{j0}(x)\\
\dot{\varepsilon}^{(4)i}(x) &= \varepsilon^{(2)i}(x) + q \,\varepsilon^{(3)k}(x) \,\epsilon_k^{\;\: ij}\,b_{j0}(x) - \varepsilon^{(4)k}(x) \,\epsilon_k^{\;\: ij}\,\omega_{j0}(x).\\
\end{aligned}
\end{align}
After using these equations \eqref{rel epsilons} in the generator \eqref{generatorGen} to eliminate the gauge parameters $\varepsilon^{(1)}$ and $\varepsilon^{(2)}$, we obtain our cherished structure in terms of the two independent gauge parameters $\varepsilon^{(3)}$ and $\varepsilon^{(4)}$,
\begin{align}
\label{generatorOur}
G = \int d^2x &\left[\left\lbrace \dot{\varepsilon}^{(3)i}(x) -  \varepsilon^{(3)k}(x) \,\epsilon_k^{\;\: ij}\,\left[ p\,b_{j0}(x)-\omega_{j0}(x) \right] + \varepsilon^{(4)k}(x) \,\epsilon_k^{\;\: ij}\,b_{j0}(x)\right\rbrace\,\pi_i^{\ 0}(x) \right.\nonumber\\
&+\left.\left\lbrace \dot{\varepsilon}^{(4)i}(x) - q \,\varepsilon^{(3)k}(x) \,\epsilon_k^{\;\: ij}\,b_{j0}(x) + \varepsilon^{(4)k}(x) \epsilon_k^{\;\: ij}\,\omega_{j0}(x) \right\rbrace \Pi_i^{\ 0}(x)  \right.\nonumber\\
&+\left. \varepsilon^{(3)i}(x)\,\bar{\mathcal{H}_i}(x) + \varepsilon^{(4)i}(x)\,\bar{\mathcal{K}_i}(x)\right].
\end{align}
On rearranging the generator and renaming the parameters as $\varepsilon^{(3)}=\varepsilon$ and $\varepsilon^{(4)}=\tau$, we obtain the generator in the form
\begin{align}
\label{generatorFinal}
\begin{aligned}
G=&\int d^2x \left[\mathcal{G}_\varepsilon(x)+\mathcal{G}_\tau(x)\right]\\
&\mathcal{G}_\varepsilon=\dot{\varepsilon}^i\,\pi_i^{\ 0} + \varepsilon^i\left[\bar{\mathcal{H}_i}- \epsilon_{ijk} \big( \omega^j_{\ 0} - p\,b^j_{\ 0}\big)\pi^{k0} + q \,\epsilon_{ijk}\,b^j_{\ 0}\Pi^{k0} \right]\\
&\mathcal{G}_\tau=\dot{\tau}^i\Pi_i^{\ 0} + \tau^i\left[\bar{\mathcal{K}_i}-\epsilon_{ijk}\big(b^j_{\ 0}\,\pi^{k0} + \omega^j_{\ 0}\,\Pi^{k0}\big)\right]\\
\end{aligned}
\end{align}
The generator thus written gives rise to gauge variations of fields in the theory. The transformations of the basic fields $b^i_{\ \mu}$ and $\omega^i_{\ \mu}$ are given by:
\begin{align}
\label{field transf gauge}
\begin{aligned}
\delta b^i_{\ \mu}(x) &:= \lbrace b^i_{\ \mu}(x),G \rbrace^* \\
&= \partial_\mu\epsilon^i(x) + \epsilon^i_{\ jk} \,\omega^j_{\ \mu}(x) \varepsilon^k(x) - p \,\epsilon^i_{\ jk} \,b^j_{\ \mu}(x) \varepsilon^k(x) + \epsilon^i_{\ jk}\,b^j_{\ \mu}(x) \tau^k(x),\\
\delta \omega^i_{\ \mu}(x) &:= \lbrace \omega^i_{\ \mu}(x),G \rbrace^* \\
&= \partial_\mu \tau^i(x) + \epsilon^i_{\ jk} \,\omega^j_{\ \mu}(x) \tau^k(x) - q \,\epsilon^i_{\ jk} \,b^j_{\ \mu}(x) \varepsilon^k(x).\\
\end{aligned}
\end{align}

We would now like to demonstrate the explicit \emph{off-shell} invariance of the action \eqref{CmbMBaction} under the above gauge transformations of the fields \eqref{field transf gauge}. The variation of the action, in general, reads:
\begin{align}
\label{gen var action}
\delta S = \delta S\Big|_\text{Einstein} + \delta S\Big|_\text{Cosmological} + \delta S\Big|_\text{Chern Simons} + \delta S\Big|_\text{Torsion}
\end{align}
Substituting our gauge transformations in the above, we observe that $\delta S$ vanishes without using any equations of motion. The cancellation of relevant terms is quite interesting and we would like to note certain features involved. An easy way of seeing the cancellation is to begin by focusing on families of similar structured terms. For example, terms containing one derivative, the parameter $\varepsilon$, $b$ and $\omega$ (where the indices have been suppressed for simplicity) may occur as $\left(~\epsilon^{\mu\nu\rho} \epsilon^i_{\ jk} ~b^j_{\ \mu} ~\partial_\nu \omega_{i\rho} ~\varepsilon^k ~\right)$ or $\left(~\epsilon^{\mu\nu\rho} \epsilon^i_{\ jk} ~\partial_\mu \varepsilon_i ~\omega^j_{\ \nu} ~b^k_{\ \rho} ~\right)$ or in other such different types. However all of them may be cast as the same term on using the properties of the levi-civita symbols and/or using partial integrals.  When all such families are identified, we see that there occur two different types of cancellation. First, many terms are identically zero or cancel algebraically, needing at most throwing of some total derivatives. Secondly, in some cases terms from different pieces of the action, with their different parameters, cancel by virtue of the relation between parameters $a, \Lambda, \alpha_3\, \&\, \alpha_4$ and the definition of the quantities `$p$' and `$q$'. We now demonstrate this for one particular family. The terms containing $(\omega\,b\ b~\varepsilon)$ can be collected from the variations of different pieces of the action \eqref{CmbMBaction}. These are written below in exactly the same order as they appear in \eqref{gen var action},
\begin{align}
\delta S\Big|_{(\omega\,b\ b\,\varepsilon) \,\text{terms}} &= -2aq \int d^2x \,\epsilon^{\mu\nu\rho} \epsilon_{ijk} \epsilon^j_{\ lm}\, b^i_{\ \mu} \omega^k_{\ \rho} b^l_{\ \nu} \,\varepsilon^m - 2\Lambda \int d^2x \,\epsilon^{\mu\nu\rho} \,\omega^j_{\ \mu} b_{j\nu} b_{k\rho} \,\varepsilon^k \nonumber\\
&\phantom{=} + 0 + \alpha_4 p \int d^2x ~\epsilon^{\mu\nu\rho} \left[\epsilon^m_{\ \ il} \epsilon_{mpq} \,b^i_{\ \mu} \omega^l_{\ \nu} b^p_{\ \rho} \,\varepsilon^q + \epsilon^i_{\ jk} \epsilon_{ilm} \,b^j_{\ \mu} b^m_{\ \rho} \omega^l_{\ \nu} \,\varepsilon^k \right] \nonumber\\
&=  2\left(-\Lambda + aq +\alpha_4 p\right) \int d^2x \,\epsilon^{\mu\nu\rho} \,\omega^j_{\ \mu} b_{j\nu} b_{k\rho} \,\varepsilon^k\nonumber\\
&= 0
\end{align}
where use of the definitions $p=\dfrac{\alpha_3\Lambda + \alpha_4 a}{\alpha_3\alpha_4 - a^2}$ and $q=-\dfrac{\alpha_4^{\ 2} +a\Lambda}{\alpha_3\alpha_4 - a^2}$ has been made, to observe that the combination $\left(-\Lambda + aq +\alpha_4 p\right)=0$. A summary of the different terms and their cancellation factors are given below. Terms that are not explicitly mentioned here, reduce to zero algebraically.
\begin{table}[ht]
\label{table:zero combs}
\caption{Cancellation of families of terms\vspace*{-1em}}
\centering
\begin{tabular}{l  c}
\\[-0.5ex]
\hline
Term & Combination of parameters giving zero\\[0.5ex]
\hline
$\omega\,b\ b~\varepsilon$ & $-\Lambda + aq +\alpha_4 p$ \\[0.5ex]
$\omega\,\omega\ b~\varepsilon$ & $\alpha_4 + q \alpha_3 + ap$ \\[0.5ex]
\hline
\end{tabular}
\end{table}

We are thus led to an intriguing situation. There are two sets of field transformations, one derived above in \eqref{field transf gauge}, and the other from the Poincare gauge gravity \eqref{Poincare3Dfieldtrans}, both of these being true symmetries of the action. We have explicitly demonstrated this by showing that variations in the action \ref{CmbMBaction}, under any of the two transformations, vanish without requiring any use of the equations of motion. Consequently these are proper gauge symmetries, i.e. they are off-shell symmetries. One would therefore expect an off-shell mapping between the two sets of parameters $\left(\varepsilon,~\tau\right)$ and $\left(\xi,~\theta\right)$. Alas, this map does not exist. Indeed, the following map, which was also mentioned in the literature \cite{Blagojevic:2004hj},
\begin{align}
\label{onshell map}
\begin{aligned}
\varepsilon^i &= -\xi^\lambda \,b^i_{\ \lambda}\\
\tau^i &= -\left(\theta^i + \xi^\lambda \,\omega^i_{\ \lambda} \right).\\
\end{aligned}
\end{align}
connects the two transformations by the identification:
\begin{align}
\label{onshell relation transfs}
\begin{aligned}
\delta_0 b^i_{\ \mu} &= \delta_{PGT} b^i_{\ \mu} - \xi^\rho\left(T^i_{\ \mu\rho}-p \,\epsilon^{ijk}\, b_{j\mu} b_{k\rho} \right)\\
\delta_0 \omega^i_{\ \mu} &= \delta_{PGT} \omega^i_{\ \mu} - \xi^\rho\left(R^i_{\ \mu\rho}-q \,\epsilon^{ijk}\, b_{j\mu} b_{k\rho} \right).
\end{aligned}
\end{align}
The terms within parentheses, which are exactly the terms destroying the mapping between $\delta_0$ and $\delta_{PGT}$, are the equations of motion \eqref{eqnmot}. So the map \eqref{onshell map} holds only \emph{on-shell}. In the next section we attempt towards a possible understanding of this point. But before that we would like to make the following comment.

\paragraph*{On the significance of (\ref{RB master 1}):} We would like to make a note on the information content of \eqref{RB master 1}, which is referred to as the ``first condition" hereafter. This equation gives the variation of the Lagrange multipliers corresponding to the primary (first-class) constraints in terms of the structure constants $\left(V^I_{\;\;\, J}\right)(x,x')$ and $\left(C^I_{\;\; JK}\right)(x,x',x'')$ defined in \eqref{StructureConsts}. However, as we show below, this equation gives us no new restrictions on the parameters. This is because the equation itself can be obtained from the properties of the total hamiltonian and the second condition -- the `master equation' \eqref{RB master 2} \cite{Banerjee:1999hu}. We now demonstrate this fact here, in the context of our theory.

We begin by calculating the time variation of the field $\displaystyle b^i_{\ 0}$ by taking its Dirac bracket\footnote{Recall that we have adopted the approach of eliminating all second-class constraints by using Dirac brackets. Hence equations of motion are given by Dirac brackets.} with the total hamiltonian \eqref{total Hamilt2}, to see that it gives the Lagrange multiplier corresponding to $\displaystyle \pi^{\ 0}_i$,
\begin{align}
\label{b lambda 3}
\dot{b}^i_{\ 0} = \lbrace b^i_{\ 0}, \int d^2x \,\mathcal{H}_T \rbrace^* = \lambda^{(1)i}_{\;\quad 0}.
\end{align}
Using this, we find the variation of the multiplier $\lambda^{(1)}$ in terms of the derivative of the field transformations,
\begin{align}
\label{field lambda 3}
\delta \lambda^{(1)i}_{\;\quad 0} = \delta \dot{b}^i_{\ 0} = \frac{d}{dt} \delta b^i_{\ 0}.
\end{align}
However, we have already calculated the transformation $\delta b^i_{\ 0}$ \eqref{field transf gauge}. Also, recall that \emph{only} equation \eqref{RB master 2} was required in deriving the generator, and so, \eqref{field transf gauge} is independent of the first condition. Now substituting these field transformations for $b^i_{\ 0}(x)$ in the last equation \eqref{field lambda 3}, and using the definitions $\varepsilon=\varepsilon^{(3)}$ and $\tau=\varepsilon^{(4)}$ introduced before in Section \ref{Cmb:GaugeGen}, we get:
\begin{align}
\label{intermediate 3}
\frac{d}{dt} \delta b^i_{\ 0} = \frac{d}{dt} \left[\partial_0 \varepsilon^{(3)i} + \epsilon^i_{\ jk}\,\omega^j_{\ 0} \varepsilon^{(3)k}-p\,\epsilon^i_{\ jk}\,b^j_{\ 0}\varepsilon^{(3)k} + \epsilon^i_{\ jk}\,b^j_{\ 0} \varepsilon^{4)k}\right].
\end{align}
This can be related with the variation of $\lambda^{(1)}$ by taking advantage of \eqref{field lambda 3}, to finally obtain
\begin{align}
\label{from theory 3}
\delta \lambda^{(1)i}_{\;\quad 0} = \frac{d}{dt} \left[ \dot{\varepsilon}^{(3)i} - \varepsilon^{(3)k}\,\epsilon_k^{\ \,ij} \left(p\,b_{j0} - \omega_{j0}\right) + \varepsilon^{(4)k}\epsilon_k^{\ \,ij}\,b_{j0} \right] = \frac{d}{dt} \varepsilon^{(1)i}.
\end{align}
Here, in the last step, we have used \eqref{rel epsilons} to express $\varepsilon^{(3)}$ and $\varepsilon^{(4)}$ in terms of $\varepsilon^{(1)}$.

Let us now return to the first condition \eqref{RB master 1}, from which it follows,
\begin{align}
\label{1st master 3}
\delta \lambda^{(1)}(x) &= \displaystyle\frac{d\epsilon^{(1)}(x)}{dt}-\int d^2x' \,\varepsilon^{(I)}(x') \,\left[ \left(V^1_{\;\;\: I}\right)(x,x')\right. \nonumber\\
&\left.\phantom{=\frac{d\epsilon^{(1)}(x)}{dt}}+\int d^2x''\,\lambda^{(B)}(x'') \,\left(C^1_{\;\; IB}\right)(x,x',x'')\right] \nonumber\\
&= \frac{d\epsilon^{(1)}(x)}{dt}
\end{align}
thereby reproducing \eqref{from theory 3}. The second term does not contribute since the structure constants $\left(V^1_{\;\;\: I}\right)$, $\left(C^2_{\;\; IB}\right)$ vanish \eqref{Cs} \& \ref{Vs}). This shows that, as claimed at the beginning of this appendix, the first condition gives us no new restrictions on the parameters. It is basically a consequence of \eqref{RB master 2}.

In our calculations above, we have used the Lagrange multiplier $\lambda^{(1)}$ corresponding to $\pi^{\ 0}_i$. However, by the same process, analogous results are obtained for the multiplier $\lambda^{(2)}$ which corresponds to $\Pi^{\ 0}_i$. The starting point of the calculation for $\lambda^{(2)}$ is now:
\begin{align}
\dot{\omega}^i_{\ 0} = \lbrace \omega^i_{\ 0}, \int d^2x \,\mathcal{H}_T \rbrace^* = \lambda^{(2)i}_{\quad\; 0}.
\end{align}
Then, going through similar steps analogous to (\ref{field lambda 3}, \ref{intermediate 3} \& \ref{from theory 3}) we get
\begin{align}
\label{from theory 4}
\delta \lambda^{(2)i}_{\quad\; 0} = \frac{d}{dt} \left[ \dot{\varepsilon}^{(4)i} - q \,\varepsilon^{(3)k}\,\epsilon_k^{\ ij}\, b_{j0} +  \varepsilon^{(4)k}\,\epsilon_k^{\ ij}\,\omega_{j0} \right] = \frac{d}{dt} \varepsilon^{(2)i}.
\end{align}
which is the analogue of \eqref{from theory 3} found above. This is nothing but the first condition \eqref{RB master 1} corresponding to  $\lambda^{(2)}$,
\begin{align}
\label{1st master 4}
\delta \lambda^{(2)}(x) &= \displaystyle\frac{d\epsilon^{(2)}(x)}{dt}-\int d^2x' \,\varepsilon^{(I)}(x') \,\left[ \left(V^2_{\;\;\: I}\right)(x,x')\right.\nonumber\\
&\left.\phantom{\frac{d\epsilon^{(2)}(x)}{dt}-\int d^2x' \,\varepsilon^{(I)}(x') \,} +\int d^2x''\,\lambda^{(B)}(x'') \,\left(C^2_{\;\; IB}\right)(x,x',x'')\right] \nonumber\\
&= \frac{d\epsilon^{(2)}(x)}{dt},
\end{align}
which follows as a consequence of the vanishing of the structure constants $\left(V^2_{\;\;\, J}\right)$ and $\left(C^2_{\;\; JK}\right)$ calculated in (\ref{Cs} \& \ref{Vs}).


\section{Comments on the lack of off-shell mapping between the transformation parameters of PGT and the independent gauge parameters}
\label{Cmb:ADM}

It has been observed in the last section that the gauge transformations \eqref{field transf gauge} of the basic fields of the theory \eqref{CmbMBaction} cannot be mapped on the transformations of the same under PGT, namely, \eqref{Poincare3Dfieldtrans} {\it{without invoking}} the equations of motion, although both sets of transformations preserve the off-shell invariance of the same action. Stated otherwise, we don't have an off-shell mapping between the two sets of parameters characterising these transformations. This, notwithstanding the facts that the number of independent parameters of the two sets match exactly and both the sets provide off-shell invariance of \eqref{CmbMBaction} as we have explicitly demonstrated above. We show  that this feature is a peculiarity of the PGT framework.

Before considering PGT, let us first analyse metric gravity theory in the second order formalism, given by the action 
\begin{align}
\label{Einsteinaction}
S=\int ~d^3x~ \sqrt{-g} R
\end{align}
where $R$ is the Ricci scalar. Here $g = {\rm{det}}g_{\mu\nu}$, $g_{\mu\nu}$ being the metric tensor. The theory is invariant under diffeomorphism
\begin{align}
\label{Diff}
x^\mu \to x^\mu + \xi^\mu.
\end{align}
The canonical analysis of the gauge transformations of the theory following the method of \cite{Banerjee:1999hu,Banerjee:1999yc} was performed in \cite{Mukherjee:2007yi}. The analysis was done in (3+1) dimensions but it can be easily adapted to ($2+1$) dimensions which is relevant here. For the canonical analysis, spacetime is foliated in spacelike two-surfaces as per the Arnowit-Deser-Misner (ADM) decomposition. The lapse variable $N^{\bot}$ represents an arbitrary variation normal to the two-surface on which the state of the system is defined whereas the shift variables $N^{\alpha}$  represent variations along the surface. They are defined by{\footnote{Note that $g^{\alpha\beta}$ is the inverse of the spatial metric $g_{\alpha\beta}$ on the two surface.}}
\begin{align}
\label{Nj}
N^{\beta} &= g^{\alpha\beta}g_{0\alpha}\\
\label{N} 
N^{\perp} &= \left(- g^{00}\right)^{-1/2}
\end{align}
These variables are not really the dynamical variables of the theory. By adding suitable divergences to the action \eqref{Einsteinaction} we can write an equivalent lagrangian \cite{Hanson:1976cn, Sundermeyer:1982gv}
\begin{align}
\label{L}
\int d^{2}x {\cal{L}} =  \int d^{2}x N^{\perp} \left(g\right) ^{1/2} \left(K_{\alpha\beta}K^{\alpha\beta} - K^{2} + R \right) 
\end{align}
where $K = K^{\alpha}{}_{\alpha} = g^{\alpha\beta} K_{\alpha\beta}$ and $R$ is the Ricci scalar on the two surface. The second fundamental form $K_{\alpha\beta}$ is defined as 
\begin{align}
\label{K}
K_{\alpha\beta} = \frac{1}{2 N^{\bot}} \left(- \dot{g}_{\alpha\beta} + N_{\alpha \mid \beta } + N_{\beta \mid \alpha} \right)
\end{align}
The ${\mid}$ indicates covariant derivative on the two-surface. The lagrangian \eqref{L} is suitable for canonical analysis because it does not contain the time derivatives of the lapse and shift variables i.e. in the canonical analysis they appear as Lagrange multipliers. Their conjugate momenta $\pi_{0}$ and $\pi_{\alpha}$ vanish weakly, providing the following primary constraints of the theory:
\begin{align}
\label{SCnew}
\begin{aligned}
\Omega_{0} &= \pi_{0}  \approx 0\\
\Omega_{\alpha}  &= \pi_{\alpha} \approx 0
\end{aligned}
\end{align}

The basic fields are $g_{\alpha\beta}$ with their conjugate momenta $\pi^{\alpha\beta}$. The canonical hamiltonian can be worked out as
\begin{align}
H_{c} &= \int d^{2}x \left(\pi_{\mu} \dot{N}^{\mu} + \pi^{\alpha\beta} \dot{g}_{\alpha\beta} - {\cal{L}}\right) \nonumber\\
\label{canonHamiltMS}
&= \int d^{2}x \left( N^{\perp}{\cal{H}}_{\perp} + N^{\alpha}{\cal{H}}_{\alpha} \right)
\end{align}
where,
\begin{align}
\label{H2} 
\begin{aligned}
{\cal{H}}_{\perp} &=  g ^{-1/2} \left( \pi_{\alpha\beta}\pi^{\alpha\beta} - \frac{1}{2} \pi^{2}\right) - \left(g\right) ^{1/2}R\\
{\cal{H}}_{\alpha} &= - 2 \pi_{\alpha}{}^{\beta}{}_{\mid \beta }.
\end{aligned}
\end{align}
The total hamiltonian is given by,
\begin{align}
\label{HTMS}
H_T = H_c + \int d^2x ~\left[\lambda^0 \Omega_0 + \lambda^\alpha\Omega_\alpha \right]
\end{align}
where $\lambda^0$, $\lambda^\alpha$ are multipliers enforcing the primary constraints $\Omega_0$, $\Omega_\alpha$. The secondary constraints, found by time conserving the primary constraints, are
\begin{align}
\label{SC} 
\begin{aligned}
\Omega_{3} &= \lbrace \pi_0, H_T \rbrace = {\cal{H}}_{\perp} \approx 0\\
\Omega_{3+\alpha} &= \lbrace \pi_\alpha, H_T \rbrace = {\cal{H}}_{\alpha} \approx 0.
\end{aligned}
\end{align}
No further constraints are generated by this iterative procedure. Note that all the constraints are first class. So, following Dirac's hypothesis \cite{Dirac:Lectures}, the gauge generator can be written as
\begin{align}
\label{GG}
G = \int d^{2}x \left(\varepsilon^{0}\Omega_{0} + \varepsilon^{\alpha}\Omega_{\alpha} + \varepsilon^{3}\Omega_{3} + \varepsilon^{3+\alpha}\Omega_{3+\alpha}\right),
\end{align}
where $\varepsilon^0$, $\varepsilon^\alpha$, $\varepsilon^3$ and $\varepsilon^{3+\alpha}$ are gauge parameters. Now using our master equation \eqref{RB master 2} we get \cite{Mukherjee:2007yi}
\begin{align}
\label{IGP}
\begin{aligned}
\varepsilon^{0} \left(x\right) &= \left[  \dot{\varepsilon}^{3} + \varepsilon^{3+\alpha} \partial_{\alpha}N^{\bot} - N^{\alpha} \partial_{\alpha}\varepsilon^{3}\right] \left(x\right)\\
\varepsilon^{\alpha} \left(x\right)&= \left[  \dot{\varepsilon}^{3+\alpha} + \varepsilon^{3+\beta} \partial_{\beta}N^{\alpha} - N^{\beta}\partial_{\beta} \varepsilon^{3+\alpha} - N^{\bot}g^{\beta\alpha}\partial_{\beta}\varepsilon^{3} + \varepsilon^{3}g^{\beta\alpha}\partial_{\beta}N^{\bot}\right] \left(x\right),
\end{aligned}
\end{align}
which shows that only three gauge parameters $\left( \varepsilon^{3}, \varepsilon^{3+\alpha}\right)$ are independent. Their number is equal to the number of primary first class constraints, in conformity with the discussion below \eqref{StructureConsts}. Also, this number matches with the number of diffeomorphism parameters $\xi^{\mu}$ (see \eqref{Diff}).

The mapping between the gauge and diff parameters is now found by comparing the variations of $N^\perp$, $N^\alpha$ and $g_{\alpha\beta}$ under both these symmetry operations. First, we consider the gauge variations which are found by Poisson bracketing with the generator, 
\begin{align}
\label{GV1}
\delta N^{\bot}\left(x \right) &= \lbrace N^\perp(x),G \rbrace = \left[\dot{\varepsilon}^{3} + \varepsilon^{3+\alpha} \partial_{\alpha}N^{\bot} - N^{\alpha} \partial_{\alpha}\varepsilon^{3}\right] \left(x\right)\\
\delta N^{\alpha}\left(x \right) &=  \left\{N^{\alpha}\left(x\right), G\right\}\nonumber\\
\label{GV2}
&= \left[  \dot{\varepsilon}^{3+\alpha} + \varepsilon^{3+\beta} \partial_{\beta}N^{\alpha} - N^{\beta}\partial_{\beta} \varepsilon^{3+\alpha} - N^{\bot}g^{\beta\alpha}\partial_{\beta}\varepsilon^{3} + \varepsilon^{3}g^{\beta\alpha}\partial_{\beta}N^{\bot}\right] \left(x\right)\\
\delta g_{\alpha\beta} \left( x \right)  &=  \left\lbrace g_{\alpha\beta} \left( x \right), G \right\rbrace \nonumber\\ 
\label{gGV}
&= -2 \varepsilon^{3} K_{\alpha\beta} +  \varepsilon^{3+\gamma} \partial_{\gamma}g_{\alpha\beta} + g_{\gamma\alpha}\partial_{\beta} \varepsilon^{3+\gamma} + g_{\gamma\beta}\partial_{\alpha} \varepsilon^{3+\gamma} 
\end{align}
The variation under general coordinate transformations or diff can be worked out after a bit of calculation \cite{Mukherjee:2007yi}. The desired variations are:
\begin{align}
\label{R1}
\delta N^{\bot}\left(x \right) &= \left(\frac{d}{dt} - N^{\alpha}\partial_{\alpha}\right) \xi^{0}N^{\bot}+\xi^{0}N^{\alpha}\partial_{\alpha}N^{\bot} + \xi^{\alpha}\partial_{\alpha}N^{\bot} \\
\label{R2}
\delta N^{\alpha}\left(x \right) &= \left(\frac{d}{dt} - N^{\beta}\partial_{\beta} \right) \left( \xi^{\alpha} + \xi^{0}N^{\alpha}\right)+ \left(\xi^{\beta} + \xi^{0}N^{\beta}\right) \partial_{\beta}N^{\alpha} - \left( N^{\bot} \right)^{2} g^{\alpha\beta} \partial_{\beta} \xi^{0} \\
\label{gR}
\delta g_{\alpha\beta} \left( x \right)  &=  \left( \xi^{0} \frac{d}{dt}  - \xi^{\gamma} \partial_{\gamma} \right) g_{\alpha\beta}  + N_{\alpha} \partial_{\beta} \xi^{0} + N_{\beta} \partial_{\alpha} \xi^{0} + g_{\gamma\alpha}\partial_{\beta} \xi^{\gamma} + g_{\gamma\beta}\partial_{\alpha} \xi^{\gamma}
\end{align}
Now comparing, for instance, \eqref{GV2} and \eqref{R2}, we can establish the mapping between the independent gauge and diffeomorphism parameters as 
\begin{align}
\label{Munattended}
\varepsilon^{3+\alpha} &= \xi^{\alpha} +  \xi^{0}N^{\alpha}\\
\label{M}
\varepsilon^{3} &= N^{\bot}\xi^{0}
\end{align}
This mapping, when substituted in the gauge variations of the basic fields $N^{\bot}$ and $g_{\alpha\beta}$ (equations \eqref{GV1} and \eqref{gGV} respectively) transforms them identically to their corresponding reparametrization variations i.e. equations \eqref{R1} and \eqref{gR}. The equivalence of the gauge and diffeomorphism symmetries is thus established. Note that this is an off-shell equivalence, established without invocation of the equations of motion.

To compare the above result with the PGT formulation, we now focus our attention on \eqref{CmbMBaction}. By chosing $a = 1$ and $\Lambda=\alpha_3=\alpha_4=0$ (for which $p=q=0$), it reduces to
\begin{align}
\label{FEinsteinaction}
S = \int \,d^3x\, \epsilon^{\mu\nu\rho} \,b^i_{\ \mu} \,R_{i\nu\rho}.
\end{align}
This is equivalent to the action \eqref{Einsteinaction} used in this section for the second order metric analysis, as can be verified by using the identity \eqref{theidentity} and the relation $g_{\mu\nu} = b^i_{\ \mu} b^j_{\ \nu} \eta_{ij}$. We wish to compare the gauge variations and the PGT variations for the theory \eqref{FEinsteinaction}. Referring back to \eqref{field transf gauge} and noting $p=q=0$, the gauge variations read:
\begin{align}
\label{field transf gauge new}
\begin{aligned}
\delta b^i_{\ \mu}(x)  &= \partial_\mu\epsilon^i(x) + \epsilon^i_{\ jk} \,\omega^j_{\ \mu}(x) \,\varepsilon^k(x) + \epsilon^i_{\ jk}\,b^j_{\ \mu}(x) \,\tau^k(x),\\
\delta \omega^i_{\ \mu}(x)  &= \partial_\mu \tau^i(x) + \epsilon^i_{\ jk}\,\omega^j_{\ \mu}(x) \,\tau^k(x).\\
\end{aligned}
\end{align}
Comparison with the PGT transformations \eqref{Poincare3Dfieldtrans} shows that there is still no off-shell correspondence between the two transformations.

In addition to our comparison here, work has also been done to compare the ADM formulation with a first order calculation of the Einstein-Hilbert action in metric $g_{\mu\nu}$ and a general (symmetric) connection $\tilde{\Gamma}^\mu_{\ \nu\rho}$ \cite{Kiriushcheva:2006gp}, leading to similar conclusions. The symmetries obtained from a first-order calculation without any ADM-type decomposition of fields, leads to symmetries that are not recognisable as the standard diffeomorphism/Poincar\'{e} symmetries. Clearly, the PGT framework is distinct from the conventional one as far as the treatment of symmetries is concerned.


\section{Discussions}
\label{Cmb:Disc}

Recently the 3D gravity models in the framework of Poincare gauge theory (PGT) have come to forefront \cite{Blagojevic:2008bn, Blagojevic:2004hj, Blagojevic:2002du, Carlip:2005zn, Park:2008yy, Grumiller:2008pr, Carlip:2008qh} in the literature. Among the various issues considered, a particularly significant one is the difference between the PGT transformations of the basic fields and the gauge variations of the same obtained in the canonical way. The two can only be mapped using the equations of motion. This fact was observed earlier \cite{Blagojevic:2004hj} but its significance was missed, principally due to the fact that the canonical gauge generator was constructed following \cite{Castellani:1981us} which maps solutions to solutions of the equations of motion. We have shown here, in the context of the Mielke-Baekler \cite{Mielke:1991nn} type topological 3D gravity with torsion, that the general gauge transformations can be obtained in the canonical way in an off-shell manner. This is done by following a method available in the literature \cite{Banerjee:1999hu, Banerjee:1999yc} that views the gauge transformations as mapping field configurations to field configurations. This naturally lends a new perspective to this issue of symmetries. In fact, we have also provided a new and compact derivation of this method of construction of the gauge generator, which highlights its off-shell nature.

The off-shell invariance of the model under PGT transformations has been explicitly verified. Then a complete canonical analysis of the model was presented. This model presents an example of a mixed constrained system with both first and second class constraints. The novelty of our approach, as against the constraint analysis done in \cite{Blagojevic:2004hj}, is the use of a reduced phase space. Using Dirac brackets, we completely eliminate the second class sector. The model then transpires to a gauge system having only first class constraints. The difference is that the symplectic structure is defined by the Dirac brackets instead of the usual Poisson brackets. We find the transformations of the basic fields by computing their Dirac brackets with the gauge generator and check by direct calculation that these gauge transformations are again  {\it{off-shell}} invariances of the action. The gauge transformations of the basic fields are then compared with the analogous transformations under PGT. There exists no off-shell map between them, though the two agree on-shell. Thus, at this point, we are still in an acute conundrum. We see that there are (apparently) two different sets of symmetries of the same action, each complete and independent of each other. This violates all expectations from the point of view of a proper gauge theory, as was discussed earlier.

To put our findings in a proper perspective, we carry out a similar analysis for 2+1 dimensional Einstein gravity in the usual metric formulation. In this case we prove an exact off-shell equivalence of the general coordinate (diff) transformations with the gauge transformations found by a canonical (hamiltonian) approach. This clearly manifests the peculiarity of the PGT vis-a-vis a standard gauge theory.

As a final remark, we mention that the methods developed here may be applied to other 3D gravity models \cite{Blagojevic:2008bn, Park:2008yy, Grumiller:2008pr, Carlip:2008qh}. In the next chapter, we will analyse the symmetries for a massive gravity model, one which has propagating degree of freedom unlike the present case. A fact that may be stressed once again here, is that, the hamiltonian method gives symmetries that are very much model dependent. However, the PGT symmetries are model-independent.


%% file: C5_bht.tex
\chapter{Symmetries in a massive theory of gravity}
\label{C:bht}

\lettrine[lraise=0.0, loversize=0.3, findent=3pt, nindent=0pt]{I}n the previous chapter, we had shown how gauge symmetries constructed out of first-class constraints of a Mielke-Baekler type 3D gravity model with cosmological constant turned out to be algebraically distinct from the underlying Poincar\'{e} symmetries of the theory. The result being a puzzle, due to two complete distinct sets of gauge symmetries appearing for the same action, we take up analysis of another model here in this chapter, adopting identical methodology. We are using a completely off-shell algorithm to construct a hamiltonian gauge generator out of the first-class constraints of the theory. The particular model that we adopt is a Poincar\'{e} gauge theory (PGT), first-order version of the 3D `new massive gravity' (NMG) action, proposed recently \cite{Bergshoeff:2009hq, Bergshoeff:2009aq}.\footnote{Taking cue from \cite{Blagojevic:2010ir}, we call this the BHT model after the original proponents Bergshoeff, Hohm and Townsend.} The PGT formulation was given in \cite{Blagojevic:2010ir}. It differs from the Mielke-Baekler type model, from the previous chapter, in that it has free local degrees of freedom \cite{Blagojevic:2010ir}.


\section{Poincar\'{e} gauge theory formulation of new massive gravity}
\label{Cbht:intro}

A unitary, renormalizable theory of gravity with propagating degree(s) of freedom is a long sought goal towards our understanding of gravitation. Recently, such a proposal (`new massive gravity' or `BHT gravity' \cite{Bergshoeff:2009hq, Bergshoeff:2009aq}) with massive propagating modes in 3 spacetime dimensions has generated much interest \cite{Deser:2009hb, Oda:2009ys, Clement:2009gq, Gullu:2010sd, Gullu:2010pc, Blagojevic:2010ir, Perez:2011qp, Jatkar:2011ue, Ahmedov:2010em}, particularly with emphasis on its symmetries \cite{Blagojevic:2010ir}. A massive spin 2 description is quite standard. At the linearised level of Einstein gravity, we have the non-interacting Fierz-Pauli (FP) model \cite{Fierz:1939ix} in any dimension. It is unitary, and in 3D has two massive degrees of freedom. Also in 3D, addition of a Chern-Simons term (in the connection variables) gives the topologically massive gravity (TMG) model \cite{Deser:1982vy,Deser:1981wh}. This theory violates parity and has one propagating degree of freedom. The BHT gravity is unitary and can give an interacting theory at the non-linear level, unlike the FP model. It is given by the action
\begin{align}
\label{action BHT simple}
S = \frac{1}{\kappa^2} \int d^3x ~\sqrt{g}\left[ R + \frac{1}{m^2} K \right],
\end{align}
where $R$, the Ricci scalar is the contraction of the Ricci tensor ($R^\mu_{\ \mu}$) and $K=R_{\mu\nu}R^{\mu\nu} - \tfrac{3}{8}R^2$. The BHT model can be both motivated as a non-linear generalization of the FP model, or a soldering of two TMG massive modes \cite{Dalmazi:2009es, Bergshoeff:2009fj, Banerjee:2012kq}. The interesting point to note is that the model is unitary in spite of fourth order derivatives present in the action, through the $K$ term. 

Now, interest in 3D gravity is fuelled by studies on the  the AdS/CFT correspondence and towards understanding fundamental problems such as entropy of the (BTZ-type) black-hole solutions. The BHT action \eqref{action BHT simple} can incorporate a cosmological term to give the action
\begin{align}
\label{action BHT cosmo}
S = \frac{1}{\kappa^2} \int d^3x ~\sqrt{g}\left[\sigma R + \frac{1}{m^2} K - 2 \lambda m^2 \right].
\end{align}
The unitarity and stability of this model depends, in general, on the choice of parameters, and unitarity has been studied in different regions of the parameter space (see, for example, \cite{Bergshoeff:2009aq}).

A consistent canonical constraint analysis of BHT gravity has been carried out in \cite{Sadegh:2010pq,Blagojevic:2010ir}. In \cite{Blagojevic:2010ir}, this was done in the first-order formulation through the Poincar\'{e} gauge theory (PGT) construction \cite{Utiyama:1956sy, Kibble:1961ba,1962rdgr.book..415S, Hehl:1976kj,Blagojevic:2002du}. However the gauge generator (from which transformation of the basic fields are obtained) is constructed by an on-shell algorithm due to Castellani \cite{Castellani:1981us}. As we have noted before, this view of symmetries is restricted. It views symmetries as on-shell maps, between solutions to solutions of the equations of motion, rather than as a map between field configurations.

Here, we construct the {\em off-shell} gauge generator of cosmological BHT gravity \cite{Bergshoeff:2009aq} through a hamiltonian algorithm following \cite{Banerjee:1999yc,Banerjee:1999hu}, based on the total hamiltonian approach.\footnote{See \cite{Henneaux:1990au} for a treatment of symmetries using the extended hamiltonian formulation.} This procedure has been used recently in the context of diffeomorphism symmetry in string theory \cite{Banerjee:2004un,Banerjee:2005bb}, second order metric gravity \cite{Mukherjee:2007yi}, interpolating formulation of bosonic string theory \cite{Gangopadhyay:2007gn}, and also in topological gravity with torsion in (2+1)-dimensions \cite{Banerjee:2009vf}, described in the previous chapter.

The derivation of the generator in the present case of BHT gravity is more subtle. The theory contains second-class constraints which have not been removed completely through Dirac brackets. This is in complete contrast to earlier examples where the theories either comprised of first-class constraints and/or second-class constraints which were totally removed through Dirac brackets \cite{Banerjee:2009vf}.

After constructing the generator, we will give explicit expression of the symmetries of the basic fields. It will be shown that these symmetries can be mapped to the underlying Poincar\'{e} symmetries through a field dependent map between gauge parameters. This mapping is only possible \emph{on-shell}, i.e. upon imposition of the equations of motion. In particular, we show that the symmetry of the triad field $b^i_{\ \mu}$, which is related to the metric through $g_{\mu\nu}=b^i_{\ \mu}b^j_{\ \nu}\,\eta_{ij}$, is identifiable with the Poincar\'{e} symmetries upon imposition of an equation of motion that implies zero torsion. This is interesting, as it is precisely the condition of zero torsion that takes us from the Riemann-Cartan spacetime of PGT to the Riemannian spacetime in which BHT was originally formulated, adopting a usual metric formalism. We will also show that though this map is field dependent, it can be used in the generator both after and before computing symmetries, equivalently, to relate the two sets of symmetries.

\paragraph{Conventions:}Latin indices refer to the local Lorentz frame and the Greek indices refer to the coordinate frame. The beginning letters of both alphabets $(a,b,c,\ldots)$ and $(\alpha,\beta,\gamma,\ldots)$ run over the space part (1,2) while the middle alphabet letters $(i,j,k,\ldots)$ and $(\mu,\nu,\lambda,\ldots)$ run over all coordinates (0,1,2). The totally antisymmetric tensor $\varepsilon^{ijk}$ and the tensor density $\varepsilon^{\mu\nu\rho}$ are both normalized so that $\varepsilon^{012}=1$. The signature of space-time adopted here is $\eta=(+,-,-)$.


\section{Canonical analysis of the model}
\label{Cbht:canAnalysis}

We begin our analysis with first order form of BHT massive gravity, written in accordance with the PGT formalism, where the basic variables are the triads $b^i_{\ \mu}$ and spin connections $\omega^i_{\ \mu}$ \cite{Blagojevic:2010ir}. The formulation of PGT starts on a globally flat space (here $3D$) with a local set of orthogonal coordinates at each point. Any global field $A^\mu$ is written in terms of these local coordinates $A^i$ by a set of vielbein fields `$b$' (triads) as $A^\mu(x) = b^i_{\ \mu}(x) A_i(x)$. Let us recall from Chapter \ref{C:pgt}, that the Lagrangian is made invariant under local Poincar\'{e} transformations by covariant derivatives $\nabla_\mu = \partial_\mu + \texttt{Conn}_\mu$, using compensating connection fields `$\texttt{Conn}$'. The respective field strengths, defined through the commutator of the covariant derivatives, are the Riemann tensor $R^i_{\ \mu\nu}$ and the torsion $T^i_{\ \mu\nu}$:
\begin{align}
\label{PGT R T}
\begin{aligned}
R^i_{\mu\nu} &= \partial_\mu \omega^i_{\ \nu} - \partial_\mu \omega^i_\nu + \epsilon^i_{\ jk}\omega^j_{\ \mu}\omega^k_{\ \nu} \\
T^i_{\ \mu\nu} &= \nabla_\mu b^i_{\ \nu} - \nabla_\nu b^i_{\ \mu}.
\end{aligned}
\end{align}
Here the covariant derivative of the triad is defined as $\nabla_\mu  b^i_{\ \nu} = \partial_\mu b^i_{\ \nu} + \epsilon^i_{\ jk}\omega^j_{\ \mu}b^k_{\ \nu}$, with $\omega^j_{\ \mu}$ being the `spin connections' arising out of the connection part $\texttt{Conn}$ of the covariant derivative. The spacetime naturally occurring in this construction is thus the {\em Riemann-Cartan} spacetime with non-zero torsion. We also recall the transformation of the basic fields under the Poincar\'{e} transformations:
\begin{align}
\label{PGT deltas}
\begin{aligned}
\delta_{\scriptscriptstyle PGT} b^i_{\ \mu} &= -\epsilon^i_{\ jk}b^j_{\ \mu}\theta^k - \partial_\mu \xi^\rho \,b^i_{\ \rho} - \xi^\rho\,\partial_\rho b^i_{\ \mu} \\
\delta_{\scriptscriptstyle PGT} \omega^i_{\ \mu} &= -\partial_\mu \theta^i - \epsilon^i_{\ jk}\omega^j_{\ \mu}\theta^k - \partial_\mu\xi^\rho\,\omega^i_{\ \rho} - \xi^\rho\,\partial_\rho\omega^i_{\ \mu}.
\end{aligned}
\end{align}
In the above symmetries, the parameter describing local Lorentz transformations is $\theta^i(x)$ and that describing general coordinate transformations is $\xi^\mu$, both transformations being of infinitesimal order. It is to be noted that the nature of these transformations depend on the behaviour of a field under the action of the Lorentz group. Thus any field having the general nature of the triad field $b^i_{\ \mu}$, i.e. which transforms as a vector in both spaces, should have the same transformations as given above in \eqref{PGT deltas}. In particular, we list the transformations of two fields `$\lambda$' and `$f$' which will be required later in this article,
\begin{align}
\label{PGT deltas lambda f}
\begin{aligned}
\delta_{\scriptscriptstyle PGT} \lambda^i_{\ \mu} &= -\epsilon^i_{\ jk}\lambda^j_{\ \mu}\theta^k - \partial_\mu \xi^\rho \,\lambda^i_{\ \rho} - \xi^\rho\,\partial_\rho \lambda^i_{\ \mu} \\
\delta_{\scriptscriptstyle PGT} f^i_{\ \mu} &= -\epsilon^i_{\ jk}f^j_{\ \mu}\theta^k - \partial_\mu \xi^\rho \,f^i_{\ \rho} - \xi^\rho\,\partial_\rho f^i_{\ \mu}.
\end{aligned}
\end{align}

The first-order BHT model we work with, to begin with, contains the usual Einstein-Hilbert piece along with a cosmological term. Now, PGT is formulated on the Riemann-Cartan spacetime, with both curvature and torsion, whereas BHT gravity was originally formulated in the Riemann spacetime with zero torsion. To be able to enforce this condition, torsion is included in the action via coupling to a Lagrange multiplier field $\lambda^i_{\ \mu}$. The distinctive term of the BHT theory which contains the square of curvature is incorporated into the action with the help of an auxiliary field, such that the action is rendered linear in curvature. On imposition of the equation of motion for $f^i_{\ \mu}$, the curvature squared term of original BHT is recovered. The lagrangian, with all the above described terms and their individual coupling parameters, take the following form:
\begin{align}
\label{lagrangian}
\mathcal{L}=a\epsilon^{\mu\nu\rho}\left( \sigma b^i_{\ \mu}R_{i\nu\rho}-\frac{\Lambda}{3}\epsilon_{ijk}b^i_{\ \mu}b^j_{\ \nu}b^k_{\ \rho} \right) + \frac{a}{m^2} \mathcal{L}_K + \frac{1}{2} \epsilon^{\mu\nu\rho}\lambda^i_{\ \mu}T_{i\nu\rho}.
\end{align}
Here $R_{i\nu\rho}$ and $T_{i\nu\rho}$ are the Riemann tensor and torsion defined earlier, while $\mathcal{L}_K$ is defined as:
\begin{align}
\label{L_K}
\mathcal{L}_K&=\frac{1}{2} \epsilon^{\mu\nu\rho}f^i_{\ \mu}R_{i\nu\rho}-b\mathcal{V}_K\nonumber\\
\mathcal{V}_K&=\frac{1}{4}\left(f_{i\mu}f^{i\mu}-f^2\right),
\end{align}
where $b$ denotes the determinant of the basic triad field $b^i_{\ \mu}$. The equations of motion corresponding to variations with respect to the basic variables $b^i_{\ \mu}$, $\omega^i_{\ \mu}$, $f^i_{\ \mu}$ and $\lambda^i_{\ \mu}$, respectively, are given below:
\begin{subequations}
\label{EOM}
\begin{align}
\label{EOM b}
& a \epsilon^{\mu \nu \rho} \left( \sigma R_{i\nu\rho} - \Lambda \epsilon_{ijk}b^j_{\ \nu}b^j_{\ \rho} \right) - \frac{ab}{m^2} \,\mathcal{T}_i^{\ \mu} + \epsilon^{\mu \nu \rho} \nabla_\nu \lambda_{i\rho} - \frac{ab}{2m^2}\,\Theta_{ij}\left( f^{j\mu}-b^{j\mu} \right) = 0\\
\label{EOM omega}
& \epsilon^{\mu\nu\rho} \left[ a\sigma T_{i\nu\rho} + \frac{a}{m^2} \nabla_\nu f_{i\rho} + \epsilon_{ijk} b^j_{\ \nu} \lambda^k_{\ \rho} \right] = 0\\
\label{EOM f}
& \frac{a}{2m^2} \left[ \epsilon^{\mu\nu\rho}R_{i\nu\rho} - b\left( f_i^{\ \mu} - f b_i^{\ \mu} \right) \right] = 0 \\
\label{EOM lambda}
& \frac{1}{2} \,\epsilon^{\mu\nu\rho} \, T_{i\nu\rho} = 0.
\end{align}
\end{subequations}
Here $f=f^h_{\ \rho}\,b_h^{\ \rho}$ is the trace of the field $f^i_{\ \mu}$ and $\mathcal{T}_i^{\ \mu}$ is defined as:
\begin{align}
\mathcal{T}_i^{\ \mu} = b_i^{\ \mu} \mathcal{V}_K - \frac{1}{2} \left( f_{ik}f^{k\mu} - f f_i^{\ \mu} \right).
\end{align}
The term $\Theta_{ij} = f_{ij} - f_{ji}$ is proportional to the antisymmetric part of the field $f^i_{\ \mu}$. Similarly, we can define an antisymmetric combination from $\lambda^i_{\ \mu}$ as $\Psi_{ij}=\lambda_{ij}-\lambda_{ji}$. The equations of motion however show that both fields $f_{ij}$ and $\lambda_{ij}$ are symmetric \cite{Blagojevic:2010ir}. Hence $\Theta_{ij} = \Psi_{ij} = 0$. Later, in this section itself, we see that $\Theta_{ij}$ and $\Psi_{ij}$ appear as constraints of the theory. Thus the symmetry of the auxiliary fields is also a result of the constraint structure and does not involve a true equation of motion (involving accelerations).

Next, we summarize the hamiltonian description of the theory along with a proper identification of the constraints {\em \`{a} la} Dirac, following \cite{Blagojevic:2010ir}. The analysis is done by treating the second-class sector in a mixed manner. First, a set of Dirac brackets is employed to eliminate a sector of the second-class constraints that arise from the primary sector, thus going into a reduced phase phase and eliminating some of the momenta. Subsequently, Lagrange multipliers corresponding to the remaining second class sector are fixed, in the reduced phase space.

To start with, the momenta corresponding to the basic fields are defined in the standard manner $p=\frac{\partial \mathcal{L}}{\partial \dot{q}}$.
\begin{table}[ht]
\centering
\begin{tabular}{l|cccc}
\hline\hline
Basic Field & $b^i_{\ \mu}$ & $\omega^i_{\ \mu}$ & $f^i_{\ \mu}$ & $\lambda^i_{\ \mu}$ \\[0.5ex] \hline
Conjugate Momenta & $\pi_i^{\ \mu}$ & $\Pi_i^{\ \mu}$ & $P_i^{\ \mu}$ & $p_i^{\ \mu}$ \\[0.5ex]\hline\hline
\end{tabular}
\caption{The basic fields and their corresponding momenta}
\label{Tab:Momenta}
\end{table}
Now, it turns out that all the momenta give rise to primary constraints. These are listed below:
\begin{subequations}
\label{primary}
\begin{align}
\phi_i^{\ \mu} &:= \pi_i^{\ \mu} - \epsilon^{0\alpha\beta} \lambda_{i\beta}\;\delta_\alpha^\mu \approx 0\\
\Phi_i^{\ \mu} &:= \Pi_i^{\ \mu} - 2a\epsilon^{0\alpha\beta} \left( \sigma b_{i\beta} + \frac{1}{2m^2}f_{i\beta} \right) \: \delta_\alpha^\mu \approx 0 \\
P_i^{\ \mu} &\approx 0 \; ; \qquad\qquad p_i^{\ \mu} \approx 0.
\end{align}
\end{subequations}
Among the above primary constraints, the set $\,X := \left( \pi_i^{\ \alpha}, \Pi_i^{\ \alpha}, P_i^{\ \alpha}, p_i^{\ \alpha} \right)$ are readily seen to be second-class in nature, and are eliminated by introducing an appropriate set of Dirac brackets, before going into the secondary stage. Consequently, the analysis is carried in a {\em reduced phase space} with a modified algebra, given below:
\begin{subequations}
\label{algebra1}
\begin{align}
\lbrace b^i_{\ \alpha}, \lambda^j_{\ \beta}\rbrace^* &= \eta^{ij}\epsilon_{0\alpha\beta}\\
\lbrace \omega^i_{\ \alpha}, f^j_{\ \beta} \rbrace^* &= \left( \frac{m^2}{a} \right) \eta^{ij} \epsilon_{0\alpha\beta}\\
\lbrace \lambda^i_{\ \alpha}, f^j_{\ \beta} \rbrace^* &= \left( -2m^2\sigma \right) \eta^{ij} \epsilon_{0\alpha\beta}.
\end{align}
\end{subequations}
The other brackets in this new algebra turn out to be same as the corresponding Poisson brackets. In particular, we note the following brackets, derivable using \eqref{algebra1} and using the inverse property of the triad field $b^i_{\ \mu}b_j^{\ \mu}=\delta^i_j$,
\begin{align}
\label{algebra2}
\begin{aligned}
\lbrace b_i^{\ \mu}, \pi_j^{\ \nu} \rbrace^* &= b_j^{\ \mu}b_i^{\ \nu}\\
\lbrace b_i^{\ \mu}, \lambda^j_{\ \nu} \rbrace^* &= \epsilon_{0\alpha\beta} \, b_i^{\ \beta}b^{j\mu}\:\delta^\alpha_\nu.
\end{aligned}
\end{align}
Since no Poisson brackets are employed in our analysis and {\em all} our brackets correspond to this reduced space algebra, we will henceforth drop the starred bracket notation and indicate the changed algebra with usual braces, i.e. $\lbrace,\rbrace^*:=\lbrace,\rbrace$.

So now, we are left with the primary momenta $\left( \pi_i^{\ 0}, \Pi_i^{\ 0}, P_i^{\ 0}, p_i^{\ 0} \right)$, all of which are primary constraints. Also, we are working in a reduced phase space with a changed algebra given by \eqref{algebra1}. The canonical hamiltonian, defined as $\mathcal{H}_C=p\dot{q}-\mathcal{L}$, after some rearrangements is given by:
\begin{align}
\label{H_C}
\mathcal{H}_C = b^i_{\ 0}\hat{\mathcal{H}}_i + \omega^i_{\ 0}\mathcal{K}_i + f^i_{\ 0}\hat{\mathcal{R}}_i + \lambda^i_{\ 0}\mathcal{T}_i + \partial_\alpha \mathcal{D}^\alpha,
\end{align}
where the relevant quantities are defined below:-
\begin{align}
\label{H_C defns}
\begin{aligned}
\hat{\mathcal{H}}_i &= \mathcal{H}_i + \frac{a}{m^2}\: b \: \mathcal{T}_i^{\ 0} \\
\mathcal{H}_i &= \epsilon^{0\alpha\beta} \left( a\sigma R_{i\alpha\beta} - a\Lambda\:\epsilon_{ijk}b^j_{\ \alpha}b^k_{\ \beta} + \nabla_\alpha \lambda_{i\beta} \right)\\
\mathcal{K}_i &= -\epsilon^{0\alpha\beta} \left( a\sigma T_{i\alpha\beta} + \frac{a}{m^2}\:\nabla_\alpha f_{i\beta} + \epsilon_{ijk} b^j_{\ \alpha} \lambda^k_{\ \beta} \right)\\
\mathcal{R}_i &= -\frac{a}{2m^2}\,\epsilon^{0\alpha\beta}\,R_{i\alpha\beta}\\
\hat{\mathcal{R}}_i &= \mathcal{R}_i + \frac{a}{2m^2} b \left( f_i^{\ 0} - f b_i^{\ 0} \right) \\
\mathcal{T}_i &= -\frac{1}{2}\:\epsilon^{0\alpha\beta}\,T_{i\alpha\beta} \\
\mathcal{D}^\alpha &= \epsilon^{0\alpha\beta} \left[ \omega^i_{\ 0}\left( 2a\sigma b_{i\beta} + \frac{a}{m^2}f_{i\beta} \right) + b^i_{\ 0}\lambda_{i\beta} \right].
\end{aligned}
\end{align}
The total hamiltonian density in the reduced phase space may be defined as the canonical hamiltonian plus all primary constraints that have not been eliminated, i.e.
\begin{align}
\label{H_T 2}
\mathcal{H}_T &= \mathcal{H}_C + u^i_{\ 0}\phi_i^{\ 0} + v^i_{\ 0} \Phi_i^{\ 0} + w^i_{\ 0}p_i^{\ 0} + z^i_{\ 0}P_i^{\ 0}.
\end{align}
Subsequent analysis yields constraints till the quartic stage and fixing of the multipliers $w^i_{\ 0}$ and $z^i_{\ 0}$. The final classified constraint structure in our reduced space is presented in Table \ref{CbhtTab:Constraints}
\begin{table}[ht]
\caption{Constraints under the modified algebra classified.}
\label{CbhtTab:Constraints}
\centering
\begin{tabular}{l c c}
\hline\hline\
& First Class & Second class \\[0.2ex] \hline\\[-1.9ex]
Primary & $\Sigma_{(1) i} = \pi''{}_i^{\ 0}\;, \;\;\Sigma_{(2) i} = \Pi_i^{\ 0}$ & $p_i^{\ 0}$, $P_i^{\ 0}$ \\[0.4ex]
Secondary & $\Sigma_{(3) i} = \bar{\mathcal{H}}_i\;, \;\;\Sigma_{(4) i} = \bar{\mathcal{K}}_i$ & $\mathcal{T}_i$, $\hat{\mathcal{R}'}_i$ \\[0.4ex]
Tertiary & & $\Theta_{ij}$, $\Psi_{ij}$ \\[0.4ex]
Quartic & & $\chi$, $\varphi$ \\[0.4ex]
\hline\hline
\end{tabular}
\end{table}
and the required quantities are defined below:
\begin{align}
\label{Constrnt defns}
\begin{aligned}
\pi''{}_i^{\ 0} &:= \pi_i^{\ 0} + f_i^{\ l}P_l^{\ 0} + \lambda^l_{\ i}\,p_l^{\ 0}\\
\bar{\mathcal{H}}_i &:= \hat{\mathcal{H}}_i + f^l_{\ i}\hat{\mathcal{R}}_l + \lambda^l_{\ i}\mathcal{T}_l + b_i^{\ \rho}(\nabla_\rho \lambda_{jk})b^k_{\ 0}\,p^{j0} + b_i^{\ \rho}(\nabla_\rho f_{jk})b^k_{\ 0}\,P^{j0}\\
\bar{\mathcal{K}}_i &:= \mathcal{K}_i - \epsilon_{ijk}\left( \lambda^j_{\ 0}\,p^{k0} - b^j_{\ 0}\lambda^k_{\ l}\,p^{l0} \right) - \epsilon_{ijk}\left( f^j_{\ 0}\,P^{k0} - b^j_{\ 0} f^k_{\ l}\,P^{l0} \right)\\
\varphi &:= \sigma f + 3\Lambda_0 + \frac{1}{2m^2}\:\mathcal{V}_K\\
\chi &:= \lambda^i_{\ \mu}b_i^{\ \mu} = \lambda.
\end{aligned}
\end{align}
Terms with determined multipliers can now be added to the canonical hamiltonian density $\mathcal{H}_C$ to form a new quantity, often denoted as $\mathcal{H}^{(1)}$
\begin{align}
\label{H1 defn}
\mathcal{H}^{(1)} &:= \mathcal{H}_C + \text{sum of primary second-class constraints with determined multipliers}\nonumber\\
&= b^i_{\ 0}\bar{\mathcal{H}}_i + \omega^i_{\ 0}\bar{\mathcal{K}}_i
\end{align}
The total hamiltonian density now becomes:
\begin{align}
\label{H_T 3}
\mathcal{H}_T &= \mathcal{H}^{(1)} + \text{sum of all primary first-class constraints with arbitrary multipliers}\nonumber\\
&= b^i_{\ 0}\bar{\mathcal{H}}_i + \omega^i_{\ 0}\bar{\mathcal{K}}_i + u^i_{\ 0}\pi''{}_i^{\ 0} + v^i_{\ 0}\Pi_i^{\ 0}.
\end{align}

The degrees of freedom can now be counted. We note that our reduced phase space has $(2N)' = 2 \times 36 - 24 = 48$ variables after fixing the $24$ momenta in sector $X$. There are $n_1 = 12$ first-class constraints and the number of second-class constraints is $n_2 = 20$. Thus the number of phase space degrees of freedom is
\begin{align}
\#\text{DoF} &= (2N)' - 2 \times n_1 - n_2 \nonumber\\
&= 48 - 24 - 20 = 4,
\end{align}
which makes the configuration space degree of freedom to be $2$ locally.

In the next section we see, that, it is the modified hamiltonian density $\mathcal{H}^{(1)}$ which becomes useful in our construction of symmetry generators of this mixed-model, with both first and second-class sectors. It plays a part analogous to that played by $\mathcal{H}_C$ in systems with only first-class sector or systems where the second-class sector is {\em completely} eliminated using Dirac brackets.


\section{Gauge generator and off-shell symmetries}
\label{Cbht:Generator}

In this section we proceed to systematically construct an off-shell generator of the model \eqref{lagrangian} following the method shown in \cite{Banerjee:1999yc, Banerjee:1999hu}. Let us denote the relevant (first-class) constraints in our theory (see Table \ref{CbhtTab:Constraints}) as:
\begin{align}
\label{CbhtRB const1}
\Sigma_{(I)} = \left[\Sigma_{(A)};\Sigma_{(Z)}\right],
\end{align}
where $A=1,2$ are primary (first class) constraints, $Z=3,4$ secondary (first class) constraints and $I=1,2,3,4$ constitute all (first class) constraints. The total hamiltonian density \eqref{H_T 3} may then be written as
\begin{align}
\label{H_T chi}
\mathcal{H}_T = \mathcal{H}^{(1)} + \chi^{(A)} ~\Sigma_{(A)},
\end{align}
with the notation $\chi^{(1)}=u^i_{\ 0}$ and $\chi^{(2)}=v^i_{\ 0}$.

By a gauge generator we mean a field dependent quantity $G$, such that for any quantity $F$ which is a function of the basic fields, the bracket $\lbrace F, G \rbrace$ gives the variation $\delta F$ consistent with the variations of the basic fields. In particular we then have
\begin{align}
\label{G q defn}
\delta q = \lbrace q, G \rbrace.
\end{align}
Now, the Dirac prescription for the generator is to consider a linear combination of {\em all} first class constraints
\begin{align}
\label{gen G}
G = \int d^2x ~\varepsilon^{(I)}\Sigma_{(I)}
\end{align}
where $\varepsilon^{(I)}$ are the gauge parameters. However, not all of these are independent. We have to now eliminate the dependent parameters and write the generator in terms of the independent gauge parameters alone.

We start by noting that the gauge variations are not completely arbitrary, but must commute with time derivatives i.e.
\begin{align}
\label{dotdelta}
\left( \delta \bullet \frac{d}{dt} \right) q \equiv \left( \frac{d}{dt} \bullet \delta \right) q,
\end{align}
where $\dfrac{d}{dt} q = \lbrace q, \int \mathcal{H}_T\rbrace$. Both sides of \eqref{dotdelta} can be evaluated separately using the generator \eqref{gen G} and the total hamiltonian \eqref{H_T 3}. The generator is composed of the first class constraints and the total hamiltonian density is the sum of $\mathcal{H}^{(1)}$ and the primary first class constraints. So the algebrae required will be those in-between the first class constraints and that of the first class sector with $\mathcal{H}^{(1)}$. For calculation, we introduce some structure functions and calculate these required algebrae.

By a theorem due to Dirac \cite{Dirac:Lectures} the first-class constraints must close amongst themselves, i.e.
\begin{align}
\label{C_defn}
\left\lbrace\Sigma_{(I) i}(x),\Sigma_{(J) j}(x')\right\rbrace = &\int d^2x''\,{\left(C^K_{\;\;\, IJ}\right)}_{ijk}(x'',x,x') ~\Sigma_{(K)}^{\quad k}(x'').
\end{align}
Also note that $\mathcal{H}^{(1)}$ \eqref{H1 defn} is a first-class quantity as the total hamiltonian must be first-class. This is analogous to the first-class nature of the canonical hamiltonian in a system with only first-class constraints. So we must have
\begin{align}
\label{V_defn}
\left\lbrace \int \mathcal{H}^{(1)},\Sigma_{(I)i}(x)\right\rbrace = &\int d^2x'\,{\left(V^J_{\;\;\: I}\right)}_{ik}(x',x)~\Sigma_{(J)}^{\quad k}(x').
\end{align}
Using the above definitions \eqref{C_defn} and \eqref{V_defn} in \eqref{dotdelta}, we reach the following set of equations relating the gauge parameters (see Section \ref{Cmb:ConstraintsRev} and \cite{Banerjee:1999hu, Banerjee:1999yc}):
\begin{align}
\label{CbhtRB master 1}
\delta\chi^{(A)}(x) = \displaystyle\frac{d\varepsilon^{(A)}(x)}{dt} - \int d^2x' ~\varepsilon^{(I)}(x') \,&\left[ \left(V^A_{\ I}\right)(x,x')\right.\nonumber\\
&\left.\ +\int d^2x''\,\chi^{(B)}(x'') ~\left(C^A_{\ IB}\right)(x,x',x'')\right]\\
\label{CbhtRB master 2}
0 = \displaystyle\frac{d\varepsilon^{(Z)}(x)}{dt} - \int d^2x' ~\varepsilon^{(I)}(x') &\left[ \left(V^Z_{\ I}\right)(x,x')\right.\nonumber\\
&\left.\ +\int d^2x''\,\chi^{(B)}(x'') \,\left(C^Z_{\ IB}\right)(x,x',x'')\right].
\end{align}
Among them, the second condition makes it possible to choose $(A)$ independent gauge parameters from the set $\varepsilon^{(I)}$ and express the generator $G$ (\ref{gen G}) entirely in terms of them. This shows that the number of independent gauge parameters is equal to the number of independent, primary first-class constraints \cite{Gomis:1989vy}. As for the first condition, it does not impose any new condition on the gauge parameters $\varepsilon$. It is actually a consistency check of the whole scheme as it can be independently derived, using the second equation and the generator constructed \cite{Banerjee:1999yc,Banerjee:1999hu}. We will demonstrate this explicitly in the case of our model, later.

Note that the derivation of (\ref{CbhtRB master 2}) is based only on the relation between the velocities and the canonical momenta, namely, the first of the Hamilton's equations of motion \cite{Banerjee:1999hu, Banerjee:1999yc}. The full dynamics, implemented through the second of Hamilton's equations $\left(\frac{dp}{dt}=\lbrace p, H \rbrace\right)$, involving accelerations, is not required to impose restrictions on the gauge parameters. Since this is the only input in our method of abstraction of the independent gauge parameters, we note that our analysis is off-shell.

\subsection{Required algebrae and finding the structure functions}

Before we begin, let us recall that all brackets are computed in the reduced phase space where a sector (second-class) of the original primary constraints has been eliminated by modifying the Poisson algebra. The algebra thus being used was presented in \eqref{algebra1} and its corollary \eqref{algebra2}.

\paragraph*{\it Algebrae within primary first-class sector:} The algebrae in this sector can be calculated directly with the definition of $\pi''{}_i^{\ 0}$ given in \eqref{Constrnt defns}. They all turn out to be either zero, or negligible square of constraint type terms (composed of {\em all} constraints, first and second class).
\begin{align}
\label{algebra PP}
\begin{aligned}
\lbrace \pi''{}_i^{\ 0}, \pi''{}_j^{\ 0} \rbrace &= 0\\
\lbrace \pi''{}_i^{\ 0}, \Pi_j^{\ 0}\rbrace &= 0\\
\lbrace \Pi_i^{\ 0}, \Pi_j^{\ 0} \rbrace &= 0.
\end{aligned}
\end{align}

\paragraph*{\it Algebrae within secondary first-class sector:} This may also be calculated using the basic algebra \eqref{algebra1} and the definitions \eqref{Constrnt defns}. We list these below \cite{Blagojevic:2010ir}
\begin{align}
\label{algebra SS}
\begin{aligned}
\lbrace \bar{\mathcal{H}}_i, \bar{\mathcal{H}}_j \rbrace &= - \epsilon_{ijk} \left( f^{kn}-f\eta^{kn} \right)\\
\lbrace \bar{\mathcal{H}}_i, \bar{\mathcal{K}}_j \rbrace &= - \epsilon_{ijk}\,\bar{\mathcal{H}}^k\\
\lbrace \bar{\mathcal{K}}_i, \bar{\mathcal{K}}_j \rbrace &= - \epsilon_{ijk}\,\bar{\mathcal{K}}^k.
\end{aligned}
\end{align}

\paragraph*{\it Algebrae between primary and secondary first-class:} Note that there are two forms of total hamiltonian; $\mathcal{H}_T$ defined in \eqref{H_T 2} with the Lagrange multipliers for the primary second-class undetermined, and the other $\hat{\mathcal{H}}_T$, with Lagrange multipliers corresponding to the primary second-class fixed \eqref{H_T 3}. These are equal upto terms which are square in constraints  \cite{Blagojevic:2010ir} and hence the difference is ignored. Now we have $\lbrace \mathcal{H}_T, \pi''{}_i^{\ 0} \rbrace = \bar{\mathcal{H}}_i$, and thus
\begin{align}
\lbrace \hat{\mathcal{H}}_T, \pi''{}_i^{\ 0} \rbrace = \bar{\mathcal{H}}_i.
\end{align}
Using the definition of $\mathcal{H}_T$ given in \eqref{H_T 3}, and after performing some manipulations, we arrive at:
\begin{align}
\label{something1}
b^j_{\ 0}\,&\lbrace \bar{\mathcal{H}}_j, \pi''{}_i^{\ 0} \rbrace + \omega^j_{\ 0}\,\lbrace \bar{\mathcal{K}}_j, \pi''{}_i^{\ 0} \rbrace = 0.
\end{align}
Note that the brackets in \eqref{something1} involve the first-class algebra within itself, which is closed. Terms linear in constraints must come from some constraint out of Table \ref{CbhtTab:Constraints}. An inspection of the same table reveals that there exist no combination of constraints such that one multiplied by $b^j_{\ 0}$ cancels out the other multiplied with $\omega^j_{\ 0}$. Thus the brackets in question must themselves be zero. Similarly, the bracket $\lbrace \mathcal{H}_T, \Pi_i^{\ 0} \rbrace = \bar{\mathcal{K}}_i$ results in the other set of brackets (between $\Pi_i^{\ 0}$ and $\bar{\mathcal{H}}_j$ or $\bar{\mathcal{K}}_j$) to also be equal to zero. We list the results below:
\begin{align}
\label{algebra PS}
\begin{aligned}
\lbrace \bar{\mathcal{H}}_j, \pi''{}_i^{\ 0} \rbrace &= 0 \\
\lbrace \bar{\mathcal{K}}_j, \pi''{}_i^{\ 0} \rbrace &= 0 \\
\lbrace \bar{\mathcal{H}}_j, \Pi{}_i^{\ 0} \rbrace &= 0 \\
\lbrace \bar{\mathcal{K}}_j, \Pi{}_i^{\ 0} \rbrace &= 0.
\end{aligned}
\end{align}

\paragraph*{\it Structure functions of the algebrae within first-class:} We can now collect and list the $C^I_{\;\;JK}$'s defined in \eqref{C_defn} from the results of all the previous algebrae calculated in this section. Only the non-vanishing ones are explicitly written.
\begin{align}
\label{CbhtCs}
\begin{aligned}
{\left(C^4_{\;\; 33}\right)}_{ijk}(x'',x,x') &= -\epsilon_{ijn}\left( f^n_{\ k} - f \delta^n_k \right) \,\delta(x-x'') \delta(x''-x')\\
{\left(C^3_{\;\; 34}\right)}_{ijk}(x'',x,x') &= -\epsilon_{ijk} \,\delta(x-x'') \delta(x''-x')\\
{\left(C^4_{\;\; 44}\right)}_{ijk}(x'',x,x') &= -\epsilon_{ijk} \,\delta(x-x'') \delta(x''-x')\\
\end{aligned}
\end{align}
In particular, we see that structure functions for algebrae within the primary first-class vanishes. Also, since the algebrae between any two first-class constraints can be expressed in terms of only the secondary first-class, all the $C^A_{\;\; IJ}$ turn out to be zero.

\paragraph*{\it Algebrae between $\mathcal{H}^{(1)}$ and first-class constraints:} The other other set of required algebrae \eqref{V_defn} can now be calculated using the definition $\mathcal{H}^{(1)}=b^i_{\ 0}\bar{\mathcal{H}}_i + \omega^i_{\ 0}\bar{\mathcal{K}}_i$. We note that this is just a combination of the secondary first-class constraints. So we use the appropriate algebrae between first-class constraints in the calculations.
\begin{align}
\label{algebrae H1_1stCl}
\begin{aligned}
\lbrace \mathcal{H}^{(1)}, \pi''{}_i^{\ 0} \rbrace &= \bar{\mathcal{H}}_i\\
\lbrace \mathcal{H}^{(1)}, \Pi_i^{\ 0} \rbrace &= \bar{\mathcal{K}}_i\\
\lbrace \mathcal{H}^{(1)}, \bar{\mathcal{H}}_i \rbrace &= \epsilon_{ijk} \omega^j_{\ 0} \bar{\mathcal{H}}^k + \epsilon_{ijk} b^j_{\ 0} \left( f^{kn} - f \eta^{kn} \right)\bar{\mathcal{K}}_n\\
\lbrace \mathcal{H}^{(1)}, \bar{\mathcal{K}}_i \rbrace &= \epsilon_{ijk} b^j_{\ 0} \bar{\mathcal{H}}^k + \epsilon_{ijk} \omega^j_{\ 0} \bar{\mathcal{K}}^k
\end{aligned}
\end{align}

\paragraph*{\it Structure functions of $\mathcal{H}^{(1)}$ with first-class sector:} The set $V^I_{\ J}$ defined in \eqref{V_defn} can be read off from the algebrae \eqref{algebrae H1_1stCl} calculated above. We list the non-vanishing ones below:
\begin{align}
\label{CbhtVs}
\begin{aligned}
{\left(V^3_{\;\;\, 1}\right)}_{ik}(x',x) &=  \eta_{ik}\,\delta(x-x')\\
{\left(V^3_{\;\;\, 3}\right)}_{ik}(x',x) &=  \epsilon_{ijk}\,\omega^j_{\ 0}\,\delta(x-x')\\
{\left(V^3_{\;\;\, 4}\right)}_{ik}(x',x) &=  \epsilon_{ijk} \, b^j_{\ 0}\,\delta(x-x')\\
{\left(V^4_{\;\;\, 2}\right)}_{ik}(x',x) &=  \eta_{ik}\,\delta(x-x').\\
{\left(V^4_{\;\;\, 3}\right)}_{ik}(x',x) &=  \epsilon_{ijl} \, b^j_{\ 0} \left( f^l_{\ k} - f \delta^l_k \right)\,\delta(x-x')\\
{\left(V^4_{\;\;\, 4}\right)}_{ik}(x',x) &=  \epsilon_{ijk} \,\omega^j_{\ 0}\,\delta(x-x')\\
\end{aligned}
\end{align}

\subsection{The generator}

Having found all the required structure functions, we can now construct the relations between the gauge parameters $\varepsilon^{(I)}$ given through the master equation \eqref{CbhtRB master 2}.
\begin{align}
\label{Cbhtrel epsilons}
\begin{aligned}
\dot{\varepsilon}^{(3)i} &= \varepsilon^{(1)i} - \varepsilon^{(3)k}\epsilon^i_{\ jk}\omega^j_{\ 0} - \varepsilon^{(4)k}\epsilon^i_{\ jk}b^j_{\ 0}\\
\dot{\varepsilon}^{(4)i} &= \varepsilon^{(2)i} - \varepsilon^{(4)k}\epsilon^i_{\ jk}\omega^j_{\ 0} - \varepsilon^{(3)k}\epsilon_{klj}\left( f^{li} - f\eta^{li} \right) b^j_{\ 0} \\
\end{aligned}
\end{align}
Note that the algebrae between the primary first-class constraints with all other first-class being zero \eqref{CbhtCs}, no $C$-structure function appears in the above relations. After using these equations \eqref{Cbhtrel epsilons} in the generator \eqref{gen G} to eliminate the gauge parameters $\varepsilon^{(1)}$ and $\varepsilon^{(2)}$, we obtain the generator in terms of the two independent gauge parameters $\varepsilon^{(3)}$ and $\varepsilon^{(4)}$.
\begin{align}
\label{CbhtgeneratorOur}
G = \int d^2x & \left[\left\lbrace \dot{\varepsilon}^{(3)i} +  \varepsilon^{(3)k} \,\epsilon_k^{\ ij}\,\omega_{j0} + \varepsilon^{(4)k} \,\epsilon_k^{\ ij}\,b_{j0}\right\rbrace\,\pi''{}_i^{\ 0} \right.\nonumber\\
&+\left.\left\lbrace \dot{\varepsilon}^{(4)i} + \,\varepsilon^{(4)k} \,\epsilon_k^{\ ij}\,\omega_{j0} + \epsilon^{(3)k} \epsilon_k^{\ lj}\,b_{j0} \left( f_l^{\ i}-f\delta^i_l \right) \right\rbrace \Pi_i^{\ 0} \right.\nonumber\\
&+\left. \varepsilon^{(3)i}\,\bar{\mathcal{H}_i} + \varepsilon^{4)i}\,\bar{\mathcal{K}_i}\right].
\end{align}
The parameters can be renamed $(\varepsilon^{(3)}=\tau\,,\; \varepsilon^{(4)}=\sigma)$, and the expression \eqref{CbhtgeneratorOur} be arranged to arrive at the generator
\begin{align}
\label{CbhtgeneratorFinal}
\begin{aligned}
G=&\int d^2x \left[\mathcal{G}_\tau(x)+\mathcal{G}_\sigma(x)\right]\\
&\mathcal{G}_\tau=\dot{\tau}^i\,\pi''{}_i^{\ 0} + \tau^i\left[ \bar{\mathcal{H}}_i -\epsilon_{ijk} \omega^j_{\ 0} \pi''{}^{k0} - \epsilon_{ijk} b^j_{\ 0} \left( f^{kn} - f\eta^{kn} \right) \Pi_n^{\ 0} \right]\\
&\mathcal{G}_\sigma=\dot{\sigma}^i\Pi_i^{\ 0} + \sigma^i\left[ \bar{\mathcal{K}}_i - \epsilon_{ijk} \omega^j_{\ 0}\Pi^{k0} - \epsilon_{ijk} b^j_{\ 0} \pi''{}^{k0} \right].
\end{aligned}
\end{align}
The above generator was also reported in \cite{Blagojevic:2010ir}, where an on-shell method of constructing gauge generators following \cite{Castellani:1981us}, was used. Our's however, is an explicitly off-shell method of construction. Also note that that the number of independent gauge parameters here ($3+3=6$) is equal to the total number of (independent) primary first-class constraints (see Table \ref{CbhtTab:Constraints}), as mentioned earlier (see discussion below eq. \ref{CbhtRB master 2}).

In the next section, we construct the symmetries of the basic fields obtained by the above generator and study their relation with the underlying Poincar\'{e} symmetries of the model.


\section{The symmetries and their mapping: hamiltonian to Poincar\'{e}}
\label{Cbht:map}

The symmetries of the basic fields $\left( b^i_{\ \mu}, \omega^i_{\ \mu}, f^i_{\ \mu}, \lambda^i_{\ \mu} \right)$ can be calculated using the generator \eqref{CbhtgeneratorFinal} constructed in the previous section. The algebra used \eqref{algebra1} is that defined in the reduced space as explained earlier in Section \ref{Cbht:canAnalysis}. Thus the variation of the triad field `$b$' is
\begin{align}
\label{delta_G b}
\delta_G b^h_{\ \zeta} = \partial_\zeta \tau^h - \epsilon^h_{\ jk}\tau^j\omega^k_{\ \zeta} - \epsilon^h_{\ jk}\sigma^j b^k_{\ \zeta}.
\end{align}
It is clear from a comparison between this symmetry generated via hamiltonian gauge generator and that of the PGT symmetry \eqref{PGT deltas} of `$b$', that the Poincar\'{e} symmetries of local Lorentz rotation and general diffeomorphism cannot be identified in the set $\delta_G b^h_{\ \zeta}$. We therefore map the arbitrary gauge parameters $\tau^i$ and $\sigma^i$ to the Poincar\'{e} parameters $\xi^\mu$ and $\theta^i$ to recover the Poincar\'{e} symmetries. The map used is
\begin{align}
\label{map}
\begin{aligned}
\tau^i &= -\xi^\rho\,b^i_{\ \rho}\\
\sigma^i &= -\theta^i - \xi^\rho \,\omega^i_{\ \rho}.
\end{aligned}
\end{align}
This type of map was reported earlier in studies of topologically massless \cite{Blagojevic:2004hj,Banerjee:2009vf} as well as massive \cite{Blagojevic:2008bn} models of gravity. In \cite{Banerjee:2009vf}, it was shown explicitly that though both $\delta_G$ and $\delta_{PGT}$ generate off-shell symmetries, they can be related to each other through the map \eqref{map} only on-shell, i.e. upon imposition of the equations of motion. We will shortly see that something similar also happens here.

\noindent On using the above map in \eqref{delta_G b}, we get the following form of symmetry for the triad:
\begin{align}
\label{G PGT B}
\delta_{\scriptscriptstyle G} b^h_{\ \zeta} &= -\partial_\zeta \xi^\rho \, b^h_{\ \rho} - \xi^\rho \,\partial_\zeta b^h_{\ \rho} + \epsilon^h_{\ jk}\xi^\rho b^j_{\ \rho} \omega^k_{\ \zeta} + \epsilon^h_{\ jk}\theta^j b^k_{\ \zeta} + \epsilon^h_{\ jk}\xi^\rho\omega^j_{\ \rho} b^k_{\ \zeta} \nonumber\\
&= -\partial_\zeta \xi^\rho\,b^h_{\ \rho} - \xi^\rho\,\partial_\rho b^h_{\ \zeta} - \epsilon^h_{\ jk}b^j_{\ \zeta}\theta^k + \xi^\rho \left( \partial_\rho b^h_{\ \zeta} - \partial_\zeta b^h_{\ \rho} + \epsilon^h_{\ jk}\omega^j_{\ \rho}b^k_{\ \zeta} - \epsilon^h_{\ jk}\omega^j_{\ \zeta}b^k_{\ \rho} \right) \nonumber\\
&= \delta_{\scriptscriptstyle PGT} b^h_{\ \zeta} + \xi^\rho\, T^h_{\ \rho\zeta}.
\end{align}
We thus recover the PGT symmetry, but modulo terms which vanish on-shell. To see this, note the equation of motion \eqref{EOM lambda} corresponding to the field $\lambda$. Since torsion is antisymmetric in its Greek indices, i.e. $T^i_{\ \mu\nu} = -T^i_{\ \nu\mu}$, we have $$\epsilon^{\mu\nu\rho}\,T^i_{\ \nu\rho} = 0 \quad \Rightarrow \quad T^i_{\ \nu\rho} = 0.$$ This phenomenon -- that among all the equations of motion the imposition of vanishing torsion is required to come back to the PGT (local Lorentz + diffeomorphisms) symmetries, from the hamiltonian gauge symmetries, is remarkable. As was earlier noted, the difference between the original and PGT formulation of BHT theory lies in that the former is built on Reimannian spacetime (only curvature, zero torsion), while the latter on Riemann-Cartan spacetime (both curvature and torsion). So we do not find it surprising that the triad field $b^i_{\ \mu}$, which alone makes up the metric $g_{\mu\nu}$, is restored to its expected PGT symmetry by use of zero torsion condition.

Coming back to hamiltonian gauge symmetries, let us examine another field, the axillary field `$f$'. The gauge transformation of `$f$' generated by the generator \eqref{CbhtgeneratorFinal} is
\begin{align}
\label{delta_G f}
\delta_{\scriptscriptstyle G} f^h_{\ \zeta} &= \dot{\tau}^i f_i^{\ h}\,\delta^0_\zeta + \partial_\alpha\left( \tau^i f^h_{\ i} \right)\delta^\alpha_\zeta - \epsilon^h_{\ jk}\tau^i f^j_{\ i}\omega^k_{\ \alpha}\delta^\alpha_\zeta + \tau^i b_i^{\ \mu} \left( \nabla_\mu f^h_{\ k} \right)b^k_{\ 0}\delta^0_\zeta - \epsilon_{ijk} \tau^i \omega^j_{\ 0} f^{kh} \delta^0_\zeta \nonumber \\
&\quad - \left( \frac{m^2}{a} \right) \epsilon^h_{\ jk}\tau^j\lambda^k_{\ \alpha} \delta^\alpha_\zeta - \left( \frac{m^2}{a} \right) \epsilon^h_{\ jk} \tau^i \lambda^j_{\ i} b^k_{\ \alpha} \delta^\alpha_\zeta - \epsilon^h_{\ jk}\sigma^j f^k_{\ \zeta}.
\end{align}
Use of the map \eqref{map} in the above transformation gives
\begin{align}
\label{delta_G f mapped PGT}
\delta_{\scriptscriptstyle G} f^h_{\ \zeta} = \delta_{\scriptscriptstyle PGT} f^h_{\ \zeta} &+ \frac{m^2}{a} ~\epsilon_{\mu\rho\zeta}\,\epsilon^{\mu\nu\sigma} ~\xi^\rho \left[ \frac{a}{m^2} \nabla_\nu f^h_{\ \sigma} + \epsilon^h_{\ jk}b^j_{\ \nu}\lambda^k_{\ \sigma} \right] \nonumber\\
&+ \frac{m^2}{a} ~\epsilon_{\mu 0 \rho}\,\epsilon^{\mu 0 \sigma} ~\xi^\rho\left[ \frac{a}{m^2} \nabla_0 f^h_{\ \sigma} + \epsilon^h_{\ jk}b^j_{\ 0}\lambda^k_{\ \sigma} \right]\delta^0_\zeta - \xi^\rho\,T^i_{\ 0\rho}\,f^h_{\ i}\,\delta^0_\zeta\,.
\end{align}
As seen earlier in case of the triad, here too we see that the hamiltonian symmetry is equal to the PGT symmetries modulo the equations of motion. In this case, the equations of motion for the fields `$\lambda$' and `$\omega$' \eqref{EOM} are required to identify with the PGT symmetries. Also in the above computations, use of the constraint $\Theta^{ij}=f^{ij}-f^{ji}$ is required.

The symmetries for the other two fields `$\omega$' and `$\lambda$' also give similar results, only the algebraic nature is more involved. Thus all the fields have two sets of symmetries, the PGT symmetries and the hamiltonian gauge symmetries. Both of these are off-shell in nature. But they can be identified with each other only on-shell.

Now, a subtle issue arises in this identification of the two symmetries through the use of the map \eqref{map}. The map between the two sets of independent gauge parameters is field dependent in nature. As a result, one may wonder whether one can use this map at the level of the generator, i.e. before computation of symmetries. Once the map is used in the generator itself, it will give rise to non-trivial brackets with other fields when computing the symmetries. The proper way to frame the question would be to study the commutativity manifested in the diagram:
\begin{center}
$\begin{CD}
G[\tau,\sigma] @>>> \delta_{[\tau,\sigma]} @.\\
@VV\text{Map}V @VV\text{Map}V @.\\
G[\xi,\theta] @>>> \delta_{[\xi,\theta]} @>>\text{\phantom{www}on-shell\phantom{www}}> \delta_{PGT}
\end{CD}$
\end{center}
The issue however can be resolved on noting that the generator is nothing but a combination of (first-class) constraints multiplied by the gauge parameters, and possibly, other fields. Thus, when terms apart from the constraints (in this case - especially the parameters) gives rise to brackets, they are rendered insignificant due to multiplication with a constraint. So it is immaterial whether we use the map in the gauge symmetries generated by the hamiltonian generator, or in the mapped generator itself.

If we use the map \eqref{map} in the generator \eqref{CbhtgeneratorFinal} constructed in the previous section, we get, upto terms proportional to square of constraints
\begin{align}
\label{generator mapped}
G = &-\dot{\xi}^\mu \left[ b^i_{\ \mu}\pi_i^{\ 0} + \omega^i_{\ \mu}\Pi_i^{\ 0} + \lambda^i_{\ \mu}p_i^{\ 0} + f^i_{\ \mu}P_i^{\ 0} \right] - \xi^\mu \left[ b^i_{\ \mu}\hat{\mathcal{H}}_i + \omega^i_{\ \mu}\mathcal{K}_i + \lambda^i_{\ \mu}\mathcal{T}_i + f^i_{\ \mu}\hat{\mathcal{R}}_i \right.\nonumber\\
&\left. \qquad\qquad\qquad\qquad\qquad\quad + ~(\partial_\mu b^i_{\ 0})\pi_i^{\ 0} + (\partial_\mu \omega^i_{\ 0})\Pi_i^{\ 0} + (\partial_\mu \lambda^i_{\ 0})p_i^{\ 0} + (\partial_\mu f^i_{\ 0})P_i^{\ 0} \right]\nonumber \\
&- \dot{\theta}^i \Pi_i^{\ 0} - \theta^i\left[ \mathcal{K}_i -\epsilon_{ijk}\left( b^j_{\ 0}\pi^{k0} + \omega^j_{\ 0}\Pi^{k0} + \lambda^j_{\ 0}p^{k0} + f^j_{\ 0}P^{k0} \right) \right].
\end{align}
This generator also generates symmetries of the basic fields and these agree with those of PGT {\em on-shell} \cite{Blagojevic:2010ir}. Thus our results are in agreement with the above conclusion of commutativity of the diagram given above.

\subsection{Consistency check}

We will now finally show an internal consistency check of the algorithm given through the relation \eqref{CbhtRB master 1} obtained in section \ref{Cbht:Generator}. This relation, unlike its twin \eqref{CbhtRB master 2}, is not a new restriction on the gauge parameters as it can be independently derived through use of \eqref{CbhtRB master 2} and the generator \eqref{CbhtgeneratorFinal} as was shown in \cite{Banerjee:1999hu}. Note that the construction of the generator itself is independent of \eqref{CbhtRB master 1}. We start with an observation on the equation of motion of the field $b^i_{\ 0}$
\begin{align}
\label{something2}
\dot{b}^i_{\ 0} = \lbrace b^i_{\ 0}, \int \mathcal{H}_T \rbrace = u^i_{\ 0},
\end{align}
where in the last step, we used the total hamiltonian density given in \eqref{H_T 3}. The variation of the Lagrange multiplier $u^i_{\ 0}$ can thus be obtained from the variation of the field $b^i_{\ 0}$ calculated in \eqref{delta_G b}
\begin{align}
\label{something3}
\delta b^i_{\ 0} = \partial_0 \tau^i - \epsilon^i_{\ jk} \tau^j \omega^k_{\ 0} - \epsilon^i_{\ jk}\sigma^j b^k_{\ 0}.
\end{align}
So, we have
\begin{align}
\label{something4}
\delta u^i_{\ 0} = \frac{d}{dt} \delta b^i_{\ 0} = \frac{d}{dt}\left[ \dot{\tau}^i - \epsilon^i_{\ jk}\tau^j\omega^k_{\ 0} - \epsilon^i_{\ jk}\sigma^j b^k_{\ 0} \right] = \dot{\varepsilon}^{(1)}{}^i.
\end{align}
Use has been made of the redefinitions $(\varepsilon^{(3)}=\tau\,,\; \varepsilon^{(4)}=\sigma)$ and the relations between the gauge parameters \eqref{Cbhtrel epsilons} which were found by employing only the second relation \eqref{CbhtRB master 2}.
Turning now to the first relation \eqref{CbhtRB master 1} that also gives variations of the Lagrange multipliers, we see for $A=1$, i.e. $\chi^{(1)}=u^i_{\ 0}$,
\begin{align}
\label{something5}
\delta \chi^{(1)} = \delta u^i_{\ 0} = \frac{d \varepsilon^{(1)}{}^i}{dt} - \int d^2x ~\varepsilon^{(I)}{}^k (V^1_{\; I}){}^i_{\ k } - \int d^2x ~\varepsilon^{(I)}{}^k \int d^2x' ~\chi^{(B)}{}^j (C^1_{\;\; IB}){}^i_{\ jk}.
\end{align}
Since $V^1_{\; I}=0$ \eqref{CbhtVs} and $C^1_{\;\; IB}=0$ \eqref{CbhtCs}, we finally get
\begin{align}
\label{something6}
\delta u^i_{\ 0} = \frac{d \varepsilon^{(1)}{}^i}{dt},
\end{align}
which is nothing but \eqref{something4}. This shows the internal consistency of our scheme.


\section{Discussions}
\label{Cbht:Disc}

In this chapter, we have constructed the hamiltonian gauge generator of the cosmological BHT model described in first-order form by the action \eqref{lagrangian} using a completely off-shell method. We have explicitly found the hamiltonian gauge symmetries resulting from this generator and shown that these symmetries can be mapped to the Poincar\'{e} symmetries only on-shell, through a mapping of the gauge parameters. Remarkably the vanishing torsion condition, which takes us from the Riemann-Cartan spacetime of first order PGT formulation to the usual metric formulation in Riemann spacetime, plays an important role in this on-shell mapping. We also noted that the map used by us, and which is also quite common in the literature \cite{Banerjee:2009vf,Blagojevic:2004hj,Blagojevic:2008bn}, is field dependent. We clarify why this does not cause any problem in computation of the symmetries through the generator. It can be used both directly in the generator, i.e. before computation of symmetries and also after computation of symmetries through the generator. The two processes were shown to be equivalent. We would see more of this map, and particularly, motivate a method of construction later in Chapter \ref{C:tr}.

Finally we would like to comment that our results would be useful in finding the corresponding conserved charges of BHT gravity consistently at an off-shell level. This would in turn play an important role in the obtention of the central charges of the asymptotic symmetry.


%% file: C6_lagrangian.tex
\chapter{Lagrangian study of symmetries}
\label{C:lag}

\lettrine[lraise=0.0, loversize=0.3, findent=3pt, nindent=0pt, lhang=0.2]{T}{he} symmetries involved in PGT are motivated and derived by considering changes occurring due to local Lorentz rotations and infinitesimal coordinate translations. However, our analysis till now shows the difficulty in constructing a canonical procedure of generating this symmetry through a set of generators. This problem was initially addressed in \cite{Blagojevic:2004hj} by computing the hamiltonian generator following the approach of \cite{Castellani:1981us}. However the off-shell symmetries that were obtained as a result, were different from the PGT symmetries. The two sets matched only on-shell. This mismatch was thought to be a consequence of the approach \cite{Castellani:1981us} which, strictly speaking, is not a completely off-shell approach. An attempt to remedy this situation was given in Chapters \ref{C:mb} and \ref{C:bht} following \cite{Banerjee:2009vf, Banerjee:2011rx}. A systematic and completely off-shell analysis was done in 2+1 dimensions, taking the 3D gravity model with torsion (modelled on the Mielke-Baekler action \cite{Mielke:1991nn}) and the Dirac Hamiltonian generator was computed. Alas, this did not result in a hamiltonian first-class generator which could give back symmetries that are (algebraically) same as the original PGT symmetries. This is even after allowing suitable maps between the hamiltonian and PGT gauge parameters, which are {\it a priori} different. Here in this chapter, we adopt a totally different lagrangian approach and show how to systematically construct lagrangian generators of the PGT symmetries \cite{Banerjee:2010kd}. The symmetries obtained from these generators reproduce the PGT symmetries, off-shell.

The task of finding the symmetries of a given action is, in general, not trivial. They cannot always be found on inspection, as is possible, say, in Maxwell electrodynamics. There exist two approaches to systematically construct the symmetries inbuilt in a given action; the Hamiltonian approach and the Lagrangian approach. In the Hamiltonian approach \cite{Castellani:1981us, Banerjee:2009vf, Henneaux:1990au, Banerjee:1999hu, Banerjee:1999yc, Banerjee:1999sz, Frolov:2009wu, Kiriushcheva:2006sg}, the generators of the symmetries are obtained as some suitable combination of first-class constraints containing time derivatives of arbitrary functions. The Lagrangian method \cite{Mukunda:1974dr, Gitman:1990qh, Chaichian:1994ug, Shirzad:1998af, Samanta:2007fk, Banerjee:2009gr, Banerjee:2006jy}, on the other hand, hinges on the condition that the existence of symmetries necessarily implies the existence of certain identities involving quantities given as variations of the action w.r.t. the basic fields,  the Euler derivatives. The Lagrangian generators can then be found through comparison with the general expressions of such identities derived from a theory involving general symmetry transformation of fields in terms of arbitrary functions of time. We have used this later method of constructing generators here, after modifying it suitably for our model with dreibeins/triads and connections as basic fields, that are written both in holonomic (global coordinate) and an-holonomic (local coordinate) indices.

\begin{figure}[ht]
\centering
\includegraphics[angle=0, width=0.9\textwidth]{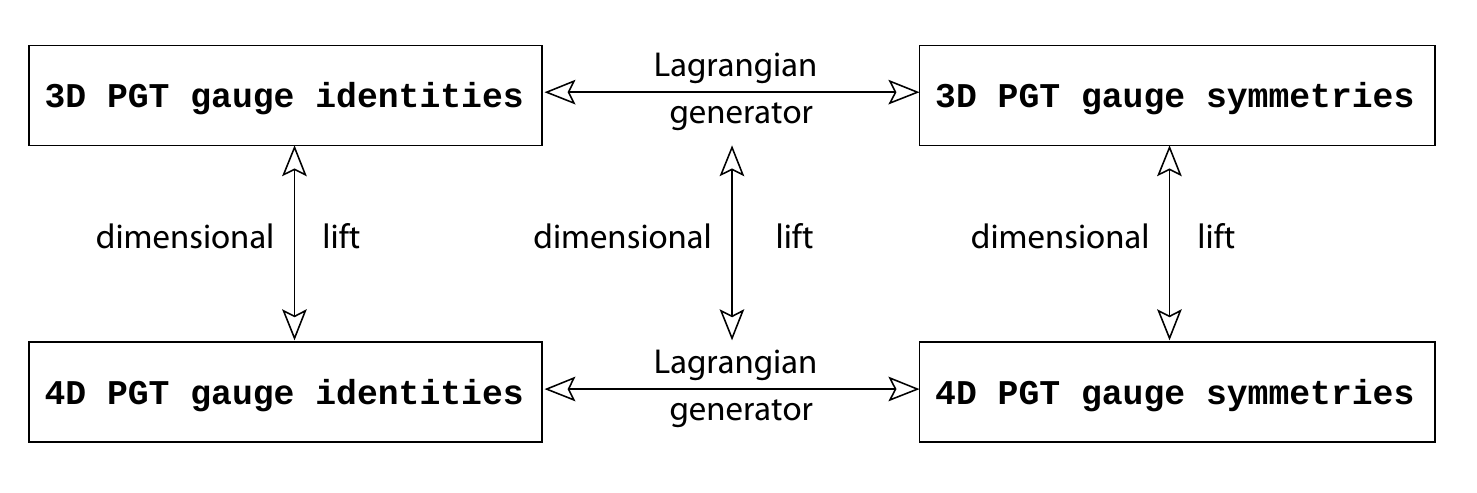}
\caption{{\it The scheme adopted for Lagrangian analysis: symmetries, corresponding gauge identities and generators in 2+1 and 3+1 dimensions.}}
\label{ClagFig1}
\end{figure}

Having constructed the generators first in 2+1 dimensions, we then repeat the process in 3+1 dimensions. The construction of symmetries through the Hamiltonian approach in 3+1 dimensions itself is a very difficult task and pure Dirac analysis and classification of constraints is non-trivial for, say, just the Einstein-Cartan theory. However it is shown that the Lagrangian version can be carried out more easily, leading to a systematic derivation of the symmetries even in 3+1 dimensions. To do this, we first lift the gauge identities corresponding to PGT symmetries from 2+1 to 3+1 dimensions and then carry out the Lagrangian analysis. However, at the end of the day, the symmetries obtained are the same as that found by directly lifting the 2+1 PGT symmetries in 3+1 dimensions. This shows that the PGT symmetries, arising out of geometrical considerations of reparametrization and local Lorentz symmetry, are beautifully consistent and useful in an extended sense. The whole scheme is summarized in Figure \ref{ClagFig1}. Besides this, we also comment on other possible applications of Lagrangian generators.

\paragraph{Conventions:}The coordinate frame or holonomic indices are written in Greek while Latin indices refer to the an-holonomic local Lorentz frame. The time and space bifurcation is indicated by choosing the beginning letters of both Greek $\left( \alpha, \beta, \ldots \right)$ and Latin $\left(a, b, \ldots\right)$ indices to run over the space indices, i.e. $1, 2, \ldots$ and choosing the middle letters of both Greek $\left( \mu, \nu, \ldots \right)$ and Latin $\left( i, j, \ldots \right)$ indices to run over both time and space indices $0, 1, 2, \ldots$. The totally antisymmetric tensor densities $\epsilon^{ijk}, \epsilon^{ijkl}, \epsilon^{\mu\nu\lambda} \text{ and } \epsilon^{\mu\nu\lambda\rho}$ are all normalized so that $\epsilon^{012} \text{ and } \epsilon^{0123}$ are unity. The spacetime signature chosen is $\left(+, -, -, \ldots \right)$. In specifying spacetime points, we have denoted by `$x$' both the time and space parts together while `$\bf{x}$' indicates only the spatial part of `$x$'. Thus $x\equiv(\mathrm{t},\bf{x})$.


\section{Lagrangian formulation: Gauge identities}
\label{Clag:GIs}

Let us first take up the lagrangian identities which we mentioned to be the key in our lagrangian analysis. We begin with the familiar example of electromagnetism. The action is,
\begin{align}
S = \int \mathcal{L}(A_\mu,\partial_\nu A_\mu) = \int F^2
\end{align}
where $F$ is the electromagnetic field tensor. The action is invariant under the gauge transformation $A_\mu \to A_\mu + \partial_\mu \Lambda$, where $\Lambda$ is the gauge transformation parameter. By Taylor expansion
\begin{align}
S[A_\mu + \partial_\mu \Lambda] = S[A_\mu] +\int \partial_\mu\Lambda \frac{\delta S}{\delta A_\mu}= S[A_\mu] -\int \Lambda \partial_\mu\frac{\delta S}{\delta A_\mu}
\end{align}
The invariance condition $S[A_\mu + \partial_\mu \Lambda] = S[A_\mu]$ leads to the gauge identity
\begin{align}
\partial_\mu\frac{\delta S}{\delta A_\mu} = \partial_\mu \partial_\nu ~F^{\nu\mu} = 0.
\end{align}
Note that this holds {\it off-shell}. In fact if we invoke the equations of motion, the gauge identity becomes a trivial $0 = 0$ statement. Note further that such a gauge identity should exist corresponding to each independent gauge parameter.

We will now proceed to construct the identities corresponding to the Poincar\'{e} gauge transformations that we had been considering from Chapter \ref{C:pgt}. Note that these symmetries are independent of any particular action.
\begin{align}
\tag{\ref{Poincare3Dfieldtrans}}
\begin{aligned}
\delta b^i_{\ \mu} &= -\epsilon^i_{\ jk}b^j_{\ \mu}\theta^k - \partial_\mu\xi^\rho b^i_{\ \rho} - \xi^{\rho}\partial_{\rho} b^i_{\ \mu}\\
\delta \omega^{i}_{\ \mu} &= -\left(\partial_\mu \theta^i + \epsilon^i_{\ jk} \omega^j_{\ \mu} \theta^k  \right) - \partial_\mu\xi^\rho \omega^{i}_{\ \rho} - \xi^{\rho}\partial_{\rho}\omega^{i}_{\ \mu}.
\end{aligned}
\end{align}
By Taylor expansion we get 
\begin{align*}
S\left[b^i_{\ \mu} , \omega^i_{\ \mu}\right] &= S\left[b^i_{\ \mu} + \delta_{\scriptscriptstyle PGT}b^i_{\ \mu}, \omega^i_{\ \mu} + \delta_{\scriptscriptstyle PGT} \omega^i_{\ \mu}\right]\nonumber\\
\Rightarrow S\left[b^i_{\ \mu} , \omega^i_{\ \mu}\right] &= S\left[b^i_{\ \mu} -\left(\epsilon^i_{\ jk} b^j_{\ \mu} \theta^k + \partial_\mu \xi^\lambda b^i_{\ \lambda} + \xi^\lambda \partial_\lambda b^i_{\ \mu} \right),\right.\nonumber\\
&\left. \phantom{S[}\omega^i_{\ \mu} -\left( \partial_\mu \theta^i + \epsilon^i_{\ jk} \omega^j_{\ \mu} \theta^k + \partial_\mu \xi^\lambda \omega^i_{\ \lambda} + \xi^\lambda \partial_\lambda \omega^i_{\ \mu} \right)\right]\nonumber\\
&= S\left[b^i_{\ \mu} , \omega^i_{\ \mu}\right] - \int d^3x ~\frac{\delta S}{\delta b^i_{\ \mu}} ~\left(\epsilon^i_{\ jk}b^j_{\ \mu}\theta^k + \partial_\mu \xi^\lambda b^i_{\ \lambda} + \xi^\lambda \partial_\lambda b^i_{\ \mu} \right)\nonumber\\
&\phantom{S[}- \int d^3x ~\frac{\delta S}{\delta \omega^i_{\ \mu}} ~\left( \partial_\mu \theta^i + \epsilon^i_{\ jk} \omega^j_{\ \mu} \theta^k + \partial_\mu \xi^\lambda \omega^i_{\ \lambda} + \xi^\lambda \partial_\lambda \omega^i_{\ \mu} \right)\nonumber\\
\end{align*}
Now collecting terms containing each parameter separately,
\begin{align}
\label{PGT taylor}
\int d^3x &\left[\frac{\delta S}{\delta b^i_{\ \mu}} \epsilon^i_{\ jk} b^j_{\ \mu} + \frac{\delta S}{\delta \omega^i_{\ \mu}} \epsilon^i_{\ jk} \omega^j_{\ \mu} -\partial_\mu\left(\frac{\delta S}{\delta \omega^k_{\ \mu}}\right) \right] \theta^k \nonumber\\
&+ \int d^3x \left[\frac{\delta S}{\delta b^i_{\ \mu}} \partial_\lambda b^i_{\ \mu} + \frac{\delta S}{\delta \omega^i_{\ \mu}} \partial_\lambda \omega^i_{\ \mu} - \partial_\mu \left(b^i_{\ \lambda}\frac{\delta S}{\delta b^i_{\ \mu}} + \omega^i_{\ \lambda} \frac{\delta S}{\delta \omega^i_{\ \mu}} \right) \right] \xi^\lambda = 0.
\end{align}
So the independent gauge identities turn out to be:
\begin{align}
\label{gauge ident PGT}
\begin{aligned}
\frac{\delta S}{\delta b^i_{\ \mu}} \varepsilon^i_{\ jk} b^j_{\ \mu} + \frac{\delta S}{\delta \omega^i_{\ \mu}} \varepsilon^i_{\ jk} \omega^j_{\ \mu} -\partial_\mu\left(\frac{\delta S}{\delta \omega^k_{\ \mu}}\right) &= 0\\
\frac{\delta S}{\delta b^i_{\ \mu}} \partial_\lambda b^i_{\ \mu} + \frac{\delta S}{\delta \omega^i_{\ \mu}} \partial_\lambda \omega^i_{\ \mu} - \partial_\mu \left(b^i_{\ \lambda}\frac{\delta S}{\delta b^i_{\ \mu}} + \omega^i_{\ \lambda} \frac{\delta S}{\delta \omega^i_{\ \mu}} \right) &=0
\end{aligned}
\end{align}

Now, the above gauge identities correspond to PGT symmetries, which are generic. Any PGT invariant lagrangian will follow these. However, the hamiltonian symmetries are not so generic, as we saw from our results in Chapter \ref{C:mb} \eqref{field transf gauge} and Chapter \ref{C:bht} \eqref{delta_G b}. For our purposes, we will adopt the Mielke-Baekler type model that we had analysed in Chapter \ref{C:mb}\footnote{This action has also been considered extensively elsewhere in literature \cite{Banerjee:2009vf, Blagojevic:2004hj, Basu:2009dy}.}, for its relative simplicity to the massive gravity model considered in Chapter \ref{C:bht}. 
\begin{align}
\tag{\ref{CmbMBaction}}
S = \int d^3x\ \epsilon^{\mu\nu\rho}&\left[a\,b^i_{\ \mu} R_{i\nu\rho} - \frac{\Lambda}{3} \epsilon_{ijk}\,b^i_{\ \mu} b^j_{\ \nu} b^k_{\ \rho} \right.\nonumber\\
&\phantom{+}\left. + \alpha_3 \left(\omega^i_{\ \mu} \,\partial_\nu\omega_{i\rho} + \frac{1}{3} \epsilon_{ijk}\,\omega^i_{\ \mu} \omega^j_{\ \nu} \omega^k_{\ \rho} \right) + \frac{\alpha_4}{2} b^i_{\ \mu} T_{i\nu\rho} \right]\nonumber.
\end{align}
The hamiltonian gauge symmetries for this action were
\begin{align}
\tag{\ref{field transf gauge}}
\begin{aligned}
\delta b^i_{\ \mu}(x) &:= \lbrace b^i_{\ \mu}(x),G \rbrace^* \\
&= \partial_\mu\epsilon^i(x) + \epsilon^i_{\ jk} \,\omega^j_{\ \mu}(x) \varepsilon^k(x) - p \,\epsilon^i_{\ jk} \,b^j_{\ \mu}(x) \varepsilon^k(x) + \epsilon^i_{\ jk}\,b^j_{\ \mu}(x) \tau^k(x),\\
\delta \omega^i_{\ \mu}(x) &:= \lbrace \omega^i_{\ \mu}(x),G \rbrace^* \\
&= \partial_\mu \tau^i(x) + \epsilon^i_{\ jk} \,\omega^j_{\ \mu}(x) \tau^k(x) - q \,\epsilon^i_{\ jk} \,b^j_{\ \mu}(x) \varepsilon^k(x).\\
\end{aligned}
\end{align}
From the invariance of \eqref{CmbMBaction} under \eqref{field transf gauge}, we find,
\begin{align*}
S\left[b^i_{\ \mu} , \omega^i_{\ \mu}\right] &= S\left[b^i_{\ \mu} + \delta_{\scriptscriptstyle\text{Gauge}}b^i_{\ \mu} , \omega^i_{\ \mu} + \delta_{\scriptscriptstyle\text{Gauge}} \omega^i_{\ \mu}\right]\nonumber\\
\Rightarrow S\left[b^i_{\ \mu} ,\omega^i_{\ \mu}\right] &= S\left[b^i_{\ \mu} + \left(\partial_\mu \varepsilon^i + \epsilon^i_{\ jk}\omega^j_{\ \mu} \varepsilon^k - p\,\epsilon^i_{\ jk}b^j_{\ \mu} \varepsilon^k + \epsilon^i_{\ jk}b^j_{\ \mu} \tau^k \right),\right.\nonumber\\
&\left. \phantom{S[} \omega^i_{\ \mu} +\left(\partial_\mu \tau^i + \epsilon^i_{\ jk}\omega^j_{\ \mu} \tau^k - q\, \epsilon^i_{\ jk}b^j_{\ \mu} \varepsilon^k \right) \right]\nonumber\\
&= S\left[b^i_{\ \mu} , \omega^i_{\ \mu}\right] + \!\int d^3x ~\frac{\delta S}{\delta b^i_{\ \mu}} \left(\partial_\mu \varepsilon^i + \epsilon^i_{\ jk}\omega^j_{\ \mu} \varepsilon^k - p\epsilon^i_{\ jk}b^j_{\ \mu} \varepsilon^k + \epsilon^i_{\ jk}b^j_{\ \mu} \tau^k \right)\nonumber\\
&+ \phantom{S[} \int d^3x ~\frac{\delta S}{\delta \omega^i_{\ \mu}} ~\left(\partial_\mu \tau^i + \epsilon^i_{\ jk}\omega^j_{\ \mu} \tau^k - q\, \epsilon^i_{\ jk}b^j_{\ \mu} \varepsilon^k \right)\nonumber\\
\end{align*}
Again, collecting terms proportional to independent gauge parameters, we have
\begin{align}
\label{Gauge gen taylor}
&\int d^3x \left[-\partial_\mu\left(\frac{\delta S}{\delta \omega^k_{\ \mu}}\right) + \frac{\delta S}{\delta b^i_{\ \mu}} \epsilon^i_{\ jk} b^j_{\ \mu} + \frac{\delta S}{\delta \omega^i_{\ \mu}} \epsilon^i_{\ jk} \omega^j_{\ \mu} \right] \tau^k \nonumber\\
&+ \int d^3x \left[-\partial_\mu\left(\frac{\delta S}{\delta b^k_{\ \mu}}\right) - q ~\frac{\delta S}{\delta \omega^i_{\ \mu}} \epsilon^i_{\ jk} b^j_{\ \mu} + \frac{\delta S}{\delta b^i_{\ \mu}} \epsilon^i_{\ jk} \omega^j_{\ \mu} -p \frac{\delta S}{\delta b^i_{\ \mu}} \epsilon^i_{\ jk} b^j_{\ \mu} \right] \varepsilon^k = 0,
\end{align}
which leads to the independent gauge identities:
\begin{align}
\label{gauge ident generator}
\begin{aligned}
-\partial_\mu\left(\frac{\delta S}{\delta \omega^k_{\ \mu}}\right) + \frac{\delta S}{\delta b^i_{\ \mu}} \epsilon^i_{\ jk} b^j_{\ \mu} + \frac{\delta S}{\delta b^i_{\ \mu}} \epsilon^i_{\ jk} \omega^j_{\ \mu} &=0\\
-\partial_\mu\left(\frac{\delta S}{\delta b^k_{\ \mu}}\right) - q ~\frac{\delta S}{\delta \omega^i_{\ \mu}} \epsilon^i_{\ jk} b^j_{\ \mu} + \frac{\delta S}{\delta b^i_{\ \mu}} \epsilon^i_{\ jk} \omega^j_{\ \mu} -p \frac{\delta S}{\delta b^i_{\ \mu}} \epsilon^i_{\ jk} b^j_{\ \mu} &=0.\\
\end{aligned}
\end{align}
Note that all the identities become trivial if invoke the equations of motion, by setting the Euler derivatives $\frac{\delta S}{\delta b^i_{\ \mu}}$ and $\frac{\delta S}{\delta \omega^i_{\ \mu}}$ to zero. Now, since we are trying to do a lagrangian analysis with the aim of understanding the PGT symmetries \eqref{Poincare3Dfieldtrans}, we will only require the identities \eqref{gauge ident PGT} in the analysis for this chapter.

In the next section, we will set up a formalism outlining how to to define and construct `lagrangian generators' corresponding to these identities, following \cite{Shirzad:1998af}. The discussion will be at a more generalised, formal level after the introduction given here through explicit examples.


\section{Lagrangian symmetry generators}
\label{Clag:lagGenGIs}

The action functional for a typical first-order theory invariant under the PGT symmetries in 2+1 dimensions, may be written in general,
\begin{align}
\label{Gen2plus1Action}
S = \int \mathrm{d}^3x ~\mathcal{L}\left[b^i_{\ \mu}(x),\omega^i_{\ \mu}(x)\right].
\end{align}
As we saw in the previous section, arbitrary variations of the basic fields give rise to variation of the action in the following form
\begin{align}
\label{2plus1GenActionVar}
\delta S=-\int \mathrm{d}^3x ~\left\lbrace\,L_i^{\ \mu}(x) \,\delta b^i_\mu(x) + \bar{L}_i^{\ \mu}(x) \,\delta\omega^i_{\ \mu}(x) \,\right\rbrace
\end{align}
where $L_i^{\ \mu}:=-\dfrac{\delta\mathcal{L}}{\delta b^i_{\ \mu}}$ and $\bar{L}_i^{\ \mu}:= -\dfrac{\delta\mathcal{L}}{\delta\omega^i_{\ \mu}}$ are the Euler derivatives. The corresponding equations of motion are just
\begin{align}
\label{GenEqnsMotion}
L_i^{\ \mu}(x)=0 \qquad\text{and}\qquad \bar{L}_i^{\ \mu}(x)=0.
\end{align}
Now let us propose the following symmetries of the fields
\begin{align}
\label{GenVarFields}
\begin{aligned}
\delta b^i_{\ \mu}(x)&=\displaystyle\sum_{s=0}^n (-1)^s\int \mathrm{d}^2{\bf z}~\left[\frac{\partial^s\xi^\sigma(z)}{\partial \mathrm{t}^s}\rho^i_{\ \mu\sigma(s)}(x,z) + \frac{\partial^s\theta^k(z)}{\partial \mathrm{t}^s}\zeta^i_{\ \mu k(s)}(x,z) \right]\\
\delta \omega^i_{\ \mu}(x)&=\displaystyle\sum_{s=0}^n (-1)^s\int \mathrm{d}^2{\bf z}~\left[\frac{\partial^s\xi^\sigma(z)}{\partial \mathrm{t}^s}\bar{\rho}^i_{\ \mu\sigma(s)}(x,z) + \frac{\partial^s\theta^k(z)}{\partial \mathrm{t}^s}\bar{\zeta}^i_{\ \mu k(s)}(x,z) \right]
\end{aligned}
\end{align}
where the functions $\rho$, $\bar{\rho}$, $\zeta$ and $\bar{\zeta}$ are known as `lagrangian generators' that generate variations in the basic fields while the six quantities $\xi^\sigma(z)$, $\theta^k(z)$ are functions of space and time serving as infinitesimal gauge parameters. Note that their number is governed by the Poincar\'{e} symmetry group, which in 2+1 dimensions has six independent symmetries. These variations of fields are symmetries in the sense that $S\left[b^i_{\ \mu}+\delta b^i_{\ \mu},\ \omega^i_\mu + \delta\omega^i_{\ \mu}\right]=S\left[b^i_{\ \mu},\omega^i_{\ \mu}\right]$, or equivalently $\delta S=0$ under these variations. Substituting the variations \eqref{GenVarFields} of $b$ and $\omega$ in \eqref{2plus1GenActionVar} yields the variation of the action as
\begin{align}
\label{ex b1}
\begin{aligned}
\delta S = -\int \mathrm{d}^2{\bf x} \int \mathrm{d}^2{\bf z} \,\displaystyle\sum_{s=0}^n (-1)^s \int \mathrm{d}\mathrm{t} &\left[\frac{\partial^s\xi^\sigma(z)}{\partial \mathrm{t}^s} \rho^i_{\ \mu\sigma(s)}(x,z) L_i^{\ \mu}(x)\right.\\
&\left.\ + \frac{\partial^s\theta^k(z)}{\partial \mathrm{t}^s} \zeta^i_{\ \mu k(s)}(x,z) L_i^{\ \mu}(x)\right]\\
-\int \mathrm{d}^2{\bf x} \int \mathrm{d}^2{\bf z} \,\displaystyle\sum_{s=0}^n (-1)^s \int \mathrm{d}\mathrm{t} &\left[\frac{\partial^s\xi^\sigma(z)}{\partial \mathrm{t}^s} \bar{\rho}^i_{\ \mu\sigma(s)}(x,z) \bar{L}_i^{\ \mu}(x)\right.\\
&\left.\ + \frac{\partial^s\theta^k(z)}{\partial \mathrm{t}^s} \bar{\zeta}^i_{\ \mu k(s)}(x,z) \bar{L}_i^{\ \mu}(x)\right].
\end{aligned}
\end{align}
To simplify the above expression, let us take a representative term -- the first term in the first line -- from \eqref{ex b1}. The other terms, having similar structure, can then be handled by a similar technique.
\begin{align}
-& \int \!\!\mathrm{d}^2{\bf x} \int \!\!\mathrm{d}^2{\bf z} \,\displaystyle\sum_{s=0}^n (-1)^s \int \!\!\mathrm{d}\mathrm{t} ~\frac{\partial^s\xi^\sigma(z)}{\partial \mathrm{t}^s} \rho^i_{\ \mu\sigma(s)}(x,z) L_i^{\ \mu}(x)\nonumber\\
= & \!\! \;-\int \!\!\mathrm{d}^2{\bf x} \int \!\!\mathrm{d}^2{\bf z} \,\left[\int \!\!\mathrm{d}\mathrm{t} ~\xi^\sigma(z) \rho^i_{\ \mu\sigma(0)}(x,z) L_i^{\ \mu}(x) - \!\int \!\!\mathrm{d}\mathrm{t} \frac{\partial\xi^\sigma(z)}{\partial \mathrm{t}} \rho^i_{\ \mu\sigma(1)}(x,z) L_i^{\ \mu}(x) + \ldots \right.\nonumber\\
& \left. \qquad\qquad\qquad\qquad \ldots + (-1)^n\int \!\!\mathrm{d}\mathrm{t} ~\frac{\partial^n\xi^\sigma(z)}{\partial \mathrm{t}^n} \rho^i_{\ \mu\sigma(n)}(x,z) L_i^{\ \mu}(x) \right].
\end{align}
We now interchange derivatives (wherever applicable) by using partial integrals, throwing away the boundary terms by assuming the fields to be well behaved at infinity. Also note that we have precisely the same number of negative signs before each term through the $(-1)^s$ factor, as required for carrying out partial integrals `$s$' times. So, simplifying the above equation yields
\begin{align}
& -\int \!\!\mathrm{d}^2{\bf x} \int \!\!\mathrm{d}^2{\bf z} \left[\int \!\!\mathrm{d}\mathrm{t} ~\xi^\sigma(z) \left\lbrace\rho^i_{\ \mu\sigma(0)}(x,z) L_i^{\ \mu}(x)\right\rbrace + \!\int \!\!\mathrm{d}\mathrm{t}~\xi^\sigma(z)\frac{\partial}{\partial \mathrm{t}} \left\lbrace\rho^i_{\ \mu\sigma(1)}(x,z) L_i^{\ \mu}(x) \right\rbrace+ \right.\nonumber\\
& \left. \qquad\qquad\qquad\qquad \ldots + \int \!\!\mathrm{d}\mathrm{t} ~\xi^\sigma(z)~\frac{\partial^n}{\partial \mathrm{t}^n} \left\lbrace\rho^i_{\ \mu\sigma(n)}(x,z) L_i^{\ \mu}(x) \right\rbrace\right]\nonumber\\
& \!\!= \;-\int \!\!\mathrm{d}^2{\bf z} \int \!\!\mathrm{d}\mathrm{t} ~\xi^\sigma(z)~\left[\displaystyle \sum_{s=0}^n\int \mathrm{d}^2{\bf x} ~\frac{\partial^s}{\partial\mathrm{t}^s} \left\lbrace \rho^i_{\ \mu\sigma(s)}(x,z) L_i^{\ \mu}(x) \right\rbrace\right].
\end{align}
Substituting this back in the variation \eqref{ex b1} of the action, we get
\begin{align}
\delta S = -\int \mathrm{d}^3z \left[ \xi^\sigma(z) \,\Lambda_\sigma(z) + \theta^k(z) \,\Xi_k(z)\right]
\end{align}
where $\Lambda$ and $\Xi$ are defined as:
\begin{align}
\begin{aligned}
\label{systemGaugeId}
\Lambda_\sigma(z) &=& \displaystyle\sum_{s=0}^n \int \mathrm{d}^2{\bf x} ~\frac{\partial^s}{\partial \mathrm{t}^s} \left\lbrace \rho^i_{\ \mu\sigma(s)}(x,z)L_i^{\ \mu}(x) + \bar{\rho}^i_{\ \mu\sigma(s)}(x,z)\bar{L}_i^{\ \mu}(x) \right\rbrace\\
\Xi_k(z) &=& \displaystyle\sum_{s=0}^n \int \mathrm{d}^2{\bf x} ~\frac{\partial^s}{\partial \mathrm{t}^s} \left\lbrace \zeta^i_{\ \mu k(s)}(x,z)L_i^{\ \mu}(x) + \bar{\zeta}^i_{\ \mu k(s)}(x,z)\bar{L}_i^{\ \mu}(x) \right\rbrace.\\
\end{aligned}
\end{align}
Since each of the gauge parameters $\xi^\sigma$ and $\theta^k$ are independent quantities, the invariance of the action $\left(\delta S=0\right)$ implies the following conditions
\begin{align}
\label{shortGaugeIds}
\begin{aligned}
\Lambda_\sigma(z) &= 0\\
\Xi_k(z) &= 0.
\end{aligned}
\end{align}
These are precisely the gauge identities that we were talking about in Section \ref{Clag:GIs}. They are identities in the sense that on substituting the Euler derivatives $L_i^{\ \mu}$ and $\bar{L}_i^{\ \mu}$ in \eqref{systemGaugeId}, all terms cancel out and we see that the relations are zero identically. Note that until now we have only used the definition of the Euler derivatives in terms of variation of fields $\left( \text{for example } L_i^{\ \mu}=-\dfrac{\delta \mathcal{L}}{\delta b^i_{\ \mu}}~\right)$, but we have not set the Euler derivative to zero, i.e. we have not used any equations of motion. In fact, using the equations of motion trivializes the gauge identities as $0=0$ relations. 

Now the algorithm for finding out the Lagrangian symmetry generators is simple. Given an action, we can easily find the Euler derivatives by varying the action w.r.t. the basic fields. Using the Euler derivatives, we can then try to build a set of independent, identically vanishing equations - the gauge identities. Alternatively, we can also explicitly check any set of identities proposed to hold as gauge identities from physical considerations. The number of these gauge identities is identical to the number of independent symmetries. In this particular case of PGT, they are six in number and are stated compactly in \eqref{shortGaugeIds}. Once we obtain a set of \emph{independent} gauge identities for the action, we then finally compare the given identities with the general form presented in \eqref{systemGaugeId}, and find out the generators $\left(\text{denoted here as } \rho, ~\bar{\rho}, ~\zeta \text{ and } ~\bar{\zeta} \,\right)$.


\section{Lagrangian generators for 2+1 dimensional PGT symmetries}
\label{Clag:2+1LagGen}

The Mielke-Baekler model which explicitly includes a torsion term, along with the Chern-Simons action and the usual Einstein-Cartan piece that we study here is given in \eqref{CmbMBaction}. This action is known to be invariant under the PGT symmetries \eqref{Poincare3Dfieldtrans}. A set of independent gauge identities corresponding to PGT symmetries for the action \eqref{CmbMBaction} is already known from our analysis in Section \ref{Clag:GIs} and are given explicitly as \cite{Banerjee:2009vf}
\begin{align}
\begin{aligned}
\label{prevGaugeId}
\Lambda_\sigma &:= -L_i^{\ \mu} \partial_\sigma b^i_{\ \mu} - \bar{L}_i^{\ \mu} \partial_\sigma \omega^i_{\ \mu} + \partial_\mu\left(b^i_{\ \sigma}L_i^{\ \mu}\right) + \partial_\mu\left(\omega^i_{\ \sigma}\bar{L}_i^{\ \mu}\right)=0\\
\Xi_k &:= -\epsilon^i_{\ jk}L_i^{\ \mu}b^j_{\ \mu} - \epsilon^i_{\ jk}\bar{L}_i^{\ \mu}\omega^j_{\ \mu} + \partial_\mu \bar{L}_k^{\ \mu}=0.
\end{aligned}
\end{align}
Here $L_i^{\ \mu}$ and $\bar{L}_i^{\ \mu}$ are the Euler derivatives obtained from the action \eqref{CmbMBaction}, and are given by,
\begin{align}
\label{Euler2plus1}
\begin{aligned}
L_i^{\ \mu} &:= -\frac{\delta \mathcal{L}}{\delta b^i_{\ \mu}} = -\epsilon^{\mu\nu\rho}\left[a R_{i\nu\rho} +\alpha_4 T_{i\nu\rho} - \Lambda\epsilon_{ijk}b^j_{\ \nu}b^k_{\ \rho}\right]\\
\bar{L}_i^{\ \mu} &:= -\frac{\delta \mathcal{L}}{\delta \omega^i_{\ \mu}} = -\epsilon^{\mu\nu\rho}\left[\alpha_3R_{i\nu\rho} + aT_{i\nu\rho} - \Lambda\epsilon_{ijk}b^j_{\ \nu}b^k_{\ \rho}\right].
\end{aligned}
\end{align}
Substituting these in \eqref{prevGaugeId} it may easily be checked that all terms cancel and they are indeed identities. Now the Lagrangian symmetry generators, which are to give us a set of symmetries of the action \eqref{CmbMBaction}, may be found by comparing these identities -- \eqref{prevGaugeId}, with the general gauge identities derived before as \eqref{systemGaugeId}.

We have employed the following strategy in comparing the two relations in question. Any sum over Greek (holonomic) indices is broken into the time and space part, {\it i.e.} say, $A^\mu B_\mu = A^0 B_0 + A^\alpha B_\alpha$; the gauge identity $\Lambda_\sigma$ is also broken into sets $\Lambda_0$ and $\Lambda_\alpha$; and finally coefficients (in general field dependant) of the Euler derivatives $L_i^{\ 0}$, $L_i^{\ \beta}$, etc. are matched between the two relations \eqref{systemGaugeId} and \eqref{prevGaugeId}.

Let us now illustrate the details for one particular term: coefficient of $L_i^{\ \beta}$ in $\Lambda_\alpha$. An inspection of \eqref{prevGaugeId} reveals that there occur terms either with zero or a single time derivative. This implies that the summation over `$s$' in \eqref{systemGaugeId} is restricted to only two values, $s=0,\, 1.$ So, the relevant terms from \eqref{systemGaugeId} are,
\begin{align}
\label{ex a1}
\int ~\mathrm{d}^2{\bf x} ~\left\lbrace \rho^i_{\ \beta\alpha(0)}(x,z)\,L_i^{\ \beta}(x) + \partial_0\rho^i_{\ \beta\alpha(1)}(x,z)\,L_i^{\ \beta}(x) + \rho^i_{\ \beta\alpha(1)}(x,z) \,\partial_0 L_i^{\ \beta}(x) \right\rbrace
\end{align}
while those from \eqref{prevGaugeId} are
\begin{align*}
-&L_i^{\ \beta}(z)\,\partial_\alpha b^i_{\ \beta}(z) + \partial_\beta \left(b^i_{\ \alpha}(z)\,L_i^{\ \beta}(z)\right).
\end{align*}
The above expression may be recast in the form
\begin{align}
\label{ex a2}
\int \mathrm{d}^2{\bf x} ~\Big\lbrace -&L_i^{\ \beta}(x)\,\partial_\alpha b^i_{\ \beta}(x)\,\delta({\bf x}-{\bf z}) - b^i_{\ \alpha}(x)\,L_i^{\ \beta}(x)~\partial_\beta^{(\bf x)}\delta({\bf x}-{\bf z}) ~\Big\rbrace\,,
\end{align}
to facilitate comparison with \eqref{ex a1}. This comparison yields the generators $\rho^i_{\ \beta\alpha(1)}=0$ and $\rho^i_{\ \beta\alpha(0)} = -\partial_\alpha b^i_{\ \beta}(x) ~\delta({\bf x}-{\bf z}) - b^i_{\ \alpha}(x)~\partial_\beta^{(\bf x)}\delta({\bf x}-{\bf z})$. The other generators may also be found in a similar manner. We list all the non-zero ones below:
\begin{align}
\label{2plus1GenSet1}
\begin{aligned}
\rho^i_{\ 0\sigma(0)}(x,z) &= -\partial_\sigma b^i_{\ 0}(x) ~\delta({\bf x}-{\bf z})\\
\rho^i_{\ \alpha \sigma(0)}(x,z) &= -\partial_\sigma b^i_{\ \alpha}(x) ~\delta({\bf x}-{\bf z}) - b^i_{\ \sigma}(x)~\partial_\alpha^{(\bf x)}\delta({\bf x}-{\bf z})\\
\rho^i_{\ 0\sigma(1)}(x,z) &= {\ } b^i_{\ \sigma}(x) ~\delta({\bf x}-{\bf z}),\\
\end{aligned}
\end{align}
\begin{align}
\label{2plus1GenSet2}
\begin{aligned}
\bar{\rho}^i_{\ 0\sigma(0)}(x,z) &= -\partial_\sigma \omega^i_{\ 0}(x) ~\delta({\bf x}-{\bf z})\\
\bar{\rho}^i_{\ \alpha\sigma(0)}(x,z) &= -\partial_\sigma \omega^i_{\ \alpha}(x) ~\delta({\bf x}-{\bf z}) - \omega^i_{\ \sigma}(x)~\partial_\alpha^{(\bf x)}\delta({\bf x}-{\bf z})\\
\bar{\rho}^i_{\ 0\sigma(1)}(x,z) &= {\ } \omega^i_{\ \sigma}(x) ~\delta({\bf x}-{\bf z}),\\
\end{aligned}
\end{align}
\begin{align}
\label{2plus1GenSet3}
\begin{aligned}
\zeta^i_{\ \sigma k(0)}(x,z) &= -\epsilon^i_{\ jk} b^j_{\ \sigma}(x) ~\delta({\bf x}-{\bf z}),\\
\end{aligned}
\end{align}
\begin{align}
\label{2plus1GenSet4}
\begin{aligned}
\bar{\zeta}^i_{\ 0k(0)}(x,z) &= -\epsilon^i_{\ jk} \omega^j_{\ 0}(x) ~\delta({\bf x}-{\bf z})\\
\bar{\zeta}^i_{\ \alpha k(0)}(x,z) &= -\epsilon^i_{\ jk} \omega^j_{\ \alpha}(x) ~\delta({\bf x}-{\bf z}) - \delta^i_k~\partial_\alpha^{({\bf x})} \delta({\bf x}-{\bf z})\\
\bar{\zeta}^i_{\ 0k(1)}(x,z) &= {\ } \delta^i_k ~\delta({\bf x}-{\bf z}).\\
\end{aligned}
\end{align}
Having obtained the Lagrangian generators, it is now possible to find the transformations of basic fields $b$ and $\omega$ through \eqref{GenVarFields}. We illustrate the process for $b^i_{\ \alpha}$.
\begin{align}
\begin{aligned}
\delta b^i_{\ \alpha}(x)=\int \mathrm{d}^2{\bf z} &\left[ \xi^0(z)\,\rho^i_{\ \alpha 0(0)}(x,z) + \xi^\beta(z)\,\rho^i_{\ \alpha\beta(0)}(x,z)\right.\\
&\left.\phantom{[ \xi^0(z)\,\rho^i_{\ \alpha 0(0)}(x,z) + \xi^\beta(z)} + \theta^k(z)\,\zeta^i_{\ \alpha k(0)}(x,z) \right]\\
-\int \mathrm{d}^2{\bf z} &\left[ \partial_0\xi^0(z)\,\rho^i_{\ \alpha 0(1)}(x,z) + \partial_0\xi^\beta(z)\,\rho^i_{\ \alpha\beta(1)}(x,z)\right.\\
&\left. \phantom{[ \partial_0\xi^0(z)\,\rho^i_{\ \alpha 0(1)}(x,z) + \partial_0\xi} + \partial_0\theta^k(z)\,\zeta^i_{\ \alpha k(1)}(x,z) \right]
\end{aligned}
\end{align}
Using the form of the generators $\rho,\,\zeta$ given in \eqref{2plus1GenSet1} and \eqref{2plus1GenSet3}, one obtains,
\begin{align}
\begin{aligned}
\delta b^i_{\ \alpha}(x)=&\int \mathrm{d}^2{\bf z} ~\left[ \xi^0(z)\,\left\lbrace -\partial_0 b^i_{\ \alpha}(x)\,\delta({\bf x}-{\bf z}) - b^i_{\ 0}(x)~\partial_\alpha^{(\bf x)}\delta({\bf x}-{\bf z}) \right\rbrace \right.\\
&\left. \qquad\;\: + \;\xi^\beta(z)\,\left\lbrace -\partial_\beta b^i_{\ \alpha}(x) ~\delta({\bf x}-{\bf z}) - b^i_{\ \beta}(x)~\partial_\alpha^{(\bf x)}\delta({\bf x}-{\bf z}) \right\rbrace  \right.\\
&\left. \qquad\;\: + \;\theta^k(z)\,\left\lbrace -\epsilon^i_{\ jk} b^j_{\ \alpha}(x) ~\delta({\bf x}-{\bf z}) \right\rbrace \right] \ - \ \int \mathrm{d}^2{\bf z} ~\left[\ 0 + 0 + 0\ \right]\\
=& -\epsilon^i_{\ jk}\,b^j_{\ \alpha}\,\theta^k - \partial_\alpha\xi^\mu\,b^i_{\ \mu} - \xi^\mu\,\partial_\mu b^i_\alpha,\\
\end{aligned}
\end{align}
which corresponds to the $\mu=\alpha$ (space) component of the PGT symmetry $\delta b^i_{\ \mu}$ given in \eqref{Poincare3Dfieldtrans}. Other PGT symmetries given in \eqref{Poincare3Dfieldtrans} are easily reproduced by this procedure.


\section{PGT construction in 3+1 dimensions and lagran\-gian analysis}
\label{Clag:3+1LagGen}

The same PGT symmetries, being constructed out of local Lorentz and general diffeomorphism symmetries, are respected by a wide class of Lagrangians \cite{Blagojevic:2002du}. In 3+1 dimensions, the Chern-Simons term of 2+1 dimensions \eqref{CmbMBaction} automatically drops out. The other terms have their counterparts in 3+1 dimensions, in addition to some other new possible terms. However, for simplicity, it suffices towards our aim of constructing the generators of PGT symmetries, to consider only the most important part of the gravitational action -- the Einstein-Cartan term. Thus we take the following action in 3+1 dimensions,
\begin{align}
\label{EC3plus1}
S=\int \mathrm{d}^4x ~b\,R
\end{align}
where $b=\text{det}\left(b^i_{\ \mu}\right)$ and the curvature scalar $R=b_i^{\ \mu}b_j^{\ \nu}R^{ij}_{\ \ \mu\nu}$. The curvature tensor and the torsion tensor are defined as:
\begin{align}
\label{3plus1CurvTor}
\begin{aligned}
T^i_{\ \mu\nu} &= \partial_\mu b^i_{\ \nu} - \partial_\nu b^i_{\ \mu} + \omega^{i}_{\ \, k\mu} b^k_{\ \nu} - \omega^{i}_{\ \,k\nu} b^k_{\ \mu}\\
R^{ij}_{\ \ \mu\nu} &= \partial_\mu \omega^{ij}_{\ \ \nu} - \partial_\nu \omega^{ij}_{\ \ \mu} + \omega^i_{\ k\mu}\omega^{kj}_{\ \ \nu} - \omega^i_{\ k\nu}\omega^{kj}_{\ \ \mu}.
\end{aligned}
\end{align}
The corresponding Euler derivatives can be found in the standard way
\begin{align}
\label{3plus1EulerDeriv}
\begin{aligned}
L_i^{\ \mu} &:= -\frac{\delta \mathcal{L}}{\delta b^i_{\ \mu}}  = -2b \left( R_i^{\ \mu} + \frac{1}{2} b_i^{\ \mu}R \right)\\
L_{ij}^{\ \ \mu} &:= -\frac{\delta \mathcal{L}}{\delta \omega^{ij}_{\ \ \mu}}  = b \left( \,b_s^{\ \mu}\,T^s_{\ ij} + b_i^{\ \mu}\,T^s_{\ js} - b_j^{\ \mu}\,T^s_{\ is} \,\right).
\end{aligned}
\end{align}

To find the appropriate gauge identities here, we will now take help of the identities found previously for 2+1 dimensions. The 2+1 dimensional model was constructed using the basic fields $b^i_{\ \mu}$ and $\omega^i_{\ \mu}$, where the latter was a dual construct of the field $\omega^{ij}_{\ \ \mu}$, valid only in 2+1 dimensions. We would now like to write the gauge identity \eqref{prevGaugeId} in terms of the fields $b^i_{\ \mu}$ and $\omega^{ij}_{\ \ \mu}$, thus getting rid of the use of special 2+1 dimensional properties. The resultant identities will then be proposed for 3+1 dimensions and a Lagrangian analysis will be carried out to find out the corresponding symmetries.

Now let us consider the $\Xi_k$ identity in \eqref{prevGaugeId}. Contracting it with the Levi-Civita symbol, we find,
\begin{align}
\label{EpsilonContr}
\Xi_{mn} = -\epsilon_{mn}^{\ \ \ k}\:\Xi_k = L_m^{\ \;\mu}\,b_{n\mu} - L_n^{\ \mu}\,b_{m\mu} + \bar{L}_m^{\ \mu}\,\omega_{n\mu} - \bar{L}_n^{\ \mu}\,\omega_{m\mu} - \epsilon_{mn}^{\ \ \ k}\,\partial_\mu \bar{L}_k^{\ \mu}.
\end{align}
Next, introducing relations between the dual fields and their corresponding counterparts through
\begin{align}
\label{LiftMaps}
\begin{aligned}
\omega^i_{\ \mu} &= -\frac{1}{2} \,\epsilon^i_{\ jk}\,\omega^{jk}_{\ \ \mu}\\
\bar{L}_i^{\ \mu} &= -\epsilon_i^{\ jk} \, L_{jk}^{\ \ \mu},
\end{aligned}
\end{align}
and using the following identity for Levi-Civita symbols
\begin{align}
\label{LeviCivitaProduct}
\epsilon_m^{\ \ hp}\,\epsilon_{njk} = \eta_{mn}\delta^h_j\delta^p_k - \eta_{mn}\delta^h_k\delta^p_j - \eta_{mj}\delta^h_n\delta^p_k + \eta_{mj}\delta^h_k\delta^p_n + \eta_{mk}\delta^h_n\delta^p_j - \eta_{mk}\delta^h_j\delta^p_n,
\end{align}
we are able to write the gauge identity completely in terms of the fields $b^i_{\ \mu}$ and $\omega^{ij}_{\ \ \mu}$. The other gauge identity $\Lambda_\sigma$ in the set of the 2+1 dim identities \eqref{prevGaugeId} can also be ridden of the duals through a similar procedure. The resultant set of identities are:
\begin{align}
\label{3plus1PGTGaugeIds}
\begin{aligned}
\Lambda_\sigma &:= -L_i^{\ \mu}\partial_\sigma b^i_{\ \mu} - L_{ij}^{\ \ \mu}\partial_\sigma\omega^{ij}_{\ \ \mu} + \partial_\mu\left(L_i^{\ \mu}b^i_{\ \sigma}\right) + \partial_\mu\left(L_{ij}^{\ \ \mu}\omega^{ij}_{\ \ \sigma}\right)=0\\
\Xi_{ij} &:= {\ } L_{i\mu}\,b_j^{\ \mu} - L_{j\mu}\,b_i^{\ \mu} + 2\left(\partial_{\nu}L_{ij}^{\ \ \nu}-L_{ik}^{\ \ \nu}\omega^k_{\ j\nu} + L_{jk}^{\ \ \nu} \omega^k_{\ i\nu}\right)=0.
\end{aligned}
\end{align}
Since these are now written independent of any dimensionally dependant dual fields, we may propose that they also hold in 3+1 dimensions. An explicit check, using the expressions for the Euler derivatives \eqref{3plus1EulerDeriv} confirms the proposition.

Expressing the action \eqref{Gen2plus1Action} in terms of basic fields, rather than the duals, we have
\begin{align}
\label{Gen3plus1Action}
S=\int \,\mathrm{d}^4x~\mathcal{L}\left[ b^i_{\ \mu},\,\omega^{ij}_{\ \ \mu} \right].
\end{align}
The symmetry transformations of this action are now given by
\begin{align}
\label{3plus1GenVarFieldsPGT}
\begin{aligned}
\delta b^i_{\ \mu}(x) &= \displaystyle\sum_{s=0}^n (-1)^s\int \mathrm{d}^3{\bf z}~\left[\frac{\partial^s\xi^\sigma(z)}{\partial \mathrm{t}^s}\rho^i_{\ \mu\sigma(s)}(x,z) + \frac{\partial^s\theta^{lk}(z)}{\partial \mathrm{t}^s}\zeta^i_{\ \mu lk(s)}(x,z) \right]\\
\delta \omega^{ij}_{\ \ \mu}(x) &= \displaystyle\sum_{s=0}^n (-1)^s\int \mathrm{d}^3{\bf z}~\left[\frac{\partial^s\xi^\sigma(z)}{\partial \mathrm{t}^s}\bar{\rho}^{ij}_{\ \ \mu\sigma(s)}(x,z) + \frac{\partial^s\theta^{lk}(z)}{\partial \mathrm{t}^s}\bar{\zeta}^{ij}_{\ \ \mu lk(s)}(x,z) \right],
\end{aligned}
\end{align}
which are the 3+1 dimensional versions of \eqref{GenVarFields}. Now adopting identical steps as in Section \ref{Clag:2+1LagGen}, we obtain the analogues of the gauge identities \eqref{systemGaugeId}
\begin{align}
\label{3plus1PGTSytemGaugeIds}
\begin{aligned}
\Lambda_\sigma(z) &= \displaystyle\sum_{s=0}^n\int \textrm{d}^3{\bf{x}} ~\frac{\partial^s}{\partial \mathrm{t}^s} \left\lbrace \rho^i_{\ \mu\sigma(s)}(x,z)L_i^{\ \mu}(x) + \bar{\rho}^{ij}_{\ \ \mu\sigma(s)}(x,z) L_{ij}^{\ \ \mu}(x) \right\rbrace\\
\Xi_{lk}(z) &= \displaystyle\sum_{s=0}^n\int \textrm{d}^3{\bf{x}} ~\frac{\partial^s}{\partial \mathrm{t}^s} \left\lbrace \zeta^i_{\ \mu lk(s)}(x,z) L_i^{\ \mu}(x) + \bar{\zeta}^{ij}_{\ \ \mu lk(s)}(x,z) L_{ij}^{\ \ \mu}(x) \right\rbrace.
\end{aligned}
\end{align}
We can compare these with the set of identities \eqref{3plus1PGTGaugeIds} term by term as explained in the discussion above eq. \eqref{ex a1}, to find out the relevant Lagrangian generators. The non-zero ones are listed below
\begin{align}
\label{3plus1GenSet1}
\begin{aligned}
\rho^i_{\ 0\sigma(0)}(x,z) &= -\partial_\sigma b^i_{\ 0}(x) ~\delta({\bf x}-{\bf z})\\
\rho^i_{\ \alpha\sigma(0)}(x,z) &= -\partial_\sigma b^i_{\ \alpha}(x) ~\delta({\bf x}-{\bf z}) - b^i_{\ \sigma}(x)~\partial_\alpha^{(\bf x)}\delta({\bf x}-{\bf z})\\
\rho^i_{\ 0\sigma(1)}(x,z) &= {\ } b^i_{\ \sigma}(x) ~\delta({\bf x}-{\bf z}),\\
\end{aligned}
\end{align}
\begin{align}
\label{3plus1GenSet2}
\begin{aligned}
\bar{\rho}^{ij}_{\ \ 0\sigma(0)}(x,z) &= -\partial_\sigma \omega^{ij}_{\ \ 0}(x) ~\delta({\bf x}-{\bf z})\\
\bar{\rho}^{ij}_{\ \ \alpha\sigma(0)}(x,z) &= -\partial_\sigma \omega^{ij}_{\ \ \alpha}(x) ~\delta({\bf x}-{\bf z}) - \omega^{ij}_{\ \ \sigma}(x)~\partial_\alpha^{(\bf x)}\delta({\bf x}-{\bf z})\\
\bar{\rho}^{ij}_{\ \ 0\sigma(1)}(x,z) &= {\ } \omega^{ij}_{\ \ \sigma}(x) ~\delta({\bf x}-{\bf z}),\\
\end{aligned}
\end{align}
\begin{align}
\label{3plus1GenSet3}
\begin{aligned}
\zeta^i_{\ \mu lk(0)}(x,z) &= {\ } \frac{1}{2}\left[\delta^i_l \,b_{k\mu}(x)-\delta^i_k \,b_{l\mu}(x)\right] ~\delta({\bf x}-{\bf z}),\\
\end{aligned}
\end{align}
\begin{align}
\label{3plus1GenSet4}
\begin{aligned}
\bar{\zeta}^{ij}_{\ \ 0lk(0)}(x,z) &= -\frac{1}{2}\left[ \delta^i_l\omega^j_{\ k0} - \delta^i_k\omega^j_{\ l0} - \delta^j_l\omega^i_{\ k0} + \delta^j_k\omega^i_{\ l0}  \right]\delta({\bf x}-{\bf z})\\
\bar{\zeta}^{ij}_{\ \ \alpha lk(0)}(x,z) &= -\frac{1}{2}\left[ \delta^i_l\omega^j_{\ k\alpha} - \delta^i_k\omega^j_{\ l\alpha} - \delta^j_l\omega^i_{\ k\alpha} + \delta^j_k\omega^i_{\ l\alpha}  \right]\delta({\bf x}-{\bf z})\\
&\phantom{= -\frac{1}{2}[ \delta^i_l\omega^j_{\ k\alpha} - \delta^i_k\omega^j_{\ l\alpha}} - \frac{1}{2}\left[ \delta^i_l\delta^j_k - \delta^i_k\delta^j_l \right] \,\partial_\alpha^{(\bf x)}\delta({\bf x}-{\bf z})\\
\bar{\zeta}^{ij}_{\ \ 0lk(1)}(x,z) &= {\ \ } \frac{1}{2} \left[ \delta^i_l\delta^j_k - \delta^i_k\delta^j_l \right] ~\delta({\bf x}-{\bf z}).\\
\end{aligned}
\end{align}
These generators will yield the symmetries of the action \eqref{EC3plus1} through the transformation \eqref{3plus1GenVarFieldsPGT}. An explicit calculation leads to the symmetries
\begin{align}
\label{3plus1PGTFieldTrans}
\begin{aligned}
\delta b^i_{\ \mu} &= \theta^i_{\ k} b^k_{\ \mu} - \partial_\mu\xi^\rho b^i_{\ \rho} - \xi^{\rho}\partial_{\rho} b^i_{\ \mu}\\
\delta \omega^{ij}_{\ \ \mu} &= \theta^i_{\ k} \omega^{kj}_{\ \ \mu} + \theta^j_{\ k} \omega^{ik}_{\ \ \mu} - \partial_\mu\theta^{ij} - \partial_\mu\xi^\rho \omega^{ij}_{\ \ \rho} - \xi^{\rho}\partial_{\rho}\omega^{ij}_{\ \ \mu}.
\end{aligned}
\end{align}
It may be easily checked that these transformations are indeed symmetries of the Einstein-Cartan action \eqref{EC3plus1} in 3+1 dimensions \cite{Blagojevic:2002du}.


We would now like to make a comparative remark on the structure of the Lagrangian generators in 2+1 and 3+1 dimensions. Let us consider the 2+1 dimensional generator $\bar{\zeta}^i_{\ \alpha k(0)}$ from \eqref{2plus1GenSet4}
\begin{align*}
\bar{\zeta}^i_{\ \alpha k(0)}(x,z) &= -\epsilon^i_{\ jk} \omega^j_{\ \alpha}(x) ~\delta({\bf x}-{\bf z}) - \delta^i_k~\partial_\alpha^{({\bf x})} \delta({\bf x}-{\bf z}).
\end{align*}
Multiplying appropriately with Levi-Civita symbols and using the map for $\omega^j_{\ \alpha}$ from \eqref{LiftMaps}, we get
\begin{align*}
\epsilon_i^{\ mn}\epsilon^k_{\ hp}\,\bar{\zeta}^i_{\ \alpha k(0)} = -\epsilon_i^{\ mn}\epsilon^k_{\ hp}\,\omega^i_{\ k\alpha}\,\delta(x-z) - \epsilon_i^{\ mn}\epsilon^i_{\ hp}~\delta_\alpha^{({\bf x})}\delta(x-z).
\end{align*}
Finally using the identity for contraction of Levi-Civita symbols \eqref{LeviCivitaProduct} and rearranging terms yield the following relation
\begin{align}
\label{ex d1}
\begin{aligned}
\bar{\zeta}^{mn}_{\ \ \ \alpha hp(0)} = &-\left[ \delta^m_h\omega^n_{\ p\alpha}-\delta^m_p\omega^n_{\  h\alpha}-\delta^n_h\omega^m_{\ p\alpha}+\delta^n_p\omega^m_{\ h\alpha} \right]\,\delta(x-z) \\
&- \left[ \delta^m_h\delta^n_p-\delta^m_p\delta^n_h \right]\,\partial_\alpha^{({\bf x})}\delta(x-z),
\end{aligned}
\end{align}
where we have defined the map
\begin{align}
\label{LagGenExMap}
\bar{\zeta}^{mn}_{\ \ \ \alpha hp(0)}(x,z) = -\frac{1}{2} ~\epsilon_i^{\ mn}\epsilon^k_{\ hp}\,\bar{\zeta}^i_{\ \alpha k(0)}(x,z).
\end{align}
Thus we have re-written the 2+1 dimensional generator $\bar{\zeta}^i_{\ \alpha k(0)}$ in terms of the original fields, getting rid of all duals. The object $\bar{\zeta}^{mn}_{\ \ \ \alpha hp(0)}$ defined in \eqref{ex d1}, however is functionally identical to the corresponding 3+1 dimensional generator \eqref{3plus1GenSet4}. Thus the map \eqref{LagGenExMap} expresses the 2+1 dimensional generator in a form that remains structurally the same, even in 3+1 dimensions.

Similarly, all the other generators from 2+1 dimensions, can be stripped off the dual fields $\omega^j_{\ \sigma}$. These dual fields were defined for the special case of 2+1 dimensions. Once having removed them, and expressed all basic fields in terms of their dimension independent form, we see that the same generators also hold in 3+1 dimensions. Below, we list all the non-trivial maps, in the sense described above, between the Lagrangian generators.
\begin{align}
\label{GeneratorLiftMap}
\begin{aligned}
\bar{\rho}^{mn}_{\ \ \ 0\sigma(0)}(x,z) &= -\epsilon_i^{\ mn}\,\bar{\rho}^{i}_{\ 0\sigma(0)}(x,z)\\
\bar{\rho}^{mn}_{\ \ \alpha \sigma(0)}(x,z) &= -\epsilon_i^{\ mn}\,\bar{\rho}^{i}_{\alpha \sigma(0)}(x,z)\\
\bar{\rho}^{mn}_{\ \ \ 0\sigma(1)}(x,z) &= -\epsilon_i^{\ mn}\,\bar{\rho}^{i}_{\ 0\sigma(1)}(x,z)\\
\zeta^i_{\ \sigma mn(0)}(x,z) &= - \frac{1}{2}\epsilon^k_{\ mn}\,\zeta^i_{\ \sigma k(0)}(x,z)\\
\bar{\zeta}^{mn}_{\ \ \ \;\sigma hp(0)}(x,z) &= -\frac{1}{2} ~\epsilon_i^{\ mn}\epsilon^k_{\ hp}\,\bar{\zeta}^i_{\ \sigma k(0)}(x,z)\\
\bar{\zeta}^{mn}_{\ \ \ \;0hp(1)}(x,z) &= -\frac{1}{2} ~\epsilon_i^{\ mn}\epsilon^k_{\ hp}\,\bar{\zeta}^i_{\ 0k(1)}(x,z)
\end{aligned}
\end{align}
Observation of this structural similarity of the generators across dimensions, from 2+1 to 3+1, indicates that in higher dimensions, similar results are expected to hold.


\section{Comments on general applications of lagrangian generators}
\label{Clag:UsesLagGen}

One might wonder about the possible applications of the lagrangian generators. The canonical generators derived in the hamiltonian formalism, apart from yielding the gauge symmetries, are often used to find the conserved charges and central terms in the Poisson algebra of spacetime symmetries \cite{Blagojevic:2008bn}. Since the hamiltonian and lagrangian formulations complement one another, it is expected that the lagrangian generators will also have a similar, though not necessarily identical, role. We now elaborate on this and related points.

The crucial ingredient in abstracting the lagrangian generators are the gauge identities. Construction of these identities can be made from physical considerations. However, there also exist systematic schemes for arriving at these gauge identities from algorithms employing lagrangian constraints \cite{Chaichian:1994ug, Shirzad:1998af}. So, given a model with some Lagrangian, we can arrive at the gauge identities and the lagrangian generators systematically. Now, some insight into these identities is gleaned from their connection with the Bianchi identities of a model \cite{Pons:2009nb, Ortin:2004ms, Weinberg:1972}. In what follows, we adopt the Einstein-Hilbert action in 3+1 dimensions for the demonstration of this connection. The calculation follows \cite{Samanta:2007fk} closely.

The Einstein-Hilbert action in 3+1 dimensions is written in terms of the basic field $g_{\mu\nu}$ -- the metric -- as:
\begin{align}
\label{App action}
S=\int \textrm{d}^4x ~\sqrt{-g}\,g^{\mu\nu}R_{\mu\nu},
\end{align}
where the Ricci tensor $R_{\mu\nu}$ is defined in terms of the Christoffel connections $\Gamma^\rho_{\mu\nu}$ in the usual way as in Einstein general relativity:
\begin{align}
\label{App RicciChrist}
\begin{aligned}
R_{\mu\nu} &= \Gamma^\lambda_{\nu\mu,\,\lambda} - \Gamma^\lambda_{\lambda\mu,\,\nu} + \Gamma^\lambda_{\nu\mu}\,\Gamma^\sigma_{\sigma\lambda} - \Gamma^\sigma_{\lambda\mu}\,\Gamma^\lambda_{\nu\sigma}\\
\Gamma^\rho_{\ \mu\nu} &= \frac{1}{2} \, g^{\rho\lambda} \Big(\, g_{\lambda\nu,\,\mu} + g_{\mu\lambda,\,\nu} - g_{\mu\nu,\,\lambda} \,\Big).
\end{aligned}
\end{align}
Varying the action \eqref{App action} w.r.t. the metric $g_{\mu\nu}$ we get the Euler derivative $L^{\mu\nu}$
\begin{align}
\label{App VariedAction}
\delta S = \int \textrm{d}^4x ~L^{\mu\nu}\,\delta g_{\mu\nu}
\end{align}
where,
\begin{align}
\label{App EulerD}
L^{\mu\nu} = \sqrt{-g}\,G^{\mu\nu} = \sqrt{-g}\,\left( R^{\mu\nu} - \frac{1}{2}\,g^{\mu\nu}R \right).
\end{align}
Invariance of the action leads to the usual Einstein's equation $L^{\mu\nu}=0$. The gauge identity may be subsequently defined as
\begin{align}
\label{App GaugeId1}
\nabla_{\!\mu\,} L^\mu_{\ \nu} = 0
\end{align}
which may also be expressed as,
\begin{align}
\label{App GaugeId2}
\nabla_{\!\mu\,} G^\mu_{\ \nu} = 0.
\end{align}
Now, the Bianchi identity for Einstein general relativity is well known and is written in terms of the Riemann tensor $R_{\lambda\mu\nu\kappa}$ as:
\begin{align}
\label{App Bianchi}
\nabla_{\!\eta\,} R_{\lambda\mu\nu\kappa} + \nabla_{\!\nu\,} R_{\lambda\mu\kappa\eta} + \nabla_{\!\kappa\,} R_{\lambda\mu\eta\nu} = 0.
\end{align}
Contracting $\lambda$ with $\nu$ and $\mu$ with $\kappa$ in the above identity (the metricity condition $\nabla_{\!\rho\,}g_{\mu\nu} = 0$ holds in Einstein general relativity), we reproduce the gauge identity \eqref{App GaugeId2}. This immediately shows that the gauge identity is nothing but a suitably contracted form of the Bianchi identity in this model.

The gauge identity, or the contracted version of the Bianchi identity, plays a significant role in the obtention of the Noether central charges. As in the hamiltonian description, here too surface terms are important. If these terms are not dropped, then \eqref{App VariedAction} takes the form,
\begin{align}
\label{App VariedActionWSur1}
\delta S = \int \textrm{d}^4x ~\sqrt{-g}\, \left[-G^{\mu\nu} \delta_{{\!}_\xi} g_{\mu\nu} + \nabla_{\!\rho}\left( 2\,g^{\mu\sigma,\rho\nu}\,\nabla_{\!\mu\,} \delta_{{\!}_\xi} g_{\sigma\nu} \right) \right].
\end{align}
Explicitly, using $\delta_{{\!}_{\xi\,}} g_{\mu\nu} = \nabla_{\!(\mu\,}\xi_{\nu)}$ we obtain,
\begin{align}
\label{App VariedActionWSur2}
\delta S = \int \textrm{d}^4x ~\sqrt{-g}\,\Big[-2\,\left\lbrace \nabla_{\!\mu\,} G^{\mu\nu}\right\rbrace \xi_\nu + \nabla_{\!\rho} \left\lbrace 2\,G^{\rho\sigma}\xi_\sigma -4\, g^{\mu\sigma,\rho\nu}\,\nabla_{\!\mu} \nabla_{\!(\sigma\,}\xi_{\nu)} \right\rbrace \Big] = 0\,.
\end{align}
The first term in the integrand vanishes due to the gauge identity \eqref{App GaugeId2}. This also implies the vanishing of the second term in the integrand. Effectively, this leads to the covariant conservation of the Noether current,
\begin{align}
\label{App NoeCurr}
j_{\scriptscriptstyle N}^\rho(\xi) = 2\,R^{\,\rho\sigma}\,\xi_\sigma - 4\,g^{\mu\sigma,\rho\nu}\,\nabla_{\!\mu} \nabla_{\!(\sigma\,}\xi_{\nu)}.
\end{align}
Corresponding to each vector $\xi^\sigma$ it is now possible to construct a conserved Noether charge from \eqref{App NoeCurr}. This yields the standard Komar's integral in general relativity.\footnote{See, Chapter 6, p. 179-180 of \cite{Ortin:2004ms}.}

A comment on the surface terms might be useful. In the hamiltonian approach, these terms are determined by requiring the functional differentiability of the generators. The corresponding criterion in the present Lagrangian formulation is to retain all surface terms in the variation of the action under a general coordinate transformation, eventually leading to the gauge identity. This is clearly manifested in \eqref{App VariedActionWSur2} where the first term in the integrand yields the gauge identity while the second is the cherished surface term.

We thus observe how the gauge identity, which is directly connected with the Lagrangian generators, leads to conserved Noether charges. Also, the complementary aspects of lagrangian and hamiltonian generators, vis-$\grave{\text{a}}$-vis the construction of conserved charges gets illuminated.


\section{Discussions}
\label{Clag:disc}

In this chapter, we demonstrated the role of Lagrangian generators in investigating gauge symmetries. In particular, the Poincar\'{e} gauge theory symmetries were reproduced through lagrangian generators for the 2+1 and 3+1 dimensional Mielke-Baekler type models of gravity. The lagrangian method of finding generators, was seen to be much simpler than its hamiltonian counterpart. To begin with, we took a 2+1 dimensional model of a PGT-invariant Lagrangian which has been of recent interest \cite{Banerjee:2009vf, Blagojevic:2004hj, Basu:2009dy} -- the 3D gravity model with torsion and a cosmological term. The starting gauge identities involving the Euler derivatives that were required for the lagrangian analysis, were constructed based on the known form of the PGT symmetries. The lagrangian generators were subsequently computed and PGT symmetries were recovered using the same. We next repeated the procedure for 3+1 dimensions, where we took only the representative and most important Einstein-Cartan term in the action. We lifted the 2+1 dimensional gauge identities to 3+1 dimensions. The validity of the lifted gauge identities in 3+1 dimensions was explicitly checked. Then the same method was adopted to calculate the generators giving rise to PGT symmetries for this case. The Lagrangian generators themselves were also shown to preserve their structure across the 2+1 to 3+1 dimension transition. The PGT symmetries were shown to be consistent throughout this process as has been shown in Figure \ref{ClagFig1}.


%% file: C7_trivial.tex
\chapter{Poincare symmetries, hamiltonian symmetries and `trivial' transformations}
\label{C:tr}

\lettrine[lraise=0.0, loversize=0.3, findent=3pt, nindent=0pt, lhang=0.1]{G}{auge} symmetries in various Poincar\'{e} gauge invariant theories are important and have received our focus in this thesis, as well as in the literature. Some, among the multitude of models where gauge symmetries have been studied, are Chern-Simons gauge theory \cite{Witten:1988hc}, Einstein-Cartan gravity \cite{Blagojevic:2002du,Frolov:2009wu}, topological gravity with torsion \cite{Blagojevic:2004hj,Banerjee:2009vf} and topologically massive gravities \cite{Deser:1981wh,Blagojevic:2008bn} including Bergshoeff-Holm-Townsend (BHT) or ``new massive gravity'' \cite{Blagojevic:2010ir,Banerjee:2011rx}. And by gauge symmetries, we have stressed that we mean those transformations of the basic fields of the action, parametrised by arbitrary functions of time, that leave the action invariant off-shell \cite{Henneaux:1992ig}. Of-course we allow for appropriate boundary conditions, and the arbitrary functions of time are our gauge parameters. The form of the Poincar\'{e} gauge symmetries `$\delta_{\scriptscriptstyle PGT}$', i.e. local Lorentz rotations and translations, dosen't depend on the particular diffeomorphism invariant model being considered. Say, for example, let us first consider the Einstein-Cartan action in 3D $$ S_1 = \int \mathrm{d^3x} ~\epsilon^{\mu\nu\rho}\, b^i_{\ \mu}R_{i\nu\rho}\,,$$ and then add to it the torsion $T_{i\nu\rho}$ enforced by a parameter $\alpha_4$ $$ S_2 = \int \mathrm{d^3x} ~\epsilon^{\mu\nu\rho} \left[ b^i_{\ \mu}R_{i\nu\rho} + \frac{\alpha_4}{2}\, b^i_{\ \mu}T_{i\nu\rho}\right].$$ The Poincar\'{e} symmetry of the (for example) triad field is the same for both of these actions $$ \delta_{\scriptscriptstyle PGT} b^i_{\ \mu} = -\epsilon^i_{\ jk}b^j_{\ \mu}\theta^k - \partial_\mu \xi^\rho \,b^i_{\ \rho} - \xi^\rho\,\partial_\rho b^i_{\ \mu} $$ and as we can see, it does not involve the coupling constant $\alpha_4$. Also, this symmetry is off-shell by construction. The gauge parameters here are $\xi^\rho$ for translations and $\theta^i$ for local Lorentz rotations.

To study the hamiltonian gauge symmetries `$\delta_{\scriptscriptstyle G}$', we have carried out {\it off-shell} canonical hamiltonian analysis for two different models in Chapters \ref{C:mb} and \ref{C:bht}. The nature of the hamiltonian symmetries depend intimately on the particular model being studied, through the structural nature of the first-class constraints. However, in all of the models, the Poincar\'{e} symmetries are not algebraically identifiable with the hamiltonian gauge symmetries. For example, in the Einstein-Cartan action with torsion we get $$\delta_{\scriptscriptstyle G} b^i_{\ \mu} = \nabla_\mu\varepsilon^i + \alpha_4 \,\epsilon^i_{\ jk} \,b^j_{\ \mu} \varepsilon^k + \epsilon^i_{\ jk}\,b^j_{\ \mu} \tau^k\,,$$ where $\varepsilon^i$ and $\tau^i$ are the gauge parameters.  Note that $\delta_{\scriptscriptstyle G}$ explicitly involves the coupling constant $\alpha_4$. To compare $\delta_{\scriptscriptstyle PGT}$ and $\delta_{\scriptscriptstyle G}$ we first have to map the (arbitrary) gauge parameters of the hamiltonian symmetries $\delta_{\scriptscriptstyle G}$ to those of the Poincar\'{e} symmetries $\delta_{\scriptscriptstyle PGT}$. The gauge parameters become different as the Poincar\'{e} parameters are dictated by either geometric or group theoretic demands while the hamiltonian parameters depend on the structure of the constraints arising in the theory. The required redefinition is usually done through an ad-hoc, field-dependant map \cite{Witten:1988hc, Blagojevic:2002du, Blagojevic:2004hj, Banerjee:2009vf, Blagojevic:2008bn, Blagojevic:2010ir, Banerjee:2011rx}. However, there is no concrete algorithm to suggest this map, and it is rather given as a proposition based on intuition and experience. After such a mapping, it is seen that the hamiltonian symmetries indeed give back the Poincar\'{e} symmetries, but modulo terms proportional to equations of motion \cite{Blagojevic:2004hj,Banerjee:2009vf}. $$\delta_{\scriptscriptstyle G} b^i_{\ \mu} \sim \delta_{\scriptscriptstyle PGT} b^i_{\ \mu} + \text{\sl Eqns.  of  motion}$$ So the hamiltonian symmetries are not exactly equal to the Poincar\'{e} symmetries and it seems that we may have two independent sets of symmetries for the same action! Each of these symmetries will now give rise to their own independent Noether identities, as we had constructed in Section \ref{Clag:GIs}.

This is not a desirable situation. It leads to an increase in the total number of independent gauge  parameters over and above that found through the canonical analysis. We now have to take the Poincar\'{e} symmetry parameters in addition to the hamiltonian gauge parameters, if they are distinct. Also, we have to deal with more number of independent Noether identities than the number of Poncar\'{e} symmetries present in the model. But we know that the number of gauge parameters and Noether identities must match the total number of independent, primary, first-class constraints (see Section \ref{Cmb:ConstraintsRev} and \cite{Henneaux:1990au,Banerjee:1999yc,Banerjee:1999hu}). This creates an apparent paradoxical situation.

In this Chapter, we finally provide our resolution of this paradox by pointing out that the pair of symmetries differ only through trivial gauge transformations. These types of transformations \cite{Henneaux:1992ig} do not introduce any new arbitrary functions of time in physical solutions, and thus are not `physical' symmetries. Thus, the hamiltonian mechanism actually reproduce the Poinacr\'{e} symmetries as the only physically relevant symmetries of the theory. Such symmetries also produce no new independent Noether identity and so the total number of identities and gauge parameters match the original number of Poincar\'{e} symmetries. Also, by exploiting the Noether identities, we provide a systematic method to construct the map between the hamiltonian and Poincar\'{e} gauge parameters, which was lacking. Finally, through this work, explicit examples of trivial gauge symmetries and the role they play in many well studied field theories get highlighted.

\paragraph*{Summary of conventions:} Latin indices refer to the local Lorentz frame and the Greek indices refer to the coordinate frame. The beginning letters of both alphabets $(a,b,c,\ldots)$ and $(\alpha,\beta,\gamma,\ldots)$ run over the space part (1,2) while the middle alphabet letters $(i,j,k,\ldots)$ and $(\mu,\nu,\lambda,\ldots)$ run over all coordinates (0,1,2). The totally antisymmetric tensor $\epsilon^{ijk}$ and the tensor density $\epsilon^{\mu\nu\rho}$ are both normalized so that $\epsilon^{012}=1$. The signature of space-time adopted here is $\eta = \text{diag}(+,-,-)$.


\section{Trivial gauge symmetries: an introduction}
\label{Ctr:trivial}

Let $S[q_i]$ describe an action with the basic field variables being $q_i$ $(i=1,2,\ldots,n)$. The canonical momenta are then defined as $\pi^i=\frac{\delta S}{\delta \dot{q_i}}$ and the hamiltonian phase space is constructed out of the conjugate pair $(q_i, \pi^i)$. The standard canonical procedure \cite{Dirac:Lectures} yields all the constraints. Let us denote the first class constraints as $\Sigma_{a}$, $(a=1,2,\ldots,f)$ and the second class constraints as $\chi_{b}$ $(b=1,2,\ldots,s)$, with $P=f+s$ being the total number of constraints. The Dirac prescription gives the gauge generator as a linear combination of all first class constraints $$G= \alpha^a\,\Sigma_{a},$$ $\alpha^a$'s being arbitrary parameters in time. However, not all the parameters $\alpha^a$ are independent. We can eliminate the dependant ones systematically and write the gauge generator in terms of only the independent $\alpha^a$'s, following a completely off-shell method \cite{Henneaux:1990au, Banerjee:1999yc, Banerjee:1999hu}.\footnote{There are other methods of construction of a gauge generator like \cite{Castellani:1981us}, though it is not an off-shell one.} The final generator yields the gauge transformations of fields through a Poisson bracket\footnote{Or a Dirac bracket, if the second class sector has been eliminated through introduction of Dirac brackets.} operation with the fields. There exist two different possibilities of defining this operation $\lbrace q, G\rbrace$, results being equivalent upto terms proportional in constraints.
\begin{align}
\label{transform Poisson}
\begin{aligned}
\delta_1 q &= \lbrace q, \alpha^a\,\Sigma_a \rbrace \\
\text{or,}\quad \delta_2 q &= \alpha^a \lbrace q, \Sigma_a \rbrace.
\end{aligned}
\end{align}
These two definitions of gauge transformations $\delta_1$ and $\delta_2$ differ upto `trivial gauge transformations' \cite{Henneaux:1990au}.

Trivial gauge transformations keep the action invariant simply by a specific antisymmetric structure within them. To write explicitly, let us consider transformations of the form
\begin{align}
\label{trivial gauge gen}
\delta q_i = \Lambda_{ij}\,\frac{\delta S}{\delta q_j},\qquad \Lambda_{ij}=-\Lambda_{ji}.
\end{align}
Here $\frac{\delta S}{\delta q_j}$ is the Euler derivative corresponding to the field $q_j$ and its equation of motion is given by setting this Euler derivative to zero. Thus on-shell, i.e. after imposition of equations of motion, trivial gauge transformations vanish. However invariance of the action $(\delta S = 0)$ is achieved off-shell due to the antisymmetry of $\Lambda_{ij}$
\begin{align}
\label{var action for trivial}
\delta S &= \frac{\delta S}{\delta q_i} \delta q_i \nonumber\\
&= \frac{\delta S}{\delta q_i}\,\Lambda_{ij}\,\frac{\delta S}{\delta q_j} = 0,
\end{align}
as the product $\frac{\delta S}{\delta q_i}\,\frac{\delta S}{\delta q_j}$ is symmetric in $i \,\&\, j$.
Since these transformations vanish on-shell they imply no degeneracy in the solutions of the equations of motion; i.e. they do not map a set of solutions to any other set through arbitrary functions of time, unlike true gauge transformations. Given any action, they can always be added as symmetry transformations and the specific form of the co-efficients do not matter, as long as they are antisymmetric in the field indices. They are not generated by first-class constraints in the hamiltonian formalism and give rise to zero gauge current as they are on-shell symmetries. Thus, trivial gauge symmetries are not true gauge symmetries and are of no physical importance.

As a consequence of the above discussion, it can be anticipated that the trivial gauge symmetries do not give rise to any new Noether identities, other than those already present due to the true symmetries of the system. Given any gauge symmetry parametrised by an arbitrary time function $\sigma$ (known as the gauge parameter), $$\delta q_i = R_{i\mu} \sigma^\mu +  \tilde{R}_{i\mu}^{\ \ \nu}\,\partial_\nu \sigma^\mu$$ where $R_i$'s and $\tilde{R}_i$'s are functions of the fields $q_i$ and possibly their derivatives, the invariance of the action leads to 
\begin{align}
\label{Noether Id gen}
\delta S &= \int \frac{\delta \mathcal{L}}{\delta q_i} \delta q_i \nonumber\\
&= \int \frac{\delta \mathcal{L}}{\delta q_i} \left(R_{i\mu} \sigma^\mu +  \tilde{R}_{i\mu}^{\ \ \nu}\,\partial_\nu \sigma^\mu \right) \nonumber\\
&= \int \left[\frac{\delta \mathcal{L}}{\delta q_i}\,R_{i\mu} - \partial_\nu\left( \frac{\delta \mathcal{L}}{\delta q_i} \, \tilde{R}_{i\mu}^{\ \ \nu} \right) \right]\sigma^\mu = 0.
\end{align}
Since $\sigma$ is an arbitrary function, we can write
\begin{align}
\label{Noether Id gen contd}
\frac{\delta \mathcal{L}}{\delta q_i}\,R_{i\mu} - \partial_\nu\left( \frac{\delta \mathcal{L}}{\delta q_i} \, \tilde{R}_{i\mu}^{\ \ \nu} \right) = 0
\end{align}
which are the Noether identities of the system.\footnote{Recall our lagrangian analysis, based on these identities and in the context of the Mielke-Baekler model, in Chapter \ref{C:lag}; also, see \cite{Banerjee:2010kd}.} They imply a dependence of the Euler derivatives $\frac{\delta \mathcal{L}}{\delta q_i}$ among themselves and thus the equations of motion are not all independent. Note that each Noether identity is proportional to a gauge parameter (here $\sigma^\mu$). Thus combinations of one set of independent  Noether identities among themselves to give rise to another equivalent set of identities is reflected at the symmetry level as a redefinition of the old gauge parameters into a new set of gauge parameters.

Now trivial gauge symmetries may affect the Noether identities in many ways. In a direct manner, if $R_i$ has antisymmetric contributions like $$R_{i\mu} \rightarrow R_{i\mu} + (\Lambda_{ij})_\mu\,\frac{\delta \mathcal{L}}{\delta q_j}\,\qquad(\Lambda_{ij})_\mu=-(\Lambda_{ji})_\mu,$$ as can arise from transformations like \eqref{trivial gauge gen}, then we will have extensions of the gauge identities \eqref{Noether Id gen contd} as shown below
\begin{align}
\label{Trivial Noether Id gen}
\frac{\delta \mathcal{L}}{\delta q_i}\,R_{i\mu} - \partial_\nu\left( \frac{\delta \mathcal{L}}{\delta q_i} \, \tilde{R}_{i\mu}^{\ \ \nu} \right) + \frac{\delta \mathcal{L}}{\delta q_i} (\Lambda_{ij})_\mu\,\frac{\delta \mathcal{L}}{\delta q_j} = 0.
\end{align}
However the last term vanishes by itself, without depending on the particular structure of the Euler derivatives, through (anti)symmetry. This generates no new identities and thus the Noether identities \eqref{Noether Id gen contd} and \eqref{Trivial Noether Id gen} are infact equivalent to each other and correspond to only one {\em physical} symmetry.

In the following sections, we work with explicit models (Einstein-Cartan gravity and a Mielke-Baekler \cite{Mielke:1991nn,Baekler:1992ab} type gravity) to show the role of trivial gauge symmetries in relating hamiltonian symmetries to the Poincar\'{e} symmetries. The analysis in each case will be based on the general formalism outlined in this section.


\section{The Einstein-Cartan model}
\label{Ctr:EC}

The Einstein-Cartan formulation of gravity is a first order generalisation of Einstein's general relativity. It is constructed through a Poincar\'{e} gauge theory (PGT) construction,  \cite{Utiyama:1956sy, Kibble:1961ba, 1962rdgr.book..415S, Hehl:1976kj, Blagojevic:2002du} on a Riemann-Cartan spacetime. As we have seen (Section \ref{Cpgt:Psymms}), the PGT gravity models are constructed to be invariant under the local Poincar\'{e} transformations
\begin{align}
\tag{\ref{Poincare3Dfieldtrans}}
\begin{aligned}
\delta_{\scriptscriptstyle PGT} b^i_{\ \mu} &= -\epsilon^i_{\ jk}b^j_{\ \mu}\theta^k - \partial_\mu \xi^\rho \,b^i_{\ \rho} - \xi^\rho\,\partial_\rho b^i_{\ \mu} \\
\delta_{\scriptscriptstyle PGT} \omega^i_{\ \mu} &= -\partial_\mu \theta^i - \epsilon^i_{\ jk}\omega^j_{\ \mu}\theta^k - \partial_\mu\xi^\rho\,\omega^i_{\ \rho} - \xi^\rho\,\partial_\rho\omega^i_{\ \mu}.
\end{aligned}
\end{align}
In the above symmetries, the parameter describing local Lorentz transformations is $\theta^i$ and that describing general coordinate transformations is $\xi^\mu$ (both transformations being of infinitesimal order). Intuitively, this explains the structure of the transformations \eqref{Poincare3Dfieldtrans} where the index `$i$' transforms as a Lorentz index while `$\rho$' transforms as a general coordinate index.\footnote{For a more detailed discussion one may refer to \cite{Blagojevic:2004hj,Banerjee:2009vf}.} The number of independent Poincar\'{e} symmetries for each field $(b,\ \omega, \text{ or any other field, if present})$ is reflected in the number of independent gauge parameters. In our model in 3D, $i=0,1,2$ and $\rho=0,1,2$. So the total number is
\begin{align}
\label{counting PGT}
3 \text{ against } \xi^\rho + 3 \text{ against } \theta^i = 6.
\end{align}
So we expect to find $6$ independent gauge parameters and $6$ independent Noether identities in our model and no more.

The Einstein-Cartan theory in 3D Riemann-Cartan spacetime gives back the standard Einstein gravity on imposition of the zero torsion condition. The action, in 3D, is
\begin{align}
\label{action EC}
S = \int \textrm{d$^3$x} ~a\,\epsilon^{\mu\nu\rho}\,b^i_{\ \mu}\,R_{i\nu\rho}.
\end{align}
The basic variables of the theory are $b^i_{\ \mu} \text{ and } ~\omega^i_{\ \mu}$ with the corresponding conjugate momenta being denoted by  $\pi_i^{\ \mu} \text{ and } ~\Pi_i^{\ \mu}$ respectively. The variational equations of motion are given by setting the Euler derivatives $\frac{\delta S}{\delta b^i_{\ \mu}}$ and $\frac{\delta S}{\delta \omega^i_{\ \mu}}$ to zero.
\begin{align}
\label{EOM EC}
\begin{aligned}
\frac{\delta S}{\delta b^i_{\ \mu}} &= a\,\epsilon^{\mu\nu\rho}\,R_{i\nu\rho} = 0\\
\frac{\delta S}{\delta \omega^i_{\ \mu}} &= a\,\epsilon^{\mu\nu\rho}\,T_{i\nu\rho} = 0
\end{aligned}
\end{align}
\begin{table}[h]
\centering
\begin{tabular}{l | c c}
\hline\hline\
& First Class & Second class \\[0.2ex] \hline\\[-1.9ex]
Primary & $\phi_i^{\ 0}\;, \Phi_i^{\ 0}$ & $\phi_i^{\ \alpha}$, $\Phi_i^{\ \alpha}$ \\[0.4ex]
Secondary & $\bar{\mathcal{H}}_i\;, \bar{\mathcal{K}}_i$ & \\[0.4ex]
\hline\hline
\end{tabular}
\caption{Constraints of the EC theory.} \label{Tab:Constraints EC}
\end{table}
A Dirac canonical analysis leads to the constraint structure \cite{Blagojevic:2004hj,Banerjee:2009vf} as given in Table \ref{Tab:Constraints EC}.
The relevant quantities in Table \ref{Tab:Constraints EC} are defined below:
\begin{align}
\label{Rel qtys EC}
\begin{aligned}
\phi_i^{\ \mu} &= \pi_i^{\ \mu}\\
\Phi_i^{\ \mu} &= \Pi_i^{\ \mu} - 2a\,\epsilon^{0\alpha\beta}\,b_{i\beta}\delta^\mu_\alpha\\
\bar{\mathcal{H}}_i &= \left[ -a\,\epsilon^{0\alpha\beta}R_{i\alpha\beta} \right] - \nabla_{\!\alpha}\phi_i^{\ \alpha}\\
\bar{\mathcal{K}}_i &= \left[ -a\,\epsilon^{0\alpha\beta}T_{i\alpha\beta} \right] - \nabla_{\!\alpha}\Phi_i^{\ \alpha} - \epsilon_{ijk}\,b^j_{\ \alpha}\phi^{k\alpha}\\
\end{aligned}
\end{align}
Once we have the constraints, we can construct the generator through an explicitly off-shell method \cite{Banerjee:1999yc,Banerjee:1999hu}. For Einstein-Cartan gravity, it turns out to be \cite{Banerjee:2009vf}
\begin{align}
\label{Gen EC}
G = {\ } &\dot{\varepsilon}^i\, \pi_i^{\ 0} + \varepsilon^i\left[ \bar{\mathcal{H}}_i - \epsilon_{ijk}\,\omega^j_{\ 0}\pi^{k0} \right] \nonumber\\
{\ } +\,& \dot{\tau}^i\, \Pi_i^{\ 0} + \tau^i\left[ \bar{\mathcal{K}}_i - \epsilon_{ijk} \left( b^j_{\ 0}\pi^{k0} + \omega^j_{\ 0} \Pi^{k0} \right) \right].
\end{align}
The hamiltonian gauge symmetries are calculated from the generator $G$, adopting the second among the definitions in \eqref{transform Poisson}
\begin{align}
\label{symm G EC}
\begin{aligned}
\delta_{\scriptscriptstyle G} b^i_{\ \mu} &= \nabla_\mu \varepsilon^i + \epsilon^i_{\ jk}b^j_{\ \mu} \tau^k\\
\delta_{\scriptscriptstyle G} \omega^i_{\ \mu} &= \nabla_\mu \tau^i .
\end{aligned}
\end{align}
Now the generator \eqref{Gen EC} is constructed as a linear combination of the products of first class constraints with gauge parameters. Looking at the first-class constraints in Table \ref{Tab:Constraints EC}, we see that they all have one local index as their free-index. This fixes the structure of the gauge parameters $\varepsilon^i$ and $\tau^i$ in the hamiltonian formulation and they turn out to be different from the Poincar\'{e} gauge parameters $\xi^\rho$ and $\theta^i$, translations and local Lorentz rotations, seen in \eqref{Poincare3Dfieldtrans}. However, to compare between the two symmetries $\delta_{\scriptscriptstyle G}$ and $\delta_{\scriptscriptstyle PGT}$ we must first have structurally similar set of gauge parameters in both sets of symmetries. This is achieved by introducing a field dependant map between the hamiltonian and Poincar\'{e} gauge parameters \cite{Blagojevic:2002du,Blagojevic:2004hj,Banerjee:2009vf}
\begin{align}
\label{map1}
\varepsilon^i = -\xi^\rho\,b^i_{\ \rho} \qquad\&\qquad \tau^i = -\theta^i - \xi^\rho \omega^i_{\ \rho}.
\end{align}
But this map is usually proposed arbitrarily and there is no process to generate this map from physical considerations. Using this map in the symmetries \eqref{symm G EC}, and after a bit of manipulations, we arrive at
\begin{align}
\label{symm mapped EC}
\begin{aligned}
\delta_{\scriptscriptstyle G} b^i_{\ \mu} &= \delta_{\scriptscriptstyle PGT}b^i_{\ \mu} + \frac{1}{2a}\,\xi^\rho\,\epsilon_{\mu\nu\rho}\,\frac{\delta S}{\delta \omega_{i\nu}} \\
\delta_{\scriptscriptstyle G} \omega^i_{\ \mu} &= \delta_{\scriptscriptstyle PGT}\omega^i_{\ \mu} + \frac{1}{2a}\,\xi^\rho\,\epsilon_{\mu\nu\rho}\,\frac{\delta S}{\delta b_{i\nu}},
\end{aligned}
\end{align}
where the Euler derivatives are defined in \eqref{EOM EC}. So the two sets of symmetries are different, and match only on-shell. Consequently, they also give rise to two sets of Noether identities.

The Noether identities corresponding to the PGT symmetries \eqref{Poincare3Dfieldtrans} can be found by proceeding along the route leading to \eqref{Noether Id gen contd}. Explicitly, they are \cite{Banerjee:2009vf}
\begin{subequations}
\label{Noether PGT EC}
\begin{align}
\label{Noether PGT EC 1}
P_k &= \frac{\delta S}{\delta b^i_{\ \mu}} \varepsilon^i_{\ jk} b^j_{\ \mu} + \frac{\delta S}{\delta \omega^i_{\ \mu}} \varepsilon^i_{\ jk} \omega^j_{\ \mu} -\partial_\mu\left(\frac{\delta S}{\delta \omega^k_{\ \mu}}\right) = 0\\
\label{Noether PGT EC 2}
R_\rho &= \frac{\delta S}{\delta b^i_{\ \mu}} \partial_\rho b^i_{\ \mu} + \frac{\delta S}{\delta \omega^i_{\ \mu}} \partial_\rho \omega^i_{\ \mu} - \partial_\mu \left(b^i_{\ \rho}\frac{\delta S}{\delta b^i_{\ \mu}} + \omega^i_{\ \rho} \frac{\delta S}{\delta \omega^i_{\ \mu}} \right) = 0.
\end{align}
\end{subequations}
The total number is $3+3=6$, as expected. Those corresponding to the hamiltonian gauge transformations \eqref{symm G EC} are, similarily,
\begin{subequations}
\label{Noether gauge EC}
\begin{align}
\label{Noether gauge EC 1}
A_k &= -\partial_\mu\left(\frac{\delta S}{\delta \omega^k_{\ \mu}}\right) + \frac{\delta S}{\delta b^i_{\ \mu}} \varepsilon^i_{\ jk} b^j_{\ \mu} + \frac{\delta S}{\delta \omega^i_{\ \mu}} \varepsilon^i_{\ jk} \omega^j_{\ \mu} = 0\\
\label{Noether gauge EC 2}
B_k &= -\partial_\mu\left(\frac{\delta S}{\delta b^k_{\ \mu}}\right) + \frac{\delta S}{\delta b^i_{\ \mu}} \varepsilon^i_{\ jk} \omega^j_{\ \mu} = 0
\end{align}
\end{subequations}
and are also $3+3=6$ in number. We would like to emphasise at this point that these identities are to be dealt with {\em off-shell}, without imposition of equations of motion, i.e. without setting the Euler derivatives to be zero. The identities in-fact become tautological $0=0$ statements on-shell as they are comprised of the Euler derivatives.

Now the question that we want to address is whether the sets of identities \eqref{Noether PGT EC} and \eqref{Noether gauge EC} are independent, or can they be shown to be actually the same. A comparison shows that among the two sets, \eqref{Noether PGT EC 1} and \eqref{Noether gauge EC 1} are already identical, i.e. $P_k \equiv A_k$. We want to check the possibility of expressing $R_\rho$ as some linear combination of $P_k$ and $R_k$. Comparing the structure of the free indices and the derivative terms among \eqref{Noether PGT EC 1} and \eqref{Noether gauge EC 1} we see that the combination $-b^k_{\ \rho}B_k - \omega^k_{\ \rho}A_k$ gives us
\begin{align}
\label{PGT Gauge Inv EC}
-b^k_{\ \rho}B_k - \omega^k_{\ \rho}A_k = &-R_\rho\nonumber\\
&+ \frac{\delta S}{\delta b^i_{\ \mu}} \left(\frac{1}{2a}\,\eta^{ij}\,\epsilon_{\mu\nu\rho}\right) \frac{\delta S}{\delta \omega^j_{\ \nu}} +  \frac{\delta S}{\delta \omega^i_{\ \mu}}\left(\frac{1}{2a}\,\eta^{ij}\,\epsilon_{\mu\nu\rho}\right) \frac{\delta S}{\delta b^j_{\ \nu}}=0
\end{align}
The last two terms in the above equation, proportional to square of Euler derivatives, cancel out due to antisymmetry of their coefficients {\em without requiring} the particular form of the Euler derivatives \eqref{EOM EC}. The net identity obtained in the process is just the second Noether identity corresponding to the Poincar\'{e} symmetries. Thus, we show that there exists only one set of true, independent Noether identities in the system. The total number of these are $3+3=6$, i.e. equal to the total number of gauge symmetries in the system.

The Noether identities are obtained, as shown in \eqref{Noether Id gen} and \eqref{Noether Id gen contd}, from collecting coefficients of the independent gauge parameters from a variation of the action through functional Taylor expansion
\begin{align}
\label{mapping and Noether EC 1}
\delta S = \int \left( \theta^k P_k + \xi^\rho R_\rho \right) &= 0 \qquad & \text{Poincar\'{e} symmetries.}\\
\label{mapping and Noether EC 2}
\delta S = \int \left( \varepsilon^k A_k + \tau^k B_k \right) &= 0 \qquad & \text{hamiltonian symmetries.}
\end{align}
The combinations $R_\rho = -b^k_{\ \rho}B_k - \omega^k_{\ \rho}A_k$ and $P_k = -A_k$, when substituted in \eqref{mapping and Noether EC 1}, gives
\begin{align}
\label{map derivation final result}
\int \left[ (-\theta^k - \xi^\rho\omega^k_{\ \rho})\, A_k\, + \,(-b^k_{\ \rho}\xi^\rho)\, B_k \right] = 0.
\end{align}
Comparing this with \eqref{mapping and Noether EC 2} gives us the required map \eqref{map1} between the two sets of gauge parameters. So the Noether identities help us to generate the required map between different sets of gauge parameters.

It is desirable to point out that, in the above analysis, we have not used any connection between the Noether identities and equations of motion. A literal application of the dependence of Euler-Lagrange equations due to Noether identities, mentioned below \eqref{Noether Id gen contd}, may lead to incorrect results.\footnote{This point was brought to our notice by the referee who also suggested, in this context, the original classic works of Hilbert on general relativity.} Here we have compared the Noether identities arising from the PGT and hamiltonian approaches to motivate the map \eqref{map1}. Also, all the Noether identities were explicitly verified.

The structure of the antisymmetric terms obtained in \eqref{PGT Gauge Inv EC}, when compared with those that arise in the case of trivial gauge symmetries as outlined in \eqref{Trivial Noether Id gen}, hints at the presence of trivial gauge symmetries within the hamiltonian formalism. The general form of trivial gauge transformations in this model would read
\begin{align}
\label{trivial gauge EC}
\begin{aligned}
\delta b^i_{\ \mu} &= \Lambda_{\left( b^i_{\ \mu},\, b^j_{\ \nu} \right)} \,\frac{\delta S}{\delta b^j_{\ \nu}} + \Lambda_{\left( b^i_{\ \mu},\, \omega^j_{\ \nu} \right)} \,\frac{\delta S}{\delta \omega^j_{\ \nu}} \\
\delta \omega^i_{\ \mu} &= \Lambda_{\left( \omega^i_{\ \mu},\, b^j_{\ \nu} \right)} \,\frac{\delta S}{\delta b^j_{\ \nu}} + \Lambda_{\left( \omega^i_{\ \mu},\, \omega^j_{\ \nu} \right)} \,\frac{\delta S}{\delta \omega^j_{\ \nu}},
\end{aligned}
\end{align}
where $\Lambda$ is antisymmetric (see \eqref{trivial gauge gen}). Here $\delta \equiv \delta_{\scriptscriptstyle G} - \delta_{\scriptscriptstyle PGT}$ is the apparently extra symmetry present within the hamiltonian symmetries. Comparing this with \eqref{symm mapped EC} we find the $\Lambda$ matrix defining the trivial gauge symmetry to be
\begin{align}
\label{Lambda EC}
\begin{aligned}
\Lambda_{\left( b^i_{\ \mu},\, b^j_{\ \nu} \right)} &= 0 \qquad &
	\Lambda_{\left( b^i_{\ \mu},\,\omega^j_{\ \nu} \right)} &= \frac{1}{2a}\,\eta^{ij}\,\xi^\rho\,\epsilon_{\mu\nu\rho} \\
\Lambda_{\left( \omega^i_{\ \mu},\, b^j_{\ \nu} \right)} &= \frac{1}{2a}\,\eta^{ij}\,\xi^\rho\,\epsilon_{\mu\nu\rho} \qquad &
	\Lambda_{\left( \omega^i_{\ \mu},\, \omega^j_{\ \nu} \right)} &= 0
\end{aligned}
\end{align}
The antisymmetry of $\Lambda$ in the diagonal ($b-b$ or $\omega-\omega$) entries is obvious. For the off-diagonal entry,
\begin{align}
\label{antisymm check EC}
\Lambda_{\left( b^i_{\ \mu},\,\omega^j_{\ \nu} \right)} &= \ \frac{1}{2a}\,\eta^{ij}\,\xi^\rho\,\epsilon_{\mu\nu\rho} \nonumber\\
&= - \frac{1}{2a}\,\eta^{ji}\,\xi^\rho\,\epsilon_{\nu\mu\rho} \nonumber\\
&= - \Lambda_{\left( \omega^j_{\ \nu},\, b^i_{\ \mu} \right)}.
\end{align}
Thus the $\Lambda$ matrix is antisymmetric in its field indices and this renders the action off-shell invariant. Indeed, in this case it is easy to write down explicitly the variation of the action under the hamiltonian symmetries \eqref{symm mapped EC}, obtained after using the map \eqref{map1}
\begin{align}
\label{trivial act inv proof EC}
\delta_G S = \delta_P S + \left(\frac{\delta S}{\delta b^i_{\ \mu}}\right) \frac{1}{2a}\,\xi^\rho\,\epsilon_{\mu\nu\rho}\,\left(\frac{\delta S}{\delta \omega_{i\nu}}\right) + \left(\frac{\delta S}{\delta \omega^i_{\ \mu}}\right) \frac{1}{2a}\,\xi^\rho\,\epsilon_{\mu\nu\rho}\,\left(\frac{\delta S}{\delta b_{i\nu}}\right).
\end{align}
The last two terms, in the above variation, cancel each other due to the antisymmetric nature of the coefficients and the action is actually off-shell invariant. We have thus shown that the difference between the hamiltonian and Poincar\'{e} symmetries is just a trivial gauge  transformation. The total number of true physical symmetries remain $3+3=6$ as both $\delta_{\scriptscriptstyle G}$ and $\delta_{\scriptscriptstyle PGT}$ are now physically equivalent.


\section{Mielke-Baekler type 3D gravity}
\label{Ctr:MB}

In this section, we study a 3D gravity model based on the Mielke-Baekler (MB) action \cite{Mielke:1991nn,Baekler:1992ab} added with a cosmological term. The action describing this topological 3D gravity model with torsion and a cosmological term is
\begin{align}
\label{action TMG}
S = \int \textrm{d$^3$x}\,\epsilon^{\mu\nu\rho} &\left[ab^i_{\ \mu}R_{i\nu\rho} - \frac{\Lambda}{3} \epsilon_{ijk}b^i_{\ \mu}b^j_{\ \nu}b^k_{\ \rho} \right.\nonumber\\
&\left. \phantom{[ab^i_{\ \mu}R_{i\nu\rho} } + \alpha_3\!\left(\!\omega^i_{\ \mu}\partial_\nu\omega_{i\rho}  + \frac{1}{3} \epsilon_{ijk}\,\omega^i_{\ \mu}\omega^j_{\ \nu}\omega^k_{\ \rho} \right) + \frac{\alpha_4}{2}b^i_{\ \mu}T_{i\nu\rho} \right]
\end{align}
Basic variables here, are $b^i_{\ \mu} \text{ and } \omega^i_{\ \mu}$ and the corresponding momenta are denoted as, $\pi_i^{\ \mu} \text{ and } \Pi_i^{\ \mu}$ respectively. The equations of motion are obtained by setting to zero the various Euler derivatives,
\begin{align}
\label{EOM MB}
\begin{aligned}
\frac{\delta S}{\delta b^i_{\ \mu}} &= \epsilon^{\mu\nu\rho} \left[ a\,R_{i\nu\rho} + \alpha_4\, T_{i\nu\rho} - \Lambda\, \epsilon_{ijk}b^j_{\ \nu}b^k_{\ \rho} \right] = 0 \\
\frac{\delta S}{\delta \omega^i_{\ \mu}} &= \epsilon^{\mu\nu\rho} \left[ \alpha_3\, R_{i\nu\rho} + a\, T_{i\nu\rho} + \alpha_4\, \epsilon_{ijk}b^j_{\ \nu}b^k_{\ \rho} \right] = 0
\end{aligned}
\end{align}
All the momenta turn out to be primary constraints in this first order theory. The consistency process ends at the secondary level itself and the constraints can be classified \cite{Blagojevic:2008bn} as given in Table \eqref{Tab:Constraints MB}.
\begin{table}[h]
\centering
\begin{tabular}{l | c c}
\hline\hline\
& First Class & Second class \\[0.2ex] \hline\\[-1.9ex]
Primary & $\phi_i^{\ 0}\;, \Phi_i^{\ 0}$ & $\phi_i^{\ \alpha}$, $\Phi_i^{\ \alpha} $ \\[0.4ex]
Secondary & $\bar{\mathcal{H}}_i\,, \bar{\mathcal{K}}_i$ &  \\[0.4ex]
\hline\hline
\end{tabular}
\caption{Constraints of the MB type 3D gravity theory.} \label{Tab:Constraints MB}
\end{table}
The relevant quantities used are defined below:
\begin{align}
\label{Rel qtys MB}
\begin{aligned}
\phi_i^{\ \mu} &= \pi_i^{\ \mu} - \alpha_4\,\epsilon^{0\alpha\beta}\, b_{i\beta}\,\delta^\mu_\alpha \\
\Phi_i^{\ \mu} &= \Pi_i^{\ \mu} - \epsilon^{0\alpha\beta} \left( 2 a\, b_{i\beta} + \alpha_3\, \omega_{i\beta} \right) \delta^\mu_\alpha\\
\bar{\mathcal{H}}_i &= - \left[ \epsilon^{0\alpha\beta}\!\left( a\, R_{i\alpha\beta} + \alpha_4\, T_{i\alpha\beta} - \Lambda \epsilon_{ijk} b^j_{\ \alpha} b^k_{\ \beta} \right)\right]\\
&\qquad\qquad\qquad - \nabla_{\!\alpha} \phi_i^{\ \alpha} + \epsilon_{ijk}\, b^j_{\ \alpha} \left( p\,\phi^{k\alpha} + q\, \Phi^{k\alpha} \right) \\
\bar{\mathcal{K}}_i &= -\left[ \epsilon^{0\alpha\beta} \left( a\,T_{i\alpha\beta} + \alpha_3\,R_{i\alpha\beta} + \alpha_4\,\epsilon_{ijk} b^j_{\ \alpha} b^k_{\ \beta} \right) \right]\\
&\qquad\qquad\qquad - \nabla_{\!\alpha} \Phi_i^{\ \alpha} - \epsilon_{ijk}\, b^j_{\ \alpha} \phi^{k\alpha}\\
p &= \frac{\alpha_3\Lambda + \alpha_4a}{\alpha_3 \alpha_4 - a^2}\,;\qquad q = -\frac{\alpha_4^2 + a\Lambda}{\alpha_3\alpha_4 - a^2}
\end{aligned}
\end{align}
Here the terms within square brackets in the definitions of the constraints $\bar{\mathcal{H}}_i$ and $\bar{\mathcal{K}}_i$, are themselves secondary in nature. The classified constraints in Table \eqref{Tab:Constraints MB} are suitable combinations of the primary and secondary constraints.

Using these constraints and an explicitly off-shell method \cite{Banerjee:1999yc,Banerjee:1999hu}, the hamiltonian generator of gauge symmetries can be constructed \cite{Banerjee:2009vf}. There are two (indexed) gauge parameters $\varepsilon^i$ and $\tau^i$ and they are (again) different from the Poincar\'{e} gauge parameters $\xi^\rho$ and $\theta^k$. The generator `$G$' can be written as a sum of two parts -- $\mathcal{G}_\varepsilon$ and $\mathcal{G}_\tau$, as shown below
\begin{align}
\label{Gen MB}
\begin{aligned}
G=&\int d^2x \left[\mathcal{G}_\varepsilon(x)+\mathcal{G}_\tau(x)\right]\\
&\mathcal{G}_\varepsilon = \dot{\varepsilon}^i\,\pi_i^{\ 0} + \varepsilon^i\left[\bar{\mathcal{H}_i}- \varepsilon_{ijk} \big( \omega^j_{\ 0} - p\,b^j_{\ 0}\big)\pi^{k0} + q \,\varepsilon_{ijk}\,b^j_{\ 0}\Pi^{k0} \right]\\
&\mathcal{G}_\tau = \dot{\tau}^i\Pi_i^{\ 0} + \tau^i\left[\bar{\mathcal{K}_i}-\varepsilon_{ijk}\big(b^j_{\ 0}\,\pi^{k0} + \omega^j_{\ 0}\,\Pi^{k0}\big)\right]
\end{aligned}
\end{align}
Symmetries of the basic fields can be computed from this generator through the second definition among \eqref{transform Poisson}
\begin{align}
\label{symm MB}
\begin{aligned}
\delta b^i_{\ \mu} &= \nabla_\mu\varepsilon^i - p \,\epsilon^i_{\ jk} \,b^j_{\ \mu} \varepsilon^k + \epsilon^i_{\ jk}\,b^j_{\ \mu} \tau^k ,\\
\delta \omega^i_{\ \mu} &= \nabla_\mu \tau^i - q \,\epsilon^i_{\ jk} \,b^j_{\ \mu} \varepsilon^k.\\
\end{aligned}
\end{align}
The hamiltonian symmetries contain the coupling constants $\Lambda, \,\alpha_3 \text{ and } \alpha_4$ through the parameters $p\, \&\, q$ defined earlier. These, they inherit from the action through the structure of the constraints. To compare with Poincar\'{e} symmetries, we take recourse to the map \eqref{map1} relating the hamiltonian gauge parameters to the Poincar\'{e} gauge parameters. After some rearrangements and remembering the Euler derivatives from \eqref{EOM MB}, we arrive at
\begin{align}
\label{symm mapped MB}
\begin{aligned}
\delta_{\scriptscriptstyle G} b^i_{\ \mu} = \delta_{\scriptscriptstyle PGT} b^i_{\ \mu} &+ \frac{\alpha_3}{2(\alpha_3\alpha_4-a^2)}\,\eta^{ij}\,\xi^\rho\,\epsilon_{\mu\nu\rho} \,\frac{\delta S}{\delta b^j_{\ \nu}} \\
&- \frac{a}{2(\alpha_3\alpha_4-a^2)}\,\eta^{ij}\,\xi^\rho\,\epsilon_{\mu\nu\rho} \,\frac{\delta S}{\delta \omega^j_{\ \nu}} \\
\delta_{\scriptscriptstyle G} \omega^i_{\ \mu} = \delta_{\scriptscriptstyle PGT} \omega^i_{\ \mu} &- \frac{a}{2(\alpha_3\alpha_4-a^2)}\,\eta^{ij}\,\xi^\rho\,\epsilon_{\mu\nu\rho} \,\frac{\delta S}{\delta b^j_{\ \nu}} \\
&+ \frac{\alpha_4}{2(\alpha_3\alpha_4-a^2)}\,\eta^{ij}\,\xi^\rho\,\epsilon_{\mu\nu\rho} \,\frac{\delta S}{\delta \omega^j_{\ \nu}} \\
\end{aligned}
\end{align}
It is again clear that the hamiltonian and Poincar\'{e} symmetries become (algebraically) identical only on-shell.

Let us now investigate the Noether identities in this model. The identities corresponding to the PGT symmetries remain the same as \eqref{Noether PGT EC}, since the form of the Poincar\'{e} symmetries do not depend upon the form of the lagrangian, as long as the lagrangian is diffeomorphism invariant in nature (and contains the same fields in construction of the action). The hamiltonian gauge symmetries \eqref{symm MB} give rise to the following identities \cite{Banerjee:2009vf}
\begin{subequations}
\label{Noether gauge MB}
\begin{align}
\label{Noether gauge MB 1}
A'_k &= -\partial_\mu\left(\frac{\delta S}{\delta \omega^k_{\ \mu}}\right) + \frac{\delta S}{\delta b^i_{\ \mu}} \varepsilon^i_{\ jk} b^j_{\ \mu} + \frac{\delta S}{\delta \omega^i_{\ \mu}} \varepsilon^i_{\ jk} \omega^j_{\ \mu} = 0\\
\label{Noether gauge MB 2}
B'_k &= -\partial_\mu\left(\frac{\delta S}{\delta b^k_{\ \mu}}\right) + \frac{\delta S}{\delta b^i_{\ \mu}} \varepsilon^i_{\ jk} \omega^j_{\ \mu} -p\, \frac{\delta S}{\delta b^i_{\ \mu}} \epsilon^i_{\ jk} b^j_{\ \mu} - q\, \frac{\delta S}{\delta \omega^i_{\ \mu}} \epsilon^i_{\ jk} b^j_{\ \mu} = 0.
\end{align}
\end{subequations}
Once again we see that one of the identities among the hamiltonian gauge \eqref{Noether gauge MB} and Poincar\'{e} ones \eqref{Noether PGT EC}, $A_k$ and $P_k$, match each other. And the combination $-\omega^k_{\ \rho} A'_k + -b^k_{\ \rho} B'_k$ leads to
\begin{align}
\label{PGT Gauge Inv MB}
\begin{aligned}
&-R_\rho \\
&\ \ + \frac{\delta S}{\delta b^i_{\ \mu}} \left(\frac{\alpha_3}{2(\alpha_3\alpha_4 - a^2)}\,\eta^{ij} \epsilon_{\mu\nu\rho}\right) \frac{\delta S}{\delta b^j_{\ \nu}} + \frac{\delta S}{\delta b^i_{\ \mu}} \left(\frac{-a}{2(\alpha_3\alpha_4 - a^2)}\,\eta^{ij}\, \epsilon_{\mu\nu\rho}\right) \frac{\delta S}{\delta \omega^j_{\ \nu}} \\
&\ \ +  \frac{\delta S}{\delta \omega^i_{\ \mu}}\left(\frac{-a}{2(\alpha_3\alpha_4 - a^2)}\,\eta^{ij}\, \epsilon_{\mu\nu\rho}\right) \frac{\delta S}{\delta b^j_{\ \nu}} + \frac{\delta S}{\delta \omega^i_{\ \mu}}\left(\frac{\alpha_4}{2(\alpha_3\alpha_4 - a^2)}\,\eta^{ij}\, \epsilon_{\mu\nu\rho}\right) \frac{\delta S}{\delta b^j_{\ \nu}}\\
&\qquad\qquad\qquad = 0.
\end{aligned}
\end{align}
The last four terms, proportional to square of Euler derivatives, cancel each other due to antisymmetry of their coefficients. The part surviving is just the missing Poincar\'{e} identity $R_\rho=0$ \eqref{Noether PGT EC 2}. So, only one set of independent Noether identities exist.

The antisymmetric terms in the Noether identities \eqref{PGT Gauge Inv MB} again point toward presence of trivial gauge symmetries. To check explicitly, we first write down the general trivial gauge symmetry structure appropriate for the MB model
\begin{align}
\label{trivial gauge MB}
\begin{aligned}
\delta b^i_{\ \mu} &= \Lambda_{\left( b^i_{\ \mu},\, b^j_{\ \nu} \right)} \,\frac{\delta S}{\delta b^j_{\ \nu}}\ +\ \Lambda_{\left( b^i_{\ \mu},\, \omega^j_{\ \nu} \right)} \,\frac{\delta S}{\delta \omega^j_{\ \nu}}\\
\delta \omega^i_{\ \mu} &= \Lambda_{\left( \omega^i_{\ \mu},\, b^j_{\ \nu} \right)} \,\frac{\delta S}{\delta b^j_{\ \nu}} \ +\ \Lambda_{\left( \omega^i_{\ \mu},\, \omega^j_{\ \nu} \right)} \,\frac{\delta S}{\delta \omega^j_{\ \nu}}\\
\end{aligned}
\end{align}
Comparing this with \eqref{symm mapped MB}, we write can down the $\Lambda$ matrix below
\begin{align}
\label{Lambda MB}
\begin{aligned}
\Lambda_{\left( b^i_{\ \mu},\, b^j_{\ \nu} \right)} &= \frac{\alpha_3}{2(\alpha_3\alpha_4 - a^2)}\,\eta^{ij}\,\xi^\rho \epsilon_{\mu\nu\rho} \quad &
	\Lambda_{\left( b^i_{\ \mu},\,\omega^j_{\ \nu} \right)} &= \frac{-a}{2(\alpha_3\alpha_4 - a^2)}\,\eta^{ij}\,\xi^\rho \epsilon_{\mu\nu\rho} \\
\Lambda_{\left( \omega^i_{\ \mu},\, b^j_{\ \nu} \right)} &= \frac{-a}{2(\alpha_3\alpha_4 - a^2)}\,\eta^{ij}\,\xi^\rho \epsilon_{\mu\nu\rho} \quad &
	\Lambda_{\left( \omega^i_{\ \mu},\, \omega^j_{\ \nu} \right)} &= \frac{\alpha_4}{2(\alpha_3\alpha_4 - a^2)}\,\eta^{ij}\,\xi^\rho \epsilon_{\mu\nu\rho}
\end{aligned}
\end{align}
The antisymmetry of this structure is easy to verify. We will just demonstrate one component
\begin{align}
\label{antisymm check MB}
\Lambda_{\left( b^i_{\ \mu},\, b^j_{\ \nu} \right)} &= \frac{\alpha_3}{2(\alpha_3\alpha_4 - a^2)}\,\eta^{ij}\,\xi^\rho \epsilon_{\mu\nu\rho}  \nonumber\\
&= - \frac{\alpha_3}{2(\alpha_3\alpha_4 - a^2)}\,\eta^{ji}\,\xi^\rho \epsilon_{\nu\mu\rho} \nonumber \\
&= - \Lambda_{\left( b^j_{\ \nu},\, b^i_{\ \mu} \right)} .
\end{align}
So the two symmetries $\delta_{\scriptscriptstyle G}$ and $\delta_{\scriptscriptstyle PGT}$ differ only by a trivial gauge symmetry which is of no physical importance. The Poincar\'{e} transformations are indeed recovered by the hamiltonian mechanism. An important point to be noted from the analysis of this model is that the hamiltonian symmetries \eqref{symm MB} of this model were different from those of the Einstein-Cartan theory \eqref{symm G EC}. However we could nevertheless recover the Poincar\'{e} symmetries from both of these. The particular difference in details between the models (various terms in the action along with their coupling constants) got manifested only through trivial gauge symmetries.


\section{Discussions}
\label{Ctr:disc}

In this Chapter, we have shown that the Dirac hamiltonian construction indeed reproduces the Poincar\'{e} symmetries in different models of gravity. We have analysed the Einstein-Cartan action and a more generalised form of a Mielke-Baekler type action with a cosmological term, both in 3-dimensions. The Noether identities corresponding to the two sets of symmetries, hamiltonian gauge and Poincar\'{e}, were shown to be the same, modulo antisymmetric cancelling terms proportional to square of Euler derivatives. Using these Noether identities, we derived a map between the two sets of gauge parameters. After using the map, we demonstrated that the difference in the hamiltonian gauge symmetries and the Poincar\'{e} symmetries was just trivial gauge transformations, characterised by coefficients antisymmetric under exchange of fields. We have explicitly found out the coefficient matrices for both Einstein-Cartan and its Mielke-Baekler type generalisation.

Since trivial gauge symmetries are of no physical importance, the Poincar\'{e} symmetries are indeed recovered through the canonical procedure. This feature should persist in all the different diffeomorphism invariant theories of interest and shows the importance of understanding and handling trivial gauge symmetries.

We have shown how the lagrangian and hamiltonian formulations complement each other and how their unified application is of great importance. Analysis of the Noether identities arising in the lagrangian formulation helps us to construct the map between gauge parameters present in the hamiltonian and Poincar\'{e} gauge transformations. This map, at the hamiltonian level, can only be guessed through an (in general case, a rather difficult) exercise of inspection and trial. In the lagrangian procedure, however, the process is much more straightforward and systematic. It is noteworthy that the map is model independent, i.e. it is the same in both examples studied here. This universal nature reveals a unifying feature among the hamiltonian gauge symmetries, a fact that is not otherwise transparent. Indeed, contrary to Poincar\'{e} gauge transformations, the structure of hamiltonian gauge transformations are distinct for distinct models.

Finally, let us recall the role of trivial gauge transformations at the quantum level. This is relevant since gauge symmetries are important in the process of quantisation. The classical gauge symmetries of the action are now replaced by the quantum (Becchi-Rouet-Stora-Tyutin or BRST) symmetries of the quantum effective action $(\Gamma)$. For general gauge theories it was shown \cite{Alexandrov:1998uy} that the set of local symmetries of $\Gamma$ comprise of the quantum gauge transformations, trivial gauge transformations and transformations induced by background fields. Taking a linear combination of all three symmetries, it is possible to find a simple or a standard form. Indeed, adopting this approach the classical gauge transformations for Yang Mills theory were reproduced in \cite{Alexandrov:1998uy}.


%% file: C8_conclusion.tex
\chapter{Comments and discussions}
\label{C:conc}

The importance of a Minkowski flat spacetime where special relativity holds is para\-mount from the point of view of our empirical knowledge of the known particles and matter fields. On the other hand, evidence of curvature and its understanding in relation to matter, as prescribed by general relativity, is now evident in our day to day lives -- say for example, through implementation in `global positioning systems' \cite{Ashby:2003vja}. In this thesis, we have explored some aspects of gravity starting from flat spacetime. Our endeavour was mainly focused along two directions:
\begin{itemize}
\item {\em Accelerated observers in flat spacetime:} The equivalence principle allows us to replace uniform acceleration with uniform gravity. So, study of uniformly accelerating frames gives us a window to the possible behaviour of fields in the presence of gravity. We calculated the temperature and spectrum corresponding to the Unruh effect from quantum tunnelling. This study is closely linked to the Hawking effect in black holes.
\item {\em Canonicity of gauge symmetries in Poincar\'{e} gauge theory:} PGT is obtained by gauging the global Poincar\'{e} symmetry of flat spacetime. This leads to additional fields that describe gravity. We established that Poincar\'{e} gauge symmetries are indeed canonical in nature modulo `trivial symmetries' (which do not give rise to any new gauge symmetry) through a completely off-shell analysis.
\end{itemize}
Let us now give a detailed summary of our results.

\section{Summary of results}
\label{Cconc:summary}

\paragraph*{Chapter 2} We first set up a scheme to accommodate arbitrary parametrisations of a family of accelerated observers with all possible accelerations between $\pm\infty$, following hyperbolic paths, so that we obtained a `{\em generalised Rindler metric}' \eqref{genRind}. This can reproduce the various different forms of the Rindler metric used in literature, as shown in Appendix \ref{App:Unruh}. Then we identified a proper analytic extension of the coordinates from the physical wedge $\mathtt{I}$ into the black hole like wedge $\mathtt{II}$ (see Fig. \ref{D:wedge}). This introduced an imaginary factor relating both set of coordinates \eqref{1_2relation} which comes due to the interchange of the nature of time and space coordinates across the `accelerated horizon.' We constructed complete set of modes for both bosonic (Klein-Gordon modes \eqref{phix}) and fermionic (Dirac modes \eqref{ABSolns}) particles in Rindler spacetime. Near the horizon, both modes reduced to plane wave like solutions. By adopting a quantum tunnelling approach, we then demanded continuity of modes across the horizon in the coordinates of the physical observer, who resides in wedge $\mathtt{I}$. This produced a set of {\em outcoming} particles -- a phenomenon classically prohibited -- suppressed by an exponentially decaying factor. Subsequently we constructed a normalised density matrix of particles out of these modes, containing pairs of ingoing and outgoing modes \eqref{scalDens}. Tracing over ingoing modes of this, we obtained a reduced density matrix \eqref{scalDensOut} which gave rise to thermal spectrum: Bose-Einstein \eqref{BoseSpect} for scalars and Fermi-Dirac \ref{Fermspect} for fermions. The Unruh temperature $T_{\scriptscriptstyle U}=\frac{\hbar \alpha}{2 \pi}$ proportional to the acceleration $\alpha$ was the characteristic temperature of these spectra, as expected. In Section \S \ref{Cunruh:gems} we studied the embedding of a Schwarzschild black hole in a 6-dimensional Minkowski flat spacetime (see \eqref{GEMSmetric} and \eqref{GEMStrans}) and studied a mapping of the Unruh effect with the Hawking effect. This procedure, acronymd as GEMS (Global Embedding Minkowski Spacetime) by the proponents Deser and Levin \cite{Deser:1998bb}, brought an interesting way of seeing both effects from flat spacetime. {\em This chapter was based on our work} \cite{Roy:2009vy}.

\paragraph*{Chapter 3} In this chapter, we mainly presented a comparison of gauging of the Lorentz group as done by Utiyama \cite{Utiyama:1956sy} and then the Poincar\'{e} group as done by Kibble \cite{Kibble:1961ba}. This clarifies our point of view regarding the setup of Poincar\'{e} gauge theory (PGT) and in what sense the transformations of the spin-connections \eqref{PoincareAconnTransfms} and triads \eqref{Poincarebframetransfms} are  gauge transformations. We also gave a brief, selective overview of the literature covering gravity theories discussed within PGT. This was done to set the stage for the models suited to our later applications. In Section \S \ref{Cpgt:geometrical} we explicitly calculated the PGT symmetry of the triad $b^i_{\ \mu}$ starting with the known diffeomorphism transformation of the metric \eqref{metricvar} and the relation of the metric with the triads \eqref{PoincareTransMetric}. This presents another view on the PGT transformations. {\em This part of the chapter was based on our work} \cite{Banerjee:2009vf}.

\paragraph*{Chapter 4} We took up canonical constrained analysis of a PGT gravity model in this chapter. But first, in Section \S \ref{Cmb:ConstraintsRev}, we re-derived the master equations for elimination of dependent gauge symmetry parameters \eqref{revChMasterEqn1} \& \eqref{revChMasterEqn2} presented originally in \cite{Banerjee:1999hu, Banerjee:1999yc}. The present derivation was a simple affair which highlighted the off-shell nature of the algorithm for construction of a gauge-generator. The model we then chose is the Mielke-Baekler type 3-D gravity model with torsion. We reproduced the constraint structure (Table \ref{table:constraints}) present in literature \cite{Blagojevic:2004hj}. However, our analysis differed in two important aspects. We employed Dirac brackets to eliminate the second-class sector. Also, we used our off-shell algorithm mentioned above, to compute the gauge generator \eqref{generatorFinal}. This is in contrast to the on-shell method \cite{Castellani:1981us} used previously. We next derived the symmetries generated \eqref{field transf gauge} and explicitly verified that these are off-shell, by computing the variation of the action. Then we confirmed that these symmetries, after the usual re-mapping of gauge parameters \eqref{onshell map}, are indeed algebraically distinct from the PGT symmetries, modulo equations of motion \eqref{onshell relation transfs}. This difference, which is puzzling as there cannot be two independent set of gauge symmetries (each having all the number of symmetries expected in the 3D Poincar\'{e} group) for the same action, is to be later explained in Chapter \ref{C:tr}. Meanwhile, we compared the above difference with an ADM analysis of the usual general relativistic action (in 3D) where no such difference between the canonical and diffeomorphism symmetries was observed. {\em This chapter was based on our work} \cite{Banerjee:2009vf}.

\paragraph*{Chapter 5} The results of the previous chapter being puzzling in nature, we took up another model here to see whether the discrepancy between canonical and PGT symmetries is model independent. This time, our model was a first-order formulation of the `new massive gravity' model of much recent interest, presented originally by Bergshoeff, Hohm and Townsend \cite{Bergshoeff:2009hq, Bergshoeff:2009aq}. The first-order formalism we adopted was presented by Blagojevic in \cite{Blagojevic:2010ir}. We reconstructed the constraint structure of this model (Table \ref{CbhtTab:Constraints}) which contains constraints upto tertiary level and computed the canonical symmetries for some of the fields (see \eqref{delta_G b} \& \eqref{delta_G f}). Once again, upon mapping of gauge parameters we could only reproduce the PGT symmetries modulo terms proportional to equations of motion. Now the mapping used is a field dependent map. So we also verified some subtleties regarding use of the map. We showed that use of such field dependent map is legitimate and we can use it both at the level of the generator or at the level of symmetries, to get identical results. {\em This chapter was based on our work} \cite{Banerjee:2011rx}.

\paragraph*{Chapter 6} After having done a purely hamiltonian analysis in the previous chapters, we turned to a lagrangian analysis in this chapter. At the heart of this analysis lies the Noether identities corresponding to each of the two different sets of symmetries, the canonical and PGT symmetries. We constructed these identities (\eqref{gauge ident generator} and \eqref{gauge ident PGT}) and used them to obtain (following \cite{Shirzad:1998af}) lagrangian generators (\eqref{2plus1GenSet1} - \eqref{2plus1GenSet4}) of PGT symmetries in 3D. We followed this up in 4D to find similar lagrangian generators from a `dimensional uplift' of the gauge identities. We also showed the connection \cite{Ortin:2004ms} of the gauge identities with the Bianchi identities, in the general relativistic Einstein-Hilbert action, to motivate a possible utility of this lagrangian description. {\em This chapter was based on our work} \cite{Banerjee:2010kd}.

\paragraph*{Chapter 7} This chapter finally concluded our canonical analysis by successfully resolving the puzzle regarding the symmetries that was elaborated earlier; taking cues from both hamiltonian as well as lagrangian analysis and bringing out the role of `trivial symmetries.' Since the `puzzle' was a model independent feature we took up the simplest model of PGT gravity: the Einstein-Cartan model. The PGT and canonical symmetries generated by an off-shell first-class generator (under suitable mapping of gauge parameters) still differed modulo terms proportional to equations of motion \eqref{symm mapped EC}. Now carefully comparing the Noether identities (\eqref{Noether PGT EC} \& \eqref{Noether gauge EC}) corresponding to both sets of symmetries gave us two important results: (i) {\em The two sets of identities are not independent.} In-fact, a linear field dependent recombination of one set gave rise to the other, as shown in \eqref{PGT Gauge Inv EC}. (ii) {\em Algorithm for the map.} The map \eqref{map1} between gauge parameters -- used repeatedly in comparing the two symmetries -- could be identified from the particular linear recombination of the canonical gauge identities that reproduced the PGT gauge identities, as shown in \eqref{map derivation final result}. Now, the first of the afore mentioned points showed us that while showing the inter-dependence of the Noether identities, it was crucial that certain terms proportional to square of Euler derivatives with antisymmetric coefficients dropped out \eqref{PGT Gauge Inv EC}. This motivated the establishment of the result that the action remains {\em off-shell} invariance under the re-mapped canonical symmetries \eqref{symm mapped EC}, as shown in \eqref{trivial act inv proof EC}. The terms proportional to Euler derivatives drop out due to antisymmetric co-efficients. This is a manifestation of `trivial symmetries' \cite{Henneaux:1990au}: symmetries that are not generated by first-class constraints and which give rise to no new gauge symmetries. Thus the two sets of symmetries are canonically equivalent, in spite of differing algebraically, as their difference in canonically trivial. The feature was also checked in the full Mielke-Baekler 3D gravity model. {\em This chapter was based on our publications} \cite{Banerjee:2011cu, Banerjee:2012jn, Roy:2012Pr, Roy1:2012MGM}.


\section{Future perspectives}
\label{Cconc:futura}

The results we obtained cover a wide arena and can be used to understand/extend a variety of results. We indicate only few of them, based on personal curiosity. In the case of the Unruh effect, it was evident that the Unruh temperature was same for both fermions and bosons. Now further investigations, based on ideas of quantum entanglement, shows a difference in the characterisation of the Unruh effect through bosons and fermions \cite{Alsing:2003es, 2006PhRvA..74c2326A}. This would be an interesting point of departure to bring in features of quantum entanglement within our framework of quantum tunnelling, which in its use of a density matrix of particles tunnelling out, is already well-suited for study of entangled systems.

In the direction of PGT models of gravity, the role of surface terms would be an interesting thing to study. In the context of general relativity, it is known \cite{Padmanabhan:2010zzb} that the action can be written (in a non-covariant form) by discarding the $\partial \Gamma$ (where $\Gamma^\mu_{\nu\rho}$ is the Christoffel) terms as surface terms. The importance of such surface terms in gravitational actions can't be overemphasised \cite{Miskovic:2009kr}. Variation of the action however gives back the covariant Einstein equations. The implication of this result would be interesting to study from our point of view, especially its canonical significance.


%% file: Capp.tex
\begin{appendices}
\addappheadtotoc

\chapter{Some standard Rindler metrics}
\label{App:Unruh}

\lettrine[lraise=0.0, loversize=0.3, findent=3pt, nindent=0pt]{T}{he} Generalised Rindler metric described by \eqref{genRind}
\begin{align*}
ds^2 = -a^2 F(x)^2 dt^2 + F'(x)^2 dx^2 + dy^2 + dz^2
\end{align*}
can yield all the various forms used in the literature through specific choice of the function $F(x)$ and the constant $a$. Some examples are discussed below. For all these metrics, the corresponding Minkowski Space variables are defined through \eqref{transformationsI}.
\vspace*{-1em}
\paragraph*{Metric 1}
\begin{align}
ds^2 = e^{2 \alpha x} \left(-dt^2 + dx^2\right).
\label{carRind}
\end{align}
For this metric taken from \cite{Carroll:2004st}, we have to use $a=\alpha$ and $F(x)=\frac{1}{a}e^{a x}$.
\vspace*{-1em}
\paragraph*{Metric 2}
\begin{align}
ds^2 = - x^2 \, dt^2 + dx^2.
\label{rindRind}
\end{align}
For this metric taken from \cite{Rindler:1966zz}, we have to use $a=1$ and $F(x)=x$.
\vspace*{-1em}
\paragraph*{Metric 3}
\begin{align}
ds^2 = - x \, dt^2 + \frac{dx^2}{x}.
\label{unruhRind}
\end{align}
For this metric taken from \cite{Unruh:1976db}, we have to use $a=1/2$ and $F(x)=2\sqrt{x}$.

\chapter{3D gravity with torsion: The Poisson algebra of constraints}
\label{App:Poisson}

\lettrine[lraise=0.0, loversize=0.3, findent=3pt, nindent=0pt]{T}{he} basic non-zero Poisson brackets of the theory \eqref{CmbMBaction} are given below.
\begin{align}
\label{fieldPoisson}
\begin{aligned}
\lbrace b^i_{\ \mu} (x), \pi^{\ \nu}_j (x') \rbrace &= \delta^i_j ~\delta^\nu_\mu ~\delta(x-x')\\
\lbrace \omega^i_{\ \mu} (x), \Pi^{\ \nu}_j (x') \rbrace &= \delta^i_j ~\delta^\nu_\mu ~\delta(x-x')\\
\end{aligned}
\end{align}
Also, we give below a list of the Poisson brackets of the quantities $\mathcal{H}$ and $\mathcal{K}$, constructed out of the the basic fields in (\ref{canon Hamilt}), with the primary constraints.
\begin{align}
\begin{aligned}
\label{App:SRel PAlgebra}
\lbrace\phi_i^{\phantom{i} \alpha}(x),\mathcal{H}_j(x')\rbrace &= 2 \varepsilon^{0\alpha\beta}\left[\alpha_4 \,\eta_{ij}\, \partial^{(x)}_\beta \delta(x-x')-\varepsilon_{ijk}\left(\alpha_4\,\omega^k_{\ \beta}-\Lambda \,b^k_{\ \beta} \right)\delta(x-x') \right]\\
\lbrace\phi_i^{\phantom{i} \alpha}(x),\mathcal{K}_j(x')\rbrace &= 2 \varepsilon^{0\alpha\beta}\left[a \,\eta_{ij} \,\partial^{(x)}_\beta \delta(x-x')-\varepsilon_{ijk}\left(a\,\omega^k_{\ \beta} + \alpha_4 \,b^k_{\ \beta} \right)\delta(x-x') \right]\\
\lbrace\Phi_i^{\phantom{i} \alpha}(x),\mathcal{H}_j(x')\rbrace &= 2 \varepsilon^{0\alpha\beta}\left[a \,\eta_{ij} \,\partial^{(x)}_\beta \delta(x-x')-\varepsilon_{ijk}\left(a\,\omega^k_{\ \beta} + \alpha_4 \,b^k_{\ \beta} \right)\delta(x-x') \right]\\
\lbrace\phi_i^{\phantom{i} \alpha}(x),\mathcal{K}_j(x')\rbrace &= 2 \varepsilon^{0\alpha\beta}\left[\alpha_3 \,\eta_{ij} \,\partial^{(x)}_\beta \delta(x-x')-\varepsilon_{ijk}\left(\alpha_3\,\omega^k_{\ \beta} + a \,b^k_{\ \beta} \right)\delta(x-x') \right].
\end{aligned}
\end{align}

We now calculate the non-trivial Poisson algebra of the constraints, by using the algebra among basic variables  (\ref{fieldPoisson}). The algebra (\ref{App:SRel PAlgebra}) comes in handy at this step (as well as in the following calculations).
\begin{align}
\label{App:PP PAlgebra}
\begin{aligned}
\lbrace\phi_i^{\ \alpha}(x),\phi_j^{\ \beta}(x')\rbrace &= -2 \,\alpha_4 \,\varepsilon^{0\alpha\beta}\,\eta_{ij}\,\delta(x-x')\\
\lbrace\Phi_i^{\ \alpha}(x),\Phi_j^{\ \beta}(x')\rbrace &= -2 \,\alpha_3 \,\varepsilon^{0\alpha\beta}\,\eta_{ij}\,\delta(x-x')\\
\lbrace\phi_i^{\ \alpha}(x),\Phi_j^{\ \beta}(x')\rbrace &= -2 \,a \,\varepsilon^{0\alpha\beta}\,\eta_{ij}\,\delta(x-x')\\
\end{aligned}
\end{align}
Observe that the Poisson algebra (\ref{App:PP PAlgebra}) between the primary constraints does not close, implying the existence of second-class constraints.

The Poisson algebra between primary and secondary constraints are:
\begin{align}
\begin{aligned}
\label{App:PS PAlgebra}
\lbrace\phi_i^{\ \alpha}(x),\bar{\mathcal{H}_j}(x')\rbrace &= \varepsilon_{ijk} \left(p\,\phi^{k\alpha} + q\,\Phi^{k\alpha} \right) \delta(x-x')\\
\lbrace\phi_i^{\ \alpha}(x),\bar{\mathcal{K}_j}(x')\rbrace &= -\varepsilon_{ijk}\,\phi^{k\alpha} \,\delta(x-x')\\
\lbrace\Phi_i^{\ \alpha}(x),\bar{\mathcal{H}_j}(x')\rbrace &= -\varepsilon_{ijk}\,\phi^{k\alpha} \,\delta(x-x')\\
\lbrace\Phi_i^{\ \alpha}(x),\bar{\mathcal{K}_j}(x')\rbrace &= -\varepsilon_{ijk}\,\Phi^{k\alpha} \,\delta(x-x'),\\
\end{aligned}
\end{align}
while the algebra among the secondary constraints are:
\begin{align}
\begin{aligned}
\label{App:SS PAlgebra}
\lbrace\bar{\mathcal{H}_i}(x),\bar{\mathcal{H}_j}(x')\rbrace &= \varepsilon_{ijk}\left(p\,\bar{\mathcal{H}^k}+q\,\bar{\mathcal{K}^k}\right)\delta(x-x')\\
\lbrace\bar{\mathcal{K}_i}(x),\bar{\mathcal{K}_j}(x')\rbrace &= - \varepsilon_{ijk}\,\bar{\mathcal{K}^k}\,\delta(x-x')\\
\lbrace\bar{\mathcal{H}_i}(x),\bar{\mathcal{K}_j}(x')\rbrace &= - \varepsilon_{ijk}\,\bar{\mathcal{H}^k}\,\delta(x-x').\\
\end{aligned}
\end{align}
We see that both sets (\ref{App:PS PAlgebra}, \ref{App:SS PAlgebra}) close.

\end{appendices}